\documentclass{aa}  

\usepackage{graphicx}
\usepackage[varg]{txfonts}
\usepackage{lscape}

\newcommand\be{\begin{equation}}
\newcommand\en{\end{equation}}

\begin{document}

\title{Reading between the lines:}

\subtitle{Disk emission, wind, ad accretion during the Z~CMa NW outburst\thanks{Based in part on observations made at Observatoire de Haute Provence (CNRS), France. Based on observations obtained at the Canada-France-Hawaii Telescope (CFHT), which is operated by the National Research Council of Canada, the Institut National des Sciences de l'Univers of the Centre National de la Recherche Scientifique of France, and the University of Hawaii.  Table B.1 is available in electronic form
at the CDS via anonymous ftp to cdsarc.u-strasbg.fr (130.79.128.5)
or via http://cdsweb.u-strasbg.fr/cgi-bin/qcat?J/A+A/
}}

\author{A. Sicilia-Aguilar\inst{1}\and J. Bouvier\inst{2}\and C. Dougados\inst{2}\and K. Grankin\inst{3}\and J. F. Donati\inst{4}}

\institute{\inst{1} SUPA, School of Science and Engineering, University of Dundee, Nethergate, Dundee DD1 4HN, UK \\
        \email{a.siciliaaguilar@dundee.ac.uk}\\
\inst{2} Universit\'{e} Grenoble Alpes, CNRS, IPAG, 38000 Grenoble, France\\
\inst{3} Crimean Astrophysical Observatory, 298409 Nauchny, Crimea\\
\inst{4} IRAP  Observatoire Midi-Pyr\'{e}n\'{e}es, 14 Avenue E Belin, 31400 Toulouse, France\\
}

\date{Submitted May 25, 2020. Accepted August 31, 2020}

 
  \abstract
{}
   {We use optical spectroscopy to investigate the disk,
wind, and accretion during the  2008 Z~CMa NW outburst.}
   {Emission lines were used to constrain the locations, densities, and temperatures
 of the structures around the star.}
{More than 1000 optical emission lines reveal accretion, a variable, multicomponent wind, and double-peaked lines of
disk origin. 
The variable, non-axisymmetric, accretion-powered wind has slow ($\sim $0 km s$^{-1}$), intermediate ($\sim -$100 km s$^{-1}$), and
fast ($\geq -$400 km s$^{-1}$) components. The fast components are of stellar origin and
disappear in quiescence, while the slow component is less variable and could be related 
to a disk wind.  The changes in the optical depth of the lines between
outburst and quiescence  reveal that increased accretion is responsible for the
observed outburst. We derive an accretion rate of 10$^{-4}$ M$_\odot$/yr in outburst.
The Fe I and weak Fe II lines arise from an irradiated, 
flared disk at $\sim$0.5-3  $\times$M$_*$/16 M$_\odot$ au with asymmetric upper layers, revealing 
that the energy from the accretion burst is deposited at scales below 0.5 au. 
Some line profiles have redshifted asymmetries,
but the system is unlikely to be sustained by magnetospheric accretion, especially in outburst. The
accretion-related structures extend over several stellar radii and, like the wind, are 
likely to be non-axisymmetric. The stellar mass may be $\sim$6-8 M$_\odot$, lower than previously thought ($\sim$16 M$_\odot$).}
{ Emission line analysis
is found to be a powerful tool to study the innermost regions and accretion in stars within a very large range
of effective temperatures. The density ranges in the disk and accretion structures are higher than in late-type stars, but the 
overall behavior, including the innermost disk emission and variable wind, is very similar for stars with different spectral types. Our work suggests a common outburst behavior for stars
with spectral types ranging from M type to intermediate mass.  }
 

\authorrunning{Sicilia-Aguilar et al.}

\titlerunning{Z~CMa NW in outburst: Disk emission and wind}

\maketitle

%

\section{Introduction \label{intro}}

Z~CMa is a binary intermediate-mass star known for its strong photometric and 
spectroscopic variability \citep{covino84,shevchenko99}. The two components are 
separated by 0.1", or about 100 au \citep{barth94},
for a distance of 1 kpc \citep{shevchenko99,kaltcheva00}.
The SE companion 
is classified
as a FU Orionis (FUor) object \citep[][]{hartmann89,hessman91}, while
the NW component is an embedded, variable intermediate-mass star, initially labeled
as an infrared companion \citep{koresko91}. The NW component was detected later in the 
optical \citep{barth94,thiebaut95} and classified as a 16 M$_\odot$, B0 III star 
based on its de-reddened photometric colors \citep{vandenancker04},
although its radius and luminosity are disputed \citep[e.g.,][]{monnier05}. 
A B0 spectral type would place the NW component of 
Z~CMa among the earliest-type stars known to have a disk \citep[e.g., see the reviews by][]{zinnecker07,beltran16}
and to undergo variable accretion in a way not dissimilar to lower-mass objects, 
although disks similar to those of solar-type
stars are also increasingly found around massive stars
\citep[e.g.,][]{bik04,alonso09,kraus10, fedriani20}.

Z~CMa has a complex light curve with 
anomalous variability \citep{shevchenko99}. Highly complex, short-timescale, aperiodic variations
were confirmed by MOST \citep[Microvariability and Oscillations of Stars;][]{siwak13}.
Although historically the main variability was attributed to FUor, further
observations, including polarization and scattered light, have suggested that the NW component is
responsible for the most dramatic changes \citep{lamzin98,szeifert10} and also dominates the 
polarization observed in the system \citep{fischer98}.
The photometric evolution of Z~CMa NW is uncertain since the first optical observations.  
Early works focussed on  the FUor, but there are several reports of anomalous
behavior that may be related to the NW component. Z~CMa was always presented as an
atypical FUor, in part for its long rising time
\citep{hartmann89,hessman91,hartmann96}, but also because of its bumpy
light curve. Several authors found the IR object
to be brighter than expected in the optical since the 1990s, although 
\citet{thiebaut95} argued that this could be 
due to scattered light of the central object on the wall of a low-inclination bipolar 
cavity. The same conclusion of the NW component being optically brighter was reached by 
\citet{barth94} from observations obtained three years after \citet{thiebaut95}. 
Subsequent interferometric observations revealed a $\sim$4 au dusty ring around Z~CMa NW
\citep{monnier05} and a larger mid-infrared structure ($\sim$68 mas$\times$41 mas, $\sim$40-70 au) 
dominated by Z~CMa NW but extended toward the FUor companion \citep{monnier09}.
The FUor companion also has a ring that is similar in size \citep{millangabet06}.

\begin{table}
\caption{Summary of spectroscopy observations.}             
\label{obs-table}      
\centering                                     
\begin{tabular}{l c  c l}       
\hline\hline                        
MJD & Date & Instrument & Status \\
\hline                                  
54813.114       & 2008-12-12 &  Sophie/OHP & Outburst \\
54820.047       & 2008-12-19 &   Sophie/OHP & Outburst \\
54839.43        & 2009-01-08 &    ESPaDOnS & Outburst \\
54840.446       & 2009-01-09 &    ESPaDOnS & Outburst \\
54841.438       & 2009-01-10 &    ESPaDOnS & Outburst \\
54843.363       & 2009-01-12 &    ESPaDOnS & Outburst \\
54844.356       & 2009-01-13 &    ESPaDOnS & Outburst \\
54845.501       & 2009-01-14 &    ESPaDOnS & Outburst \\
55121.170       & 2009-10-16 &   Sophie/OHP & Quiescence \\
55122.181       & 2009-10-18 &   Sophie/OHP  &  Quiescence  \\  
55123.182       & 2009-10-19  & Sophie/OHP  &    Quiescence  \\ 
\hline                                             
\end{tabular}
\end{table}

The classification of the FUor object emphasized its double-peaked 
absorption lines, a telltale of strongly accreting sources \citep{hartmann96}. 
The spectra revealed a radial velocity $\sim$30 km s$^{-1}$, and double-peaked
absorption lines consistent with a self-luminous disk 6 au in size and with maximum velocity 
$\sim$120 km s$^{-1}$ \citep{hartmann89}. 
In addition, Z~CMa showed lines with strong P Cygni profiles
\citep[][]{hartmann89,vandenancker04} that were enhanced during the episodes of anomalous, bumpy variability \citep{hessman91}.
\citet{covino84} pointed out significant spectral variations since observations
began in the 1920s, and several authors have suggested that the origin of the strong line emission lies in the NW component
\citep{garcia99,benisty10, szeifert10,bonnefoy17}.

There is little
information on the accretion mechanisms and accretion evolution 
in intermediate-mass stars. Unlike T Tauri stars, Herbig Be (HBe) stars are
not expected to have magnetospheric accretion as their magnetic fields are too weak \citep{alecian13} as well as very small magnetospheres \citep{cauley15}. 
This would lead to a different accretion mechanism compared 
to solar-type stars \citep{cauley16}, such as a boundary layer \citep{popham93,vink02,eisner04,mendigutia11,wichittanakom20}, a
hot inner disk
scenario \citep[e.g.,][]{fairlamb15,mendigutia15}, 
or non-axisymmetric radial flows \citep{mendigutia17}. 
 Therefore, exploring the
small-scale accretion structures is especially important in objects such as Z~CMa NW.

A new outburst started in January 2008 \citep{grankin09}, ending in October 2009. 
High spatial resolution observations  
confirmed that the intermediate-mass Z~CMa NW was responsible for the increased brightness, and also resolved the
near-infrared (near-IR) spectra of both components, revealing the expected characteristic lines of FUor objects for
the SE companion, and a large number of emission lines reminiscent of EXor variables for Z~CMa NW \citep{bonnefoy17}. An enhanced
bipolar wind with velocities  up to 700 km s$^{-1}$ was also detected during the outburst \citep{benisty10,szeifert10}, which is a
further sign that variability episodes are related to increased accretion.

In this paper, we present the spectroscopic analysis of the 2008-2009 outburst and return to
quiescence of Z~CMa NW, exploring the physical structure and properties in 
the wind and the innermost disk with the aim of understanding accretion in high-mass
stars. Observations, data reduction, and 
emission line classification are presented in Section \ref{obs}. The analysis of the emission lines
is presented in Section \ref{analysis}, while the implications for the outbursts of massive versus low-mass stars
are discussed
 in Section \ref{discussion}. Our
results are summarized in Section \ref{conclusions}.

\section{Observations and data reduction \label{obs}}

\subsection{Spectroscopic data\label{spectra-obs}}

Two series of spectra were obtained for Z CMa at the Observatoire de Haute-Provence (OHP) using the SOPHIE spectrograph 
\citep{perruchot08}. The first set of two spectra was obtained on 2008 December 12 and 19 during the outburst. An additional set of 
three spectra was obtained with the same setup from 2009 October 16 to 19 as the system had returned to quiescence. OHP/SOPHIE spectra are automatically reduced online at the telescope to provide a continuum normalized 1D spectrum ranging from 382 to 693 nm at a spectral resolution of 39,000. 
With an integration time of 1 hour, and depending on weather conditions, the spectra have a signal-to-noise ratio (S/N) ranging from 40 to 150 at 600 nm.
Another series of spectra were obtained with the Echelle SpectroPolarimetric Device
for the Observation of Stars \citep[ESPaDOnS;][]{donati03}  in spectroscopic mode at the Canada-France-Hawaii Telescope (CFHT) 
from 2008 January 8 to 14 as the system was reaching the peak of the outburst. A total of 
six spectra were obtained over seven nights, covering from 370 nm to 1000 nm at a spectral resolution of 
65,000. An exposure time of 4800 s yielded a S/N of 500-600 at 600 nm. Spectra were reduced 
with the Libre-Esprit software \citep{donati97} to provide continuum normalized 1D spectra. 
Table \ref{obs-table} summarizes the available spectroscopic data.

Our data are not spatially resolved, so
the spectra contain emission from both Z~CMa NW and from the FUor 
companion. We assume that the FUor contribution can be neglected during
outburst because the luminosity is at least 8 times higher during outburst as in quiescence,
the FUor outburst seems to have ended (or substantially weakened) after 1995 (see Appendix \ref{historical-data}), and the
wind component has been estimated to be at least 20 times stronger for Z~CMa NW in outburst as for the
FUor \citep{antoniucci16}. The quiescence spectra do not reveal any
features typical of a FUor. Even though there is a larger relative
contribution from the FUor companion, the spectra resemble those of a typical, strongly accreting intermediate-mass 
star. The FUor features such as 
double-peaked absorption components that were present before \citep[e.g.,][]{hartmann89,hartmann96} are no
longer observed, indicating the evolution of the FUor outburst
in the last $\sim$20 years. We thus suspect 
that the quiescence data may also be dominated by Z~CMa NW, although a significant
contribution from the companion cannot be excluded.

\begin{figure*}
\centering
\begin{tabular}{cc}
\includegraphics[height=4.4cm]{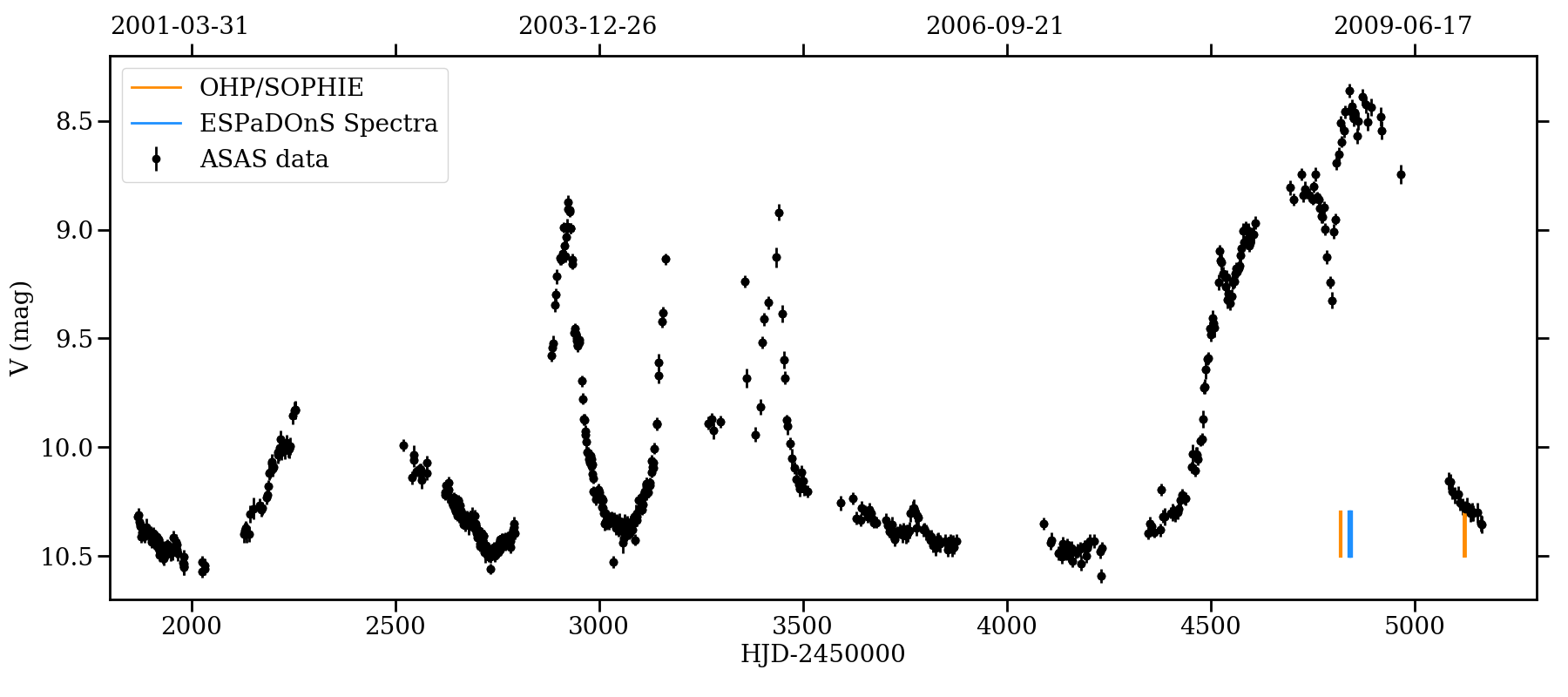} & 
\includegraphics[height=4.4cm]{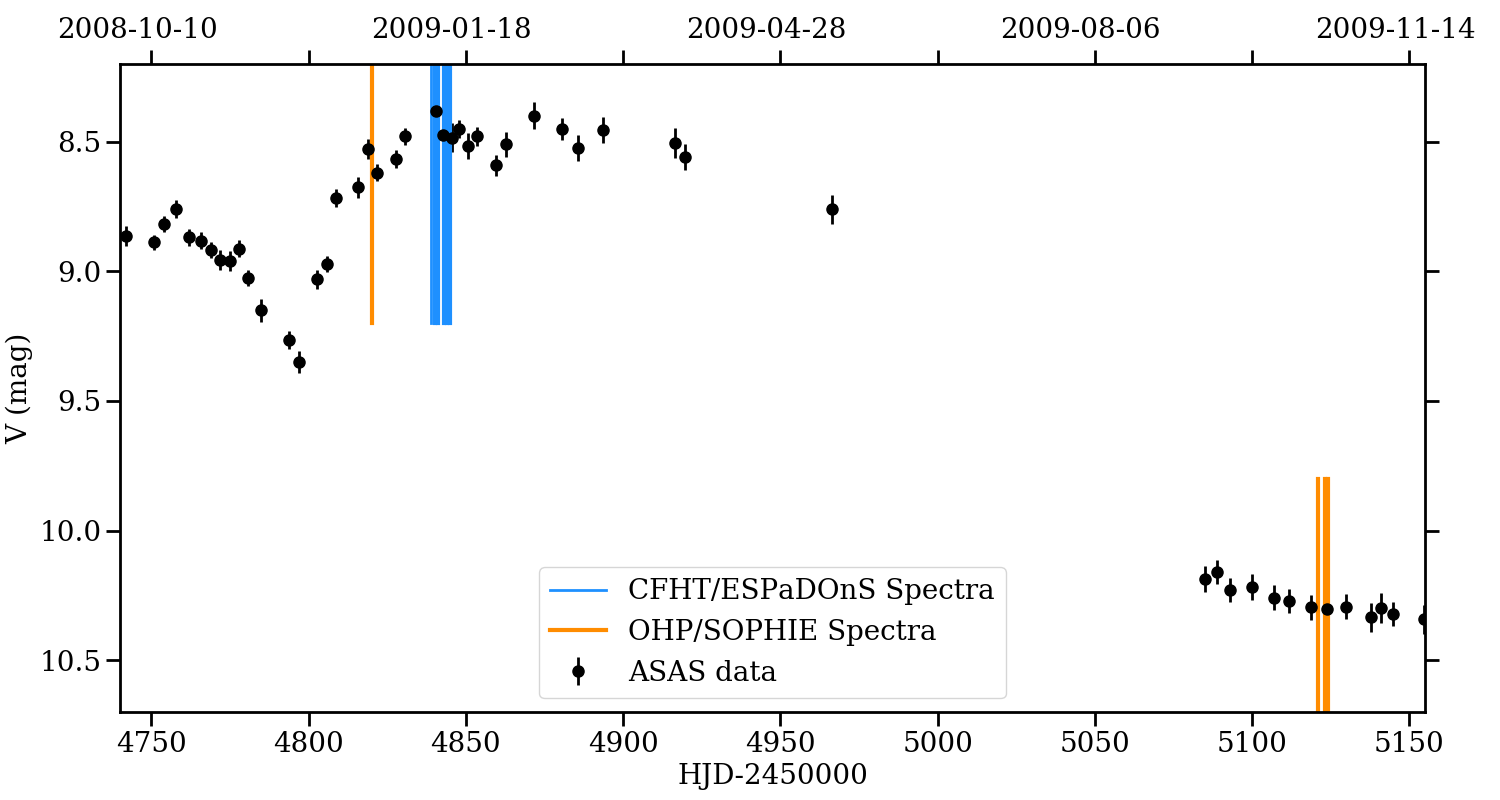}  \\
\end{tabular}
\caption{Full light curve (left) from ASAS and zoom around the times of the spectroscopic observations (right). 
The times at which the spectra were taken are indicated as blue for outburst spectra and
orange for quiescence spectra.  A complete light curve is given in Appendix \ref{historical-data}, Figure \ref{historical-fig}.}
\label{lightcurve-fig}
\end{figure*}

\subsection{Photometric observations\label{photo-obs}}

Z~CMa was observed by the All Sky Automated Survey \citep[ASAS;][]{pojmanski03}
between December 2000 and December 2009 in the V filter. The photometry is calculated using 
five different apertures.
We chose the smallest (2 pix, where the native pixel is 15"\footnote{http://www.astrouw.edu.pl/asas/explanations.html}) to minimize 
the amount of nebular emission included, although the differences 
between apertures are negligible, especially considering the variability range. Only the quality A data are used, 
containing 637 photometric data points spread over nine years 
(Figure \ref{lightcurve-fig}). The ASAS data include the full outburst, 
rising from quiescence levels by more than two magnitudes and coming back to quiescence,
and offer a good baseline to explore the status of the object at the time 
of the spectroscopic observations.
The ASAS data also reveal that, although the 2009 outburst is significantly longer and
very bright, the source has suffered several bursts, rising to nearly the same magnitude for shorter times,
since their records started. 

The historical data from the American Association of Variable Stars Observers
(AAVSO; see Appendix \ref{historical-data}) show that the smooth
FUor outburst profile is punctured by brief bursts \citep[similar to the bumps cited by][]{hessman91}
since the early 1980s. The overall shape of the light curve suggests that the NW component has 
become increasingly active and likely brighter and less extincted in the last $\sim$40 years.
The FUor outburst
had faded substantially by 1995. More recent AAVSO observations show that the bumpy behavior 
recorded by ASAS continues until the present day (see Figure \ref{historical-fig}), so that
the outbursts described in this work are recurrent.

Analysis of the ASAS data does not reveal any significant periodic signature on short 
(days--months) or long (months--years) timescales. A low-significance 249 d quasi-period can be derived from
 the whole V data (ASAS plus the AAVSO, see Appendix \ref{historical-data}), suggesting
that outbursts happen roughly every 249 days.
Such a period is similar to that of a body in Keplerian rotation at a distance
2.0$\times$(M$_*$/16~M$_\odot)^{1/3}$ au.
Regular perturbations of the disk by companions may trigger
outbursts  \citep{lodato04}, although the results have low significance and 
need follow up for confirmation.
The ASAS data also show a brief dip or occultation event that happened 
just before the first
spectrum was acquired. Although the dip was not completely over at the time of the first observation,
we do not find any significant spectroscopic signature linked to this event.

\subsection{Emission line classification \label{findlines}}

\begin{figure*}
\centering
\begin{tabular}{cccc}
\includegraphics[width=4.3cm]{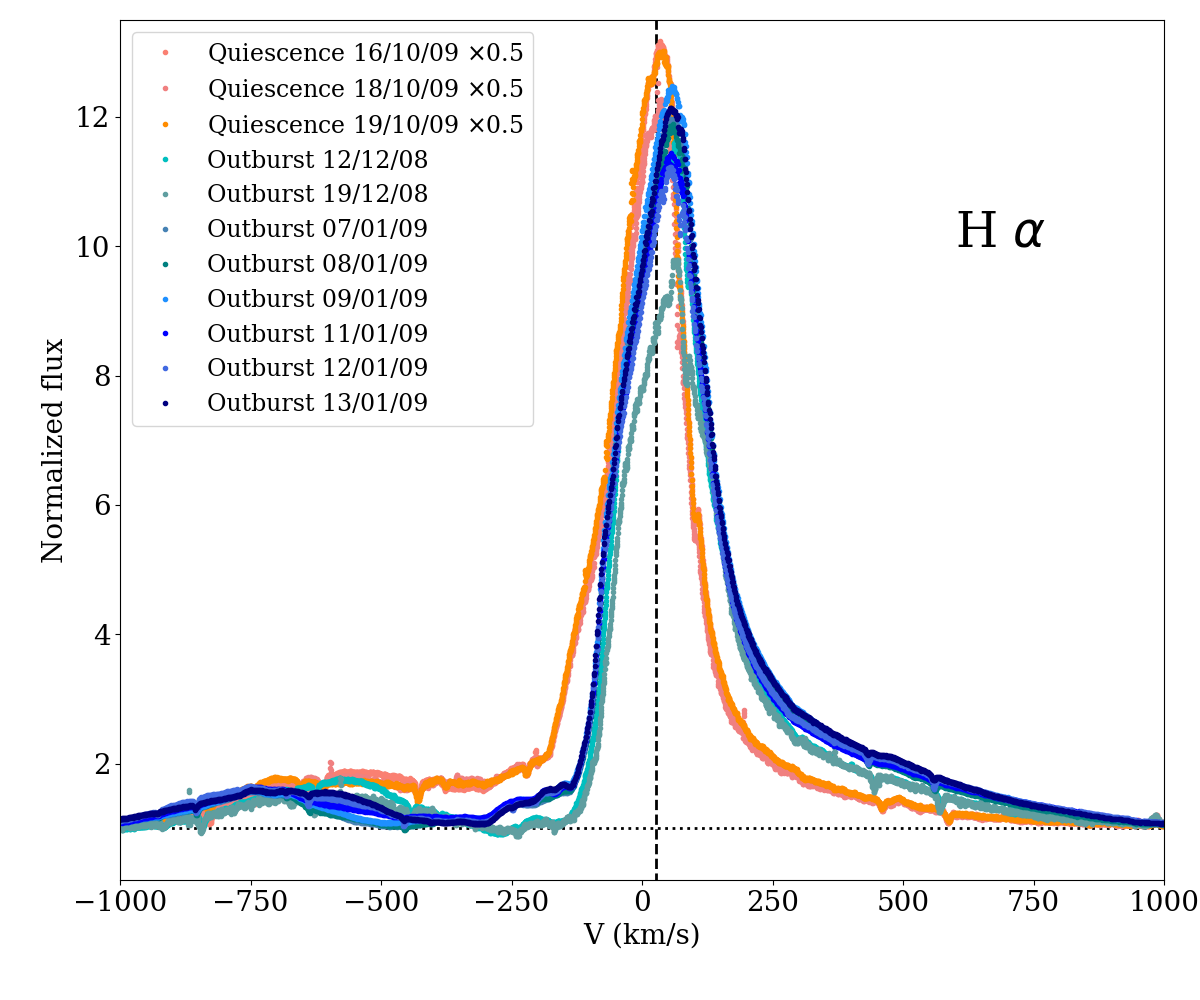} &
\includegraphics[width=4.3cm]{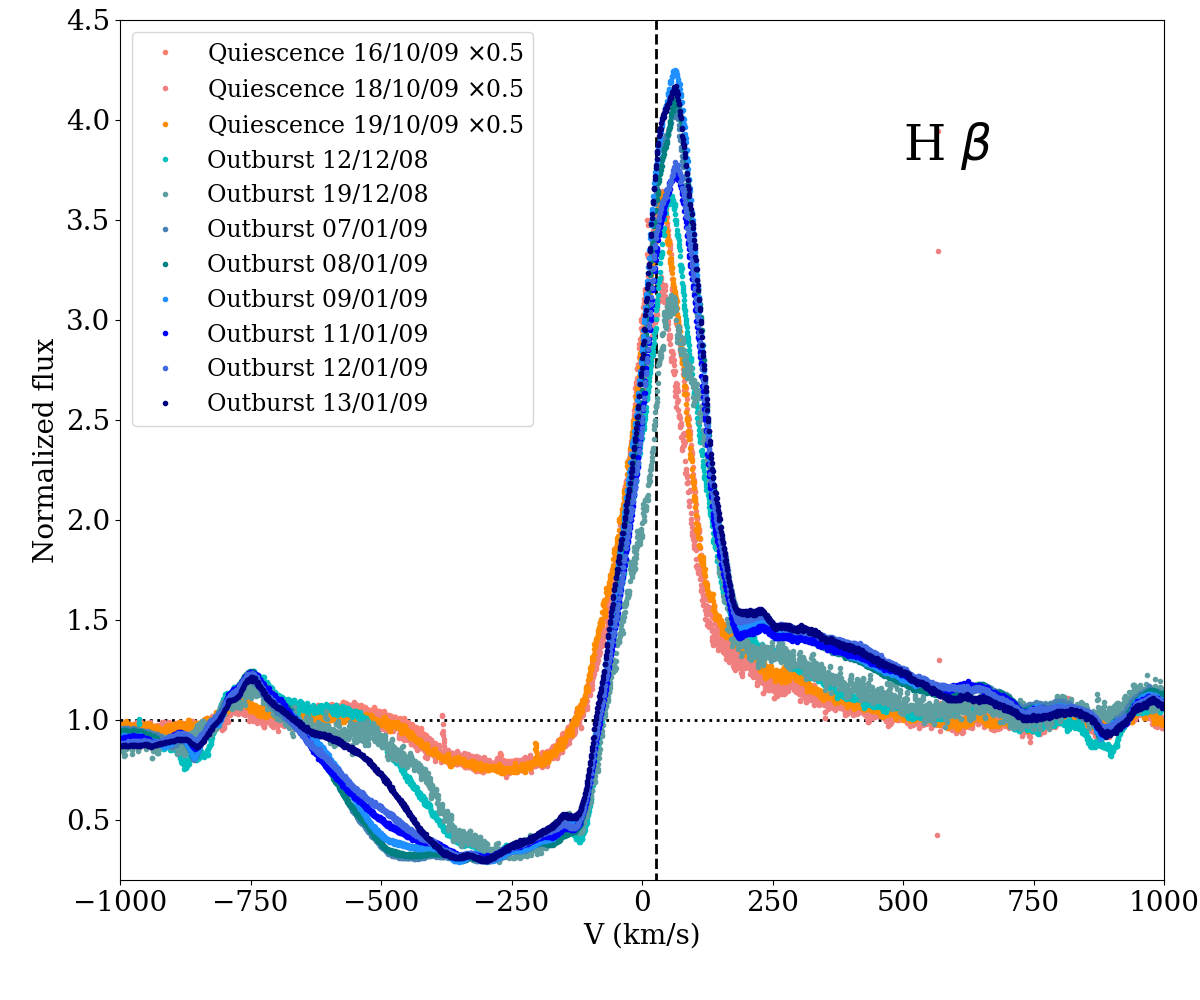} &
\includegraphics[width=4.3cm]{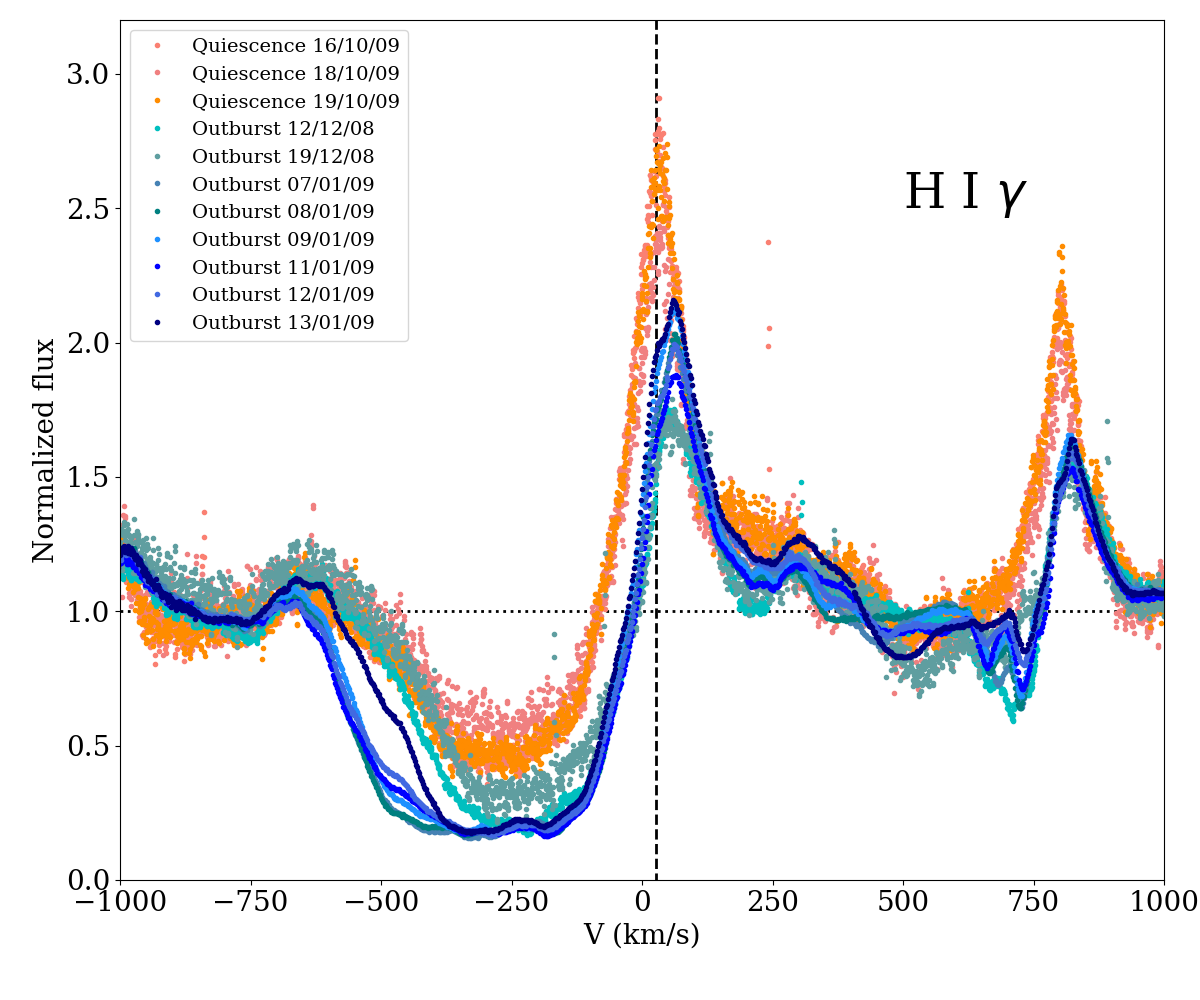} &
\includegraphics[width=4.3cm]{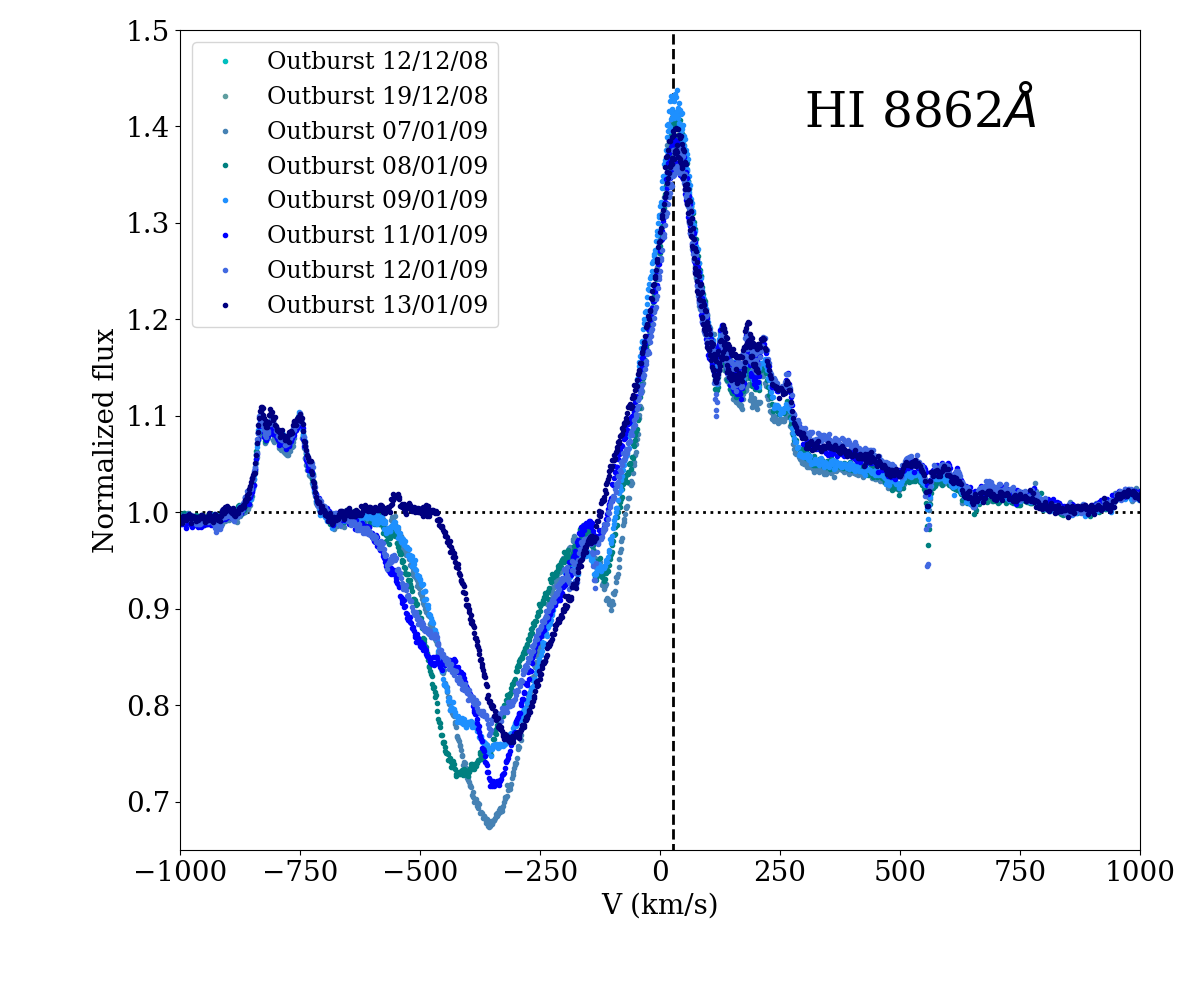}\\
\includegraphics[width=4.3cm]{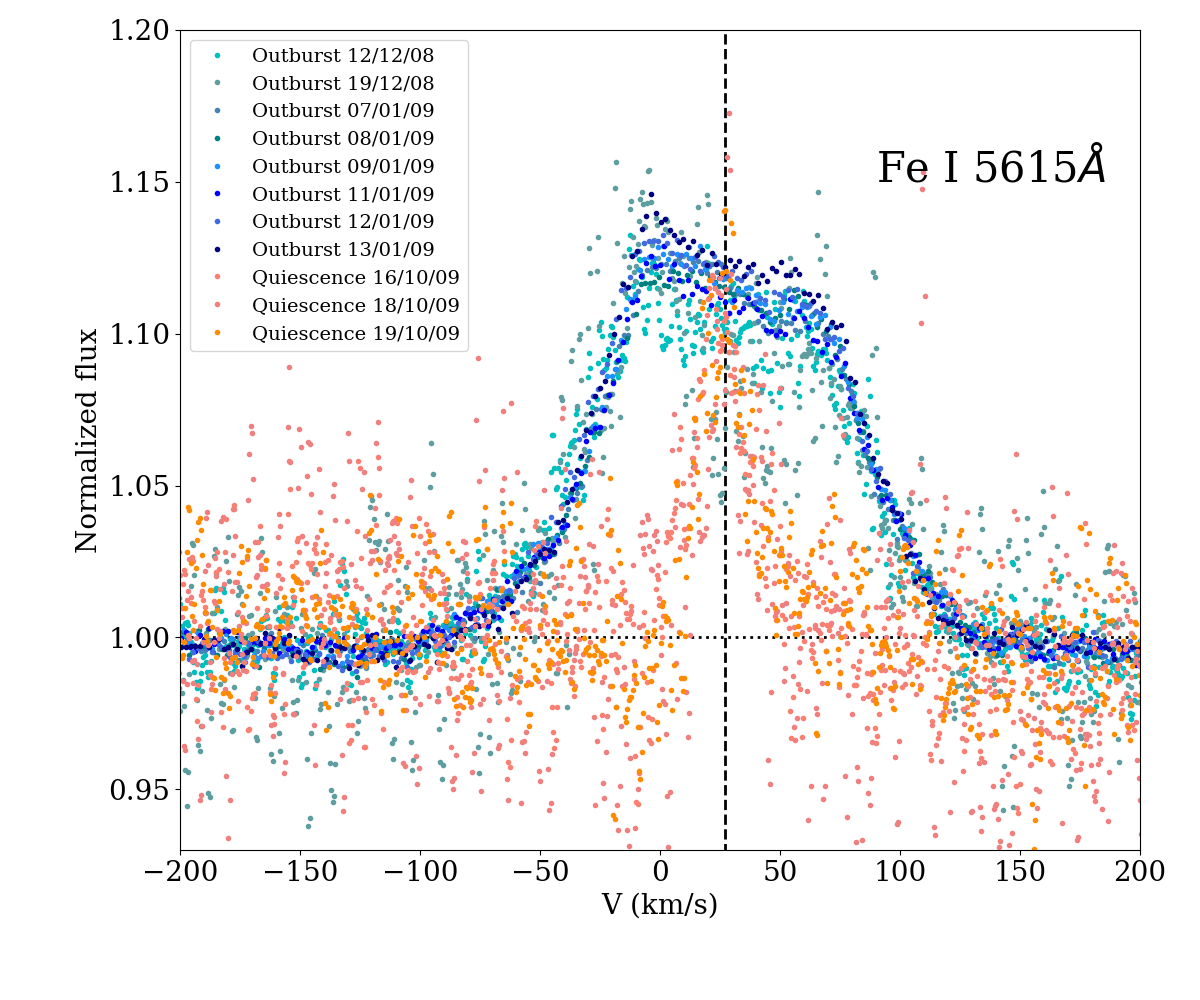} &  
\includegraphics[width=4.3cm]{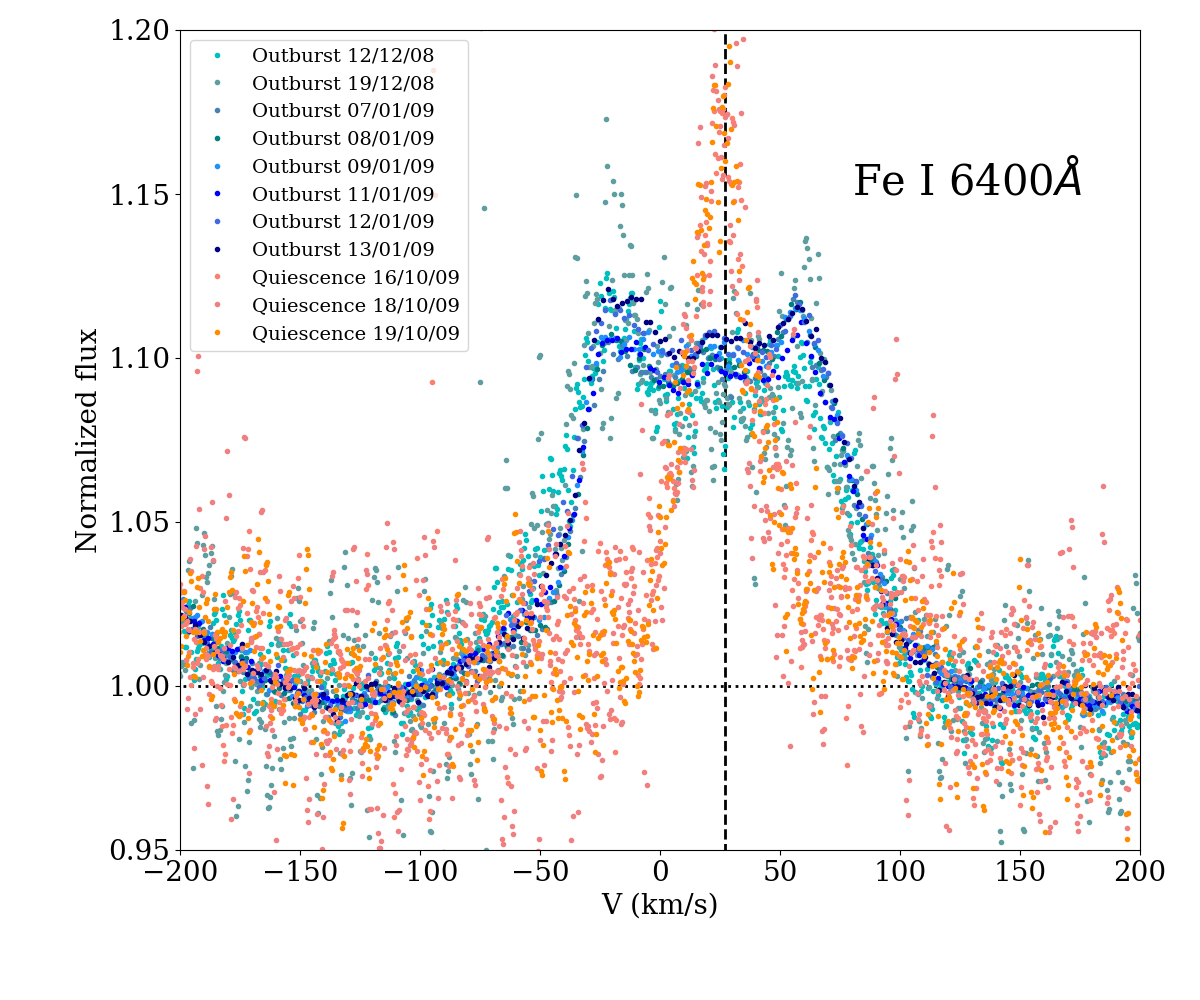} & 
\includegraphics[width=4.3cm]{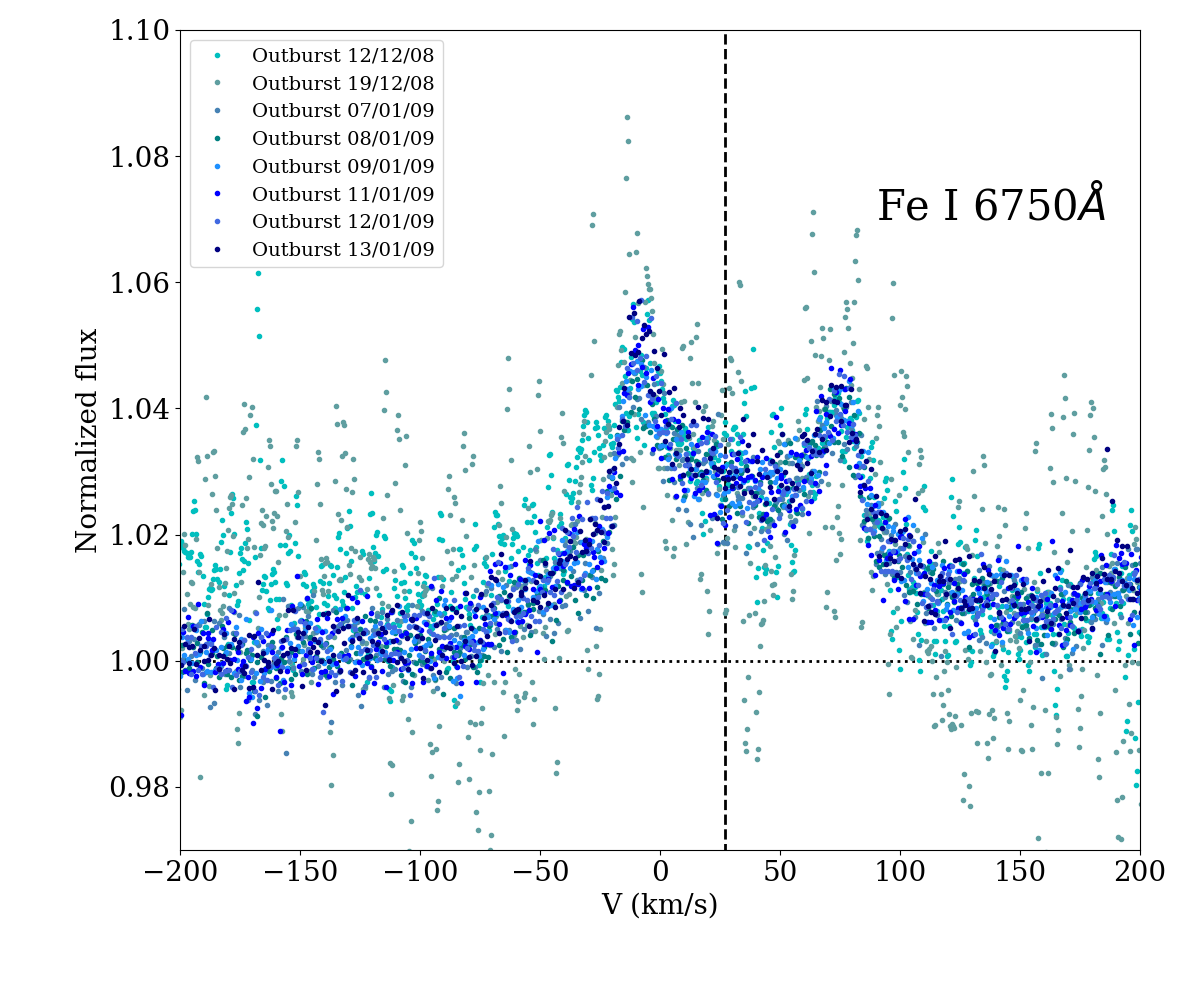} &
\includegraphics[width=4.3cm]{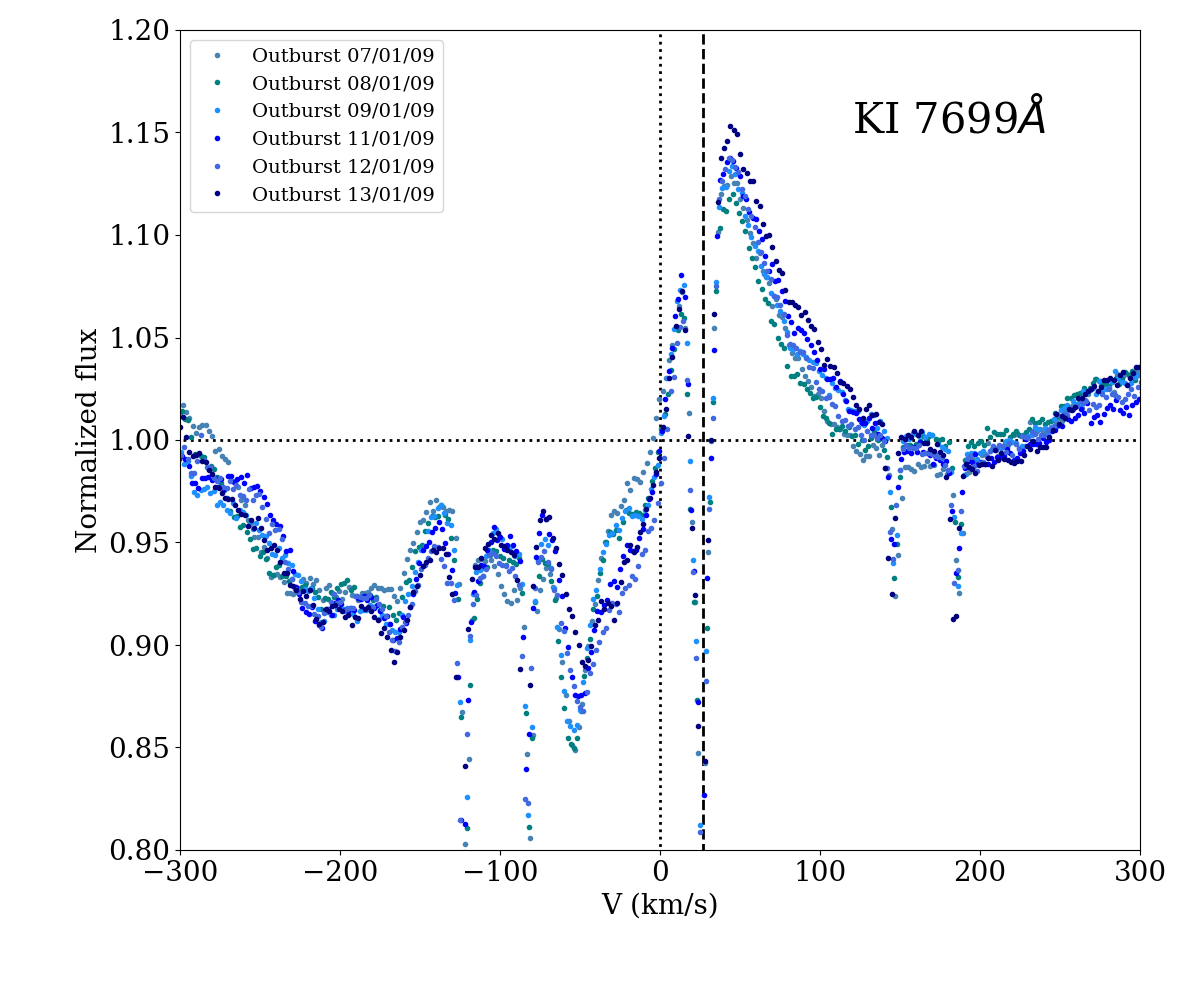}\\
\includegraphics[width=4.3cm]{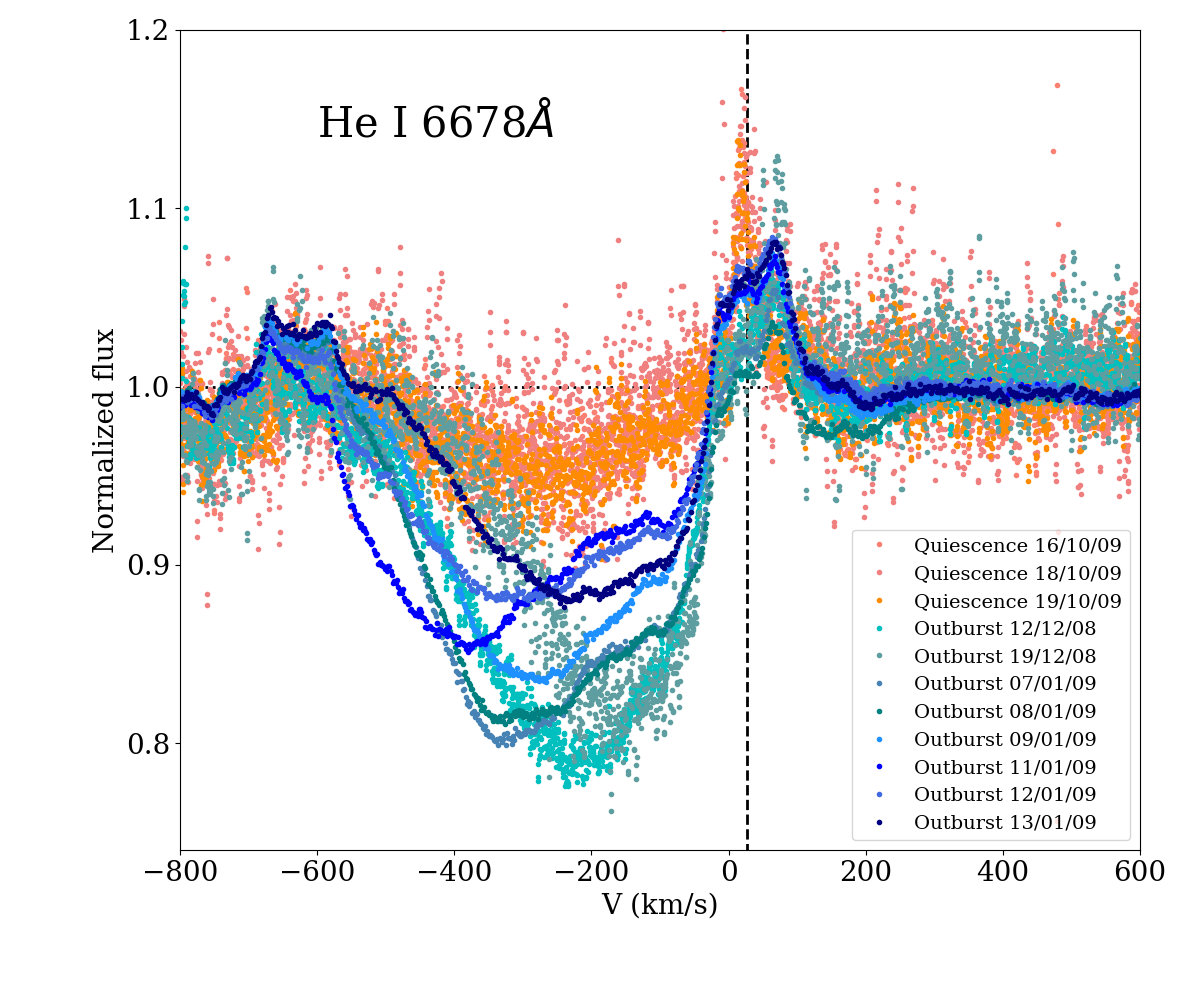} & 
\includegraphics[width=4.3cm]{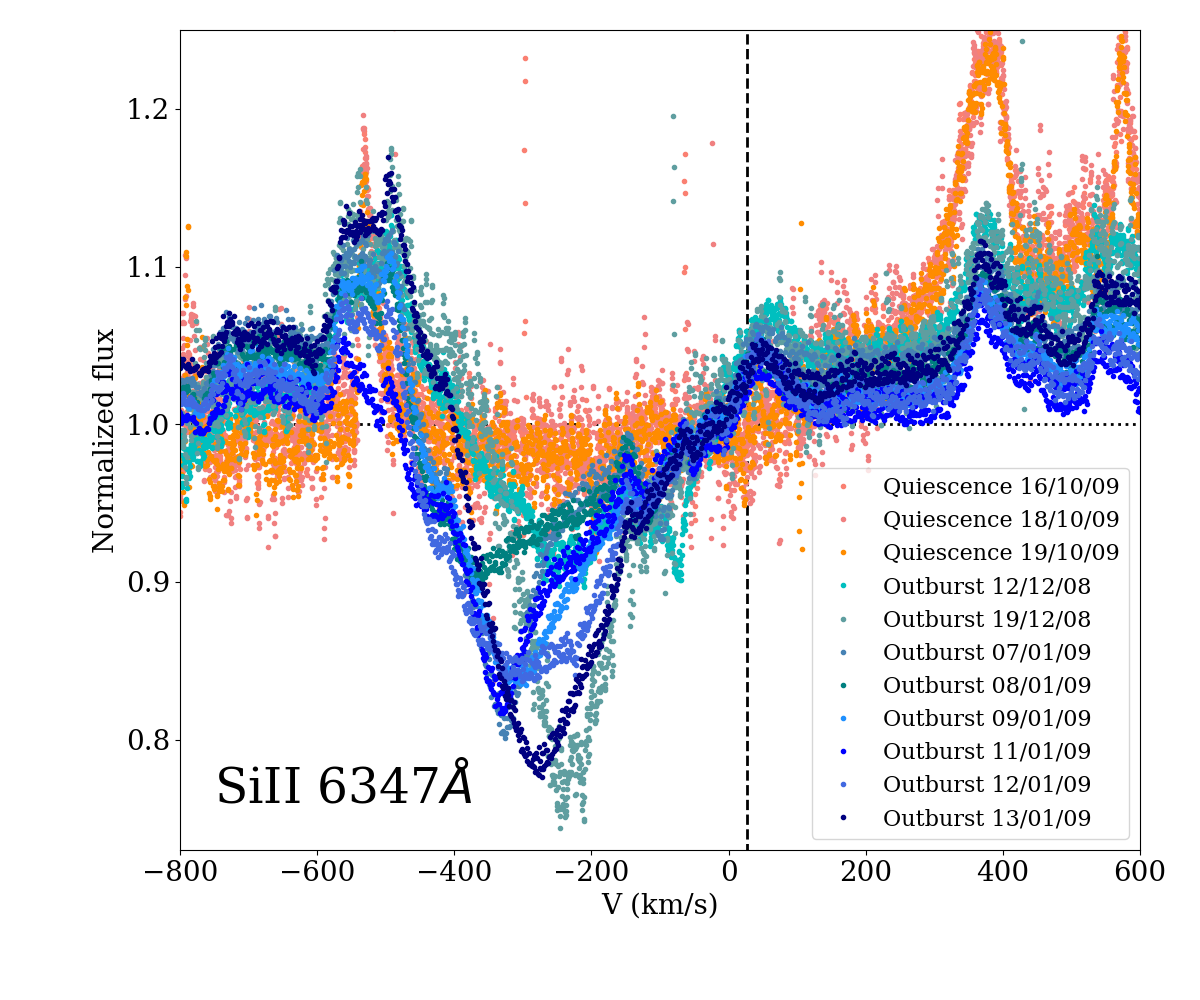} &
\includegraphics[width=4.3cm]{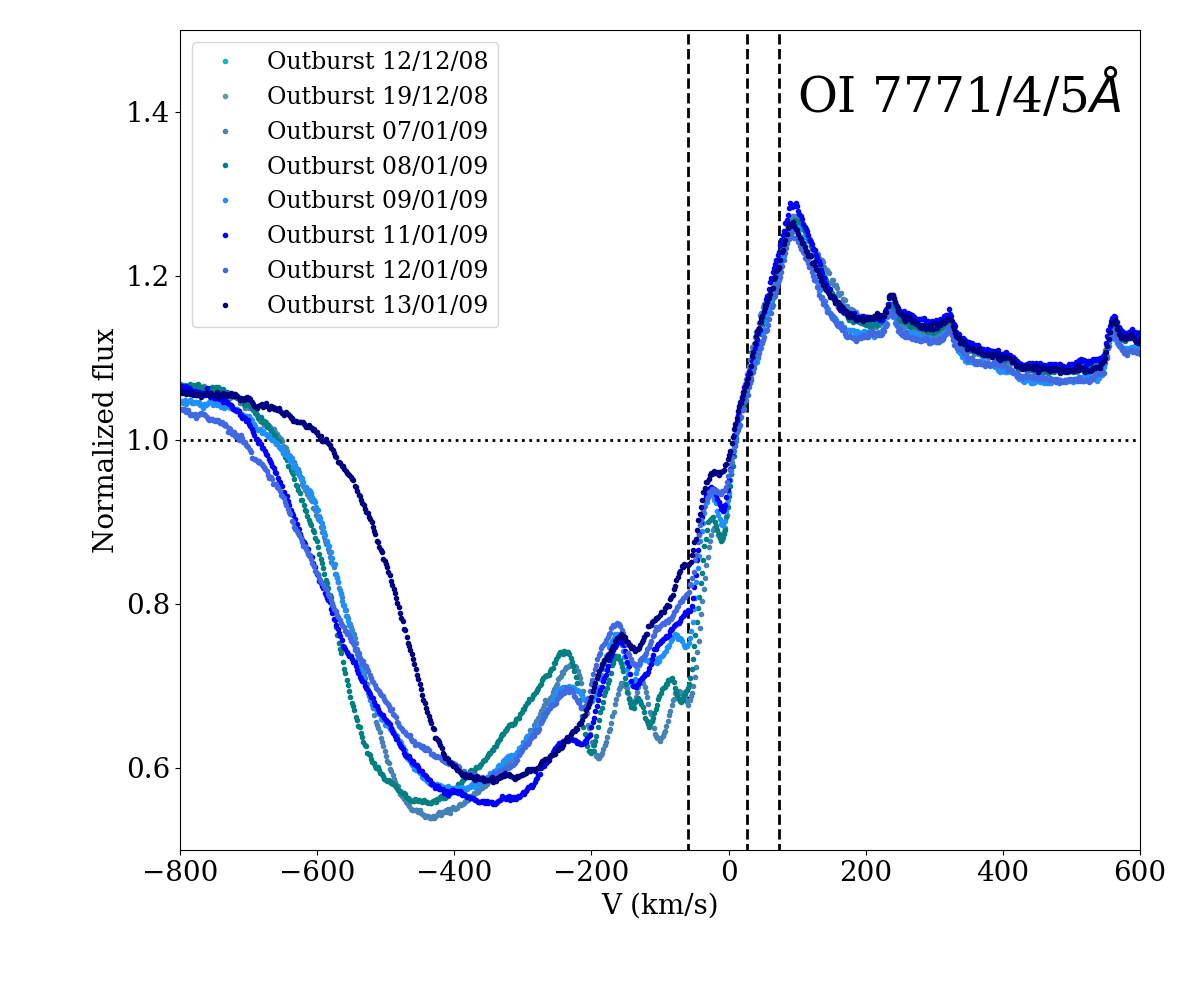} &
\includegraphics[width=4.3cm]{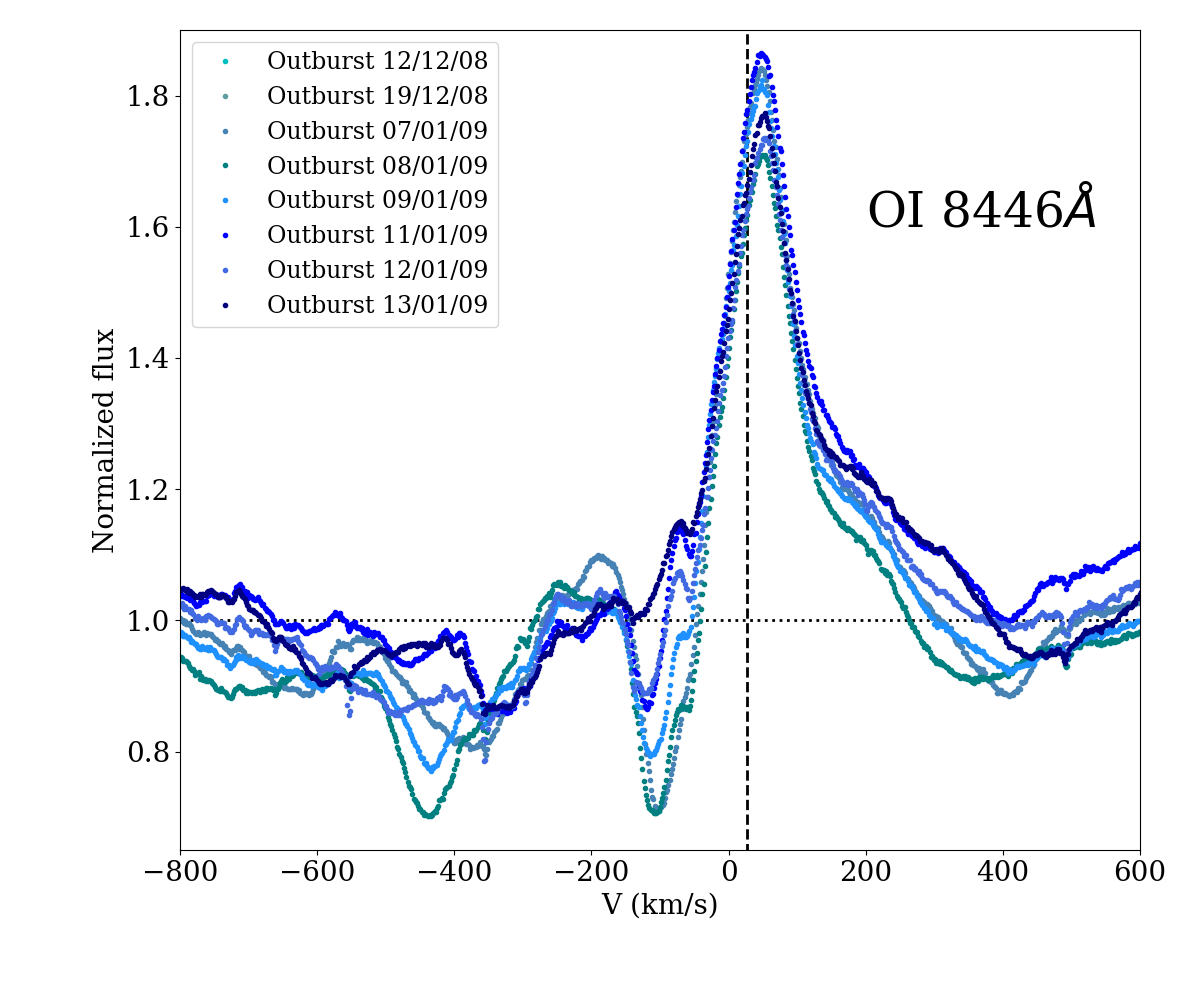}\\
\includegraphics[width=4.3cm]{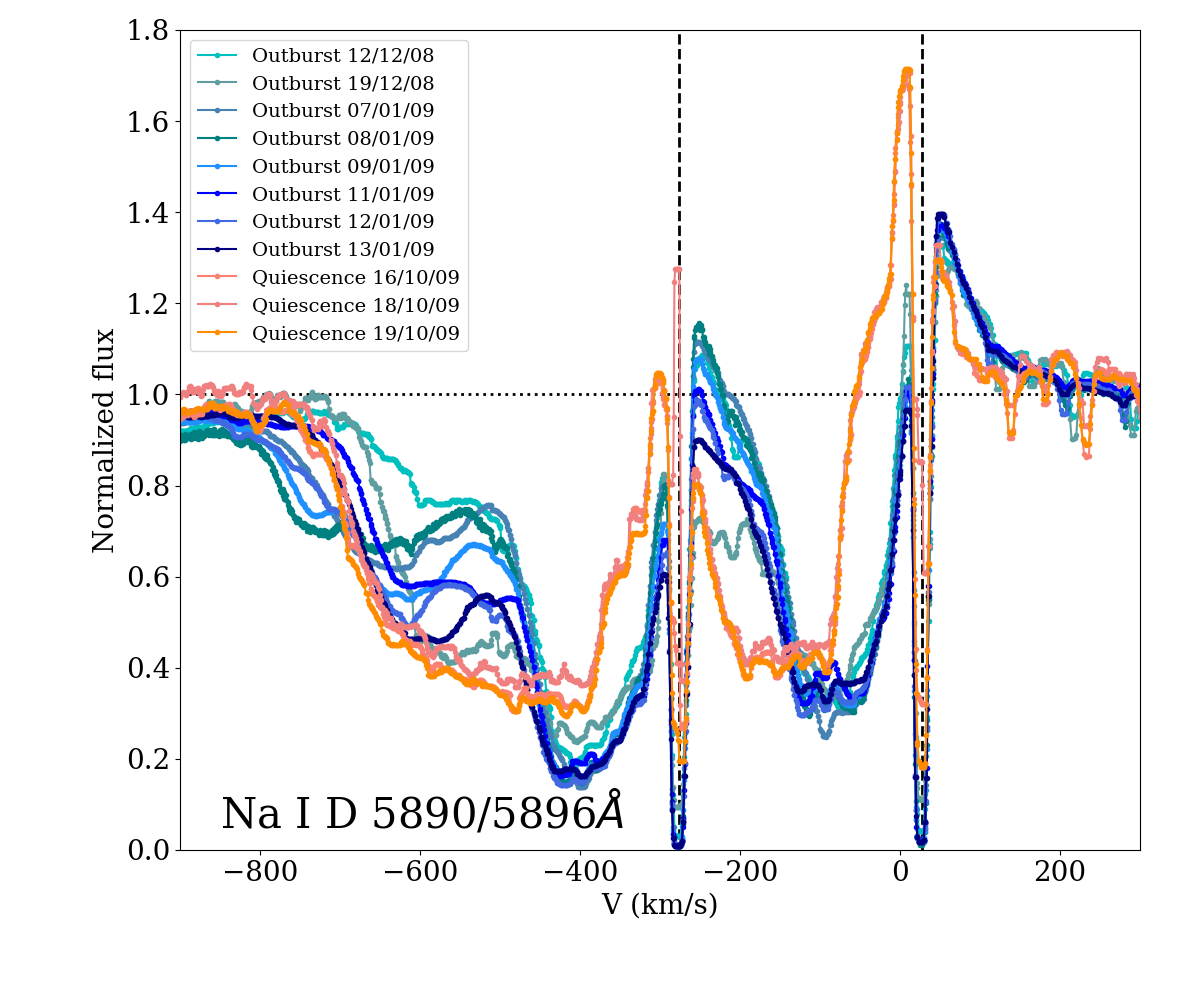} &
\includegraphics[width=4.3cm]{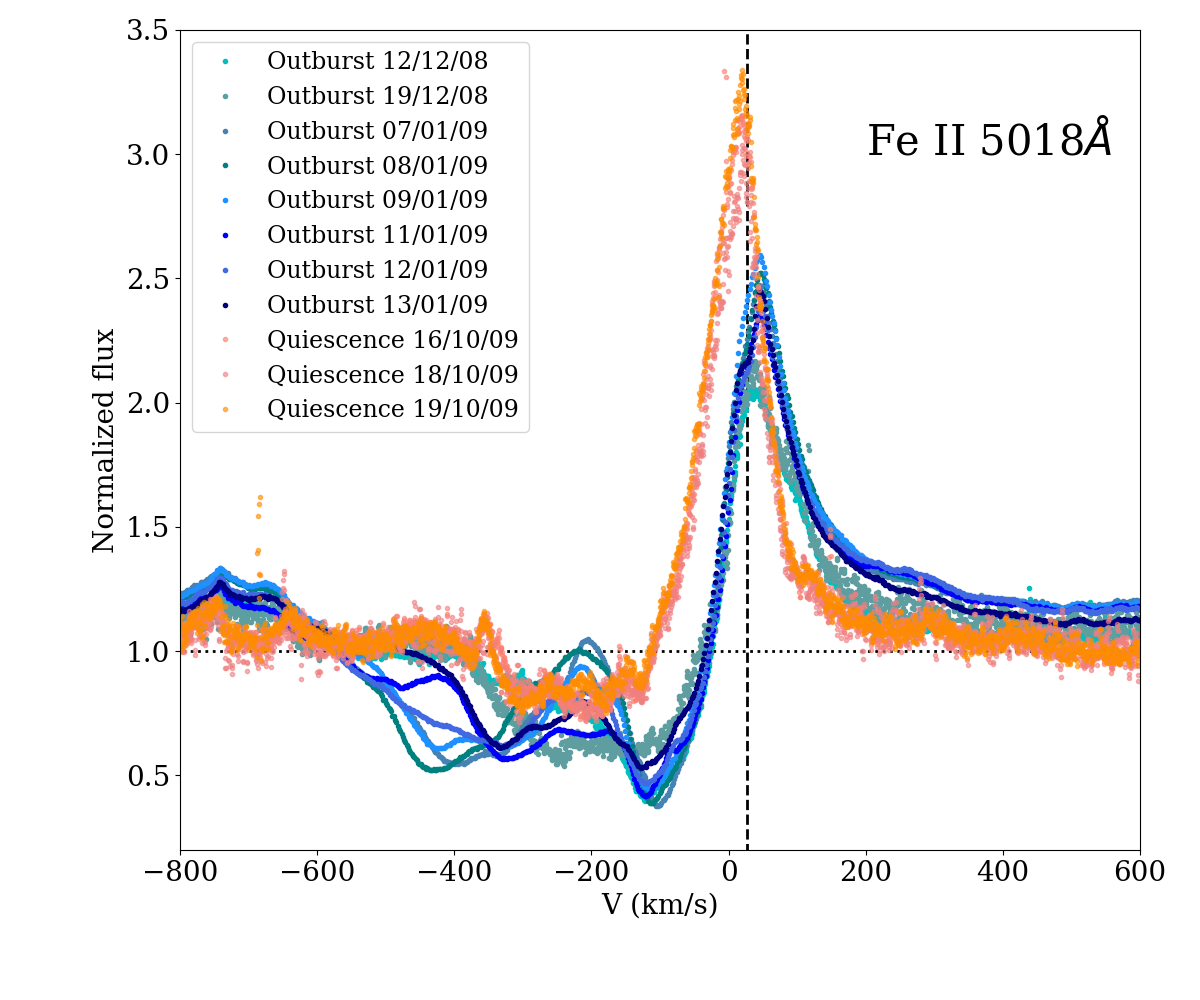} &
\includegraphics[width=4.3cm]{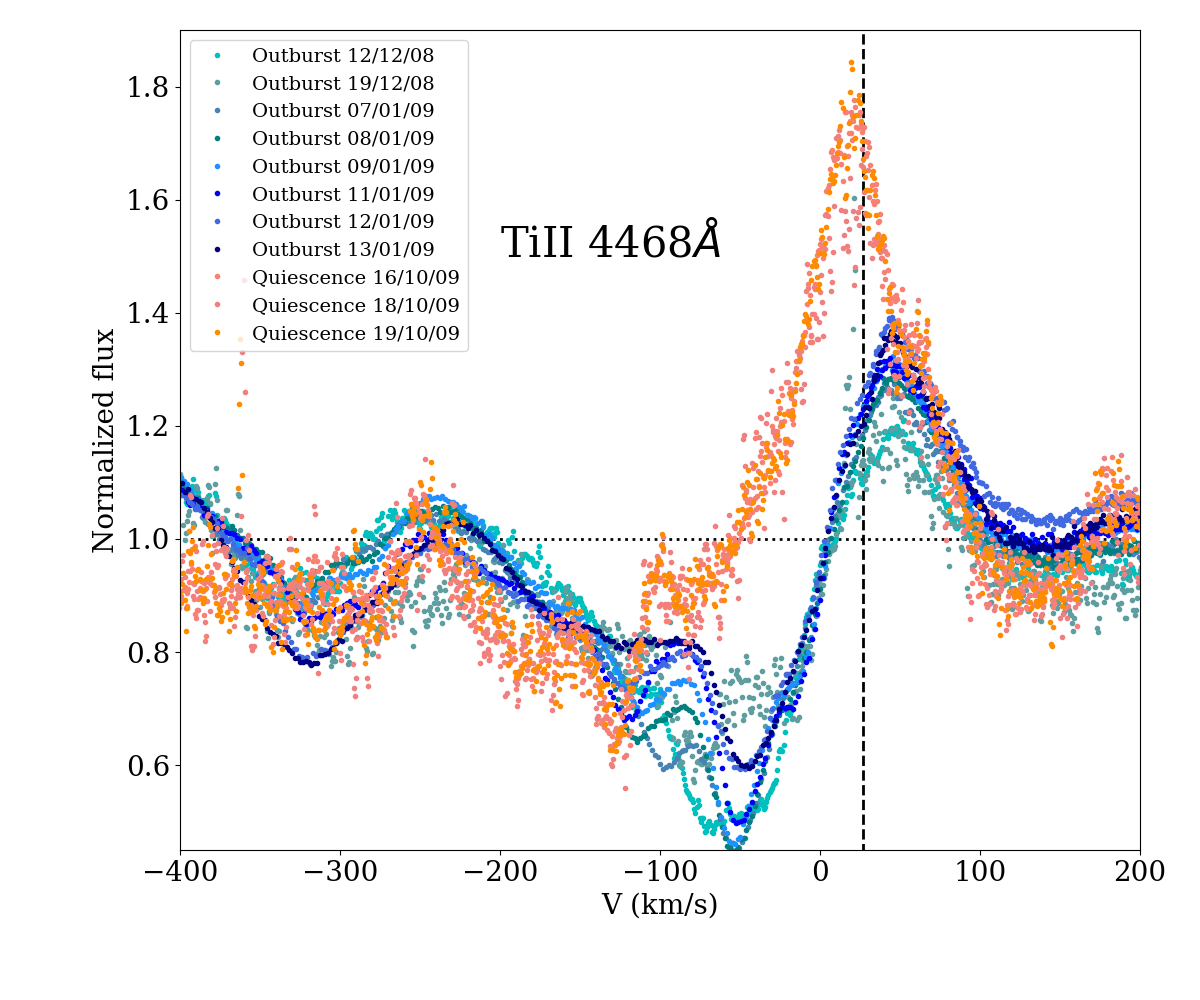}&
\includegraphics[width=4.3cm]{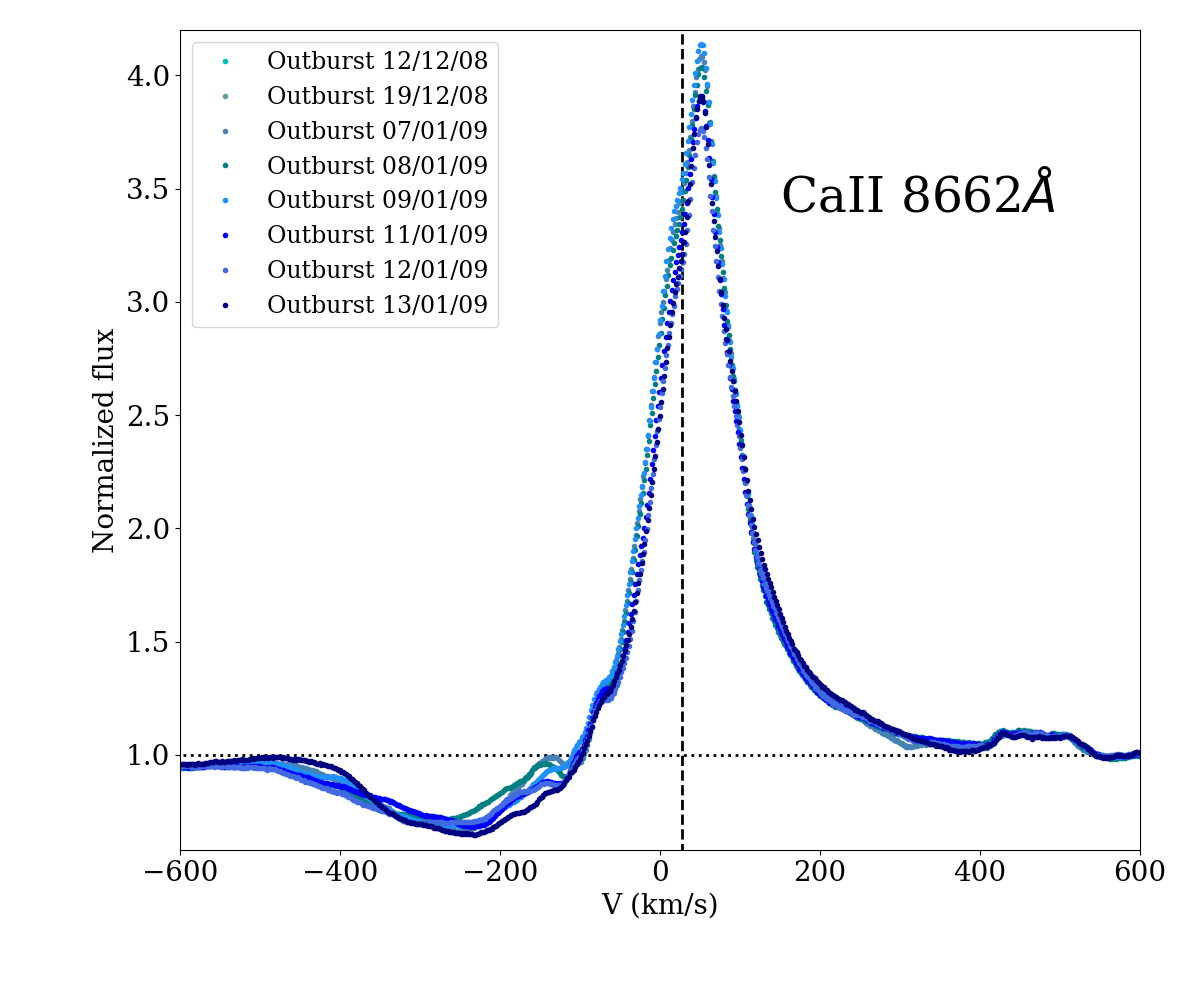}\\
\end{tabular}
\caption{Examples of lines observed for Z~CMa.
The lines are normalized to the local
continuum. 
In this and the following figures, the quiescence
data are plotted in various shades of orange, while the outburst spectra are shown in blue. The dashed line shows the
radial velocity of the system (27 km s$^{-1}$) and the dotted line indicates the continuum level.  First row: Some H Balmer lines and an IR H line. 
The quiescence spectra of H$\alpha$ and H$\beta$
are scaled by $\times$0.5 for better display. Second row: Some of the Fe I lines with disk-like profiles and a K I resonance line.
Third  and fourth row: Some lines with various types of P Cygni profiles.
The quiescence spectra do not cover the range beyond 6940 \AA.}
\label{lines-fig}%
\end{figure*}

The emission lines were visually selected and classified using the NIST database
\citep{ralchenko10,kramida18} and the list of lines observed in Z~CMa and other outbursting objects 
\citep{vandenancker04,sicilia12,sicilia15,sicilia17}. The narrow neutral metallic emission lines in quiescence reveal 
a radial velocity $\sim$27$\pm$3 km s$^{-1}$, which is consistent with
previous estimates \citep[$\sim$30 km s$^{-1}$;][]{hartmann89}. 
This radial velocity was used to 
revise the classification of the remaining emission lines, in particular, those that were weak or not observed
in other young stars.  In total, we estimated more than 1000 emission lines in the optical outburst
spectra, but because of low S/N and blends, we only resolved 497 of these, 
among which 26 lines cannot be identified in the literature. The complete line list is presented in Appendix 
\ref{lines-app} (Table \ref{alllines-table}). The table contains all identified lines plus blends, together with
excitation potentials, transition probabilities, profile types, strength, and whether they have been observed in outburst and/or 
quiescence.

The lines include typical H, He, and Ca II emission 
plus a large number of Fe I, Fe II, Ti I, Ti II, Si II, and 
other metallic lines commonly found in 
young stars \citep{joy45,hamann92}.  Nearly all lines correspond to permitted transitions and
the only forbidden lines that can be clearly identified\footnote{There is some potential  [Fe II] and [N II] emission, but the
lines are very weak and blended and the identification is not univocal.} are the [O I] lines at 
$\lambda \lambda$6300, 6363\AA\ and the [S II] lines at $\lambda \lambda$6717, 6731\AA.
Many of these are common to both quiescence
and outburst, but their profiles are different.  Several examples of lines with various profiles are shown in Figure \ref{lines-fig}.
Although most of the lines are broad in both quiescence and outburst, the outburst lines are
significantly broader; some such as H$\alpha$ have line wings up to $\pm$1000 km s$^{-1}$.
A P Cygni profile is most common among the strong,
high-energy lines, such as H and Fe II, 
in both outburst and quiescence; 89 lines have this kind of profile. 
Very high energy lines, such as He I and Si II, have extreme P Cygni profiles in outburst,
which are entirely dominated by blueshifted absorption and with a very small emission component
(see Figure \ref{lines-fig}), being thus the only lines that show the typical profiles
of FUor objects. No such lines are 
observed in quiescence, so they are likely related to the outbursting Z~CMa NW and
not to the FUor. The blueshifted absorptions become much weaker (or disappear
completely, especially in the most energetic lines) in the quiescence spectra. 
This suggests a powerful hot wind developing during outburst that is likely related to the
increased accretion episode. 
Most of the outburst blueshifted absorption
features have several components and show a very high degree of variability, as we
discuss in Section \ref{wind-sect}.

We detected many low-energy, neutral metallic lines in emission
during both quiescence and outburst, especially Fe I.
In quiescence, most of the Fe I, Na I, and Ni I lines are narrow (FWHM $<$30 km s$^{-1}$) or absent, while in outburst, 91 of
these lines show a box-like or double-peaked profile as expected for Keplerian disk 
emission \citep[][]{horne86} and similar to the disk emission profiles observed in HAeBe
stars in CO \citep[e.g.,][]{hein16}.  The 
double-peaked profiles are very similar to those observed by \citet{hartmann89} in FUor 
objects, but they are in emission instead of in absorption. 
Although viscous dissipation often dominates disk heating in strongly accreting systems
\citep[producing a disk that is hotter in the midplane than in the surface;][]{hartmann89}, this is a signature of a temperature inversion in the
Z~CMa NW disk, which remains hotter in its upper layers
as seen in irradiated disks \citep{calvet92}. Interferometry and polarimetry  have also revealed
structures that are consistent with disks in some HBe stars \citep{eisner04,ababakr17}, which could be
similar to what we trace in the neutral metallic lines.
There are differences in the disk-like profiles, considering line strength (or transition probability) and
some asymmetries in the blue- versus redshifted parts. The former is expected since lines with various strengths saturate at
different heights over the disk and stronger lines can be produced over larger, less well-defined regions (and may thus include other components, including low-velocity ones)
compared to weaker lines, resulting in box-like rather than double-peak
profiles \citep{ferguson97}. The
latter could be related to asymmetries in the disk and/or wind- or accretion-related absorption components, which is
discussed in detail in  Sections \ref{lineratio-sect} and \ref{diskanalysis-sect}.

Redshifted absorptions are typically a clear indication of infall in the spectra of young stars. We did not observe clear
redshifted absorptions in the Z~CMa spectra, although there are several lines with redshifted asymmetries that
 could be consistent with weak absorption components at various velocities. 
Redshifted absorption asymmetries were observed at 
$\sim$200 km s$^{-1}$ in the higher Balmer lines, IR H lines, the O I line at 8446\AA\, and the Ca II IR triplet 
(Figure \ref{lines-fig}). As we discuss in Section \ref{quiescence-sect}, Fe I lines also have weak redshifted asymmetries in
outburst and quiescence that could be
related to absorption components, although the velocities are significantly lower.

\begin{figure}
\centering
\begin{tabular}{cc}
\includegraphics[height=4.5cm]{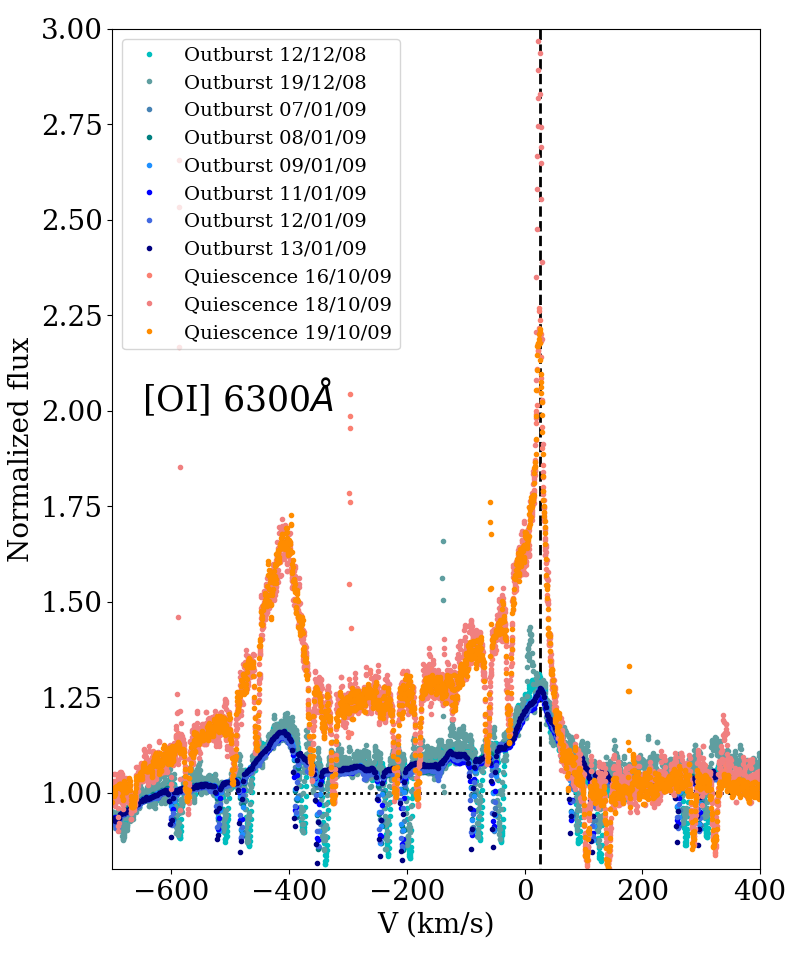} &
\includegraphics[height=4.5cm]{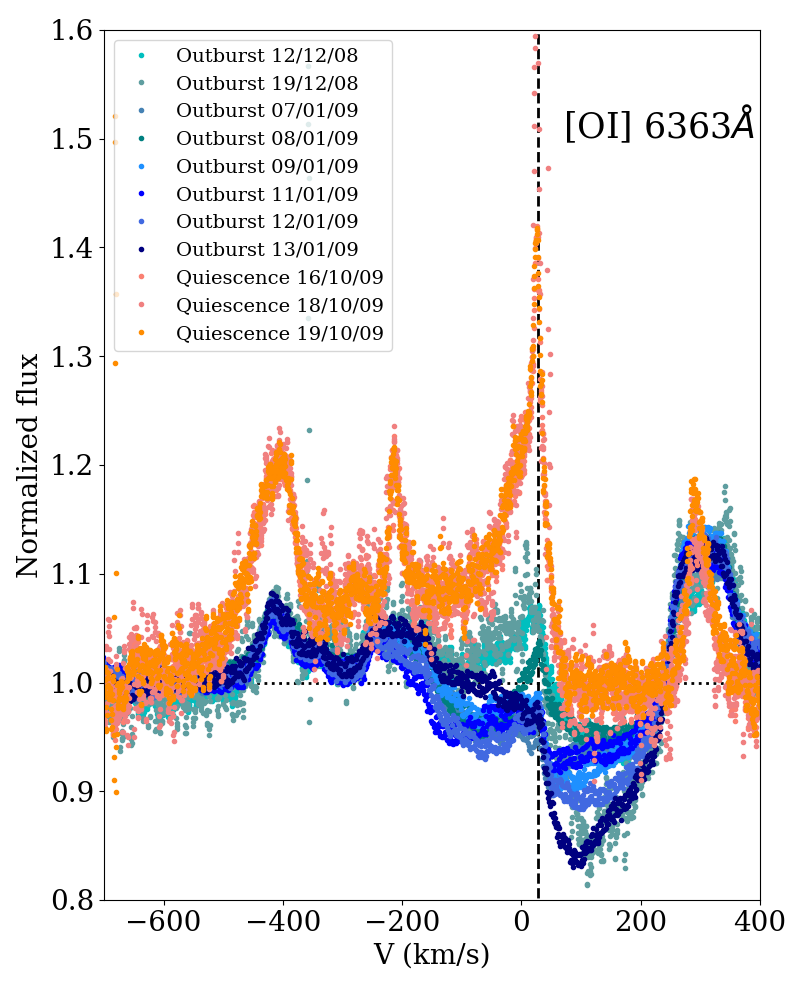}  \\
\end{tabular}
\begin{tabular}{c}
\includegraphics[height=4.5cm]{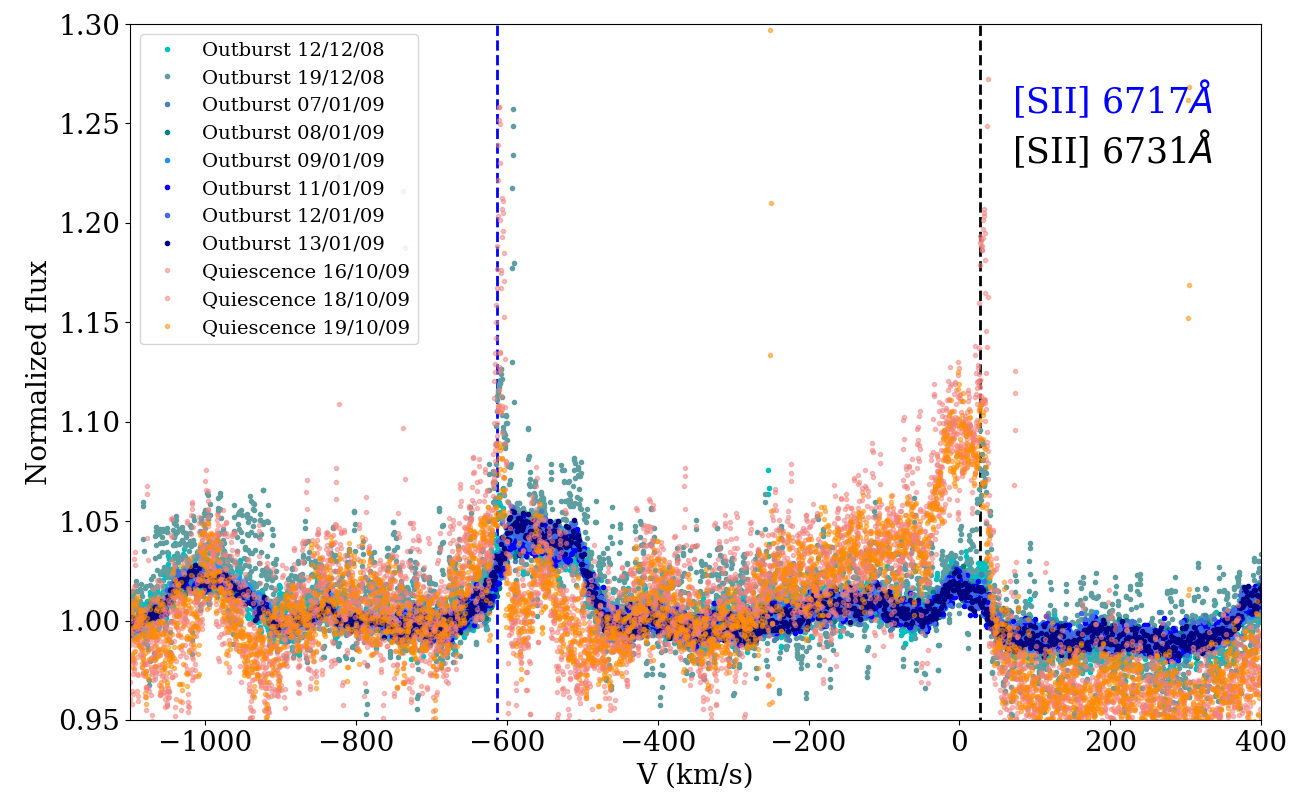}  \\
\end{tabular}
\caption{Upper: Forbidden [O I] emission. Lower: Forbidden [S II] emission.
The dotted lines indicate the the continuum level, and the dashed
lines represent the radial velocity of 27 km s$^{-1}$. The data from the outburst are plotted in 
various shades of blue, while the quiescence data are shown in orange. 
The extra lines observed in the 6363\AA\ spectrum
correspond to unrelated Fe I and Fe II lines, and the strong absorption is
due to Fe II 6369\AA. The narrow absorption lines are telluric.}
\label{forbidlines-fig}%
\end{figure}

Regarding forbidden line emission, the [O I] line is the most evident because of its strength and lack of blends with other features. 
The [O I] lines have two components with typical shock profiles (Figure \ref{forbidlines-fig}): one component slightly blueshifted with respect to the source velocity and a second,
high-velocity component at $\sim -$420 km s$^{-1}$. The high-velocity component is consistent with the high-velocity peak of the jet identified by
\citet{whelan10}, who also detected the SE source jet at velocities $-$300 to $-$100 km s$^{-1}$ (not evident in our spectra). [S II] emission at 6731 and 6717\AA\ is also observed. Its high-velocity component at -400 km s$^{-1}$ is detected in quiescence, but it is negligible in outburst, 
which could be a contrast effect combined with blending with other broad lines in the outburst spectrum. Additional [N II] emission and some 
[Fe II] lines may be also present, but because they are heavily blended with other lines (see Table \ref{alllines-table}), it is hard to confirm their presence and
to reveal any details about their velocity components.

\begin{figure}
\centering
\includegraphics[width=8.5cm]{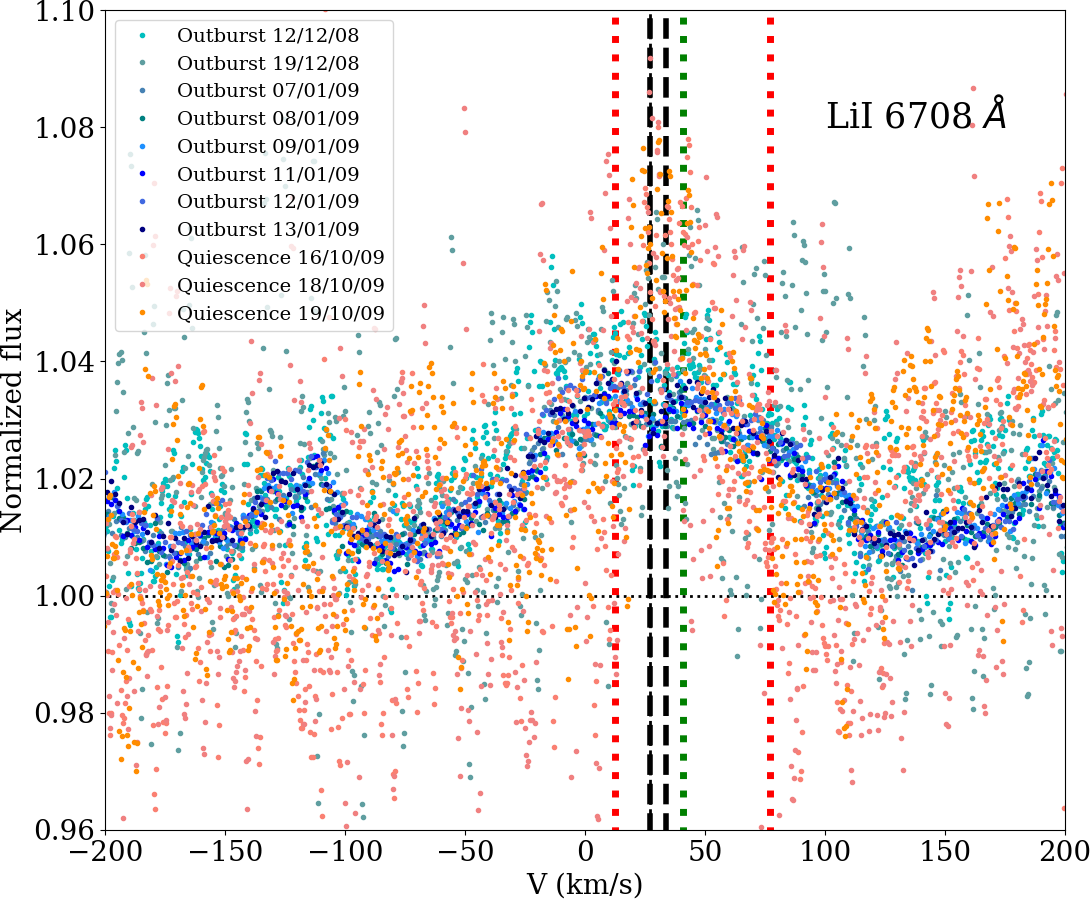}
\caption{Zoom around the Li I 6708\AA\ line. The dotted black line indicates the continuum level, the dashed black
lines indicate the Li I transitions (6707.76 and 6707.91\AA),  the vertical red dotted lines indicate 
the Fe I  6707.43\AA\ and Fe II 6708.88\AA\ lines, and the green dotted line indicates the V I line at 6708.07\AA; all of these are shifted to match a radial velocity of 27 km s$^{-1}$. 
The emission feature in quiescence is likely due to the high-velocity component of [S II]. 
}
\label{li-fig}%
\end{figure}

The Li I resonance line, which is characteristic of
young T Tauri stars \citep{white03}, is also detected  (Figure \ref{li-fig}), albeit mostly in emission, as observed for 
V1118 Ori in its 2005 outburst \citep{herbig08}.  In outburst, the line has a profile that is very similar to
that of EX Lupi in outburst with broad line wings in emission and
a
shallow absorption at the stellar rest velocity \citep[][]{sicilia12}; this could suggest a similar origin. 
It is nevertheless very hard to produce Li I emission under normal conditions and abundances \citep[e.g.,][]{shore20}, so we 
explored other possibilities. 
The profile is clearly different from disk- or box-like profiles, which
have distinct peaks or, for box-like lines, a central 
peak instead of a central absorption. As in the case of EX Lupi, 
Li I is the only line with this type of profile. There is potential contamination by nearby Fe I and Fe II lines and 
by the high-velocity component of the [S II] line at 6717\AA, which is likely
the cause of the emission feature observed in quiescence, although the wavelength of the weak
absorption feature is more consistent with Li I. It is unlikely that [S II] is responsible for the feature observed at 6708\AA\ during outburst
because the high-velocity  component of the 6731\AA\ [S II] line is negligible. 
The Fe II line is clearly more energetic than the typical Fe II emission observed, and thus unlikely to contribute.
The Fe I line is only slightly more energetic than observed Fe I emission lines, so a contribution is hard to rule out, 
although it would be blueshifted. In addition, a V I line could be also present at 6708.07\AA\ and 
has similar energetics to the V I line observed at 6643.786\AA,\ which has a disk-like
profile in outburst and an emission profile in quiescence. Nevertheless, the lack of consistency between the 
profiles of the tentative V I identifications makes the association uncertain.
Li I absorption is usually not seen in stars earlier than F0 \citep{zappala72},
which could hint at an origin in very hot and dense circumstellar
material or inner hot disk atmosphere for the small absorption feature.

Finally, a handful of absorption lines are observed, mostly in quiescence, but some of these lines are also observed
in outburst (e.g., 5781, 5798  6614\AA), which are likely diffuse interstellar bands \citep[DIBs;][]{herbig95,hobbs08}.
There is no clear identification of photospheric absorption lines in outburst nor quiescence.

\section{Analysis \label{analysis}}
 
In this section, we constrain the properties of the various components of the
star-disk system using the  emission and absorption lines. 
First, we revise the
stellar parameters and accretion in Section \ref{stellar-sect}, followed by a discussion of the wind components in Section \ref{wind-sect}. 
Then, we use several methods to extract information about the
physical conditions and velocities from the observed emission lines. 
The details of the methods are given in Appendix \ref{methods-app}.
There are many uncertainties in the structure of a
complex object such as Z~CMa~NW, so we use various techniques that work under different assumptions. These include Saha's equation, which
assumes local thermodynamic equilibrium (LTE), and ratios from lines from the same upper 
level, which do not depend on ionization equilibrium \citep[][]{beristain98}. 
We finally use velocity brightness decomposition for disk-like profiles \citep{acke06}
to introduce further constraints derived from the line profile on the velocity, temperature, and density 
structure of the system, as well as to reveal
where the observations depart from the simplified models. 
Blended lines and those that may belong to several species and/or transitions
are excluded from the analysis. 
In addition, non-LTE effects and line pumping may affect line
ratios; for instance,
due to strong UV emission lines directly feeding the upper level of a line. Lacking UV data, but knowing
that lines in the UV, as in the optical, may be very broad, we used the exhaustive
line list from \citet{herczeg05}
for the young star RU Lupi to exclude any line that could
be pumped by UV lines with $\pm$1000 km s$^{-1}$ wings.

\subsection{Constraining the stellar properties and accretion \label{stellar-sect}}

\begin{figure}
\centering
\includegraphics[width=9cm]{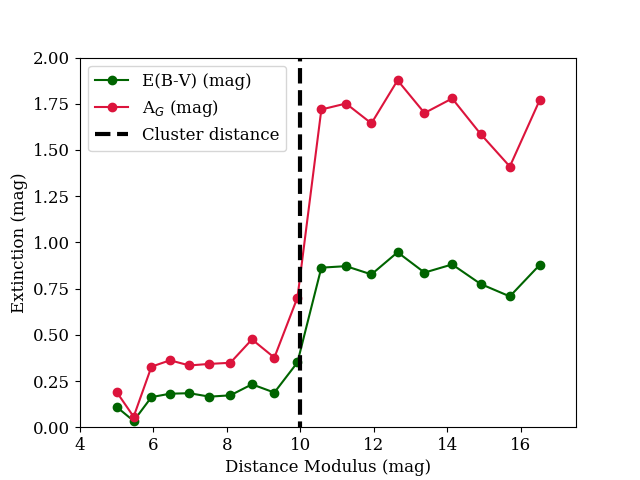} 
\caption{ Extinction in the Gaia G band and Gaia BP-RP around a 0.2 degree radius field toward Z~CMa. 
The rise of extinction is fully consistent with the
location of the Z CMa cloud at 1 kpc.}
\label{extinction-fig}%
\end{figure}

Because of the disagreements regarding the spectral type and quiescence luminosity of the star \citep{vandenancker04,monnier05},
our first step was to revise these assumptions.
First, we used Gaia DR2 data to revise the distance. Z~CMa itself is not a reliable Gaia source because it is too bright, is a binary, and
is surrounded by nebulosity. Nevertheless, a rise in extinction at a certain distance can be used to identify the location of its 
cloud \citep{green15,sicilia17}. Examining the Gaia DR2 stellar extinction for stars 0.2 degrees around Z~CMa reveals
a clear rise around 1~kpc (Figure \ref{extinction-fig}), confirming the previously assumed distance to the Z~CMa cloud, and 
thus the  stellar luminosity. 
For a  quiescence luminosity of 2400 L$_\odot$ \citep{thiebaut95} and assuming that the system has an age 
in the 0.3-3 Myr range, the CMD 3.3 stellar evolutionary 
tracks\footnote{http://stev.oapd.inaf.it/cmd} \citep{bressan12,marigo17,pastorelli19} 
suggest a mass between 6-8 M$_\odot$. The temperature would be consistent with a B-type star (16000-22000 K), albeit later than B0. 
The star could be colder ($\gtrapprox$10000 K) 
but similarly massive if it is younger than 0.3 Myr.
Assuming that half of the observed luminosity in quiescence is due to accretion, we still get a mass 
between 6 M$_\odot$ and 8 M$_\odot$ and a temperature between 13000 K and 20000 K (for ages 0.3-2 Myr) or 8000 K and 13000 K (for ages younger than 0.3 Myr). So while the object is definitely an intermediate-mass star,
the mass is likely lower and the spectral type is uncertain.

We thus adopted the quiescence luminosity of  L$_*$=2400 L$_\odot$ and give all results as
a function of stellar mass.
An effective temperature of 29200 K  (for a B0 star) results in a stellar radius of about 1.93 R$_\odot$ in quiescence,
but the radius would be larger if T$_{eff}$ is lower, especially if the star is very young. 
During outburst, the luminosity increased by a factor of 8. Assuming a blackbody-like scenario, the effective temperature for
this increased luminosity could reach 49100 K if the emission 
originated from the star. Nevertheless, we expect a range of temperatures for the star and innermost disk and accretion structures. 
Despite all these uncertainties, if we assume that the extra energy due to accretion during outburst 
is dissipated in the innermost regions (star and innermost disk), the temperature of the regions outside
this area depends on the total luminosity rather than on the actual temperature and radius of the star 
or the size of the emitting region. The temperature required for
silicate dust sublimation ($\sim$1500 K) is reached at $\sim$3.4 au during
quiescence \citep[consistent with the dusty ring at 4 au resolved by][]{monnier05}
and at $\sim$9.6 au in outburst. These distances are lower limits to the inner dusty rim
in the disk,  its radius could be larger if there is in situ disk heating. 
Large amount of silicates may have been vaporized in outburst. 
The temperatures at 1 au are 
$\sim$2800 K in quiescence and $\sim$4700 K in outburst. Orbital velocities are 120$\times (M_*/16M_\odot)^{1/2}$ km s$^{-1}$ at 1~au,
and 38$\times (M_*/16M_\odot)^{1/2}$  km s$^{-1}$ at the outburst dust sublimation radius (9.6 au).

The accretion rate (\.{M}) can be estimated from the increase in luminosity, 
as the luminosity in outburst is significantly higher than the
quiescence luminosity of the object. For 
T Tauri stars with magnetospheric accretion, 
the accretion luminosity is
\begin{equation}
L_{acc} = \frac{G M_* \dot{M}}{R_*} \bigg(1 - \frac{R_*}{R_{in}}\bigg), \label{magnacc-eq}
\end{equation}
where M$_*$ and R$_*$ are the stellar mass and radius, G is the gravitational constant,
and R$_{in}$ is the radius from which the material is accreted \citep[usually, the
magnetospheric or corotation radius in T Tauri stars; e.g.,][]{gullbring98,sicilia10,fairlamb15}. 
We assume that the infall factor $(1-R_*/R_{in})$ is about 0.5, which can be attained from material infalling from a 
minimum distance of 2 stellar radii and is also equivalent to the result
for a boundary layer 
\citep[][]{bertout88}. We
obtain \.{M}$\sim$10$^{-4}$ $\times (16 M_\odot/M_*)$ M$_\odot$/yr, which is similar to the accretion rates of
FUors \citep[][]{hartmann96}. In general, B stars are  fast rotators \citep{abt02,mueller11},
which results in small, if any, magnetospheres \citep{tambovtseva16}. A magnetosphere
extending to several times the stellar radius would result in a lower accretion rate, while detailed
boundary layer fits to B stars tend to produce higher accretion estimates \citep{wichittanakom20}.

The feasibility of free-falling material feeding accretion can be also 
estimated looking at the free-fall velocity, $v_{ff}$,
\begin{equation}
v_{ff} = \bigg(\frac{2 G M_*}{R_*}\bigg)^{1/2} \bigg(1 - \frac{R_*}{R_{in}}\bigg)^{1/2}. \label{freefallv-eq}
\end{equation} 
For Z~CMa and an infall factor of 0.5, we obtain a free-fall velocity $\sim$1200$\times (M_*/16M_\odot)^{1/2}$ km s$^{-1}$. This is
similar to the velocities observed in the H$\alpha$ line wings (up to $\sim$ 1000 km s$^{-1}$), but
H$\alpha$ can be strongly affected by Stark broadening.
The strongest metallic lines  have maximum velocities of the range of 500-600 km s$^{-1}$
and  any redshifted absorption features appear at lower velocities (see Section \ref{findlines}). 
Therefore, whether velocities can be explained by free-fall
needs further discussion after other possibilities have been considered (Section \ref{discussion}).

Some initial constraints on the density can be derived from the estimated accretion 
rate. Assuming that accreted material moves in 
at a typical velocity $v$
over one or more accretion
channels that cover a fraction $f$ of the stellar surface, we can write
\begin{equation}
\dot{M} = 4\pi R_*^2 f v \rho = 4\pi R_*^2 f v \mu m_H n,       \label{mdotdensity-eq}
\end{equation}
where $\rho$ is the density of the accreting material, which can be written in terms of the
mean atomic weight ($\mu$=1.36 is a typical value), 
the mass of the hydrogen atom ($m_H$), and the number density \citep[$n$; e.g.,][]{sicilia12}. Taking a
typical velocity $v$=500 km s$^{-1}$, we obtain a density $n\sim$1.6e14$/f$ cm$^{-3}$. For any
small value of $f$, these densities are about two orders of magnitude larger
than observed in outbursting T Tauri stars. Nevertheless, if the material is brought onto the star via an extended
structure involving part of the innermost disk, we could have $f>1$ and 
lower densities, as found, for instance, for ASASSN-13db \citep[][but note that ASASSN-13db has 
magnetospheric accretion]{sicilia17}. 

Temperature estimates also suggest that the energy is deposited over a region larger than the
stellar surface. If the accretion luminosity was released in a stellar spot at the stellar surface, the temperature of the spot could 
be unphysical, reaching (47e3/$\sqrt{f}$) K, where $f$ is the fraction of the stellar surface covered by the spot. Assuming a boundary layer scenario, 
the boundary layer temperature $T_{BL}$ can be written  as
\begin{equation}
T_{BL} = \bigg(\frac{L_{acc}}{4 \pi R_* \delta \sigma}\bigg)^{1/4} \label{tbl-eq},
\end{equation}
where $\sigma$ is Stefan's constant, $\delta$ is the scale of the boundary layer, and the rest of symbols retain their previous meanings
\citep{popham93,blondel06,mendigutia20}. The
value of $\delta$ is calculated as
\begin{equation}
\delta = \delta_0 R_* \bigg[1-\bigg(\frac{\Omega_*}{\Omega_K}\bigg)^2\bigg]^{-1/3} \label{delta-eq}
,\end{equation}
where $\delta_0$ is a scale parameter, with typical values between 0.01-0.5 \citep{popham93,blondel06}; $\Omega_*$ is the angular velocity of the star;
and $\Omega_K$ is the Keplerian angular velocity. For a rotation rate of 25\% of the breakup velocity \citep[typical for B stars;][]{abt02},
$\delta_0$=0.5 would still result in T$_{BL} \sim$56,000 K, therefore the requirement of 
a structure that extends over several stellar radii remains.  This is also consistent with the very extended boundary layers, 
often hard to distinguish from the disk,  expected in objects with very high accretion rates \citep{popham93}.

\subsection{Wind variability in outburst and quiescence\label{wind-sect}}

\begin{figure*}
\centering
\begin{tabular}{ccc}
\includegraphics[width=5.9cm]{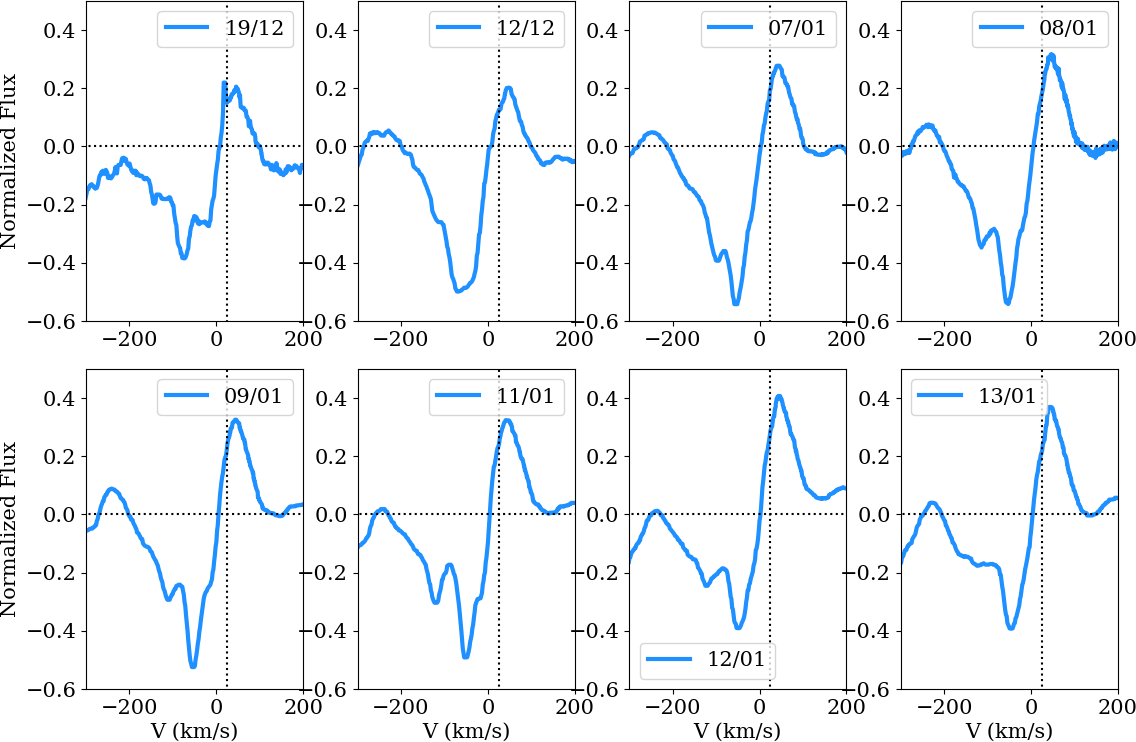} &
\includegraphics[width=5.9cm]{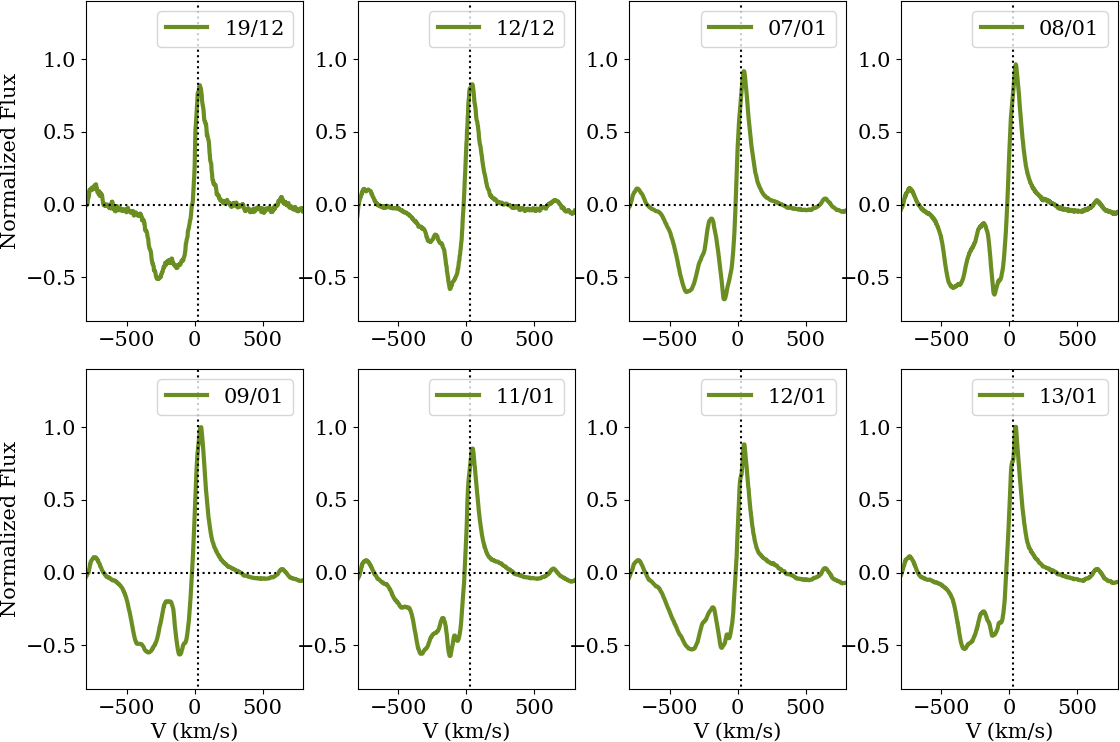} &
\includegraphics[width=5.9cm]{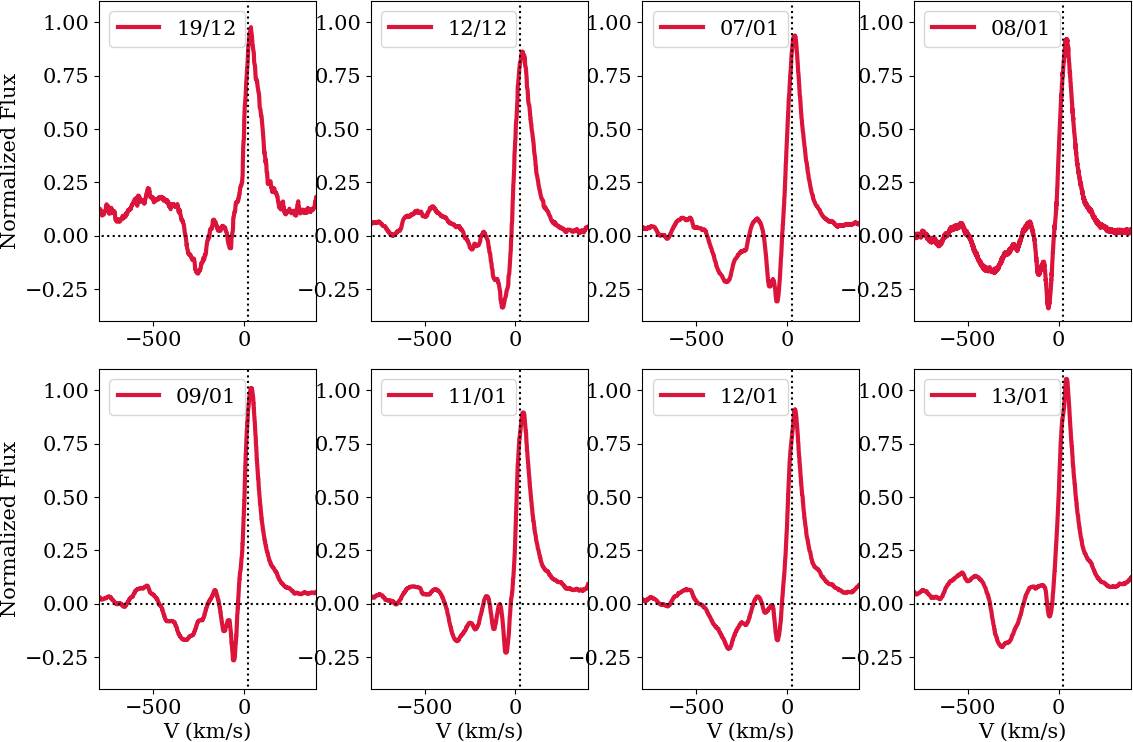} \\
\end{tabular}
\caption{Examples of the time variability observed for the Ti II 4468\AA\ line (left, blue), the Fe II 4923\AA\ line
(middle, green), and the
Fe II 5316\AA\ line (right, red). In each case, the 8 subfigures display the individual spectra as
they were obtained in time. The dotted lines indicate the continuum level and the radial velocity of the source. }
\label{Windtime-fig}%
\end{figure*}

Variable and complex, strong winds are typical of quiescent early-type stars \citep{prinja86}.
The many emission lines with P Cygni profiles are a sign of a
powerful wind that  gets stronger and more complex during outburst, which is typical of accretion outbursts and similar to what 
is observed in previous events \citep{covino84,vandenancker04}. 
At least three absorption components (at $-$900, $-$500 and $-$100 km s$^{-1}$)
were reported by \citet{vandenancker04} during the 1999 outburst; the blueshifted
absorption extended up to $-$1000 km s$^{-1}$ in the Balmer H series.
According to \citet{antoniucci16}, the wind from Z~CMa NW is at least 20 times 
stronger than the FUor wind for velocities $-$90 to $-$400 km s$^{-1}$. This is confirmed 
by our quiescence spectra, which show
much weaker absorption components despite having a higher proportional contribution 
from the FUor, and justifies our approach of treating the observed outburst lines as coming
from Z~CMa NW\footnote{ The jet of Z~CMa SE has velocities
similar to those of the intermediate and slow wind components, but if we assume that the SE component
dominates in quiescence,  it may contribute more in emission than in absorption, because
the wind features are much weaker during quiescence.}. The analysis of the quiescence wind is thus limited to
pointing out the differences with outburst, since the quiescence data have an unknown contribution of Z CMa SE.

In our 2008/2009 data,
the absorptions in the strongest Balmer lines are highly saturated 
and the wind is stronger than in the past, consistent with the 2 mag rise experienced in the 2009 outburst
compared, for instance,  to the 1 mag rise in 1999. The
wind affects preferentially the high-energy lines, which is a sign of high
wind temperature. The absorption features have complex and highly variable, high-velocity profiles,
up to $-$800 km s$^{-1}$ in certain lines. 
The velocities of the fast and intermediate-velocity winds  ($-$100 to $-$500 km s$^{-1}$) are 
similar to escape velocities in the innermost regions
of the system ($<$0.1-0.3 au), which is consistent with stellar winds; these winds are associated with innermost
accretion structures or initially poorly collimated launching regions that become the large-scale
collimated jet.  Since hotter temperatures
are more likely to originate closer to the star, the high-velocity wind is the best candidate
for a stellar or innermost-disk wind, while the low-velocity
wind could originate in a disk wind.

Inner disk winds can explain the deep and broad absorptions observed in FUors
\citep{milliner18}, which are similar to what is observed for Z~CMa in the higher energy lines. 
He I and Si II are peaked toward the fastest wind 
components at $>-$400 km s$^{-1}$
(see Figure \ref{lines-fig}), indicating a very high temperature. 
He I and energetic lines may come from a more direct view (hotter, denser, saturated) than the
Fe II and Ti II lines, which can also explain the differences in wind velocity and
maximum redshifted velocities in the emission part of the PCygni profiles.  
The O I 8446\AA\ line  shows fast and slow components similar to Fe II
lines,  such as the Fe II 5018/4923 \AA\ multiplet, 
maybe due to pumping by UV lines \citep[e.g., H$_2$ C-X0-2Q(10) line;][]{herczeg05}. 
The absorption components observed in the optical lines are also in agreement with
the large-scale jet observed around Z~CMa NW, which have velocities from $-$450 to $-$600 km s$^{-1}$ \citep{whelan10, antoniucci16}.
Ti II lines, which typically trace material at lower temperatures and densities, 
are dominated by the intermediate and slow wind.

The redshifted emission peaks of most strong ionized lines
also suggest a slower wind around $-$20 km s$^{-1}$. 
The velocity of the slow wind is comparable to the escape velocity 
at $\sim$10-30 au, which is beyond the disk region detected in emission lines,
but the orbital period at this distance is too long to expect significant day-to-day modulations,
unlike what is observed.

Variations in the velocity of the absorption
components are observed in all lines with non-saturated PC profiles, including those that have only 
a low-velocity absorption component, so that
all wind components are variable.
The day-to-day variations of the wind components could be caused by real variability, rotation, and/or small-scale occultation
events in the innermost regions.   Rapid  
variability in a high-velocity wind component has been also observed for ASASSN-13db
in outburst, caused by rotation of an accretion-powered, non-axisymmetric wind \citep{sicilia17}. 
A difference in inclination angle can cause a variation in the
absorption versus emission parts of the wind \citep[e.g.,][]{milliner18}, 
therefore we may observe material at multiple inclinations depending on the temperature structure.

\begin{figure}
\centering
\includegraphics[width=8.5cm]{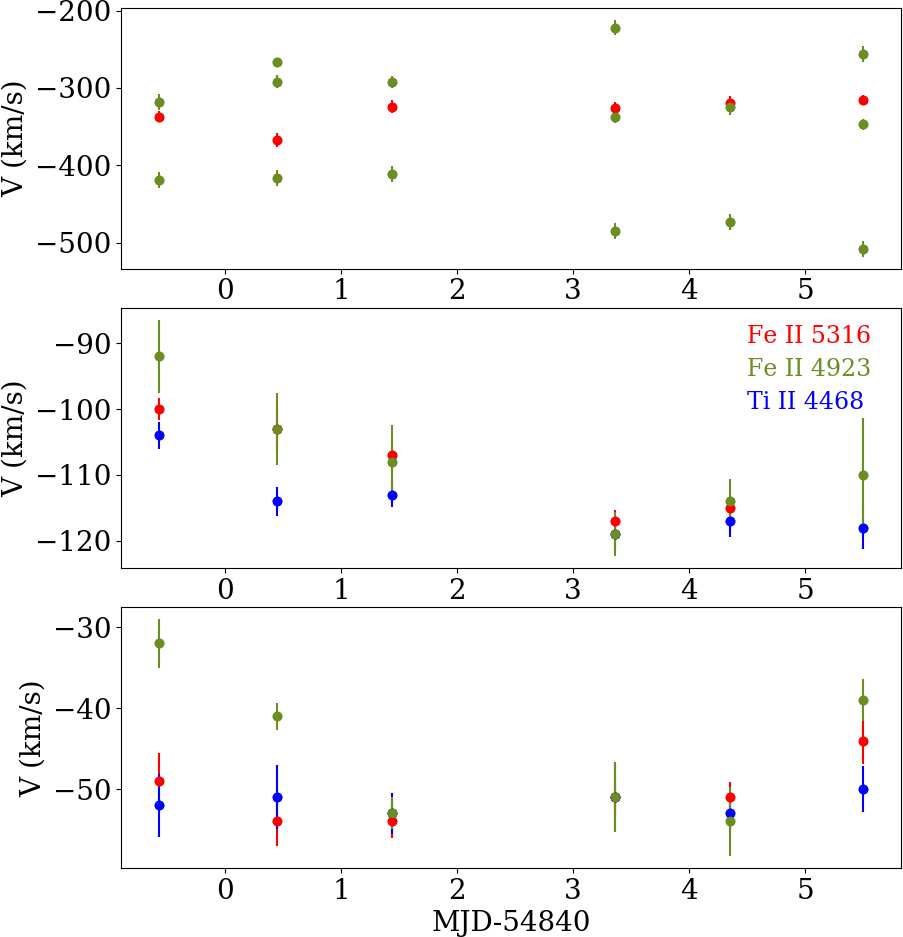} 
\caption{Day-to-day velocity variation of the wind components in 2009 January,
classified according to the central velocities as very fast (top, observed clearly only for
Fe II), fast (middle), and slow (lower). Results for the Fe II 5316 (red), Fe II 4923 (green), and Ti II 4468\AA\ (blue) lines, which have good S/N
and no signs of contaminations, are shown. The velocities are given as measured, without correcting for the radial velocity of the system. 
  The correlations between the components observed
for both intermediate velocity and slow wind lines, despite some shifts in velocity, are shown. The fast wind component appears to split into 
several components for some dates for the Fe II 4923\AA.}
\label{Windtimeline-fig}%
\end{figure}

The variability of the wind shows a clear time evolution.
This is particularly evident for the six spectra taken on
consecutive nights between 2009 January 7 and 13 (Figure \ref{Windtime-fig}). 
To examine these spectra, the various absorption components were locally fitted by
Gaussians, considering their centers as the representative velocity of each individual wind
component. This fit is a simple approximation,
not taking into account other effects (e.g., optical depth, radiative transfer).
Figure \ref{Windtimeline-fig} shows that the Fe II and Ti II absorption profiles vary in parallel, even if their central velocities
differ, which may arise from different line opacities \citep[][]{sicilia15}.
If we consider the  Fe II 5316\AA\ and Ti II 4468\AA\ lines as examples,
the Fe II very fast, broad wind component (with velocities
between $-$300/$-$380 km s$^{-1}$) is the most variable and does not display a well-defined trend (besides
not being observed in the lower energy Ti II lines). The intermediate-velocity wind 
component at $-$95/$-$120 km s$^{-1}$ and the slower wind component at $-$47/$-$55 km s$^{-1}$ are correlated
for both lines (correlation coefficient 0.90, with false-alarm-probability 0.04 using a Spearman
rank test) and vary smoothly with time, revealing the development and evolution of 
each individual component during the six observations.

\begin{table*}
\caption{Lines selected for the Saha analysis, including the NIST data quality flag.}             
\label{saha-lines}      
\centering                                     
\begin{tabular}{l c c c c c c c l}       
\hline\hline                        
Species & $\lambda$ & A$_{ki}$ & g$_k$ & g$_i$ & E$_i$ & E$_k$ &  NIST & Comments \\
 & (\AA) & (s$^{-1}$) &   &  & (eV) & (eV) & flag &  \\
\hline                                  
FeI & 5216.27 & 3.47E+05 & 5 & 5 & 1.608 & 3.984 & A & \\
FeI & 5455.61 & 6.05E+05 & 3 & 3 & 1.011 & 3.802 & B+ & \\
FeI & 5506.78 & 5.01E+04 & 7 & 5 & 0.990 & 3.241 & A & Some odd results \\
FeII & 5627.50 & 2.90E+03 & 6 & 8 & 3.387 & 5.589& C & Potential blend \\
FeII & 6084.10 & 3.00E+03 & 8 & 10 & 3.199 & 5.237 & E & \\
FeI & 6256.36 & 7.40E+04 & 9 & 9 & 2.453 & 4.435 & C & \\
FeI & 6265.13 & 6.84E+04 & 7 & 7 & 2.176 & 4.154 & B & \\
FeI & 6358.63 & 5.19E+05 & 7 & 7 & 4.143 & 6.092 & C+ & Blend likely\\
FeI & 6393.60 & 4.81E+05 & 9 & 11 & 2.433 & 4.371 & C & \\
FeI & 6400.00 & 9.27E+06 & 9 & 7 & 3.603 & 5.539 & B & \\
FeI & 6421.35 & 3.04E+05 & 5 & 5 & 2.279 & 4.209 &  B & \\
FeI & 6462.73 & 5.60E+04 & 9 & 9 & 2.453 & 4.371 & C & \\
FeI & 6750.15 & 1.17E+05 & 3 & 3 & 2.424 & 4.260 & B & \\
\hline                                             
\end{tabular}
\end{table*}

Two scenarios are possible. The first would be rotational modulation of the wind 
(either in a stellar or disk wind), as it has seen in 
the multiple system GW Ori \citep{fang14} and in the outbursting star FU Orionis \citep{powell12}
and ASASSN-13db \citep{sicilia17}. The wind modulation in GW Ori is caused by a companion,  but
any process resulting in a non-axisymmetric
wind associated with either the star or the disk can produce the desired effect. 
The wind modulation in FU Orionis is periodic, which cannot be confirmed in this work
due to the few spectra available. 
The second scenario would result from the propagation
of a clump of matter along the wind. The radial velocities could vary due to changes in
their angle with respect to the line of sight, as  also occurs in the rotation scenario, or to
changes in the physical velocity (e.g., in an accelerating wind). Both scenarios produce 
a smooth velocity evolution (as it is observed), but an accelerated wind would move toward larger
velocities, while rotation would result in changes in the angle with respect to the line of sight.
The variability ranges proportional to their velocity; the fast 
component changes by $\sim$25 km s$^{-1}$ and the slow wind
component changes by $\sim$8 km s$^{-1}$ during these six day (see Figure \ref{Windtimeline-fig}).
This is consistent with rotation and projection effects as observed in  rapidly variable, winding winds in certain B-type stars
\citep{prinja86,prinja88}, where a  rotating/spiraling, non-axisymmetric  clumpy wind, moving not only radially, but also azimuthally,
causes differences in opacity along
the line of sight and thus rapid velocity variations. 
The trend extends to the very fast wind component observed in Fe II, 
although the error in its central velocity is large because the absorption features are very broad. 
Testing whether the system shows any periodic or
quasi-periodic behavior in the wind velocity and strength can help to distinguish these scenarios.

The conclusions of this section are that Z~CMa NW develops a hot, fast wind in outburst, which is likely 
of stellar or innermost disk origin, as well as colder components that could be related to the disk. 
The wind is variable on timescales of days, and the intermediate and strong, slow wind components are clearly correlated,
betraying a small-scale structure and a clumpy nature in a rotating, non-axisymmetric wind.

\subsection{Saha's  equation constraints on temperature and density in outburst and quiescence \label{saha-sect}}

We can use the observed emission lines to put constraints on the density and temperature of the
emitting material, assuming LTE. We use Saha's and Boltzmann's equations (Appendix \ref{equations}) to derive expected
intensities for a range of temperatures and densities and compare these with the observations.
For velocity-resolved lines, it is possible 
to distinguish various gaseous structures considering their velocities. For
Z~CMa NW, where different
lines are related to different physical structures (e.g. winds,  hot inner disk, 
 accretion structures), it is crucial to classify the lines
according to their profiles. Most Fe II lines have P Cygni profiles, while Fe I lines have disk- or
box-like profiles, a sign that they are not produced by the same  
structure and cannot be treated together.
Many lines are saturated and probably originate
in optically thick regions, so a proper fit would require full radiative transfer.
As a result, the only cases for which we can apply this method are
weak Fe I and  Fe II lines with similar box- or disk-like profiles (see Table \ref{saha-lines}). There are only two clean Fe II
lines with box-like profiles, $\lambda \lambda$ 5627.497, 6084.103\AA. These lines have transition probabilities on the order of 1e3 s$^{-1}$ and are very 
weak\footnote{This 
weakness is probably the reason why the lines do not have wind signatures (self-absorptions 
and P Cygni profiles) as most Fe II lines do.}.
Among the clean Fe I lines, only
four have low enough transition probabilities for the assumption
of optically thin lines to be justified.

\begin{figure*}
\centering
\begin{tabular}{c}
\includegraphics[width=15cm]{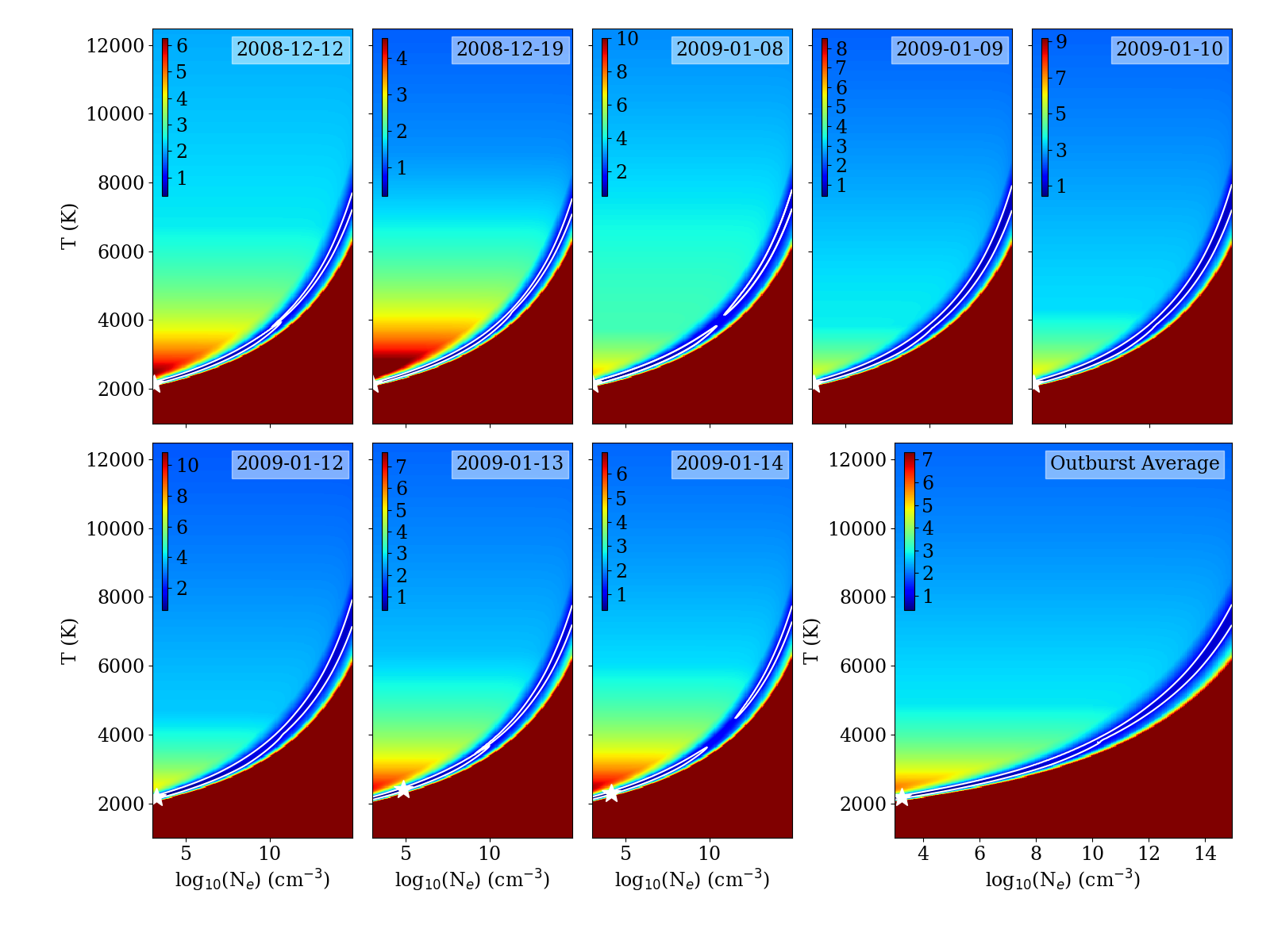}\\
\includegraphics[width=15cm]{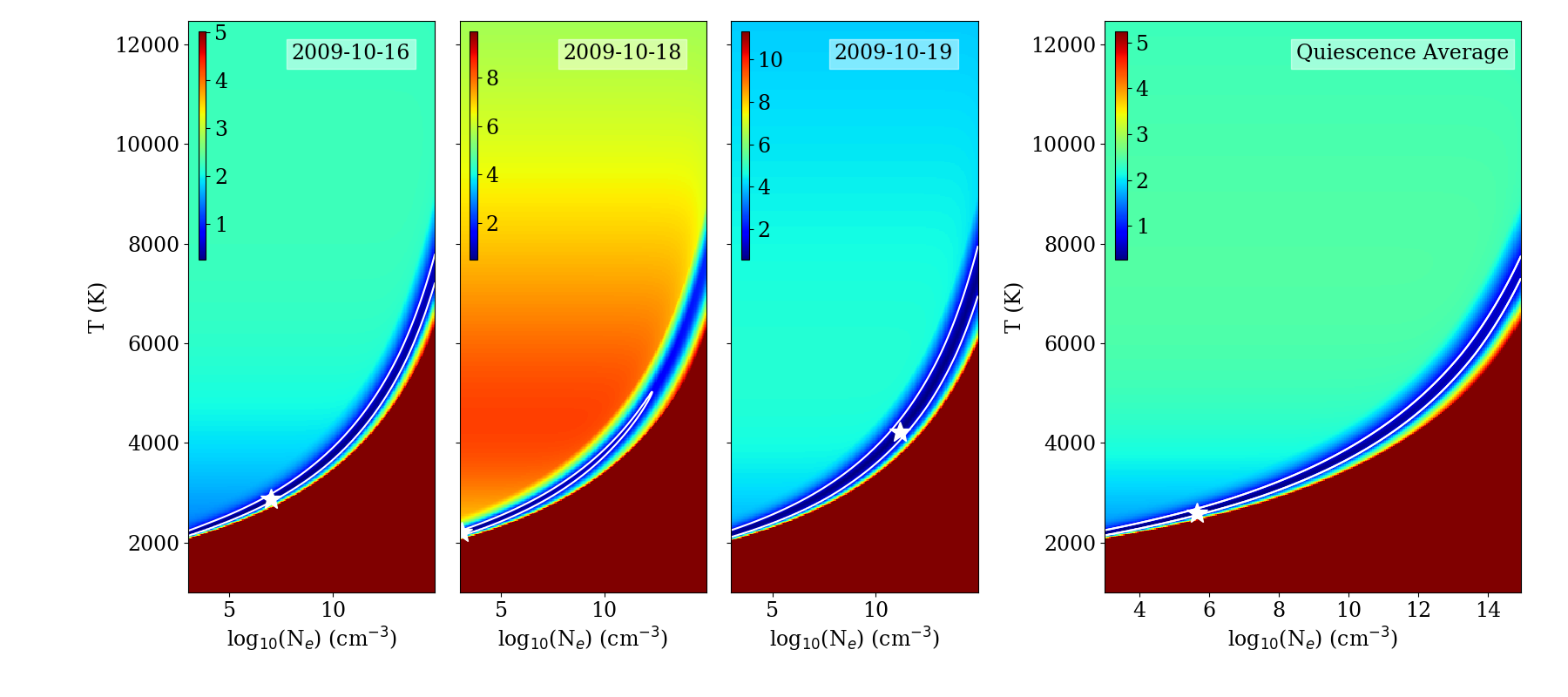} \\
\end{tabular} 
\caption{Result of Saha's equation analysis for the lines with disk-like profiles
and low transition probabilities for the observations taken during the outburst phase (upper panels)
and for the same lines as observed during quiescence (lower row).
The panels show the data corresponding to individual days in chronological order; the last panel shows the result of fitting the average spectrum. 
The color scheme reveals the relative variations in 
the line intensity ratio for each combination of temperature and electron density (such that
0.1 would correspond to 10\% average deviations). The white contour indicates the regions for which
the relative variations of the line ratio are up to 3 times the minimum value observed, which is considered as our best-fit region; a white star
denotes the best fit. }
\label{saha-fig}%
\end{figure*}

The lines were extracted from the continuum and scaled to the same arbitrary-unit scale. 
Considering a reasonable range of temperatures and electron densities, we can estimate
the relative line intensity expected depending on the physical conditions. 
Leaving the total density of Fe I as an unknown, the relative strength of Fe II lines with respect to
Fe I lines is given by Saha's equation (Eq. \ref{saha-eq}). Then, for each transition we
consider the atomic parameters so that the total line strength is given by the transition probability 
(A$_{ki}$) divided by the wavelength and multiplied by the Boltzmann distribution (Eq. \ref{boltzmann-eq}).
We estimated theoretical line ratios for temperatures from 1000 K to 12500 K and electron densities
10$^3$-10$^{15}$ cm$^{-3}$. Saha's equation has an intrinsic 
degeneracy between the electron density and the temperature, which means that
higher temperatures with higher electron densities may reproduce the same line ratios, 
but we add further constraints in the following sections.
The spectra are not flux-calibrated, and the theoretical intensity 
is calculated as a ratio of the intensities depending on the ratio of ionized to neutral elements, so
the lack of scale invalidates the $\chi^2$ statistics \citep[e.g.,][]{andrae10}. Therefore, 
the best-fitting density is obtained by minimizing the standard deviation in the ratio of the observed
to theoretical line intensities, taking the median of the ratio of the observed to theoretical 
lines as a scaling factor.  The contribution of each line is weighted taking into
account the uncertainties in its atomic parameters using the NIST data quality flag. 
To assess the uncertainties in the results, the exercise was repeated 
for each one of the eight outburst observations and for the average outburst spectrum. There are no significant differences
between dates, thus the variations are due to noise.

The exercise can be repeated for the  lines
observed in quiescence, remembering that there is no evidence that quiescence lines originate in the disk.
The data from 2009-10-18 suggest higher
temperatures and densities than for the other quiescence spectra, which could be related to an increased accretion rate in either of the
two Z~CMa components. Although the H$\alpha$ emission line is strong on that particular
day, the noise level is also higher and
there are no other significant differences  between all three quiescence spectra. Therefore, random noise cannot be fully excluded and the results from the averaged quiescence
spectra may be more representative of the typical quiescence state.

\begin{table*}
\caption{Results from the Fe I/FeII Saha line analysis.}              
\label{saha-table}     
\centering                                     
\begin{tabular}{l c c c c c l}       
\hline\hline                        
Date & T$_{best}$ & T$_{median}$ &  T$_{range}$ & n$_{e,best}$ & n$_{e,range}$ & Notes \\
        &  (K)    & (K) & (K) & (cm$^{-3}$) & (cm$^{-3}$) &  \\
\hline  
2008-12-12  & 2400 & 4050 & 2125-7725 & 7.1e4 &  1e3-1e15 & \\
2008-12-19  & 2300 & 4260 & 2125-7700 & 1.4e4 &  1e3-1e15 & \\
2009-01-08  & 2150 & 4600 & 2125-7775 & 1.3e3 & 1e3-1e15 & \\
2009-01-09  & 2150  & 4600 & 2125-7900 & 1e3 & 1e3-1e15 & \\
2009-01-10  & 2150 & 4600 & 2125-7925 & 1.3e3 &  1e3-1e15 & \\
2009-01-12  & 2150 & 4525 &  2100-7925 & 1.3e3 &  1e3-1e15 & \\
2009-01-13  & 2150 & 4625 & 2125-7750 & 1.3e3 &  1e3-1e15 & \\
2009-01-14  & 2175  & 4800  & 2125-7725 & 1.8e3 &  1e3-1e15 & \\
Outburst Average & 2175 & 4525 & 2125-775  & 1.8e3 & 1e3-1e15  & \\                             
2009-10-16 & 2875 & 4560 & 2150-7775 & 1.0e7 &  1e3-1e15  & \\
2009-10-18 & 2200 & 3075 & 2150-5025 & 1.3e3 &  1e3-2e12  & \\
2009-10-19 & 4225 & 4550 & 2125-7950 & 1.4e11 & 1e3-1e15  & Anomalous values\\ 
Quiescence Average &  2600 & 4250 & 2150-7750 & 4.5e5 & 1e3-1e15  & \\
\hline                                             
\end{tabular}
\tablefoot{Range corresponds to the span of the fits for which
the relative variations of the line ratio are up to 3 times the minimum value observed.}
\end{table*}

The explored range of  electron densities and
temperatures and the best-fitting values are shown in Figure \ref{saha-fig}. 
One warning regarding the interpretation of these results is that the deviations from the best-fit models for
different groups of lines are not random. While
neutral lines tend to favor a
relatively low temperature (on the order of 2000-3000 K), to reproduce the ionized lines at the same level
we tend to require higher temperatures and densities ($\sim$7600 K) and   the best fit tends to the
lower temperature values because the atomic parameters of the neutral lines have smaller uncertainties. There are no perfect fits
and, in particular, the FeI line at 5506\AA\ and the FeII line at 6084\AA,\ which has 
a particularly large uncertainty in its atomic parameters, are usually
badly fitted (within a factor of $\sim$10) by any combination of temperatures and electron
densities that reproduces the remaining lines. As observed in EX Lupi \citep{sicilia15}, this may indicate 
multiple components with various densities and temperatures, although for Z~CMa there are too few weak Fe II lines to
attempt a more complex model fit. 
Despite the uncertainties and degeneracy, the best fits suggest a relatively low temperature for the formation of the observed disk-shaped lines.
The constraints on the density are weaker, although the best-fitting values tend toward the lower end as well
(see Figure \ref{saha-fig}). Although we cannot use the Ti I/Ti II
lines for this exercise because all the Ti II lines are heavily 
affected by the winds, we note that the
temperature and density ranges obtained from Fe I/Fe II are consistent
with the strongest Ti I lines having rounded or narrow
profiles. These lines are very sensitive to temperature, disappearing 
for temperatures more than\ 3000 K at the observed best-fit densities. The fact that no disk-like
profiles are observed for Ti I lines suggests that they may come from regions in the disk at larger radii (and thus colder)
than the Fe I lines. Therefore, this section adds an 
additional proof of the inner disk emission being extended, which agrees with the
range of temperatures required to explain both the Fe I and Fe II emission (from 2000-3000 K to 7000 K)
and with the variety of observed disk-like profiles. The narrow quiescence lines may be linked to the accretion structures in
quiescence.

\begin{table*}
\caption{Lines that share a common upper level, among those with disk-like profiles.}              
\label{linepairs-table}     
\centering                                     
\begin{tabular}{l c c c l}       
\hline\hline                        
Species & $\lambda_0$ & A$_{ji}$ & Uncertainty & Notes \\
        &  (\AA)    & (s$^{-1}$) & & \\
\hline                                  
FeI     & 4786.8, 5262.9 & 1.03E+06, 8.70E+04 & 50\% & Second one blended/uncertain \\
FeI     & 4839.5, 7511.0 & 3.90E+05, 1.35E+07   & 18\% & Uncertain continuum (7511), blends\\
FeI     & 4939.7:, 5051.6, 5142.9 & 1.39E+04, 4.65E+04, 2.40E+04  &  40\% & Blends\\
FeI     & 4957.6, 6271.3 & 4.22E+07, 3.05E+04 & 18\% & Blends\\
FeI     & 5141.7, 6270.2 & 4.86E+05,  1.39E+05 & 50\% & Blends \\
FeI     & 5393.2, 5615.6, 5709.4 & 4.91E+06, 2.64E+07, 2.13E+06  & 18\% & Blends \\
FeI     & 5543.9, 5641.4 & 3.40E+06,  3.00E+06 & 50\% & Strength too similar, uncertain\\
FeI     & 5572.8, 5624.5, 6408.0 & 2.28E+07, 7.41E+06, 3.12E+06 & 10\% & OK first two lines \\
FeI     & 5747.9, 6103.3 & 8.80E+05, 1.52E+06 & 50\% & Blends, uncertain\\
FeI     & 6191.6, 6256.4 & 7.41E+05, 7.40E+04 & 25\% & OK \\
FeI     & 6393.6, 6462.7 & 4.81E+05, 9.70E+04 & 25\% & OK \\
FeI     & 6265.1, 6335.3 & 6.84E+04, 1.58E+05 & 18\% &  Blends \\
FeI     & 6750.2, 8674.7, 8838.4 & 1.17E+05, 6.17E+05, 3.83E+05 & 25\% & Strength too similar, uncertain \\
\hline                                             
\end{tabular}
\tablefoot{Uncertainty levels are taken from the NIST database
and reflect the uncertainty in the atomic parameters for the most
uncertain line. For lines with errors $>$40\%, the theoretical line ratio
is very uncertain, especially in cases in which the transition probabilities
are not very different.}
\end{table*}

\subsection{Line ratio constraints on the disk and wind properties \label{lineratio-sect}}

Here, we follow \citet{beristain98}  to derive the optical depth 
of the line-emitting disk material using lines
that share the same upper level. This avoids having to deal with the way the upper level is populated, so that the
resulting relative intensities depend only on the atomic parameters, temperature, and line opacity. For velocity-resolved lines, 
we can apply the Sobolev approximation to write down the line opacity, which results in the line strength
depending only on the atomic parameters, temperature, and column density-velocity gradient ($n dv/dl$, where
$n$ is the density, $v$ is the velocity,
and $l$ is the spatial scale). The details of the procedure are given in Appendix \ref{equations}. 
We first apply this method to the disk-like emission lines, 
and then examine the results for the lines with P Cygni profiles.

\subsubsection{Line ratio constraints on the inner disk \label{disklineratio}}

 There are 13 groups of lines from the same upper level among the disk-like lines
(Table \ref{linepairs-table}). Nevertheless, only 3 groups
remain for which the lines are not blended, 
are close enough to minimize the effects of the continuum variations, 
and have high-quality NIST atomic data\footnote{If the uncertainties in
the NIST transition probabilities are larger than\ 50\%, they are unusable unless they
have very different strengths.}. A fourth pair, Fe I $\lambda\lambda$ 8674.7, 8838.4 \AA,
also satisfies the above conditions, but the presumably stronger line is observed to be weaker. This is
probably caused by the 25\% uncertainty in the transition probabilities
plus the fact that both lines have a very similar strength.
Considering the uncertainties, the best line pair
is Fe I 5572/5624~\AA\footnote{There is another Fe I line at 5573.1\AA,\, but its atomic parameters suggests that
this line would be much weaker than observed in this work, thus we expect little contamination.} (10\% uncertainty or class B, compared to
25\% uncertainty for the other two pairs). The smaller difference between the transition probabilities
for the Fe I 6393/6492 \AA\ pair makes it the least constrained. 
All three line pairs are also detected in quiescence, but the Fe I 5572/5624\AA\ pair  in quiescence is
too weak to be used.

\begin{figure*}
\centering
\begin{tabular}{cc}
Outburst & Quiescence \\
\includegraphics[width=8cm]{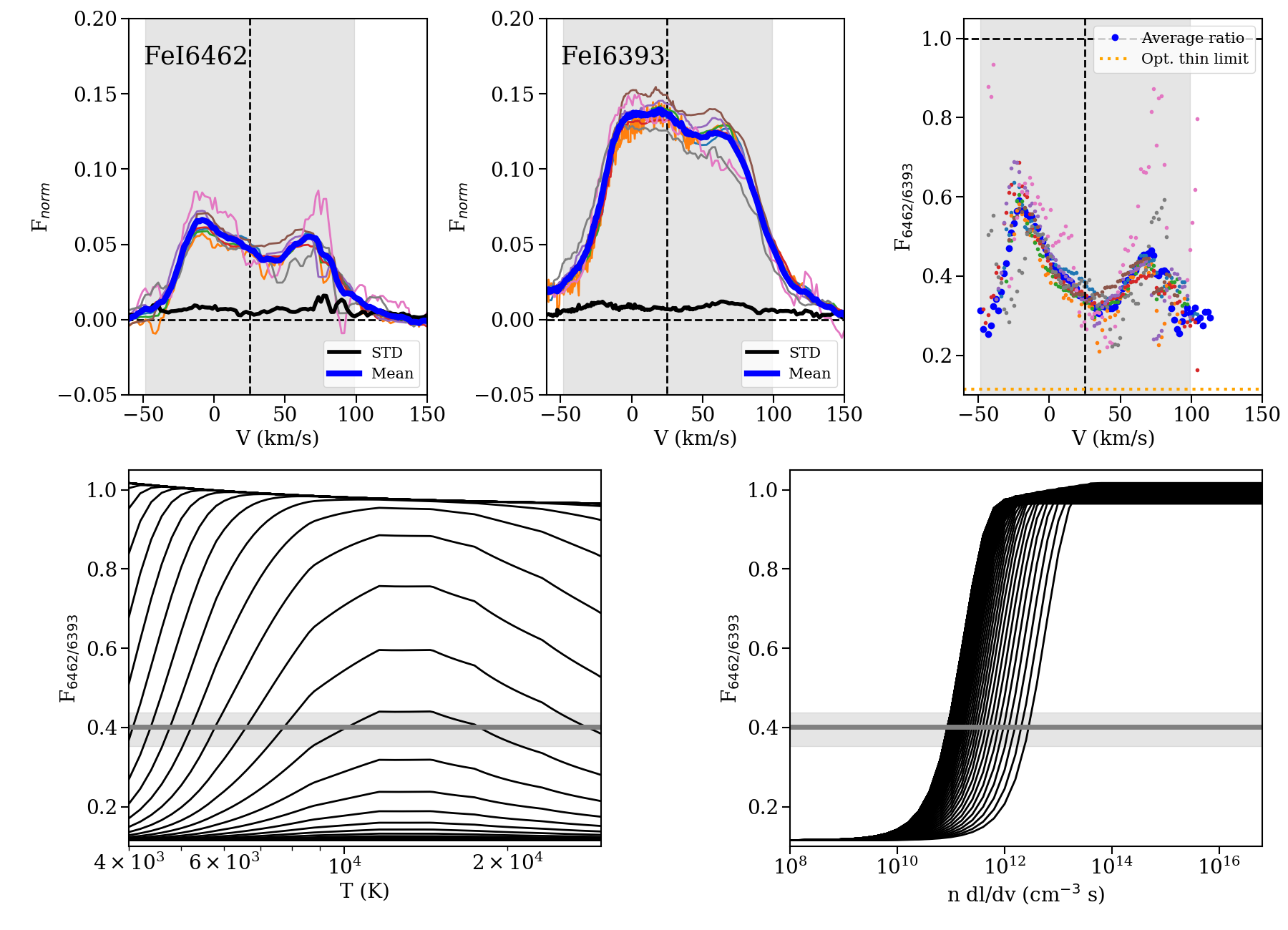} &
\includegraphics[width=8cm]{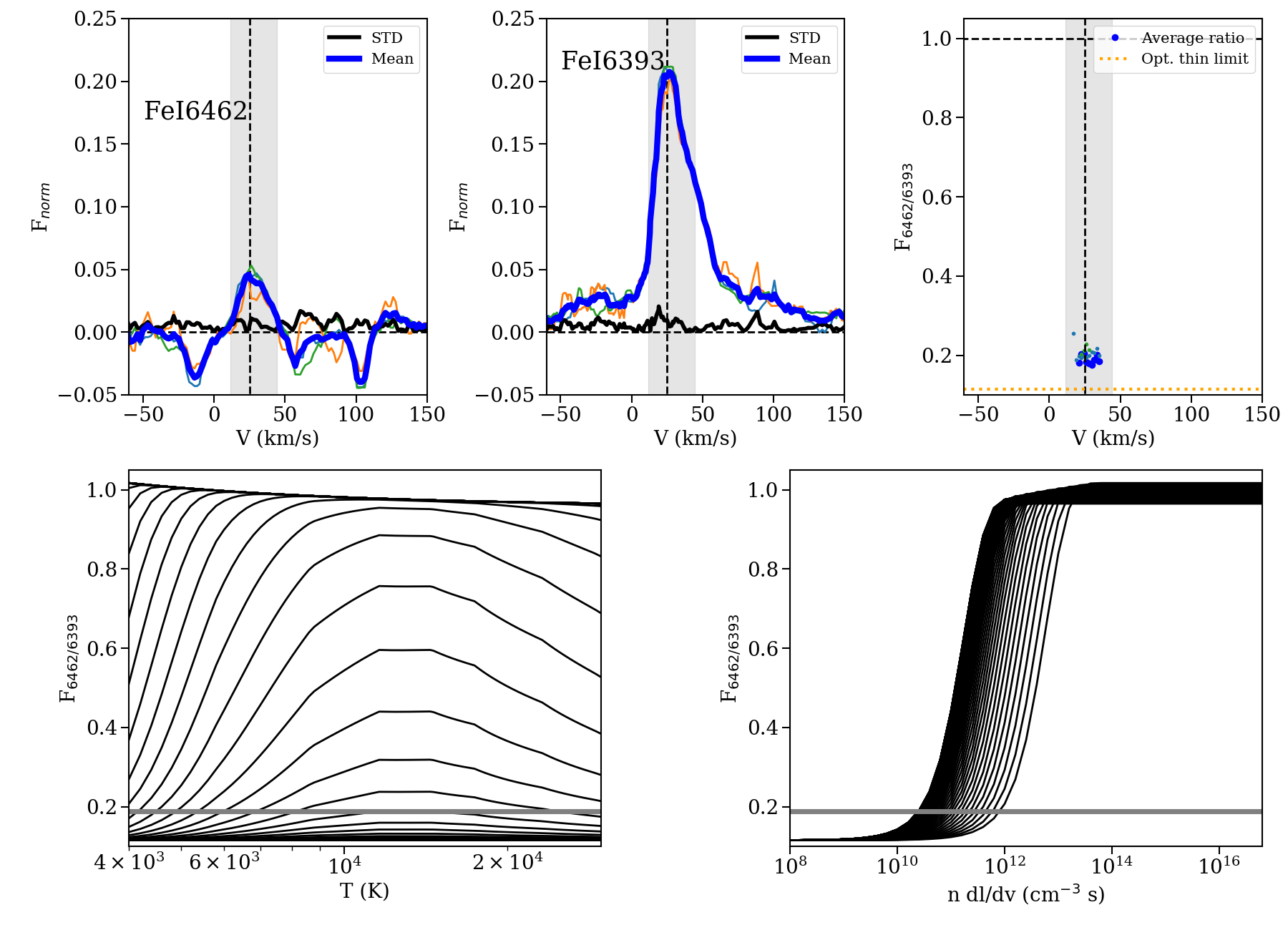} \\
\end{tabular}
\caption{Analysis of the line ratio for Fe I transitions with common
upper levels (continued in Appendix \ref{methods-app}). The upper left panels show the
line emission in outburst and the upper right panels show the quiescence data. 
The ratios observed  are compared to the theoretical calculations for a range of temperatures and column density-velocity gradients in the lower panels.
The gray area shows the region for which the line emission is $>$5$\sigma$
(in the velocity panels) and the 1$\sigma$ error in the temperature and density planes.
Quiescence lines are often very weak and thus uncertain.  }
\label{FeIratio-fig}%
\end{figure*}

To derive the ratios, we normalized every line individually.  First, the global shape of the spectrum
has to be taken into account, which for lines that have similar wavelengths
can be done assuming a linear fit between the two regions. We imposed a minimum of 5$\sigma$ emission
above the noise level
to obtain the ratio, where $\sigma$ corresponds to the standard deviation of the flux measured
in a line-free nearby region, to avoid unphysical values due to noise. 
We analyzed the outburst 
spectra individually, together with the
averaged spectrum, to estimate  the noise and variability. For comparison, we also
analyzed the quiescence data in the same manner, which is
discussed in Section \ref{quiescence-sect}.

The results are shown in Figure \ref{FeIratio-fig} and continued in Appendix \ref{methods-app}. All
figures show the velocity-by-velocity line ratio, followed by a comparison with models covering a range of
temperature and velocity-density gradient going from the conditions where both lines are optically thin, to the strong line saturating,
and finally, to the limit where the two lines are saturated \citep{beristain98}.
Leaving aside the Fe I 5572/5624\AA\ pair that is close to saturation in the line wings as well as the center, the 
central part (closer to zero velocity) appears to be optically thinner
than the high-velocity wings in the outburst lines. For a disk scenario, this can be explained if the density decreases
toward larger radii (lower velocities), which is reasonable. 
For the quiescence data, the emission is close to being optically thin in the weak line
for both the Fe I 6256/6191\AA\ and the Fe I 6462/6393\AA\ pairs.
This is significantly different from what has been observed in EX Lupi \citep{sicilia12,sicilia15} and DR Tau \citep{beristain98},
where narrow lines are optically thicker than the broad components and can 
be traced back to dense,
hot spots on the stellar surface\footnote{We note that Sobolev's approximation breaks down for very narrow lines.}, 
and suggests a disk 
origin.

\begin{figure}
\centering
\includegraphics[width=9cm]{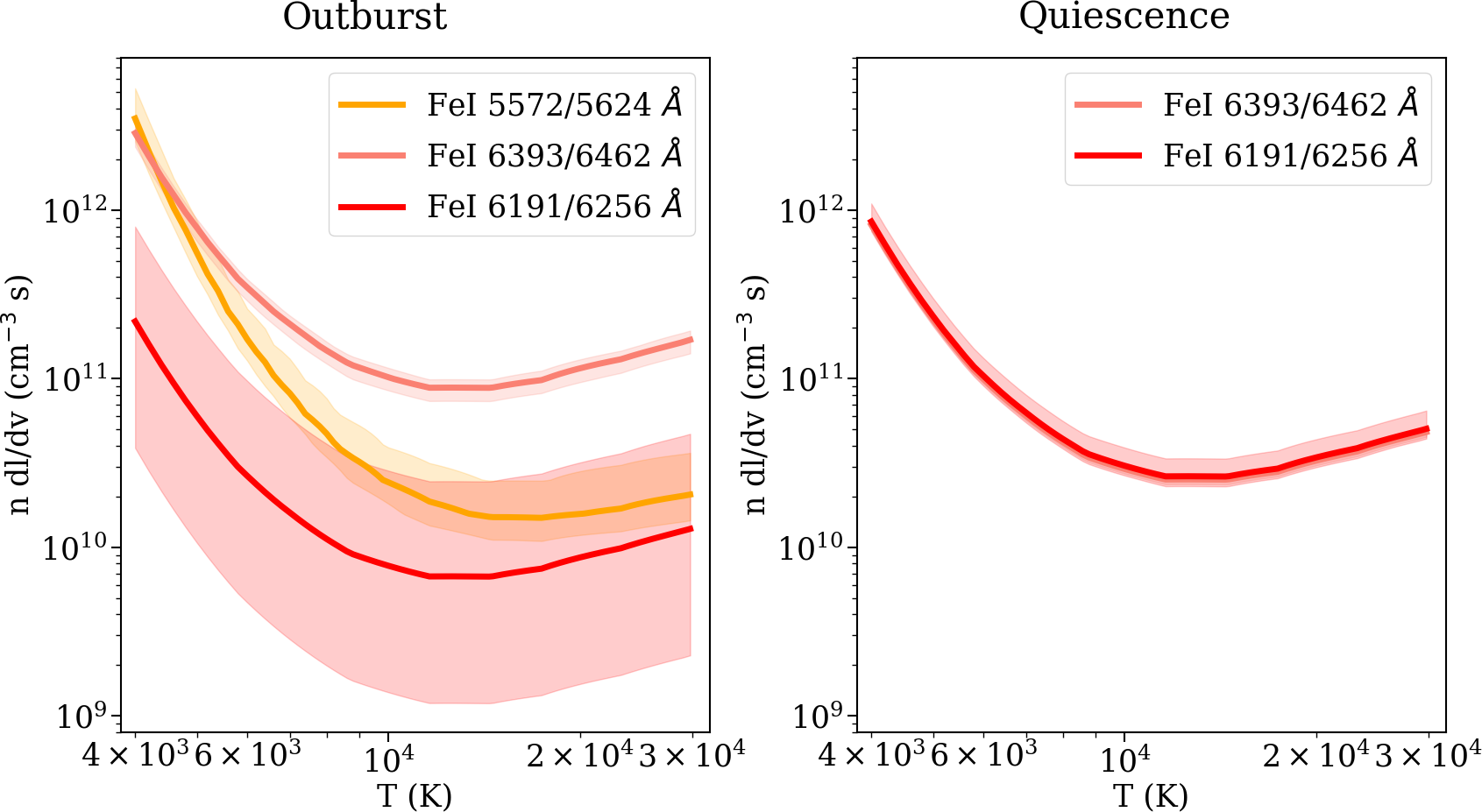}
\caption{Result of the line ratio analysis for neutral lines with disk profiles in outburst (left) and
narrow profiles in quiescence (right).}
\label{lineratiosdisk-fig}%
\end{figure}

We can combine all of the lines to 
derive a better constraint on temperature and density (see Figure \ref{lineratiosdisk-fig}). 
The temperature is not too well constrained from this exercise, but most of the lines agree with 
column density-velocity gradients ($n dl/dv$) on the order of 10$^{10}$-10$^{12}$ cm$^{-3}$s, higher than in quiescence. 
The weakness of the lines, the uncertainty of some of the
atomic parameters, and the potential breakdown of the Sobolev approximation 
for lower velocity gradients may cause part of the temperature mismatch between line pairs.
If we combine this result with
Saha's equation from the previous section, both converge toward larger densities and lower ($\sim$4000-5000 K) temperatures, which is
also consistent with the origin in a disk and the values derived from the lines with lower uncertainties. 

A limitation of this procedure is that lines with various
strengths (and line profiles) trace material originating in slightly different locations, which adds a further uncertainty. 
The line pairs examined to trace the disk emission during outburst 
have typical ratios in between the optically thin and optically thick limits.
This means that the weak line is usually optically thin, while the strong line is optically
thick. The optically thin line probes material deeper into the disk, which
explains why lines with low transition probabilities look more disk-like (instead of box-like)
than lines with high transition probabilities. In addition, narrow-line components
(originating in more compact but relatively hot and dense regions, such as 
hot spots at the stellar surface or in the accretion shock region) would be stronger 
for the lines with high transition probability because low transition probability lines would
be more susceptible to collisional deexcitation in these environments. Such departures from disk-like
profile related to optical depth are also observed in CO lines \citep[e.g., see][]{panic08,hein16,carmona17}, contributing to the box-like
appearance of optically thick lines.  As in such cases, Fe I lines with various optical depths 
can give information concerning the vertical temperature structure.
In the case of the strong lines, the box-like
features would not result from optically thick material in a flat disk \citep[e.g., as in the 
examples from][]{horne86}, but rather from material coming from various layers 
and distances in the optically thick, flared disk.
In addition, contributions from a disk wind can affect the double-peaked profiles in
optically thick lines, and can, together with optical depth, result in box-like profiles 
 \citep{murray96,murray97,ferguson97}. Wind can quickly overwhelm disk (and magnetospheric) emission due to
its larger volume \citep{tambovtseva16}, which is not observed here.

Most of the observed disk lines are systematically brighter on their blue side. This blue-red asymmetry could  
be due to inclination and flaring---the red and the
blue parts may appear to have different inclinations; \citep[][]{hein16}---a density enhancement on the
blue side, an off-center disk, or contamination by other contributors to the
line emission such as accretion structures or the disk wind, which does not need to be uniform. We further examine the location of
the emitting material based on the velocities in Section \ref{diskanalysis-sect},
but the results are consistent. A blue asymmetry, likely related to wind, is also observed, albeit in absorption, in the strongest
disk-like lines in FU Ori \citep{donati05}.

The column density-velocity gradients can be used to estimate the densities and constrain the 
values derived from Saha's equation. Considering that the maximum $dv/dl$ is given by 
approximately $v/r$ where v and r are the Keplerian velocity and radius, the best-fitting column
density-velocity gradient of 1e12cm$^{-3}$s suggests a particle density at least 3e4 cm$^{-3}$. 
This value is clearly lower than our estimate for the accretion structures (Section \ref{stellar-sect}), 
which is consistent taking into account that the disk is likely less dense than the accretion channels, especially at larger radii.
Together with the temperature derived from this procedure ($\sim$4000-5000 K), this section confirms the agreement of the 
neutral line emission with an origin in an extended disk structure around the star as well as a density (optical depth), 
radial differences between the high-velocity disk (line wings, formed closer in), and the lower-density, lower-velocity regions (related to the
outer disk).

\subsubsection{Line ratio constraints on the wind absorption and redshifted emission components \label{lineratiowind}}

\begin{table*}
\caption{Lines with PCygni profile that share a common upper level.}              
\label{linepairsPC-table}     
\centering                                     
\begin{tabular}{l c c c l}       
\hline\hline                        
Species & $\lambda_0$ & A$_{ji}$ & Uncertainty & Notes \\
        &  (\AA)    & (s$^{-1}$) & & \\
\hline                                  
HI      & 3750.151, 8750.46 & 2.83e+04, 2.02e+04 & 1\% & Complex continuum \\
HI      & 3797.909, 9015.300 & 7.1225e+04, 5.1558e+04 & 1\%  & Telluric cont. \\
TiII    & 3900.540, 5129.150 & 1.60E+07, 1.00E+06 & 50\% & OK \\
FeII    & 3938.290, 6432.680 & 6.10E+03, 8.50E+03 &  50\% & First line blend \\
FeII    & 4128.748, 4303.176, 4522.634 & 2.60E+04, 2.20E+05, 8.40E+05 & 50\% & Blends\\
FeII    & 4178.862, 4520.224, 5276.002 & & 50\% & Blends \\
FeII    & 4233.167, 4731.453 & 1.72E+05, 2.80E+04 & 50\% &  OK \\
FeII    & 4258.154, 5197.577 & 3.10E+04, 5.40E+05 & 50\% & First line blend\\
FeII    & 4273.326, 4508.288 & 9.10E+04, 7.30E+05 & 50\% & Blends \\
FeII    & 4296.652, 4489.183, 5234.625 & 7.00E+04, 5.90E+04, 2.50E+05 &  50\% & Blends \\
TiII    & 4443.800, 4450.490, 4501.270 & 1.1E+07, 2.00E+06, 9.80E+06 & 50\% & Blends \\
FeII    & 4629.339, 4666.758 &  1.72E+05, 1.30E+04 & 25\% & OK \\
TiII    & 4779.985, 4805.100 & 6.20E+06, 1.10E+07 & 50\% & OK \\
FeII    & 6238.375, 6247.562 & 7.50E+04, 1.60E+05 & 50\% & OK \\
OI      & 7771.940,7774.170, 7775.390 & 3.69e+07 & 3\% & Blends\\
\hline                                             
\end{tabular}
\tablefoot{Limitations to use the
lines ratios can be due to blends, telluric
contamination, or if the lines have very different wavelengths and underlying 
continua. The uncertainty levels are taken from the NIST database
and reflect the uncertainty in the atomic parameters for the most
uncertain line. For lines with errors $>$40\%, the theoretical line ratio
is very uncertain, especially if the transition probabilities
are similar.}
\end{table*}

\begin{figure*}
\centering
\begin{tabular}{cc}
Outburst & Quiescence \\
\includegraphics[width=6.3cm]{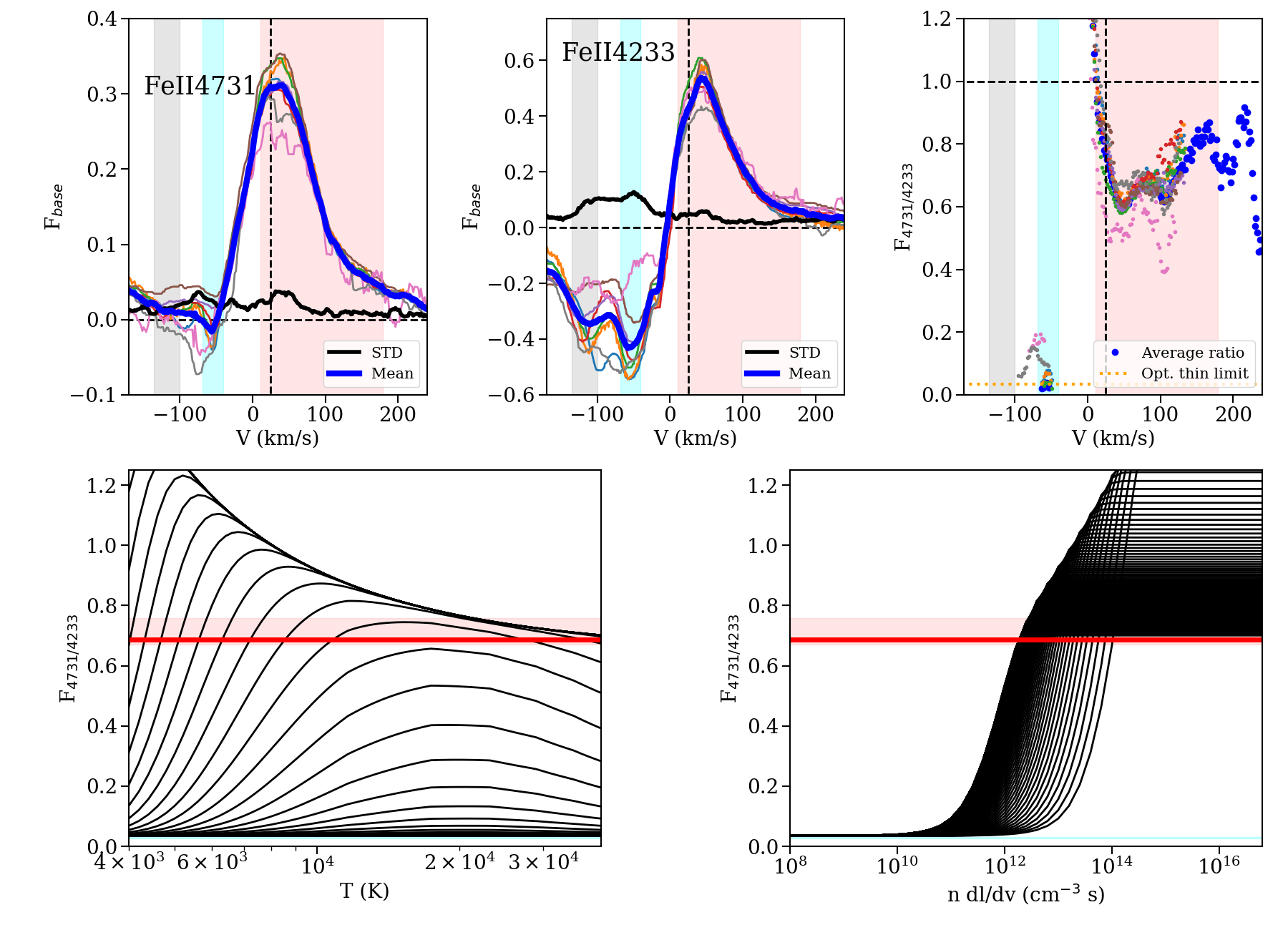} &
\includegraphics[width=6.3cm]{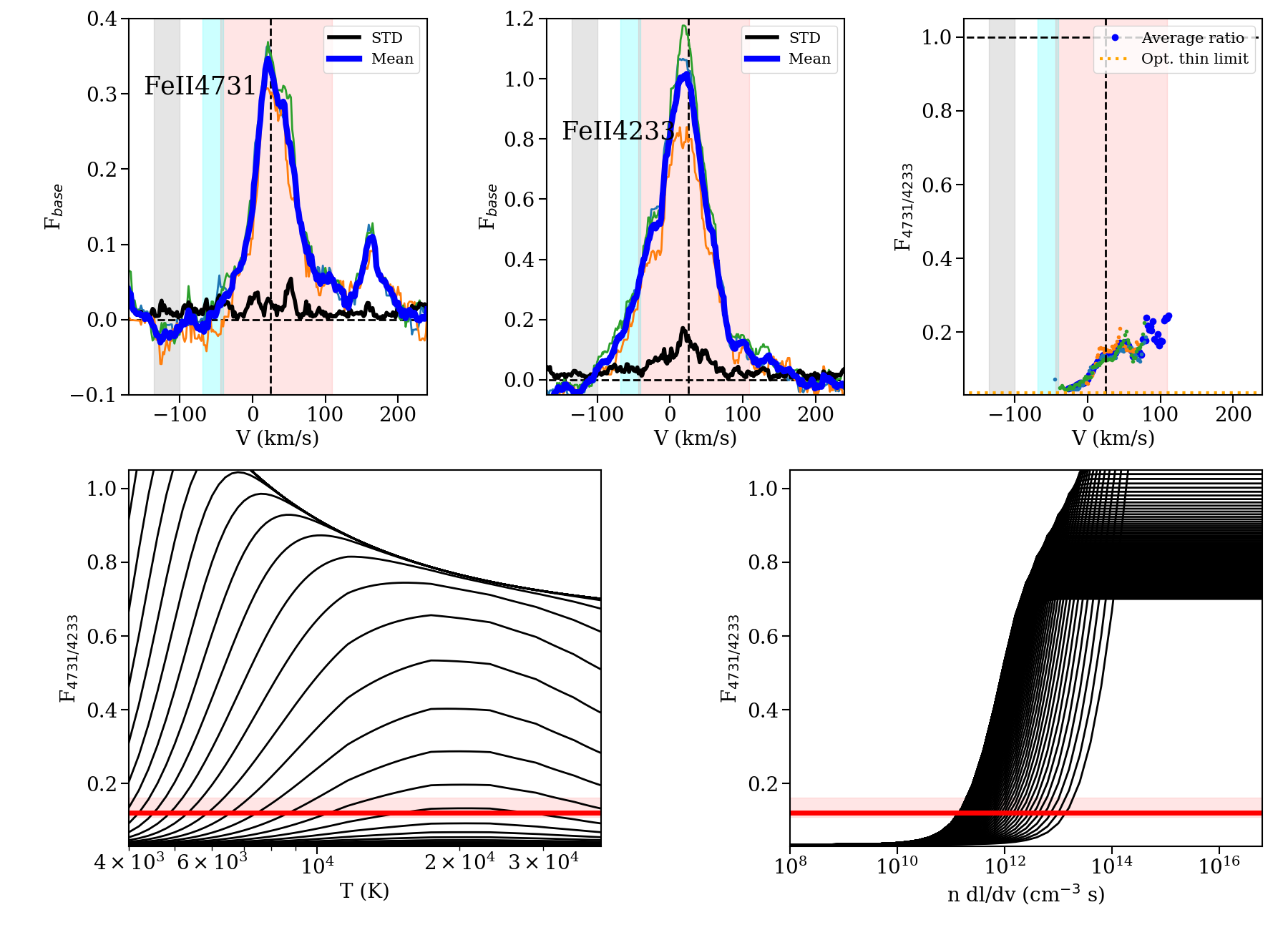} \\
\includegraphics[width=6.3cm]{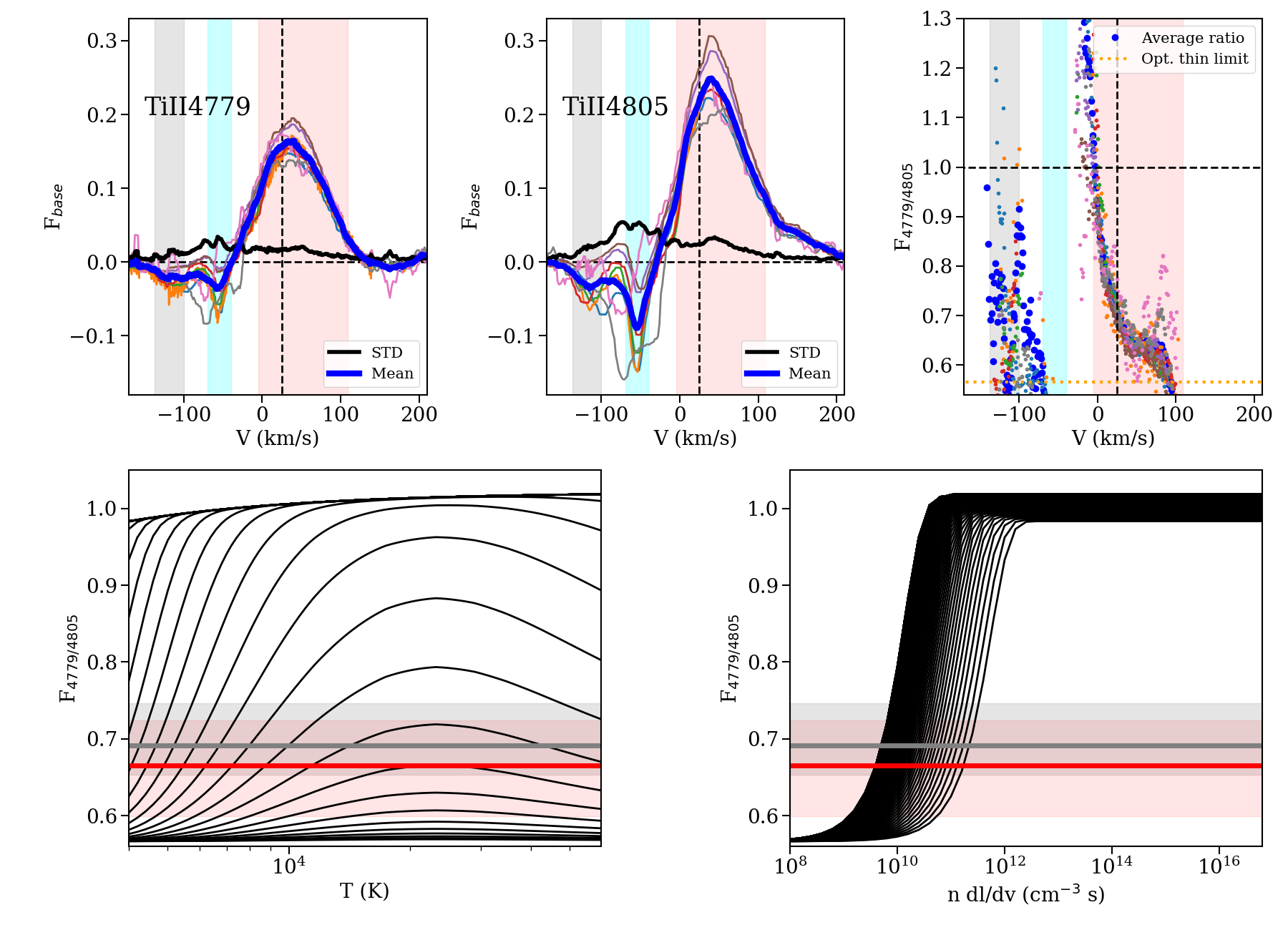} &
\includegraphics[width=6.3cm]{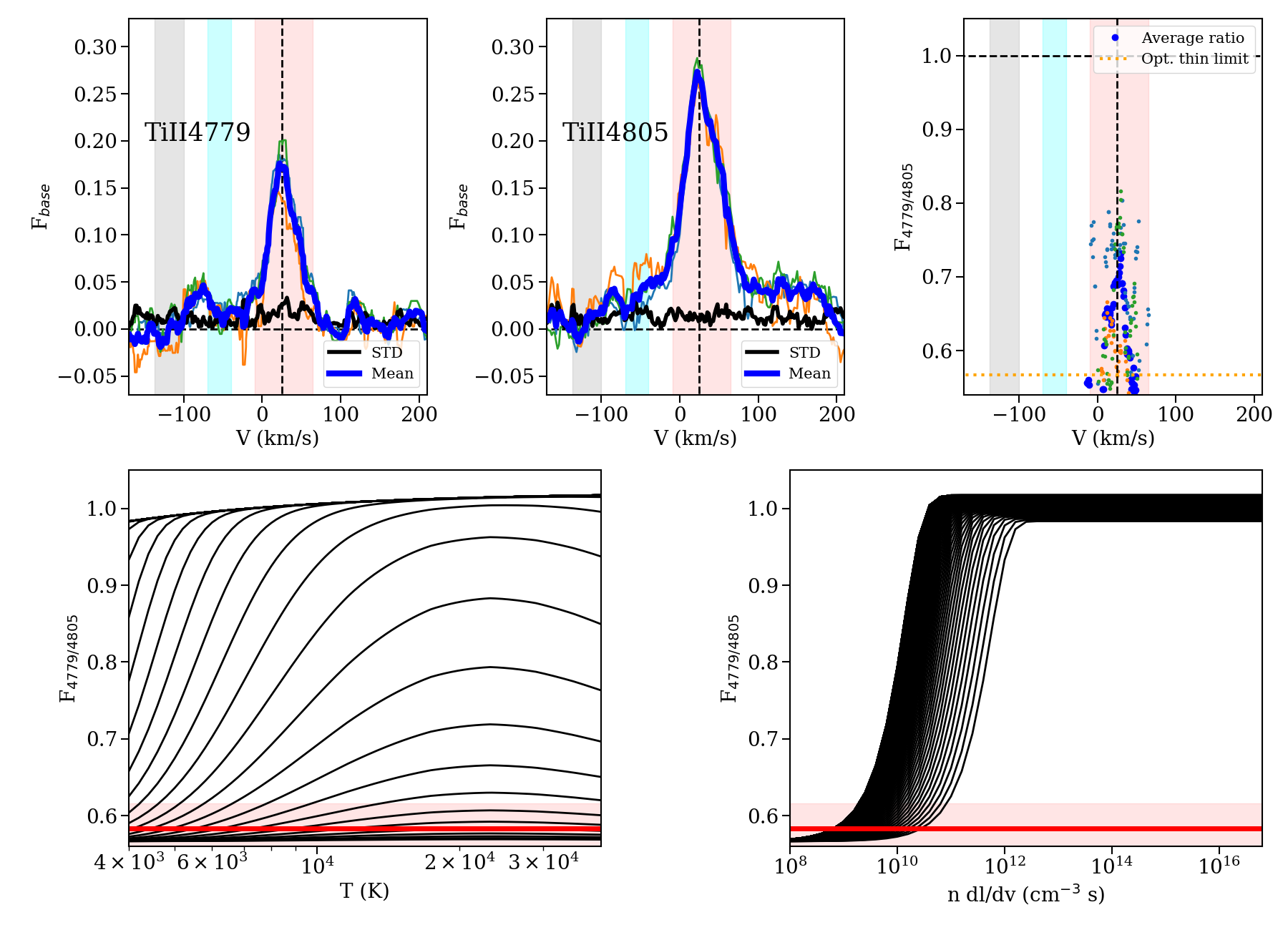} \\
\end{tabular}
\caption{Analysis of the line ratio for the FeII and TiII lines with PCygni profiles from 
transitions with common upper
levels (continued in Appendix \ref{methods-app}). The upper left panels show the
line emission in outburst and the upper right panel shows the quiescence data. 
The ratios observed  are compared to the theoretical calculations for a range of temperatures and column density-velocity gradients in the lower panels.
In the velocity panels, the red, cyan, and gray areas show the regions with significant redshifted emission and slow and fast wind absorption, respectively. The shaded colored
areas in the temperature and density planes show the regions that are compatible with the line ratios observed 
in the emission and the two wind component parts with their 1$\sigma$ error.
See text for details.}
\label{Windratio-fig}%
\end{figure*}

The same exercise of line ratios was done for the pairs of lines with wind
signatures that share an upper common level. 
We find a total of 15 pairs or triplets of lines with common upper levels 
and P Cygni profiles (see Table \ref{linepairsPC-table}), but only 5 pairs remain if we exclude pairs. 
For the line ratio analysis, we checked both the line ratio in the
emission part (measured as positive over the continuum) and in the absorption part (measured as 
negative under the continuum). The velocity space was divided in the region in which
both lines have significant ($>$5$\sigma$) emission and the regions in which both lines have
significant absorption. The ratio was only calculated where both lines were either
in emission or in absorption. 
The wind was separated in two components, the fast and the slow wind, but since the 
emission part of the lines is very broad, it is very likely that the slow
wind is contaminated by emission. This explains the unphysical ratios observed at some 
velocities. The line pairs show no significant absorption during quiescence, a signature
of the wind becoming optically thin, as expected from an
accretion-powered wind.

\begin{figure*}
\centering
\includegraphics[width=17cm]{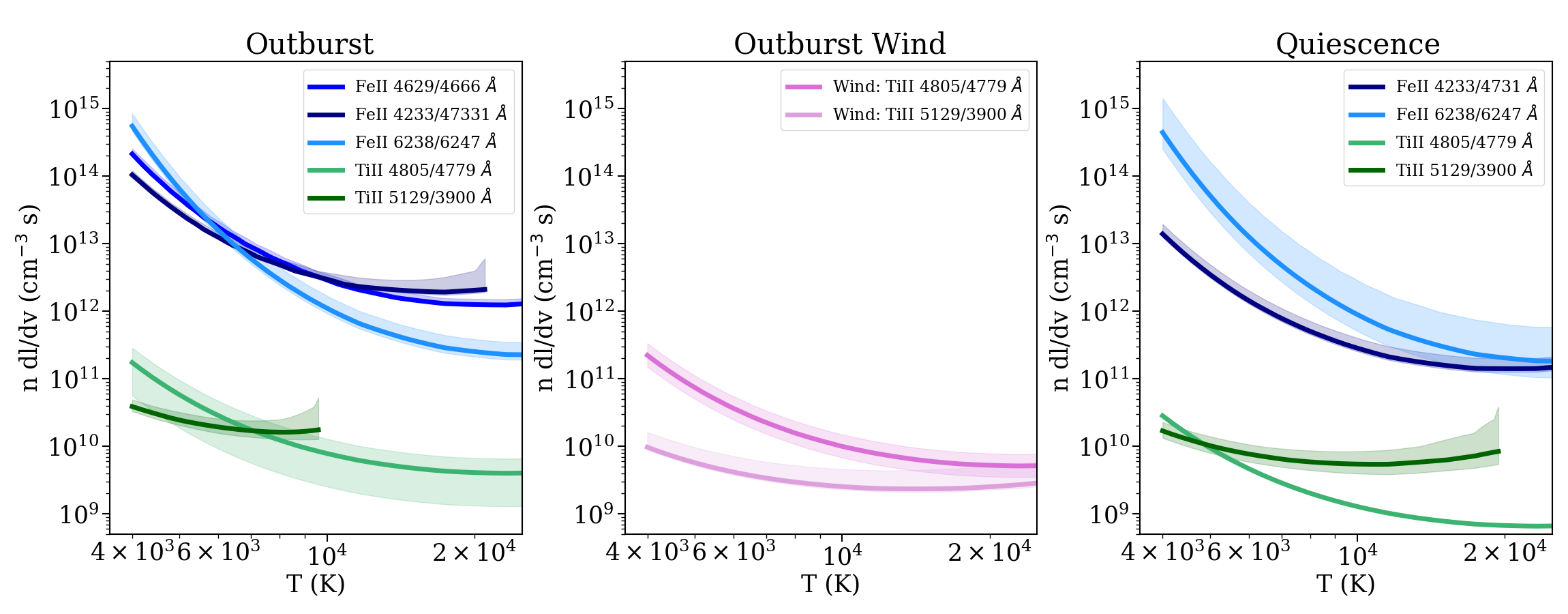} 
\caption{Result of the line ratio analysis ion pairs of lines 
with a common upper level for the emission and wind absorption parts (denoted as "wind" in
the legend). Some of the Ti II lines have sharp cuts in the n dv/dl vs. T plane due to
unphysical values that arise in regions of low S/N (see Appendix \ref{methods-app}). The results have to be regarded with care
when approaching such regions. }
\label{lineratios-fig}%
\end{figure*}

Figure \ref{Windratio-fig} shows the results. 
The wind components cover a large range of temperatures and densities. 
We can put together the column density-velocity gradient and temperature ranges favored by
each line ratio to get a better constraint for both the
emission and fast wind (Figure \ref{lineratios-fig}).
Since the wind absorptions are very weak for all the line pairs, 
derived wind properties are strongly affected by the noise. Weak lines with low transition probabilities
have very weak or even absent absorption, a signature of being optically thin. This 
constrains the maximum density allowed for a given wind temperature and the
range of densities that this method can probe. 
For the emission component of the line, both the Fe II line pairs and Ti II line pairs are consistent
with a temperature in the range 6000-7000 K, although the Ti II lines favor a column 
density-velocity gradient value three orders of magnitude lower.  Given that the line
profiles for Fe II and Ti II are not identical, differences in the physical location are 
expected. Only the two Ti II pairs give significant results for the fast
wind component, suggesting a much lower column density-velocity gradient and a very high temperature ($>$2e4 K).
Since the uncertainties in the atomic parameters for Ti II are $\sim$50\%,
the final results have uncertainties on the order of 70\%, which would
be compatible with a lower bound temperature of $\sim$ 10$^4$ K. Such large temperatures are more
consistent with a stellar, rather than a disk wind, origin. The relative values for the Ti II lines suggest 
that the column density-velocity gradient is between 1 and 2 orders of magnitude
lower for the wind than for the redshifted (potential infall-related) part.

The emission part of the line has a narrow, central component in the
weakest ionized lines, probably because the strong lines with large transition probabilities are
highly saturated. The lack of narrow components in low transition probability neutral 
lines is likely related to temperature: the
originating accretion structure, wind, or hot spots in the proximities of the star
are likely too hot to allow for strong Fe I emission. If the temperatures are low enough to 
ensure that the metals are all once-ionized and hydrogen remains neutral, the electron density would be
approximately 0.1\% of the hydrogen number density. Typical redshifted velocities in the metallic lines
are around 100 km s$^{-1}$, which is likely too low to be explained by free fall around an
intermediate-mass star, but compatible with, for example, rotation at
a few stellar radii. The 10$^{13}$ cm$^{-3}$s$^{-1}$ column density velocity gradient constrained by the line ratios would then result
in electron densities on the order of 10$^5$, which require temperatures of at least 3500 K to
avoid Fe I emission (according to Saha's equation). The temperatures from line ratios are about a factor of 2 higher (see below), 
explaining the lack of Fe I emission with accretion or wind signatures
and the scarcity of Fe II lines with disk profiles.

The temperatures of the redshifted wings derived from the line pairs are around 7000 K, which are low for an origin in the accretion columns in the 
proximity of a luminous, intermediate-mass star. This suggests that
the lines originate in a more distant region, inconsistent with
magnetospheric accretion; infall velocity onto an intermediate-mass star
would rapidly exceed the 100-200 km s$^{-1}$ ranges observed in the line wings. 
A boundary layer scenario also fails to explain the velocities in the inner disk,
which are expected to be significantly lower than observed in disk-like lines \citep{popham93},
although increased velocities due to turbulence may reach a few 100 km s$^{-1}$ \citep{bertout88}. As in the disk analysis,
because optical spectra trace relatively low-energy lines, 
we have a selection bias toward lower temperatures. This could explain their origin not too close to the star, 
so we may be indeed missing the hottest, highest velocity
emission in most of these line pairs. Much higher velocities, more consistent with infall on an intermediate-mass
star, are seen in the very extended wings of the H Balmer lines. 
A hotter and higher-velocity region 
could be explored further with  spectroscopic 
observations covering metallic UV lines.

Summarizing, line ratios reveal a highly structured wind and a complex origin for
the redshifted emission components. Both cover a broad range of densities and temperatures that can be probed by lines
with various excitation energies. The wind temperatures are consistent with a stellar (or innermost disk) origin,
while the redshifted emission could originate at a few stellar radii. The wind becomes optically thinner in
quiescence, which is consistent with being powered by accretion.

\subsection{Disk structure derived from the velocity analysis\label{diskanalysis-sect}}

\begin{table*}
\caption{Summary of the brightness decomposition of lines with disk-like profile. }              
\label{brightness-table}     
\centering                                     
\begin{tabular}{l c c c c c l}       
\hline\hline                        
Species/$\lambda_0$ & A$_{ki}$ &  E$_k$ & v$_{rad}$ & V$_{p2p}$ & FWHM & Notes \\
          (\AA)     & (s$^{-1}$)  & (eV) &  (km s$^{-1}$) &  (km s$^{-1}$)  &(km s$^{-1}$)  & \\
\hline                              
FeI 5506.8    & 1.14E+08        & 3.241  & 26.9$\pm$0.3 &  78$\pm$6 & 113$\pm$23 & Blue-dominated, no high-velocity wings. \\                                                     
FeI 5572.8 &  2.28E+07 &  5.621 &  35.8$\pm$0.3   & 72$\pm$4 & 115$\pm$2  & Very disky, no high-velocity wings, some central emission. \\
CaI 5598.5    & 4.30E+07        &  4.735 & 21.3$\pm$0.6 & 81$\pm$2 & 113$\pm$3  &  Small blue wings.\\ 
FeI 5615.6    & 2.39E+05                & 5.539  & 26.2$\pm$0.1 & 62$\pm$1 & 120$\pm$1  & Strong high-velocity wings.\\                                                     
FeI 5624.5    & 8.30E+05        & 5.621  & 30.2$\pm$0.1 &  74$\pm$2 & 105$\pm$2 & Strong blue high-velocity wing. \\                                                     
FeI 6065.5    & 1.07E+06        &  4.652 & 31$\pm$1 & 74$\pm$2 & 119$\pm$3  & Strong high-velocity wings, mostly blue.\\                                                     
NiI 6108.1    & 1.30E+05        & 3.706  & 24.5$\pm$0.8 & 55$\pm$3: & 95$\pm$2  & Very asymmetric, red side may be absorbed by nearby feature.\\ 
FeI 6256.4    & 7.40E+04        &  4.435  & 30$\pm$3 &  87$\pm$1 & 108$\pm$11  & Disk-like, but has low S/N and nearby absorption lines. \\                                                     
FeI 6265.1    & 6.84E+04        & 4.154  & 36.9$\pm$0.1 & 76$\pm$4 &108$\pm$2  &Clean, no wings. \\                                                     
FeI 6358.6    & 5.19E+05        & 6.092  & 36$\pm$1 &  81$\pm$5 & 107$\pm$14  & Low S/N and asymmetric continuum due to nearby lines. \\
FeI 6393.6    & 4.81E+05        & 4.371  & 23.5$\pm$0.5 & 62$\pm$2 & 115$\pm$1   &Asymmetric, may have an additional blueshifted absorption.\\                                                     
FeI 6400.0    & 9.27E+06        &   5.539 & 37.5$\pm$0.3 & 79$\pm$2 & 118$\pm$2  & High-velocity wings.\\   
FeI 6411.6    & 4.43E+06        & 5.587  & 41$\pm$1 & 82$\pm$17 & 107$\pm$15   & Red-dominated.\\                                                     
FeI 6421.4    & 3.04E+05        & 4.209  & 29.1$\pm$0.3 &  73$\pm$22 & 109$\pm$2  &\\    
CaI 6439.1    & 5.30E+07        & 4.103  & 27$\pm$1 &  82.5$\pm$0.8 & 111$\pm$15 &Blue-dominated, very strong high-velocity blue wing.\\                                                     
FeI 6462.7    & 5.60E+04        &  4.371 & 25.9$\pm$0.6 &  75$\pm$2  & 113$\pm$2 &\\
CaI 6717.7    & 1.20E+07        & 4.554  & 27.4$\pm$0.9 &  74$\pm$2 & 124$\pm$4  &Asymmetric continuum. \\                                                      
FeI 6750.2    & 1.17E+05        & 4.260   & 25$\pm$0.1 &  80$\pm$1 & 107$\pm$17 &Very extended high-velocity wings to red and blue parts. \\                            
NiI 6767.8    & 3.30E+05        & 3.658  & 27.9$\pm$0.8 & 79$\pm$1  & 109$\pm$2 & Blue high-velocity wing.\\                            
FeI 8674.7    & 6.17E+05        & 4.260  & 37.6$\pm$0.6 &  78$\pm$2 & 108$\pm$4  & Flat-top.\\ 
FeI 8824.2    & 3.53E+05        & 3.603  & 31.1$\pm$0.3 & 68$\pm$1 & 121$\pm$2   & High-velocity wings, central narrow emission. \\
FeI 8838.4   & 3.83E+05 & 4.260  & 34.9$\pm$0.4 & 80.1$\pm$0.7 & 114$\pm$3 &\\                                                                                                                                    
\hline                                             
\end{tabular}
\tablefoot{ 
Radial velocity estimate is affected if the line
is very asymmetric. If the radial velocity is very much off for any
reason, it introduces artificial red/blue asymmetries. Also,
if the NIST wavelength has large uncertainties, this may result in uncertain wavelengths
and thus a further error in the radial velocity and line asymmetry. The quantity
V$_{p2p}$ indicates the velocity between the double peaks of the line.
Lines with potential contamination, blends, or very irregular continuum have been removed from the list.}
\end{table*}

We used the brightness decomposition technique from \citet{acke06}
to study the velocity structure of the disk-like lines. 
The details of the procedure are given in Appendix \ref{bdecomp-app}. For lines with disk-like profiles, the highest velocities are
caused by material in the innermost disk. Starting with the highest velocity, we fit a ring that can reproduce the observed flux and
subtracted it from the line profile, repeating the procedure until the line is entirely fitted or
we reach a velocity in which emission from narrow, non-disk components
dominates. The emission scales as a blackbody with temperature given by a power law of the radius. The final emission 
is obtained by summing all rings.  All the observed lines are best 
fit using the temperature power law for a flared, irradiated disk model \citep[][]{dalessio99}.
Given that the accretion during the outburst is very high\footnote{Accretion luminosity in outburst is seven times the stellar luminosity, so the source is 
accretion-dominated.}, to account for an irradiated disk at few astronomical units 
we need the bulk of the accretion energy to be released closer to the star than the 
region that produces the disk-like lines, at a fraction of an astronomic unit or less, or
radii smaller than those inferred from the lines with disk profiles, justifying our
discussion in Section \ref{stellar-sect}.
Some viscous heating is likely present,  but because of its $r^{-3}$ dependence with the 
disk radius \citep{dalessio98,woitke15},
any viscous heating at the distances at which we observe the double-peaked lines would be
strongly suppressed, so that irradiation (by the central star, accretion spots, or even the
viscous heating in the innermost part of the disk) dominates.

Our models assume $i$=30 degrees, but the inclination is unknown. 
Since the observed velocity is v$_{obs}$ = v$_{kepler}sini$, the radius derived assuming Keplerian rotation
is proportional to sin$^2i$. If the stellar mass was lower or
the disk was closer to edge-on, the emission zone would move toward
smaller radii, constraining even more the locations at which the main source of outburst energy is released.
The results are easily scalable, noting that the temperatures,
based on irradiation/luminosity, would not change. Nevertheless, the vertical scale height and flaring could be affected
if the gravity pull is different.
Putting together the effects of inclination and stellar mass, the scaled radius (r$_{s}$) for an
inclination $i$ and a stellar mass of M$_*$ would be
\begin{equation}
r_{s} = r \frac{sin^2 (i)}{sin^2(30^o)} \frac{M_*}{16 M_\odot},
\end{equation}
where $r$ is the value derived for $i$=30 degrees and $M_*$=16M$_\odot$ shown in the figures.

The main limitations of our analysis are the presence of non-Keplerian velocity components (e.g., arising from
a boundary layer at low velocities, or from an accretion column or a wind at high velocity) 
and the fact that different lines (and different velocities) 
trace material at different depths within the disk. As in previous sections, the results
from weak lines are the most reliable.
The final result is obtained by fitting the median
data, while individual fits to each spectrum are used to determine
the uncertainties. For the noisiest lines (Fe I 6358\AA\ and Ca I 6439\AA), 
some of the individual datasets were disregarded. 
A summary of the fitted lines is shown in  Table \ref{brightness-table}. 

\begin{figure*}
\centering
\begin{tabular}{ccc}
\includegraphics[height=5.8cm,width=5.8cm]{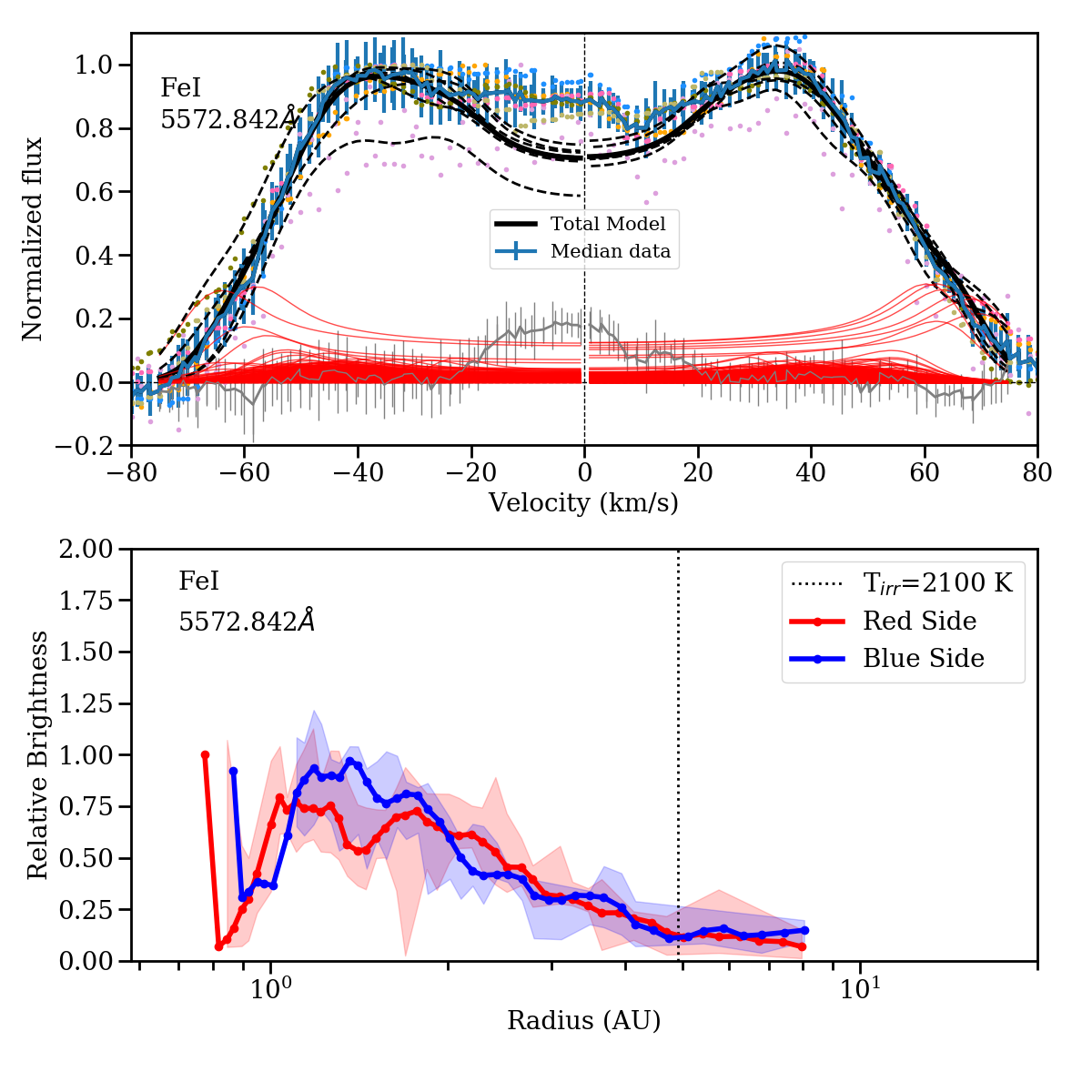}&
\includegraphics[height=5.8cm,width=5.8cm]{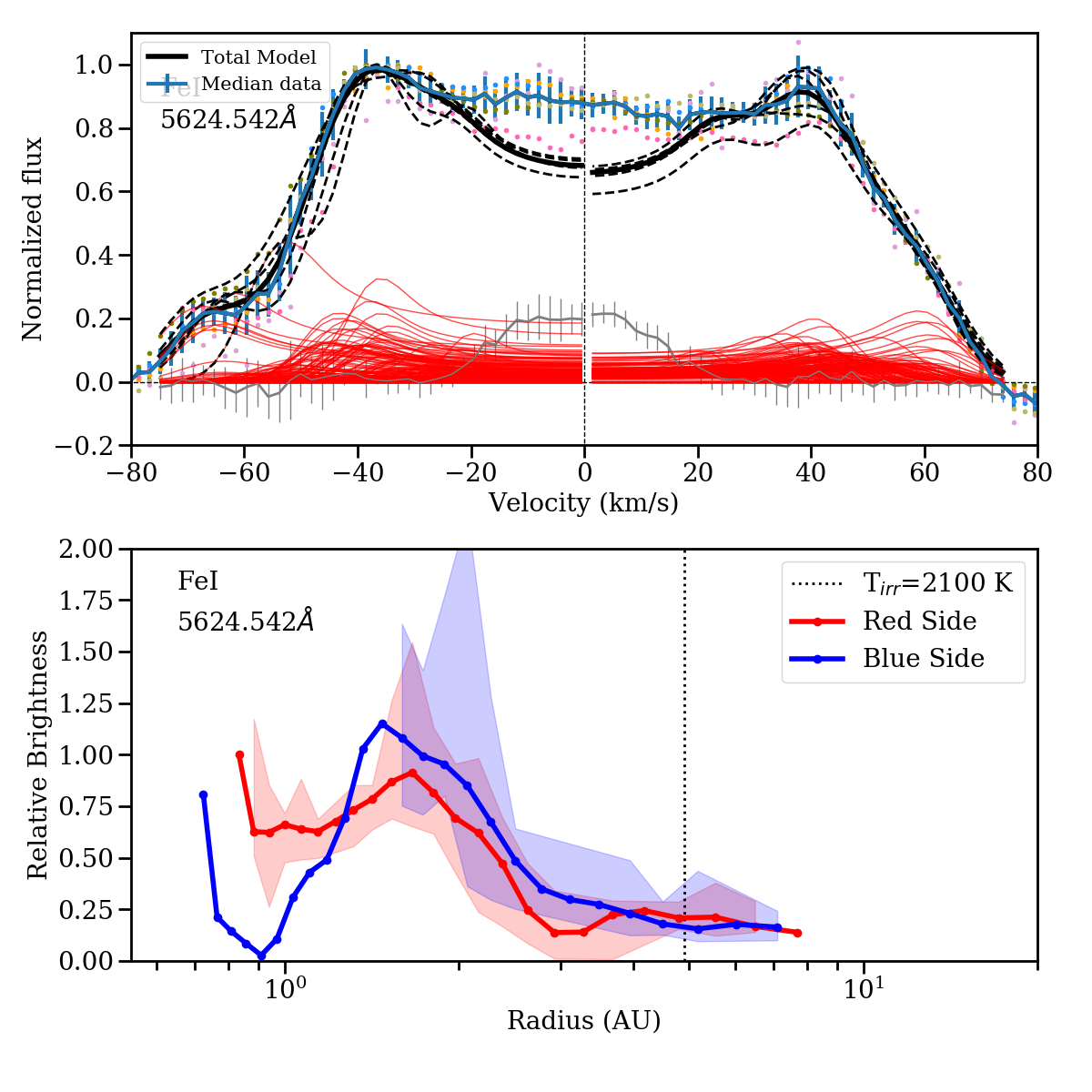} &
\includegraphics[height=5.8cm,width=5.8cm]{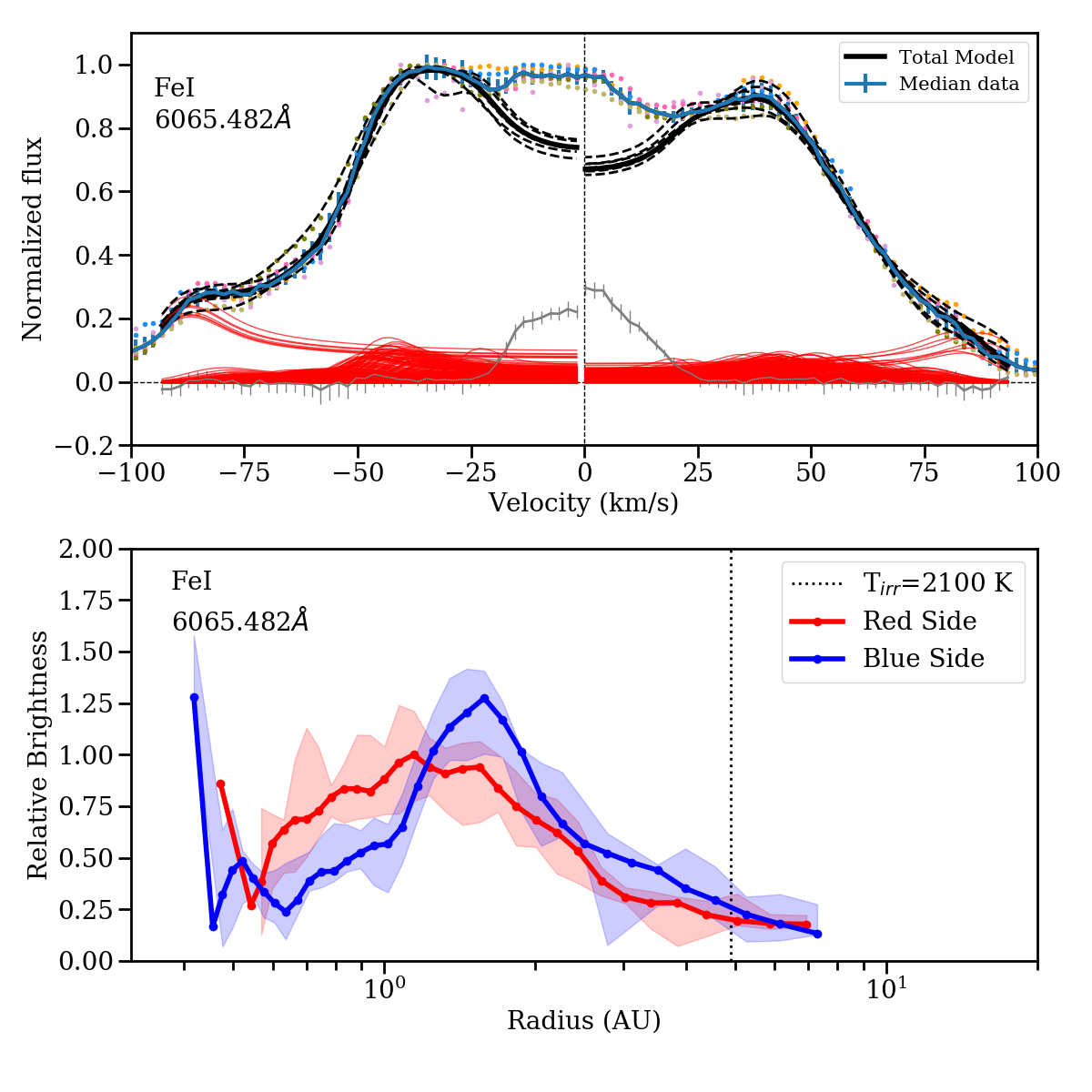} \\
\includegraphics[height=5.8cm,width=5.8cm]{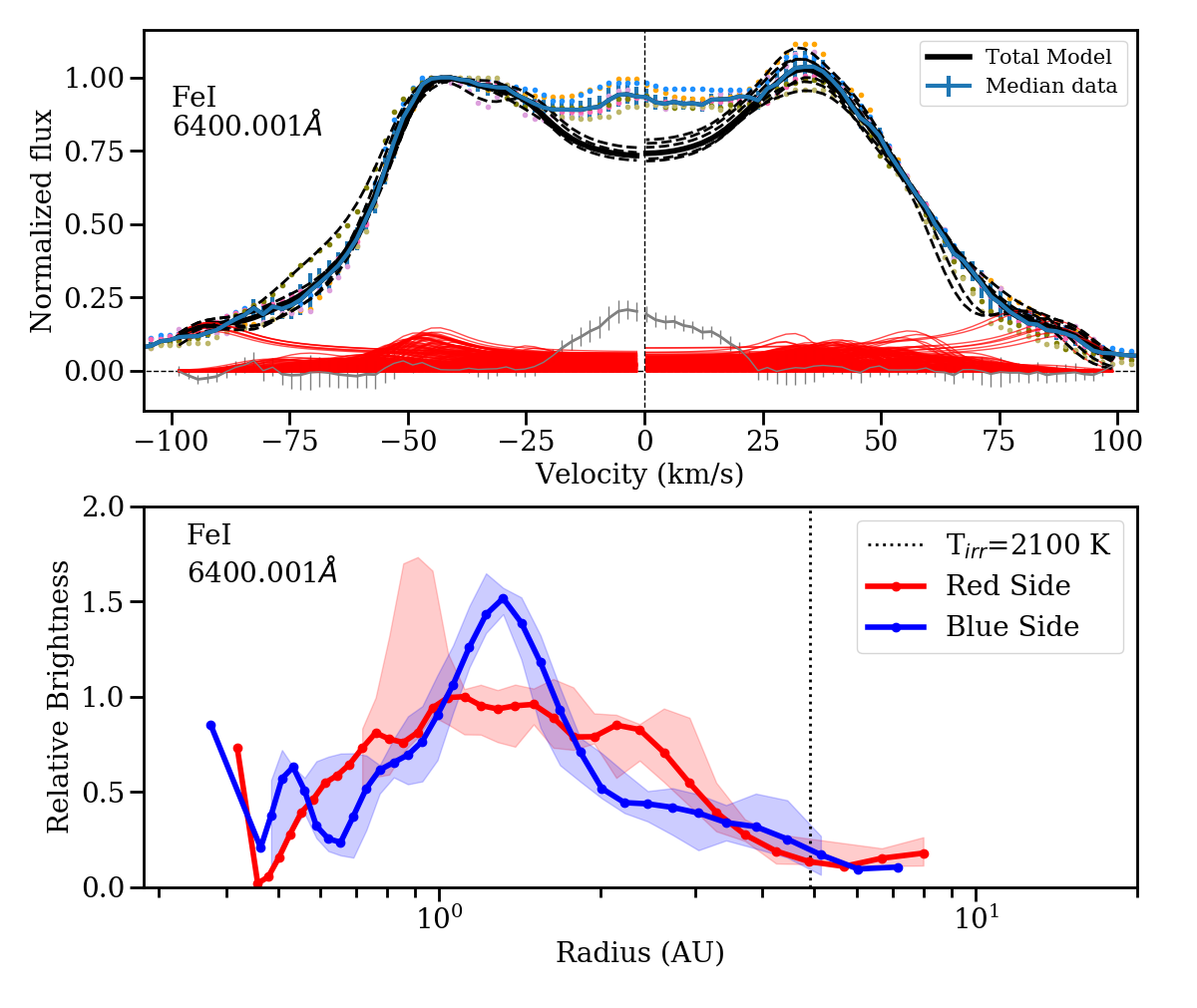} &
\includegraphics[height=5.8cm,width=5.8cm]{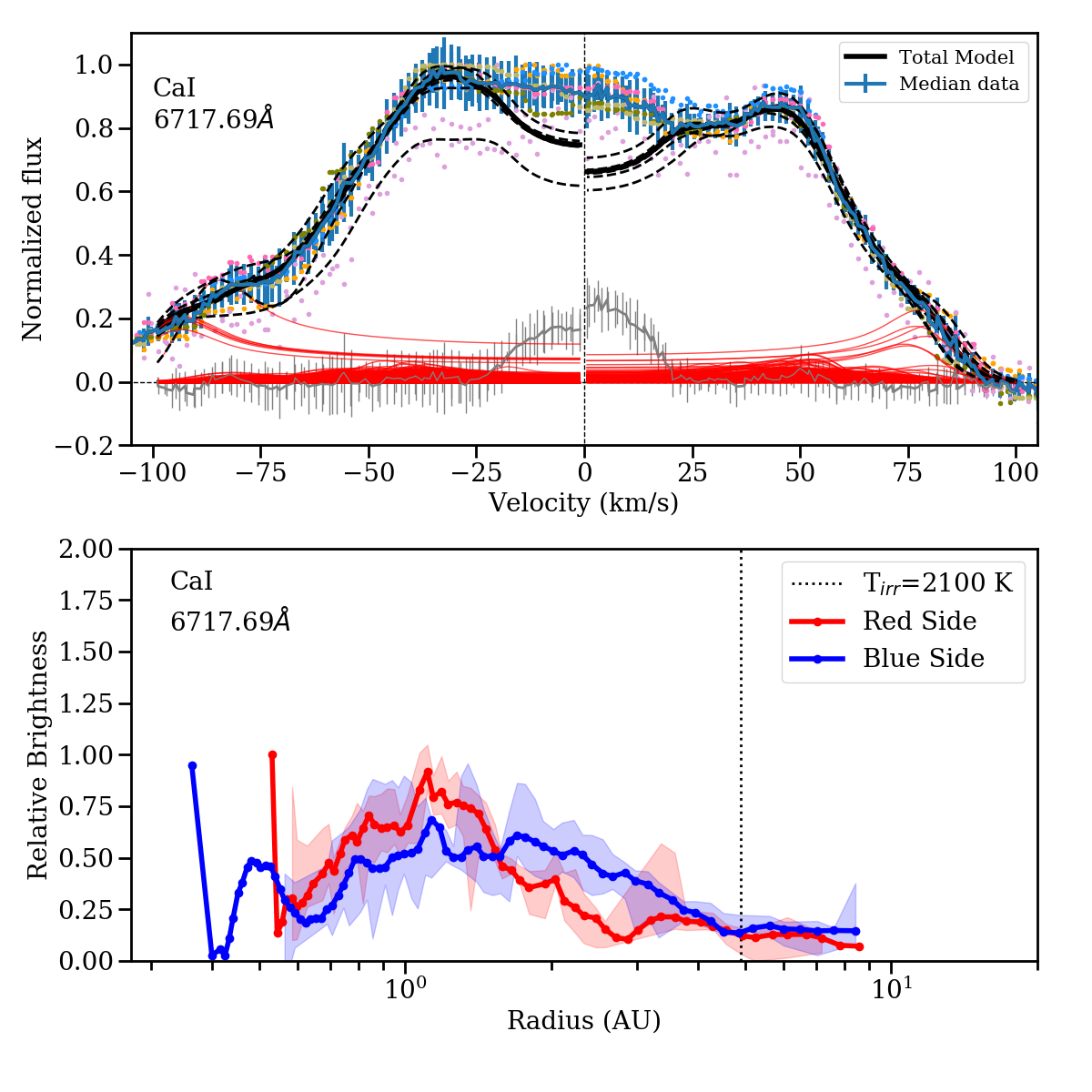} &
\includegraphics[height=5.8cm,width=5.8cm]{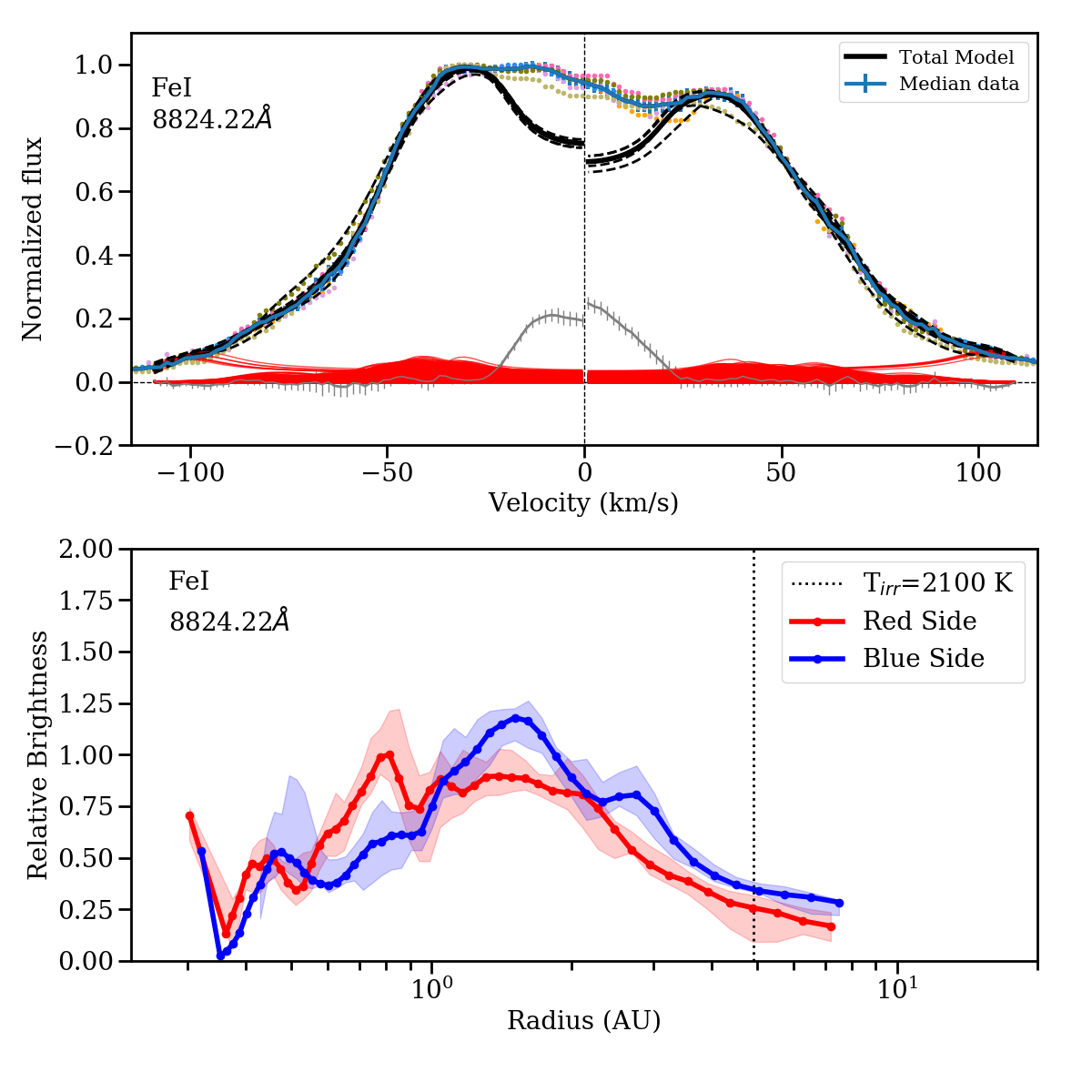} \\
\end{tabular}
\caption{Brightness profile decomposition for the disk-like lines (continued in Appendix \ref{methods-app}).
For each line, the upper panel shows the normalized, median line emission (blue), the individual
data points (colored dots), and the model fit. The black solid line is the 
best fit to the
median data, while the dashed lines are the fits to the individual datasets that are used to
compute the errors. The zero velocity (measured with respect to the symmetry axis of the line) is shown as a dotted line. 
The individual ring fits are shown by thin red lines, and the residuals
are shown in gray. Since we impose a limit of 35 au to avoid fitting
potential narrow emission within the line, the center of the line is not well fitted.  In the lower panel for each line, the red and blue thick lines show the best fit to the  median spectrum for the red and the blue sides, while the shaded areas show 
the range spanned by
the fits to the individual spectra. Since the noise in the single spectra is higher than in the median spectrum, there are no significant results at extreme velocities for some of the individual datasets. 
The vertical dotted line indicates the place where direct irradiation results in a temperature of 2100 K,
which is independent on the model assumptions and stellar mass.
All models are calculated for an inclination of 30 degrees and a stellar mass of 16 M$_\odot$. A different
inclination $i$ would change the derived radii by sin$^2 i$/sin$^2(30)$, and a different stellar mass M$_*$ would change them
by a factor of M$_*$/16M$_\odot$, but the relative brightnesses would remain the same. }
\label{brightness-fig}%
\end{figure*}

The results of the velocity decomposition are shown in Figure \ref{brightness-fig}.
The central radial velocity of the lines is $\sim$20-35 km s$^{-1}$ (average 30$\pm$5 km s$^{-1}$), 
which is consistent with but not identical to the 27$\pm$3 km s$^{-1}$ radial velocity. 
The line decomposition algorithm defines the center of the line by
symmetry, so that deviations in the line symmetry 
would alter the line radial velocity. 
The decomposition reveals that the bulk of disk emission originates at radii $r\sim$0.5-3$\times$(M$_*$/16M$_\odot$) au,
and for most lines (especially, those with well-defined profiles and high S/N), the emission peaks at 
$r\sim$1-2$\times$(M$_*$/16M$_\odot$) au; this is consistent with the radius at which the temperature originated by irradiation is
expected to be high enough to result in significant Fe I emission.

The blue and red sides are not symmetric: most of the lines 
are dominated by the blue component. This suggests that
the gaseous disk may be either non-axisymmetric \citep[as it had been observed in the metallic emission
lines of EX Lupi in outburst][]{sicilia12} and/or not fully centered around the star. Stronger lines, which are likely to
originate in the disk surface layers, tend to be more asymmetric and more variable, suggesting that the outer layers of the disk 
have a more complex structure than the midplane. The disks
of HAeBe stars often have asymmetries in the outer layers at all radii \citep{benisty15,laws20,kluska20},
so this is not unexpected. Hot spots could produce
extra emission at low velocity, although 
we do not observe any rotational modulation in the narrow components. 
Many of the lines have
high-velocity wings or tails (in the range of $-$75 to $-$90 km s$^{-1}$) 
that are clearly distinct from the main disk profile. Although
in some cases the wings are seen in both the blue and red part of the spectrum (e.g., 
Fe I 6400\AA), in most of them the blue high-velocity wing is stronger (or even the only one 
seen; e.g., Fe I 5624\AA) and certain lines are particularly 
asymmetric on their blue sides (e.g., Fe I 6750\AA, Ni I 6767\AA).
The stronger Fe I lines with higher A$_{ki}$ have stronger blue wings\footnote{Similar 
wings are also seen in some of the box-profile, low A$_{ki}$ Fe II lines.
Any disk-like emission in strong Fe II lines would be masked by the much stronger 
wind features.
This may also result in disk-like profiles being very hard to 
detect in stars with lower masses and smaller, hot inner disks, unless there is a 
large difference in temperature and density between the disk and the accretion columns and winds.}.
The high-velocity region could suggest a discontinuous, 
asymmetric disk structure rotating 
faster and closer to the star
(weak because of its smaller area).
Alternatively, the high-velocity tails and blue/red asymmetries could result from absorptions in the range of 
$-$60 to $-$40 km s$^{-1}$ related to the slow wind component (in the blue) or to infall absorption (in the red,
at positive velocities).
Nevertheless, the lack of significant variability in these high-velocity tails
contradicts what is observed in the wind components, and is more in agreement with
an origin in a stable structure (with a longer period) such as 
a radially and azimuthally asymmetric disk.

Besides the difference in line profile with the transition probability (weak lines look more disk-like
than strong ones), 
there are no significant correlations between excitation potentials,
A$_{ki}$, and the brightness decomposition. Correlations between the velocity, the full width at half maximum (FWHM) or the peak-to-peak
distance, and the atomic parameters are uncertain and noise-dominated.
Part of the noise may be caused by the larger variability observed in the line wings of the strong lines that affect their FWHM.
Leaving aside the changes that could be attributed to 
depth and inclination described before, this homogeneity 
could be a sign that the emitting region is very well constrained and ring-like 
\citep[][]{hein16}.

To summarize, this section confirms the previous results of neutral lines in an irradiated disk-like structure,
extending between 0.5-3$\times$(M$_*$/16M$_\odot$) au, and presenting some asymmetries in the surface and the
disk midplane as well as radial and azimuthal variations.
The double-peaked lines arise from very large radii compared 
to the stellar radius (and to any reasonable value of
the stellar magnetosphere or boundary layer), so these lines cannot provide much information on the accretion mechanism.
The temperature ranges inferred from the disk-like lines in the previous sections are not too different from those observed
in FUors \citep[i.e., similar to photospheres of F- or G-type stars][]{hartmann96}, although this could result from
 Fe I emission being strongly suppressed at much higher temperatures. Observations
of line profiles in the UV during an outburst would be needed to obtain information on the innermost
zones around the star and its temperature and velocity structure.
As observed in FUors \citep{hartmann96},  we also see that the lines at longer wavelengths are less
disk-like. This could respond to the same effect observed in FUors where lower temperatures, corresponding to further away
distances in the disk, have different line profiles. In this respect, it would be useful to check the profiles of near-IR lines,
but unfortunately the available near-IR spectroscopy does not resolve
the line profiles of neutral atomic lines \citep{bonnefoy17}.

\subsection{Return to quiescence \label{quiescence-sect}}

\begin{figure}
\centering
\includegraphics[width=7cm]{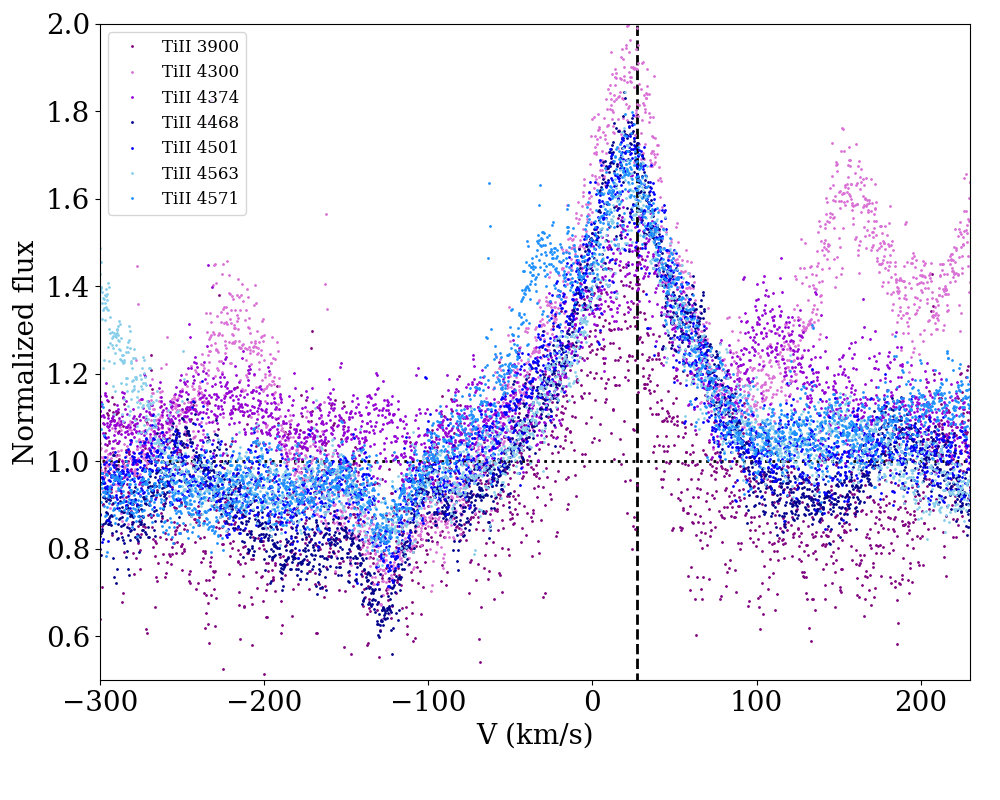}
\caption{Line profiles of the strong Ti II as observed in quiescence. The data for the various dates are all plotted together,
given that there is no significant variability. The vertical line indicates the 27 km s$^{-1}$ radial velocity, and the dotted horizontal line indicates the
continuum level.}
\label{TiIIqui-fig}%
\end{figure}

\begin{figure*}
\centering
\begin{tabular}{cccc}
Outburst & Quiescence & Outburst & Quiescence \\
\includegraphics[width=4cm]{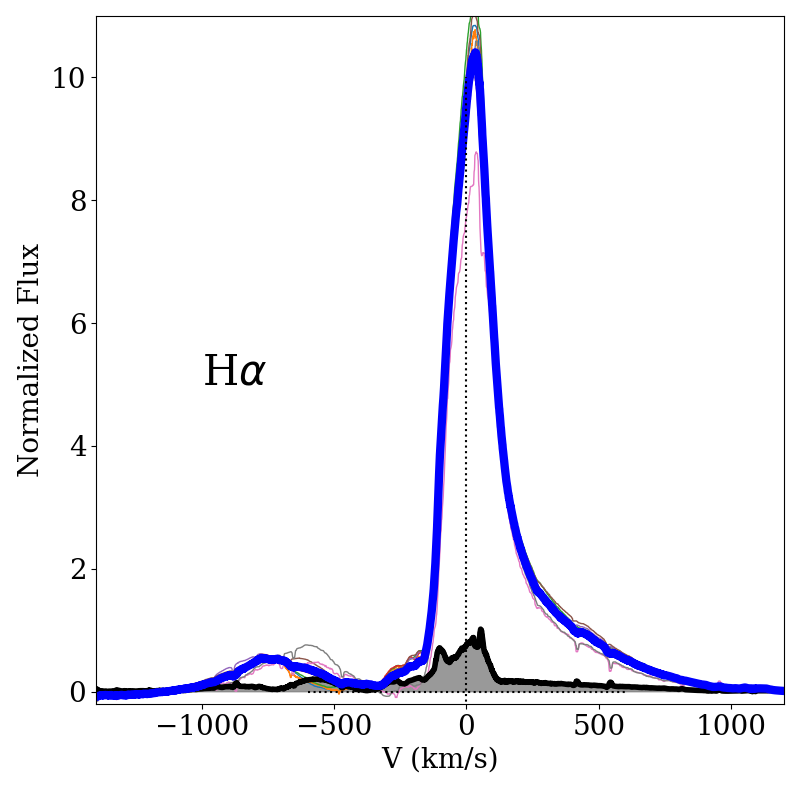} &
\includegraphics[width=4cm]{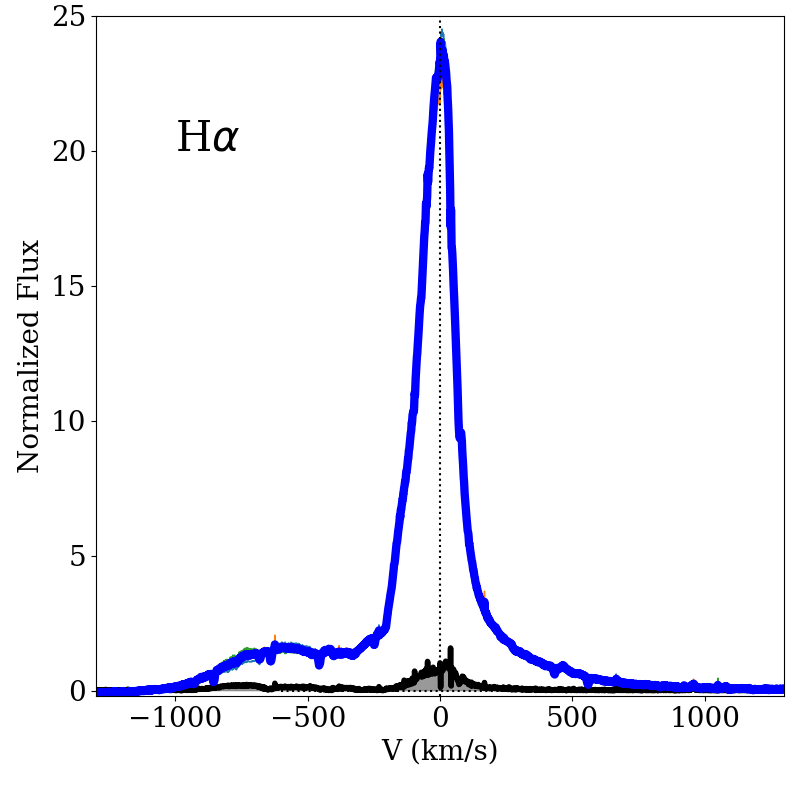} &
\includegraphics[width=4cm]{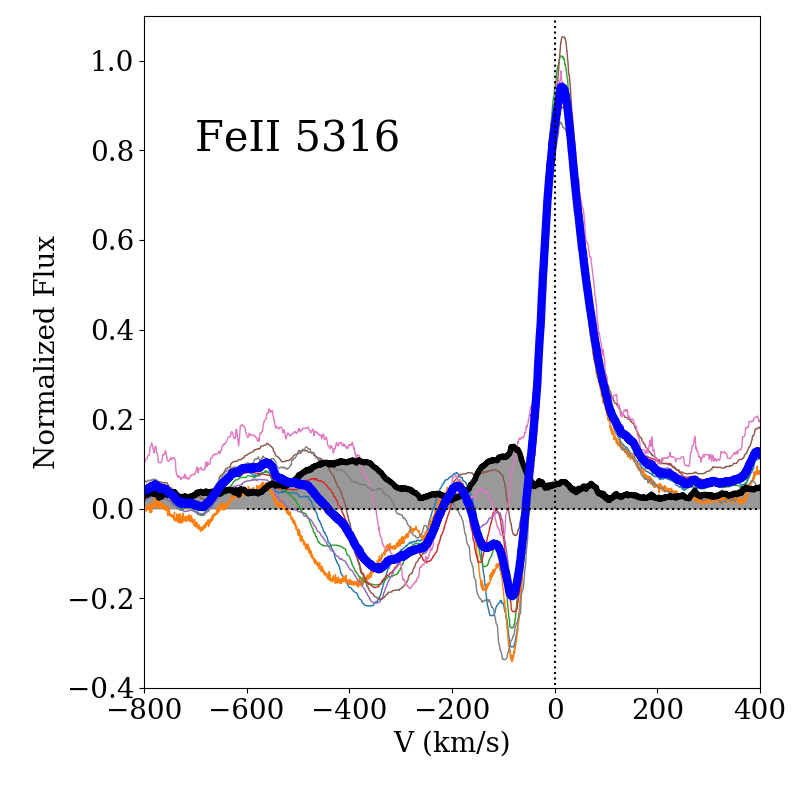} &
\includegraphics[width=4cm]{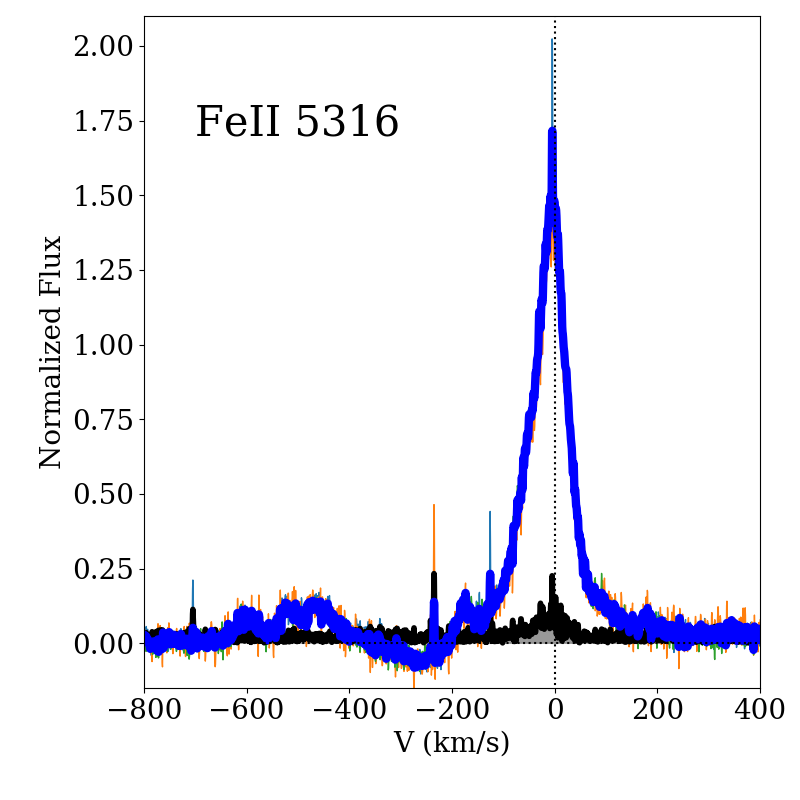} \\
\includegraphics[width=4cm]{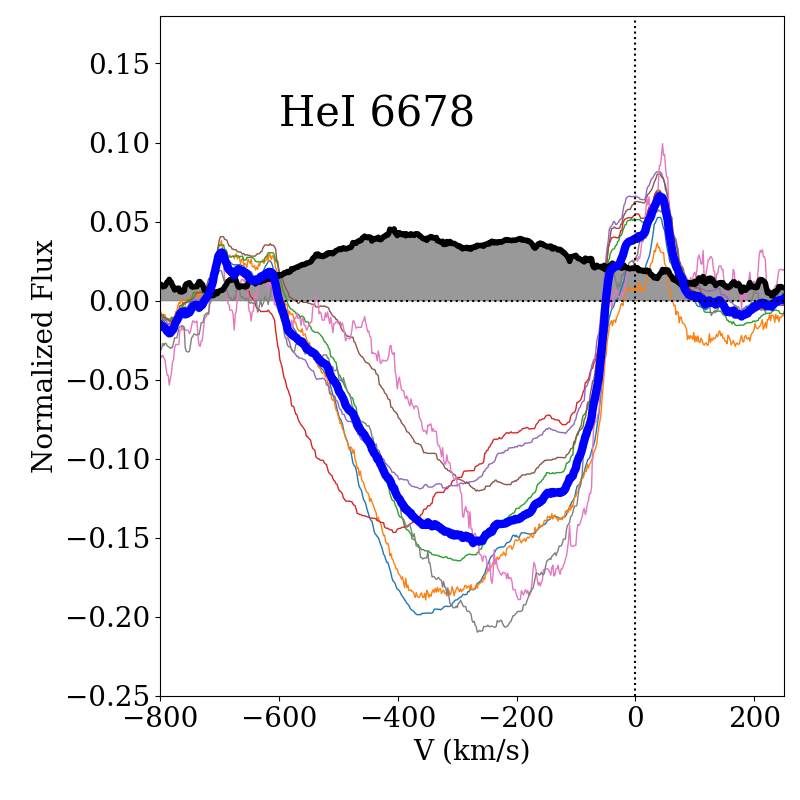} &
\includegraphics[width=4cm]{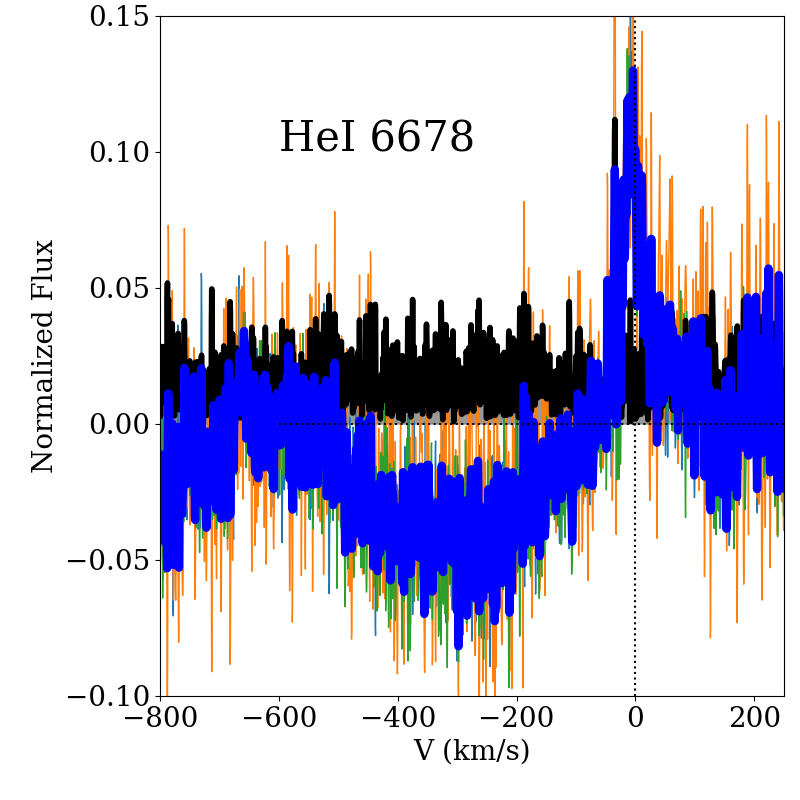} &
\includegraphics[width=4cm]{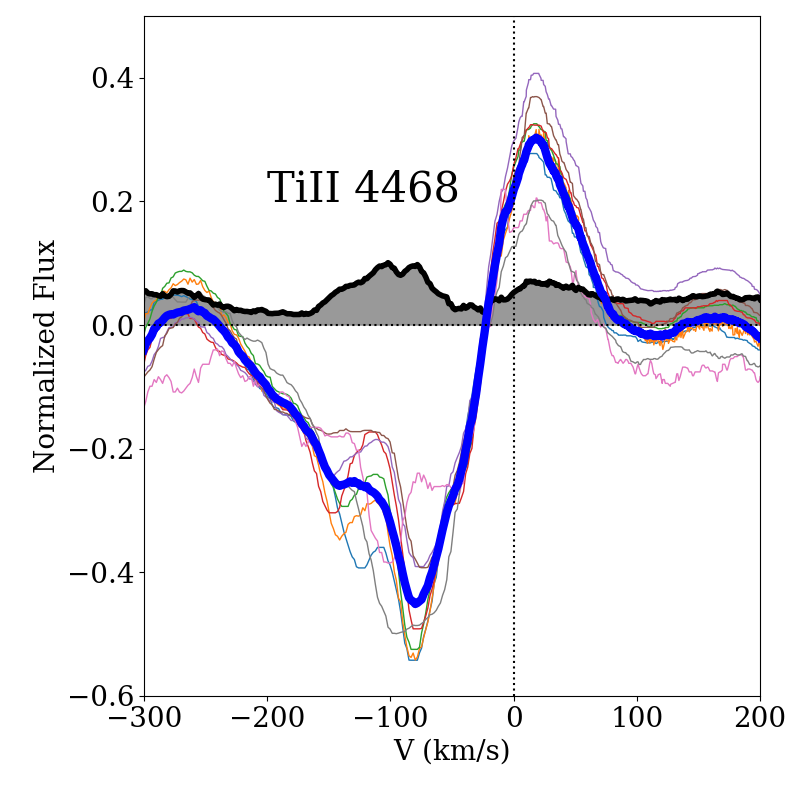} &
\includegraphics[width=4cm]{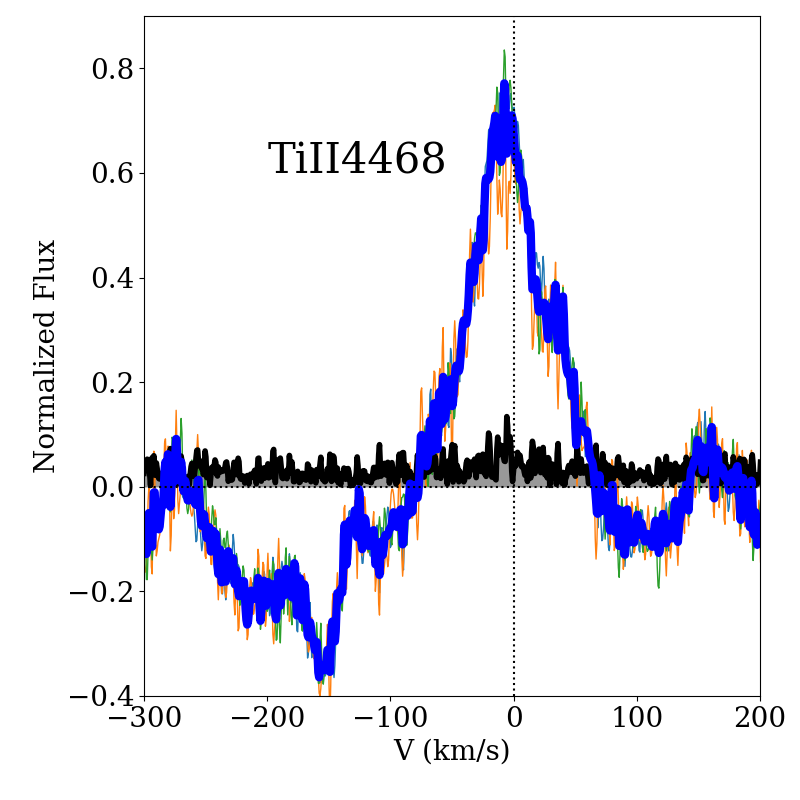} \\
\includegraphics[width=4cm]{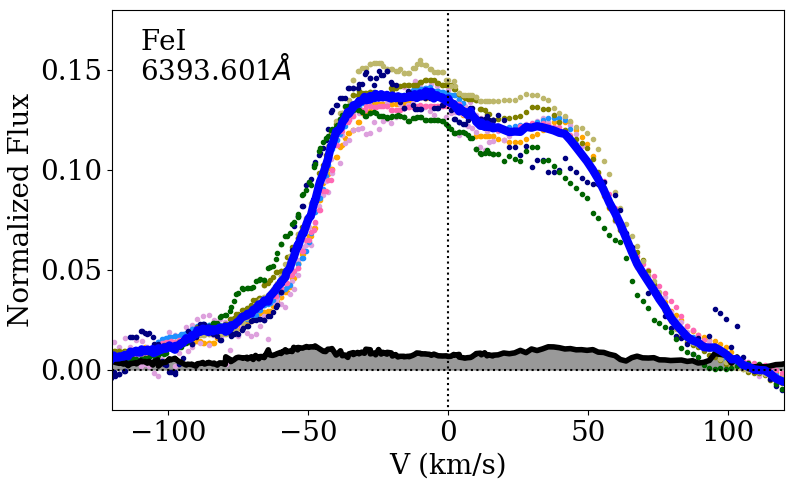} &
\includegraphics[width=4cm]{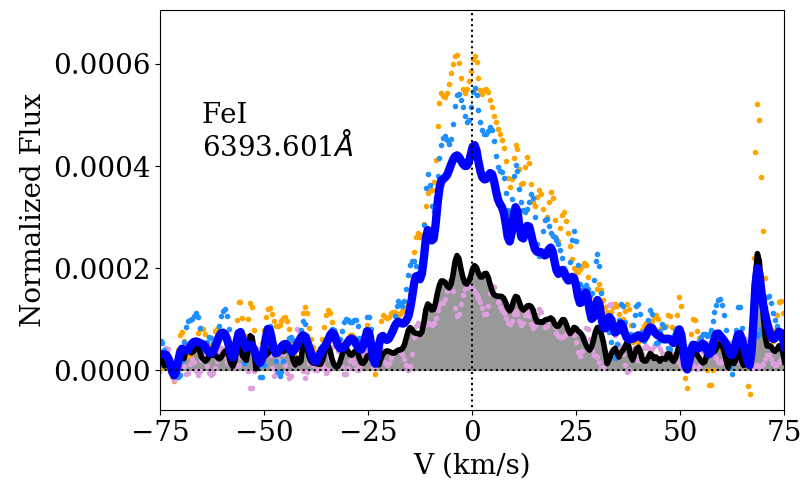} &
\includegraphics[width=4cm]{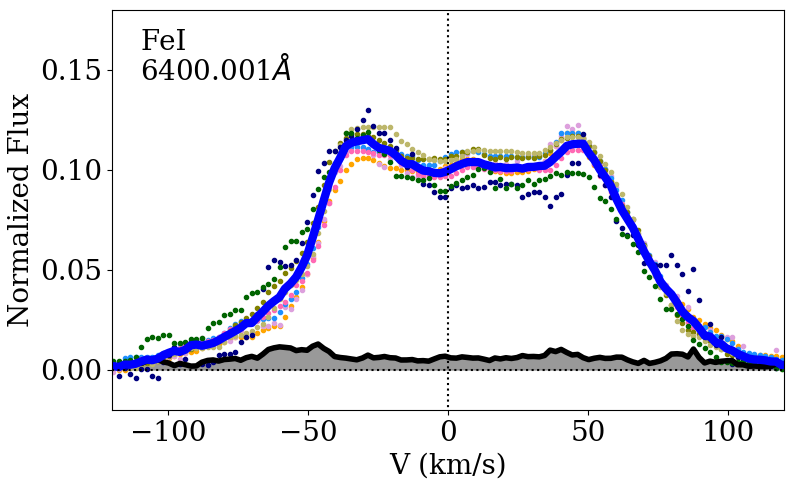} &
\includegraphics[width=4cm]{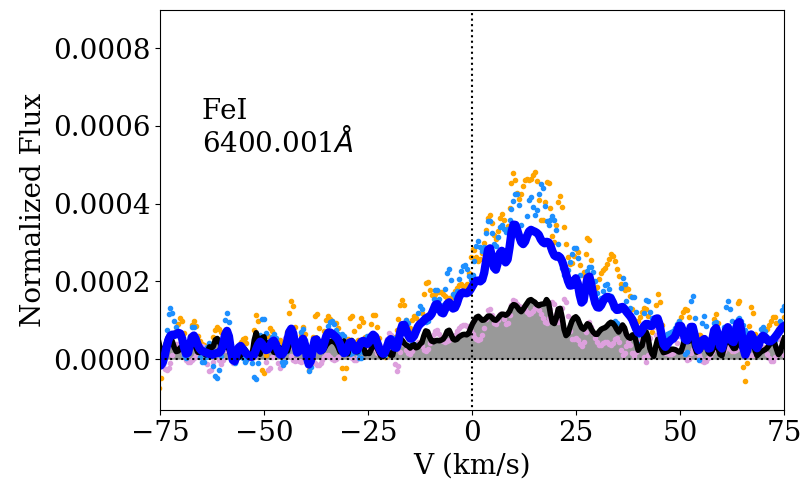} \\
\end{tabular}
\caption{Comparison of the variability observed in several lines in outburst and quiescence. The average flux is plotted in blue, while the standard deviation is shown in black with
grey shadowing. The zero velocity (dotted vertical line) is given with respect to the stellar radial velocity. Note: the Fe I 6400\AA\ line in quiescence may 
contain some contribution of a nearby forbidden [Fe I] line.}
\label{outbqui-fig}%
\end{figure*}

Z~CMa NW in quiescence may not be much brighter than its companion, so the three quiescence spectra
may contain a combination of Z~CMa SE (likely no longer in outburst; Appendix \ref{historical-data}) and the quiescent Z~CMa NW. In
the quiescence spectra, the number of emission lines is still high despite being 
lower than in outburst, but the line profiles are dominated by redshifted emission components
instead of blueshifted absorption.
The lines are weaker, but due to the changes in the continuum and in the wind absorption, 
equivalent widths tend to be higher in quiescence. There is no significant variability 
in the quiescence data (see Figure \ref{outbqui-fig}, the apparent
variability in faint atomic lines is at the noise level).
 This could be due to the short time baseline (2009-10-16 to 2009-10-19), but since the outburst spectra
show variations on consecutive days,
the rapid variations observed in outburst clearly subsided in quiescence.

The maximum velocity observed in the H$\alpha$ red wing decreases from 
$\sim$1000 km s$^{-1}$  in outburst to $\sim$750 km s$^{-1}$ in quiescence, and the wind components 
faster than -400 km s$^{-1}$ are absent from the metallic lines
in quiescence. High-energy lines such as Si II are absent in quiescence, and in the He I lines the emission
component is stronger and the absorption features are shallower in quiescence (Figure \ref{lines-fig}).
The wind during the quiescence stage is not as clumpy as in outburst, 
but dominated by a single, more symmetric, 
wind component instead of several non-axisymmetric components. 
The intermediate-velocity wind (v$_{wind} >-$100$/$-200 km s$^{-1}$) 
is still observed in the 
metallic (including Fe II  and Ti II), H I, and He I lines in quiescence, 
but the strong slow wind (v$_{wind} \sim -$50$/-$100 km s$^{-1}$) 
disappears in quiescence (Figure \ref{lines-fig}). 
This is not unreasonable if the slow wind originates in the disk and the
amount of matter transported through the disk decreases in quiescence,
affecting the wind mass and optical depth, as often found in accretion-driven outbursts.
The low-excitation material also becomes less optically thick during quiescence, which causes most
of the weak lines to become very faint or disappear altogether (see Figure \ref{FeIratio-fig}).  
The pair Fe I
5624/5572 \AA\ is too weak in quiescence to provide physical results, leaving
the temperature largely unconstrained (Figure \ref{lineratios-fig}).
Optically thinner circumstellar material in quiescence 
suggests that the disk
accretion rate has substantially decreased after the accretion burst, 
as has also been observed in CO
lines in other outbursting stars \citep{banzatti15}.

The decrease in optical thickness and column density-velocity gradient in quiescence
reveals that, even if the quiescence lines were dominated by the
FUor component, the structures accreting onto the FUor are significantly less dense than those
of the outbursting Z~CMa NW. Differences in inclination may change the observed line widths and
depths, but unless accretion 
covers a significantly larger fraction
of the FUor, the accretion rate of the FUor inferred from the densities would be
lower, by a couple of orders of magnitude, than the
estimated value of 10$^{-4}$M$_\odot$yr$^{-1}$ for Z~CMa NW in outburst. 
Since  10$^{-4}$-10$^{-5}$M$_\odot$yr$^{-1}$ is a typical
accretion rate for a FUor \citep{hartmann96}, this is an additional sign that, as of 2009, Z~CMa SE had 
returned to quiescence, although it could be still a strong accretor.
We also checked for further signatures of
redshifted absorption, characteristic of nonspherical infall or magnetospheric accretion, in the
less-blended strong Ti II lines observed in quiescence (Figure \ref{TiIIqui-fig}). 
The asymmetric redshifted side of the line is suggestive of redshifted absorption, but no
consistent trend is observed and the
main signature of the Ti II quiescence lines is a strong, narrow wind component at $-$130 km s$^{-1}$.

The [O I] components at $\sim$0 km s$^{-1}$ and $\sim -$400 km s$^{-1}$ seem stronger in quiescence
due to the lower continuum levels,
but their velocities do not change between quiescence and outburst
(see Figure \ref{forbidlines-fig}). The high-velocity component becomes visible in [S II] in quiescence. The
profiles, which are only clearly resolved in quiescence due to contrast, are typical of Herbig-Haro objects. 
The high-velocity, forbidden line 
components could be related to the main wind ($\sim -$400 km s$^{-1}$) and could correspond to
a shock in the wind emission. The second, at $\sim$ 0 km s$^{-1}$, could be related to a shock
along the disk or extended envelope, but the origin is uncertain since Z~CMa SE also drives shocks \citep{bonnefoy17}.

Many the low-excitation lines with box- or disk-like profiles in outburst
are detected in quiescence, but they are narrow  and do not have
double-peaked profiles. Since in quiescence the irradiated disk temperature is expected to decrease rapidly,
we would expect the disk emission to move to faster velocities as the disk cools down, 
but the emitting area (and thus the line strength) would also decrease rapidly. 
Strong neutral lines are often still strong and have complex profiles, suggesting collisional 
de-excitation in a still-high-density environment. Some of the strongest Fe I lines have asymmetric
profiles (e.g., Fe I 6191 and 6393\AA\ lines in Figure \ref{FeIratio-fig}), dominated by blueshifted emission. 
This redshifted asymmetry could be indicative of infall or nonspherical accretion,
although the velocities are very low ($\sim -$ 20 km s$^{-1}$). 
The typical FWHM of 20-30 km s$^{-1}$ for the metallic quiescence lines
would correspond to Keplerian rotation at $\sim$16$\times$(M$_*$/16M$_\odot$) au, at which the disk will be too cold in quiescence. 
Their wings extend out to 
$\pm$50 km s$^{-1}$ and could be related to the innermost disk (3-5$\times$(M$_*$/16M$_\odot$) au), but line width is very small compared to 
infall or rotation velocities around an intermediate-mass star.
The lower luminosity in quiescence suggests
line formation closer to the star. 
If the
lines were associated with small and localized hot spots on the stellar surface, 
we would expect to see some rotational modulation, 
unless the star is seen close to pole-on. The lack of rotational modulation
in the narrow quiescence emission lines suggests that the material
is either at a polar, always-visible spot, or extended in a rather uniform way around 
the star. This could be expected for
unstable accretion  \citep{kurosawa13} or if the star accreted through an axisymmetric 
structure or boundary layer. Stellar rotation could account for the line
broadening if the star is a slow or highly inclined rotator. Further observations over a more extended period of time
would be needed to clarify this issue and, in particular, to determine the longer-term variability during quiescence
that may betray its spatial location.
Summarizing, quiescence brings a dramatic decrease of mass accretion
and wind density, triggering the disappearance of the low-velocity disk wind and non-axisymmetric wind
components.

\section{Outbursts in high-mass versus low-mass stars \label{discussion}}

One of the most surprising results of our analysis is that the spectral behavior of Z~CMa NW in outburst
and quiescence is qualitatively not too different from what has been observed in other EXors 
\citep{herbig07, herbig08, sicilia12,sicilia17,holoien14, audard14}. 
From historical photometric evidence (see Appendix 
\ref{historical-data}), the outbursts of Z~CMa NW are not different from the
outbursts of  lower-mass EXors, repeating in time every few years and showing various patterns in 
time duration and intensity \citep{herbig01, herbig07,audard14}.
In both EX Lupi and Z~CMa NW, the accretion-related wind becomes stronger during outburst, and the
line wings of essentially all lines move toward greater (positive and negative) velocities. Emission
originating from the inner disk is also detected in both objects, although in EX Lupi the 
emitting material within the inner disk
is highly non-axisymmetric (e.g., a single accretion column or disk spiral) and very close-in, 
which results in dramatic day-to-day velocity variations in the broad components of the lines 
\citep{sicilia12}. Non-axisymmetric winds have also been observed in the M5 star ASASSN-13db during outburst \citep{sicilia17}. 
Evidence of changes in the inner disk content after bursts
has been documented  in EX Lupi \citep{banzatti15}, which is consistent with the decrease in optical thickness in
the line-emitting region observed for Z~CMa NW after outburst. The larger luminosity of this outburst agrees with the idea of a continuum of bursting stars, not only in the magnitude of the burst \citep{contreraspena17},
but also in stellar mass.

The emission lines observed in Z CMa NW, EX Lupi, and ASASSN-13db
have statistically indistinguishable distribution of transition probabilities. The
only significant difference arises between the excitation potentials of the lines observed in
Z~CMa during outburst and ASASSN-13db, thus rejecting at the 1\% level the hypothesis that both samples are drawn from
the same distribution, according to a double-sided Kolmogorov-Smirnov test.  This difference in excitation potentials likely results from the large
difference in stellar temperature (B-type vs. M5).   
The similarities point to
a rather uniform range of physical conditions for the material in the inner disk and
accretion structures. However, because  of the differences in the masses and effective temperatures
of the three stars, the regions in which this emission originates ranges from a few stellar radii 
for ASASSN-13db \citep{sicilia17}, to 0.1-0.2 au for EX Lupi \citep{sicilia12},
and to 0.5-3$\times$(M$_*$/16M$_\odot$) au for Z~CMa NW.

The accretion scenario for Z~CMa NW can be also explored through its emission  and absorption lines.
Although contributions from different origins (e.g., nonspherical winds) may affect the extremely
redshifted H$\alpha$ wings, there are some small redshifted absorption asymmetries in the Ti II and Fe I lines (in
the range of 50-150 km s$^{-1}$) and
in the high-velocity wings of accretion-related lines. These asymmetries are mainly observed during outburst in the
higher Balmer lines and in H IR lines (see Figure \ref{lines-fig}, small absorption components appear at $\sim$+200 and 
+400 km s$^{-1}$ in H$\beta$ and $\sim$250 km s$^{-1}$ in H$\gamma$).
These could indicate a high-velocity  infall scenario \citep[e.g.,][]{koenigl91,edwards94,muzerolle98} rather than a classical boundary
layer, where the velocities in the boundary would be much reduced \citep[e.g.,][]{popham93}.
Nonetheless, these velocities suggest a stellar mass much lower than 16 M$_\odot$, even if the magnetosphere was very small.
Using as the maximum infall velocity that observed in the the H$\alpha$ line wings (1000 km s$^{-1}$) 
this would result in a 2.7 R$_*$ or 5.2 R$_\odot$, which is on the small side for magnetospheric accretion
in intermediate-mass stars \citep[e.g.,][]{mendigutia17}, albeit not unexpected in a fast-rotating star with a weak magnetic field. A variation in stellar mass by a factor of two would bring the infall radius to the 
classical infall radius for a low-mass stars \citep{gullbring98}. 

Nevertheless, the corresponding field to ensure a magnetospheric cavity extending as far as the corotation radius for R$_{cor}$=2.5 R$_*$ would be on the upper limit for HAeBe stars.
The minimum magnetic field $B_{min}$ required to support magnetospheric accretion can be estimated as
\begin{equation}
B_{min} = 1.1 \bigg(\frac{M_*}{M_\odot}\bigg)^{2/3} \bigg(\frac{\dot{M}}{10^{-7} M_\odot/yr}\bigg)^{23/40} \bigg(\frac{R_*}{R_\odot}\bigg)^{-3} \bigg(\frac{P}{1d}\bigg)^{29/24} \text{kG} \label{bfield-eq} 
\end{equation}
\citep[see][]{colliercameron93,johnskrull99,mendigutia20} as a function of the rotational period of the star, P, which is a priori not known. 
A corotation radius of 2$\times$R$_*$ requires a stellar rotation period of 0.22 d (for a 16 M$_\odot$ star),
down to 0.44 d  (for a 4 M$_\odot$ star), but the minimum magnetic field needed to support the accretion rate observed in outburst
via magnetospheric accretion would
be at least 8 kG and up to 16 kG for a 4 M$_\odot$ star, since a lower stellar mass also means a higher accretion rate
for the same accretion luminosity. This is much larger than
the existing estimates for the Z CMa pair  \citep[1.2 kG;][]{szeifert10,hubrig15} and the typical fields of HAeBe stars. On the other hand,
assuming that in quiescence the accretion rate drops to the typical HAeBe rate of 10$^{-7}$M$_\odot$/yr,
a magnetic field of 0.2 kG would be able to support magnetospheric accretion for a 16 M$_\odot$ star, and
a lower-mass star would require a weaker field. 

A more detailed estimate of the equatorial magnetic field required could be derived following \citep{bessolaz08} 
\begin{equation}
\begin{aligned}
B_*= & 140 \times 2^{-7/4} m_s^{-1/2} \bigg(\frac{R_t}{R_*}\bigg)^{7/4}   \bigg(\frac{\dot{M}}{10^{-8} M_\odot/yr}\bigg)^{1./2.} \bigg(\frac{M_*}{0.8M_\odot}\bigg)^{1/4}\\
& \bigg(\frac{R_*}{2R_\odot}\bigg)^{-5/4} G,\\
\end{aligned}
\end{equation}
where $R_t$ is the truncation radius,  $m_s$ is the Mach number \citep[$\sim$0.45;][]{bessolaz08}, and the rest of symbols retain their previous meaning. This approach results in an even higher magnetic field 
during outburst (46-65 kG) for a star with a
mass between 4-16 M$_\odot$, which is completely unfeasible. The result for the quiescence accretion rate of 1e-7 M$_\odot$/yr would range 
from 1-1.5 kG for stars with masses between 4-16 M$_\odot$, which although feasible for the more magnetic HAeBe stars, remains
questionable until such a high magnetic field has been confirmed. Therefore, although magnetospheric accretion is highly unlikely in outburst, 
which would be a significant difference compared to lower-mass EXors,
it may be possible in quiescence.

\section{Summary and conclusions \label{conclusions}}

We present optical spectroscopy during the outburst and post-outburst phase of Z~CMa NW, 
using the wealth of emission lines observed to explore the causes of the outburst and the properties
of the accreting star. We confirm that the outburst of Z~CMa NW was related to an
increased accretion episode reaching \.{M}$\sim$10$^{-4}$ M$_\odot$ yr$^{-1}$. This
accretion rate is significantly higher than the current accretion rate of the FUor companion
Z~CMa SE, for which photometric and spectroscopic evidence suggest a return to quiescence by year 2000.
The historical light curve (including AAVSO data since the 1930s) and past spectroscopic activity 
\citep[e.g.,][]{hessman91} suggest that the outbursts of Z~CMa NW
are repetitive, as in low-mass EXors. We use several techniques, including
velocity decomposition to track the spatial location,
Saha's equation and line ratios of lines from the same upper level, to estimate the
density and temperature of the various regions, which include a non-axisymmetric, multicomponent
wind and extended disk emission.

The optical spectra strongly resemble EXor outburst spectra for
lower-mass stars, suggesting a continuum of
accreting behaviors for disked stars with all masses. 
The spectra of Z~CMa  NW in outburst reveal lines with multiple origins. A
 double-peaked line emission originates in a hot disk (detected in neutral atomic lines, especially Fe I). Complex wind
absorption  with several velocity components is most evident in H, He I, and in the higher-energy 
metallic ionized lines such as Fe II,
Si II, and Ti II. There is also some evidence of redshifted asymmetries that could be 
related to infall or accretion structures, which are detected preferentially in H and single-ionized elements such as
Fe II, and Ca II in outburst, and in the asymmetry of neutral lines in quiescence. 
This is not dissimilar to what is seen  in
lower-mass stars, although the velocities also suggest that the stellar mass is likely lower than 16 M$_\odot$. Age and luminosity
considerations suggest a value around 6-8 M$_\odot$. 

The magnetosphere - if present -
is expected to be very small, and it is highly unlikely that the magnetic field is strong enough to support magnetospheric accretion
during outburst, although it may be possible in quiescence if the star belongs to the most magnetic HAeBe stars.
This is significantly different from lower-mass EXors, and challenges the role of a magnetosphere in the triggering of
some outbursts \citep[e.g., as suggested for EX Lupi;][]{sicilia15}.
We do not observe a boundary 
layer in the classical sense either, but rather extended accretion structures. We also observe a more distant
Keplerian, irradiated disk. The energetics of the optical emission lines may impose a limitation to the
temperature we can reach, and UV lines may be needed to trace such a hotter, closer structure.

The disk emission 
observed in Fe I and Ca I lines is best fitted with
an irradiated, flared disk originating in the inner $\sim$0.5-3$\times$(M$_*$/16M$_\odot$) au regions, which have temperatures 
in the 2000-7000 K range and densities that decrease toward lower velocities
(larger radii). Weak Fe II lines are also consistent with disk emission, although they trace higher
temperature regions. For the disk to be dominated by
irradiation, it is necessary that the bulk of energy released in the accretion burst is
deposited at a small radius (e.g., in the innermost disk, accretion structures, or on stellar hot spots). 
There is no evidence of the temperature inversion
associated with viscous heating in strongly accreting systems, but this could happen 
if the part of the irradiated disk
that we observe in the Fe I lines is too distant from the hot, dense, accretion-powered inner disk. The disk-like features of
higher energy and ionized lines from the hotter, innermost regions in the disk 
may be masked by the strong emission and absorption profiles produced by the hot stellar wind.  
We observe a decrease in column density-velocity gradient 
per velocity bin, and for the redshifted components of the lines, decreases 
by at least one order of magnitude between outburst and quiescence, a sign of a lower mass
transport through the disk.

The blueshifted absorption features are consistent with an accretion-powered wind. The fastest 
and slowest wind components disappear
in quiescence, when the wind density decreases. The temperature in the fast wind is 
higher than in the slow wind, betraying an origin very close to the star, while the slow component
could be a disk wind. 
The changes in velocity and depth of the absorption are consistent with 
a clumpy, non-axisymmetric wind with spatially variable opacity, modulated by rotation.
This is similar to the winding winds observed in B-type
stars \citep{prinja88}, but the behavior disappears in quiescence and thus is linked to the outburst rather than to
the spectral type. Our small number of observations does not allow us to 
determine whether the velocity changes display any
periodicity, but the presence of a non-axisymmetric wind is an additional sign that accretion is 
non-axisymmetric as well.

This study, together with our previous work on low-mass stars \citep{sicilia12,sicilia15,sicilia17},
demonstrates that optical spectroscopy can be used to trace the structure, radial
distance, and physical conditions of accretion in stars with a large range of mass. Combined with
historical data, it shows that a similar outbursting behavior can be observed through a very large
range of stellar masses and spectral types (from M5 to B-type), pointing to common outburst mechanisms 
and similarities in accretion and inner disk transport within this whole range, despite potential differences in 
the structure and size of the magnetosphere. The limited energy ranges covered by the lines that can be observed in
optical spectra also point out the need for high-resolution UV and IR spectroscopy data to study the full
energetic range of this kind of objects.

\vskip 0.3truecm
{\bf Acknowledgments:} We thank Steve Shore,  Lynne Hillenbrand, Andrew Collier Cameron, Lee Hartmann, Peter Woitke, and Ignacio Mendigut\'{i}a for inspiring discussion,
as well as the anonymous referee for valuable comments.
ASA is partly supported by the STFC grant number ST/S000399/1 ({\textit "The Planet-Disk Connection: Accretion, Disk Structure, and Planet Formation"}) and by the 2017 Visiting Teacher Campaign at the University of Grenoble.
We thank the former OHP Director, Michel Boer, for granting discretionary time to perform these observations. We thank Anne Eggenberger for performing the second set of spectroscopic observations at OHP during quiescence. This project has received funding from the European Research Council (ERC) under the European Union’s Horizon 2020 research and innovation programme (grant agreement No 742095; {\textit {\it SPIDI}: Star-Planets-Inner Disk-Interactions”}, http://spidi-eu.org, and grant agreement No 740651 {\textit NewWorlds}). 
We acknowledge with thanks the variable star observations from the AAVSO International Database 
contributed by observers worldwide and used in this research, and in particular, observers from the Royal 
Astronomical Society of New Zealand (RASNZ).

%

\Online

\onecolumn

\begin{appendix}

\section{Variability: 90 years of photometric data \label{historical-data}}

\begin{figure*}
\centering
\includegraphics[width=13cm]{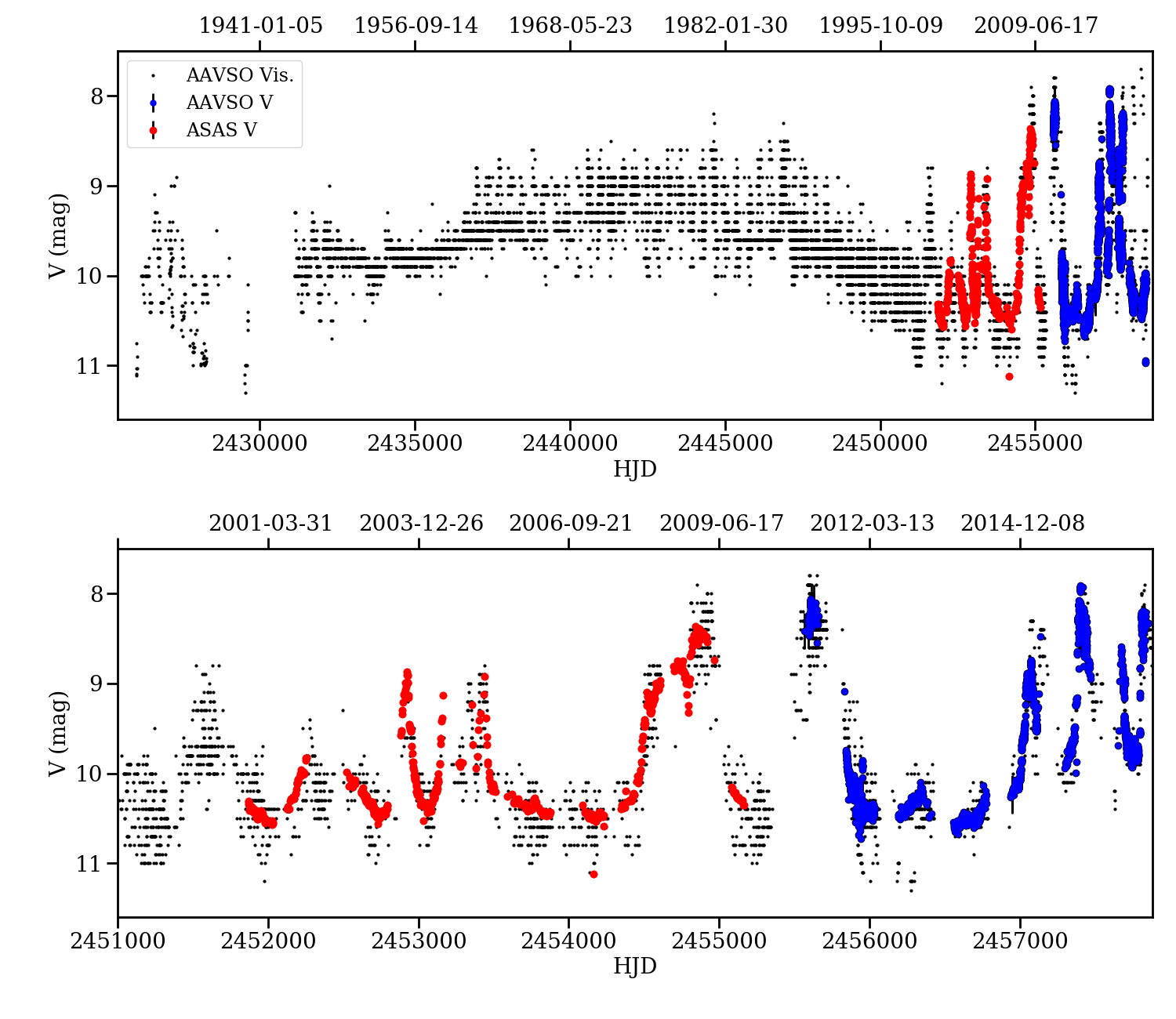} 
\caption{Light curves from AAVSO for Z~CMa since 1930s (upper) and from 1999 to 2018 (lower). Only verified observations are included. The visual and V filters are plotted together, since they are similar for the current purpose of examining the timescales of the outbursts. As a comparison, the ASAS V filter is also plotted (red), which is in excellent agreement with the amateur V data.}
\label{historical-fig}%
\end{figure*}

The AAVSO contains data on Z~CMa since the 1930s. These data provide a great opportunity to examine
the behavior of the Z~CMa pair over nearly 90 years. 
The AAVSO confirms that the light curve is anomalous compared to other FUors \citep{hartmann89,hessman91},
since the documented magnitude rise spans only 2 mags and seem to have progressed very
slowly for the first 30 years.
The light curves are shown in Figure \ref{historical-fig}, suggesting 
that the FUor outburst probably ended by 1995. Since then, the outbursts
of Z~CMa NW have been recurrent and increasing in
brightness, although both the AAVSO data and early mentions of small-scale variability over the FUor 
curve \citep{hessman91} suggest 
that the outbursts may have been occurring at least since the early 1980s, masked in 
part by the FUor outburst of the companion. Thus the P Cygni profiles observed to dominate the 
spectrum during bumps \citep{hessman91} may  have been caused by the HBe star, and not by its FUor companion. In any case, the
bursts appear to have increased in intensity since the 2000s, and deserve some
future follow-up.

\begin{figure*}
\centering
\includegraphics[width=12cm]{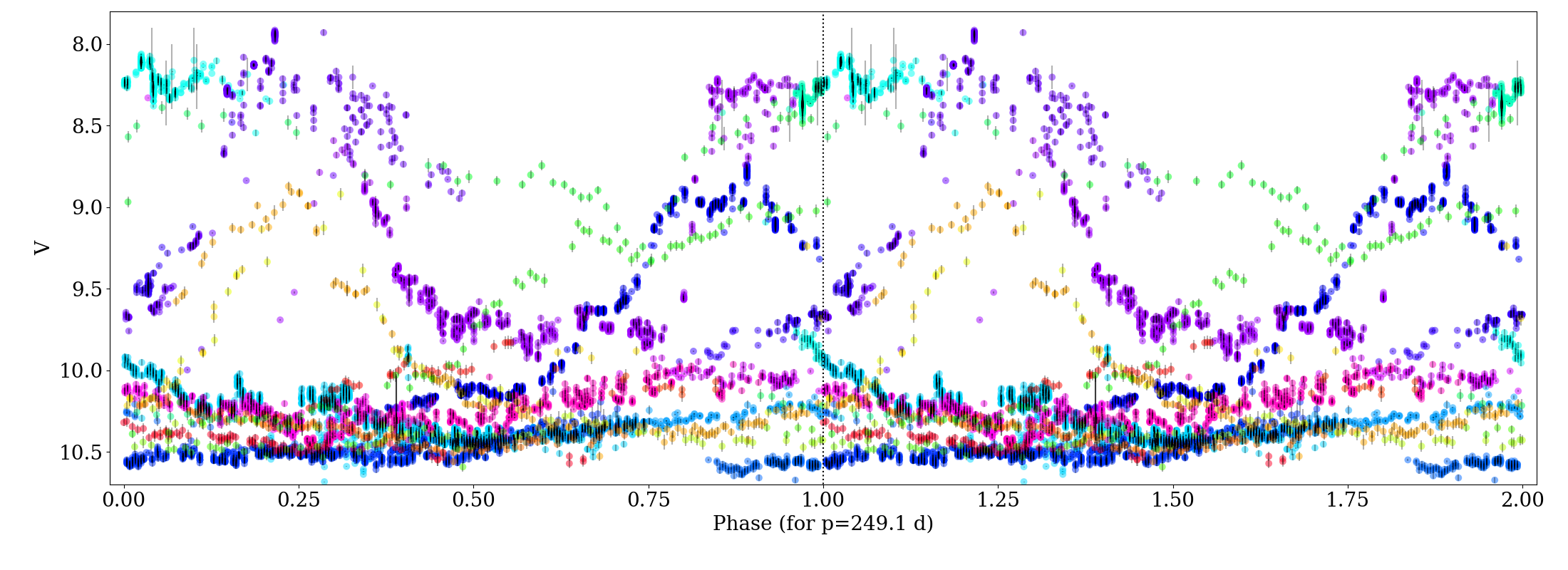} 
\caption{Data from ASAS and AAVSO V  wrapped according to a 249.1 d period. The data are colored depending
on the epoch to enable visualization.}
\label{phased-fig}%
\end{figure*}

A generalized Lomb-Scargle periodogram \citep[GLSP;][]{scargle82,horne86glsp,zechmeister09} 
reveals a low-significance quasi-period of 
249$\pm$20 d (Figure \ref{phased-fig}) when the ASAS and AAVSO V data (covering 14 years) are taken into account.
The modulation appears to change in phase, which suggests a quasi-periodic behavior in a component
that changes at about the same rate as the periodicity is observed. Using a red noise 
model \citep[as done in][]{sicilia20}, the false-alarm-probability of the modulation is $\sim$1\%.
Although this potential period needs to be confirmed in the coming years, it could be related to
repeated disturbances in the innermost disk causing the bursts. 

As seen in Figure \ref{lightcurve-fig}, the two first spectra were taken as the star emerged from
an occultation-like event, in which the rising magnitude experienced a drop of more than 0.5 magnitudes and
then recovered; the whole event lasted about 30 days.  Unfortunately, the spectra were taken
when the star had nearly recovered, so any occultation effects may be mild. These two
first datasets also seem to show distinct wind features, such as a less complex profile more 
similar to the quiescence profile. This could be caused by either a larger contribution of the FUor
spectrum if Z~CMa NW is partly occulted or suggest that part of the wind-originating location
could have been occulted by the material that caused the observed dip. The high temperatures
inferred for the fast wind are consistent with an origin closer to the star, which would be
also easier to be hidden by an occultation event than, for instance, a more extended disk wind. 
The effect is mostly observed in Fe II lines (5018, 4923, 4233 \AA), which
are also some of the most variable lines among those with P Cygni profiles. Nevertheless, since the 
first spectrum has stronger wind absorption than the second spectrum,  the wind
is by itself very variable and these observations were taken about a month before the following observations,
what we observe in the wind may be just part of the usual wind variability unrelated to the occultation event.
Figure \ref{lightcurve-fig} reveals that similar dips during previous outbursts of Z~CMa NW have occurred. They are more evident
during the last ten years, after the FUor outburst had dimmed and during the time when the
activity of Z~CMa NW seems to have increased, and could be related to variable extinction  
suggested by \citet{blondel06}.
Future follow-up may help to understand the origin of the dips and to locate the occulting material and
the regions it affects.

\section{Complete list of emission lines \label{lines-app}}

The line identification was done using the NIST database, except 
for lines denoted with WS, for which the line identification and atomic parameters have been taken from 
Wahlgren \& Shore (S. Shore, private communication) and their observations of the symbiotic star 
V407 Cyg outburst. We compare the observed lines with the lists of \citet{vandenancker04} from observations of
Z~CMa between the late 1990s and the early 2000s (including weaker outbursts) and
the lines observed in two outbursting stars, EX Lupi
\citep{sicilia12,sicilia15} and ASASSN-13db \citep{sicilia17}, for which a similar analysis is published.
The observed profiles were also used to classify a line, or at least to identify the dominant component in case of 
blends or multiple options, when more than one transition was possible. As mentioned in the text, strong Fe II lines (and also Cr II and Si 
II lines) have P Cygni profiles. Weak Fe II lines with low transition probabilities are similar to Fe I lines. 
A few forbidden lines are also observed; [O I] 6300/6363\AA\ is the strongest, followed by [S II].  Some redshifted
outflow emission could be present, but  blends make this possibility uncertain at best. Some [Fe II] lines are detected, but
they tend to be very weak and, in most cases, they appear blended with or very close to other very strong lines so 
that their profiles are hard to compare and the detections remain tentative. Some of the [Fe II] lines may also be observed in quiescence.

\section{Methodology \label{methods-app}}

This appendix contains the details of the different methods used to analyze the emission and absorption lines
 that are applied in Section \ref{analysis}.

\subsection{Deriving physical properties from the spectral lines \label{equations}}

To constrain the physical conditions of lines originating from the same location, 
 we compared the strength of the ionized and neutral lines to put constraints on the density and
temperature using Saha's equation,
\begin{equation}
\frac{n_{j+1}n_e}{n_j} = \bigg(\frac{2\pi m_e k T}{h^2}\bigg)^{3/2} \frac{2 U_{j+1}(T)}{U_j(T)} e^{-\chi_I/kT},     \label{saha-eq}
\end{equation}
where $n_{j+1}$ and $n_j$ are the densities of the ionized and neutral species, respectively; 
$\chi_I$ is the ionization potential;
$n_e$ is the electron density; $m_e$ is the electron mass; $k$ is the Boltzmann constant; 
$h$ is the Planck constant; T is the temperature; and $U_x(T)$ is the partition function 
for the x-ionized ion at a temperature T  \citep{mihalas78}. The main limitations of this approach 
are non-LTE effects and optical depth and self-absorption 
\citep[as observed in EX Lupi,][]{sicilia12}. We expect
at least some of the lines to be optically thick, which is a problem when using line ratios to derive
the physical properties. Nevertheless, the non-detection (or detection) 
of low-energy lines is a very powerful way to set a limit to the temperature for reasonable densities,
or vice versa. For this exercise, we used the atomic parameters and partition functions for different temperatures from the NIST 
spectral line database\footnote{See 
https://physics.nist.gov/PhysRefData/ASD/lines\_form.html, https://physics.nist.gov/PhysRefData/ASD/levels\_form.html} \citep{ralchenko10,kramida18} and made sure that the
line profiles of the compared transitions are very similar in velocity and line profile, to minimize the risk of including emission produced
by different structures.

A further constrain is obtained from the ratio of the transitions of the same ion. 
As a first approach, if we ignore radiative transfer effects and
assume that the lines are optically thin, 
the levels are populated according to Boltzmann statistics, so that
\begin{equation} 
\frac{n_i}{n_0} = \frac{g_i}{g_0} e^{-Ei/kT},\label{boltzmann-eq}
\end{equation}
where $n_i$ and $n_0$ are the densities in the levels $i$ and the ground state, respectively;
$g_i$ and $g_0$ are their statistical weights; and $E_i$ is the energy of the level with respect to
the ground state.

A second calculation can be performed considering the ratio of two lines that share a common upper level.
In such a case, the way the upper level is populated does not matter, and the line ratio depends only on the optical thickness of each line \citep{beristain98}. If we estimate the ratio of
the weakest to the strongest line, the optically thin limit (for which both lines are optically thin)
depends on the line properties alone. 
While the strongest line is saturated, the ratio increases until it reaches a limit once the
weakest line is also saturated. In particular, for isotropic high-velocity gradients, we can use the 
Sobolev approximation to treat the escape probability \citep{beristain98} 
so that the line ratio can be written as
\begin{equation}
r_{w/s} = \frac{A^w_{ki} \lambda_s}{A^s_{kj} \lambda_w} 
\frac{1-e^{-\tau_w}}{\tau_w}\frac{\tau_s}{1-e^{-\tau_s}},    \label{lineratio-eq}
\end{equation}
where A$_{ul}$ is the Einstein coefficient for the spontaneous transition from levels u to 
l, $\lambda$ is the wavelength, and $\tau$ is the opacity, and the indices $w$ and $s$ mark the 
weak and the strong line, respectively. The opacity in the Sobolev approximation for the transition between levels k (up) and i (low) can be written
as\begin{equation}
\tau = \frac{hc}{4\pi} \frac{n_i B_{ik}-n_j B_{ki}}{dv/dl} \approx \frac{hc}{4\pi} B_{ik}g_i \frac{e^{-E_i/kT}}{U(T)} \frac{n dl}{dv},  \label{sobolev-eq}
\end{equation}
where $B_{ik}$ and $B_{ki}$ are the Einstein coefficients for stimulated absorption and emission and 
the approximation can be obtained by neglecting the population in the upper level and assuming thermal
equilibrium \citep{mihalas78,beristain98}. The velocity 
gradient is written $dv/dl$ and $E_i$ is the energy of the lower level.
Rearranging Equations \ref{lineratio-eq} and \ref{sobolev-eq}, we can write the line ratio in terms of
the temperature and modified column density or column density-velocity gradient $\mathcal{N}=n dl/dv$.

\begin{figure*}
\centering
\begin{tabular}{cc}
Outburst & Quiescence \\
\includegraphics[width=5.5cm]{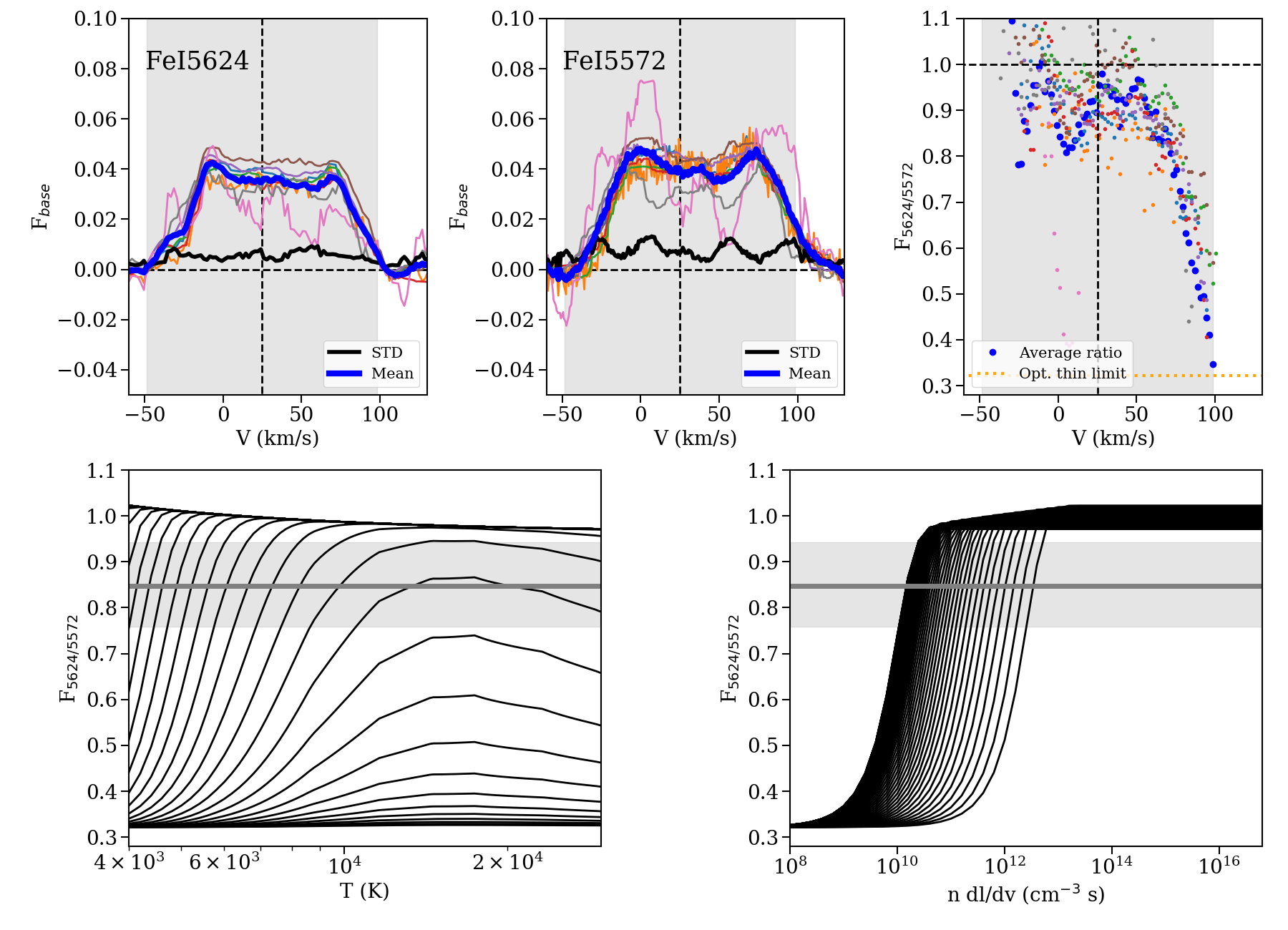} &
\includegraphics[width=5.5cm]{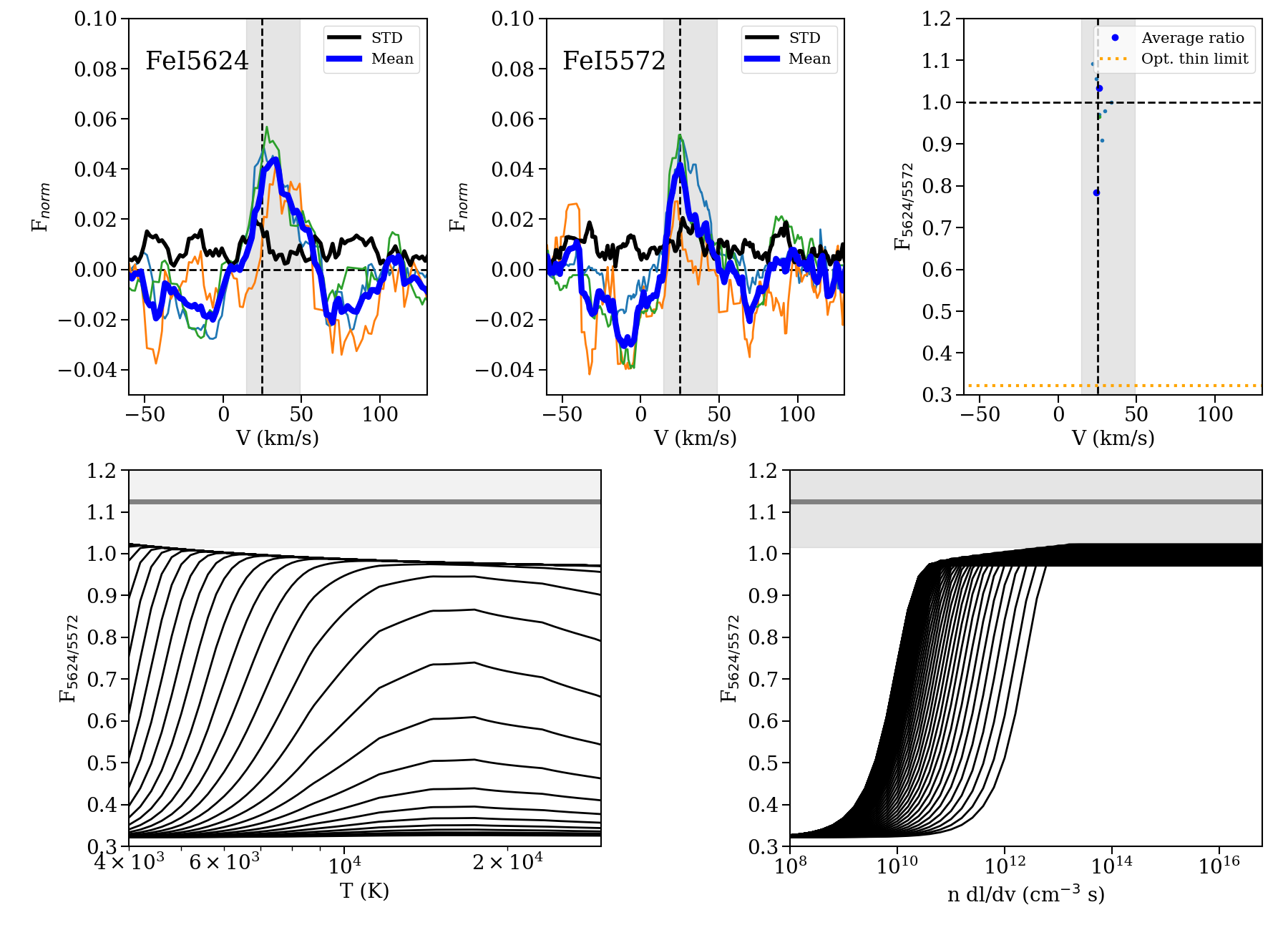} \\
\includegraphics[width=5.5cm]{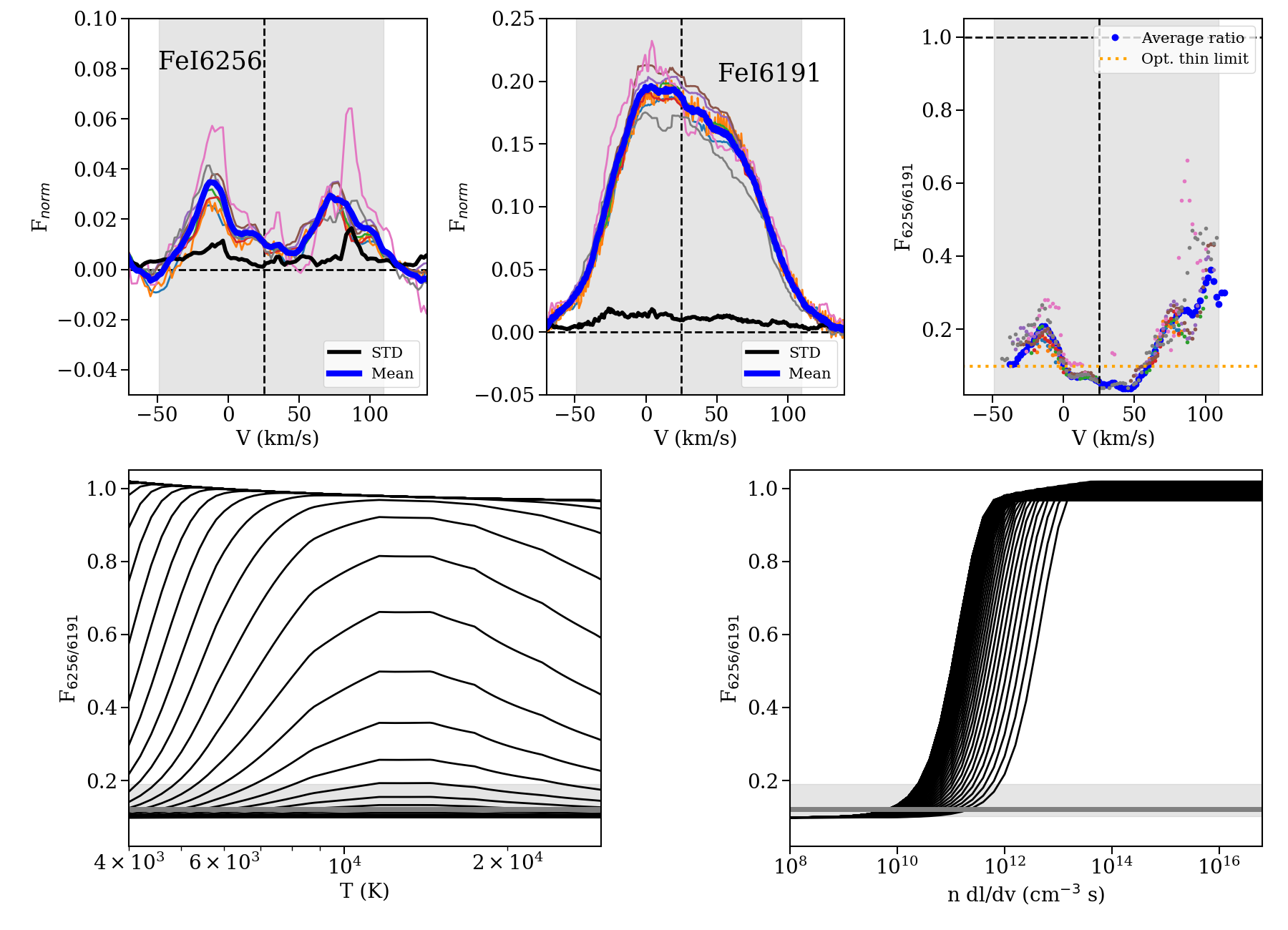} &
\includegraphics[width=5.5cm]{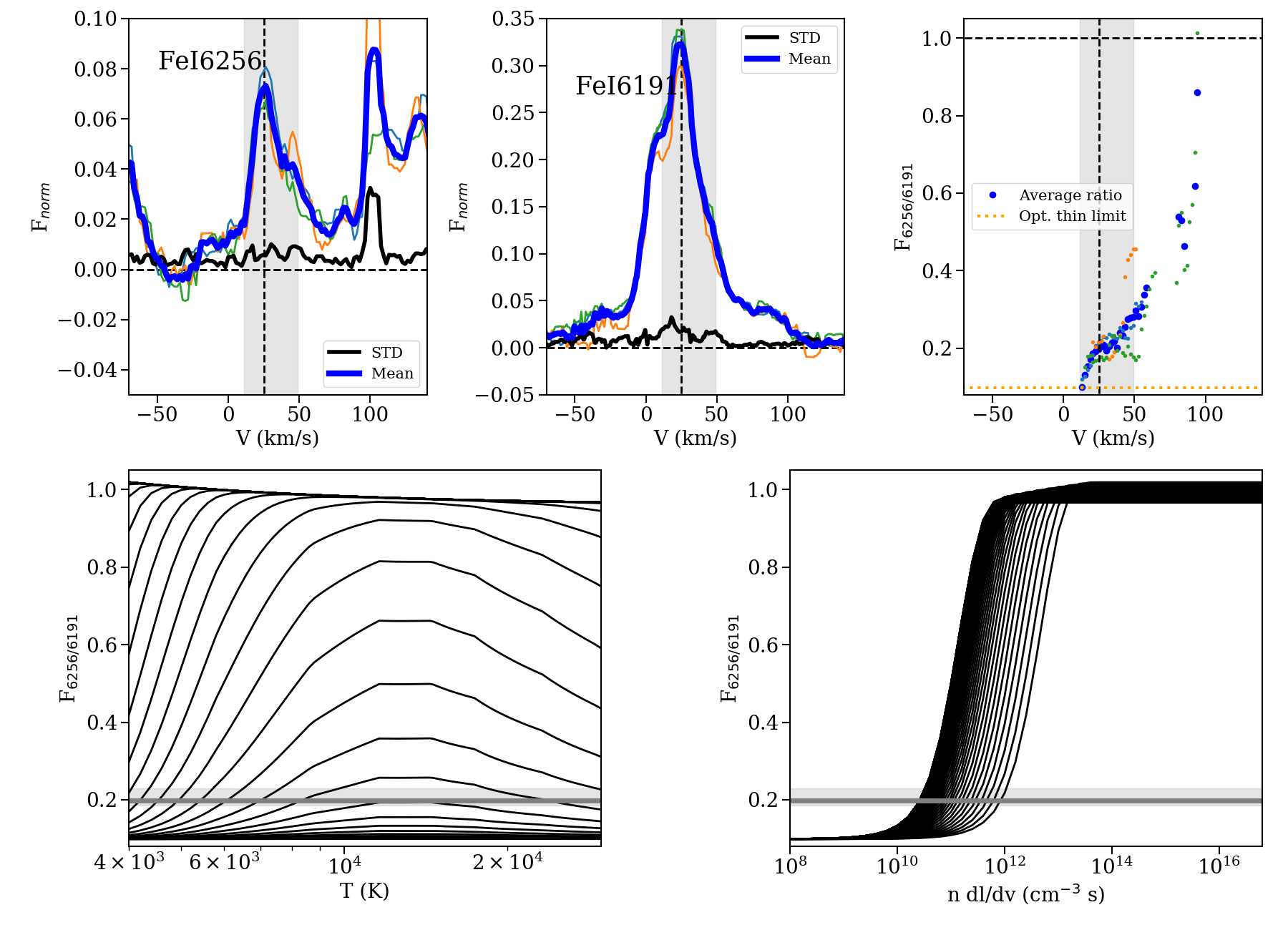} \\
\end{tabular}
\caption{Analysis of the line ratio for Fe I transitions with upper common
levels (Figure \ref{FeIratio-fig}, continued). The upper left panels show the
line emission in outburst; the upper right panel shows the quiescence data. 
The ratios observed  are compared to the theoretical calculations for a range of temperatures and column density-velocity gradients in the lower panels.
The gray area shows the region for which the line emission is $>$5$\sigma$
(in the velocity panels), and the 1$\sigma$ error in the temperature and density planes.}
\label{FeIratio2-fig}%
\end{figure*}

\begin{figure*}
\centering
\begin{tabular}{cc}
Outburst & Quiescence \\
\includegraphics[width=5.5cm]{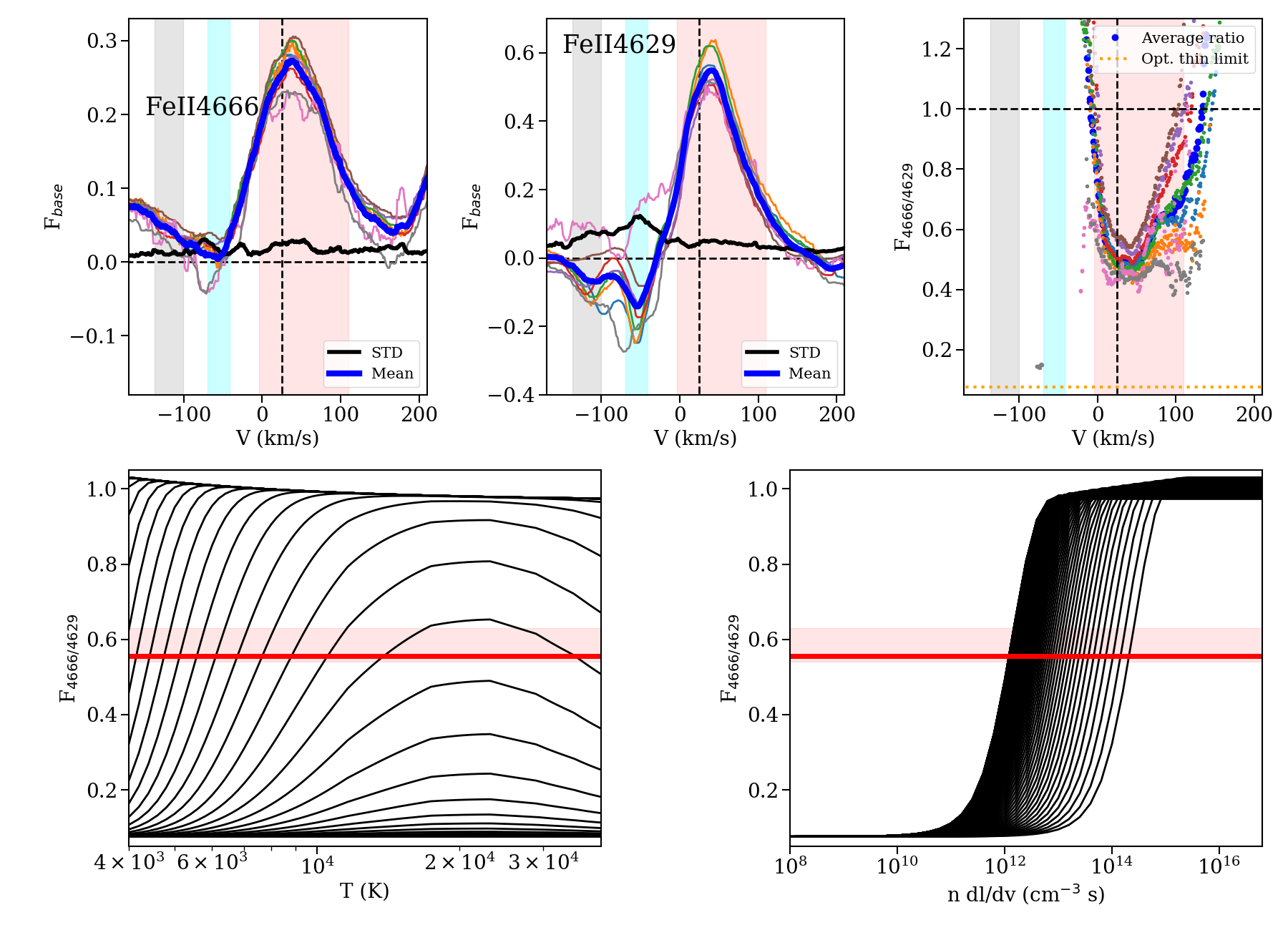} &
\includegraphics[width=5.5cm]{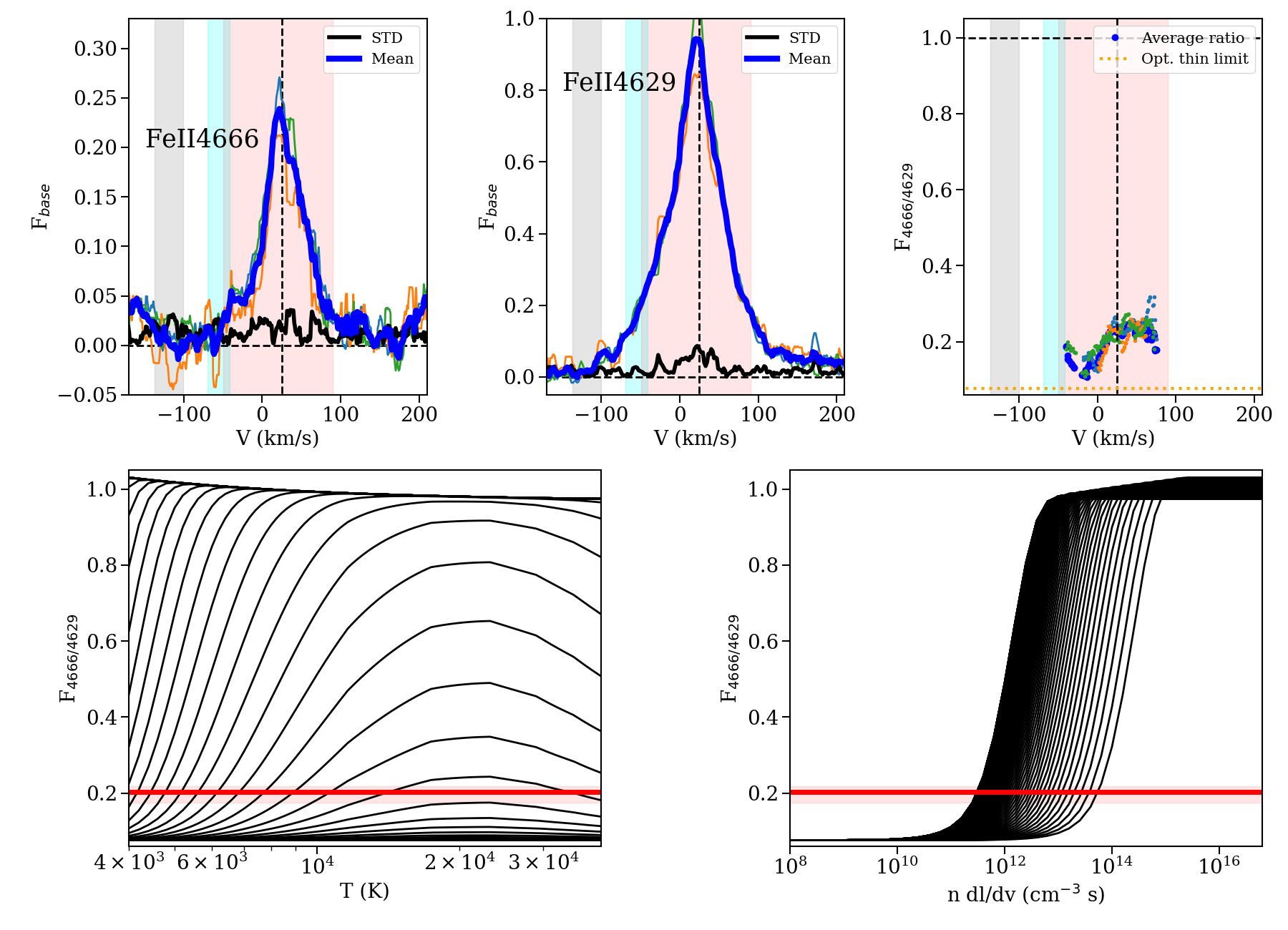} \\
\includegraphics[width=5.5cm]{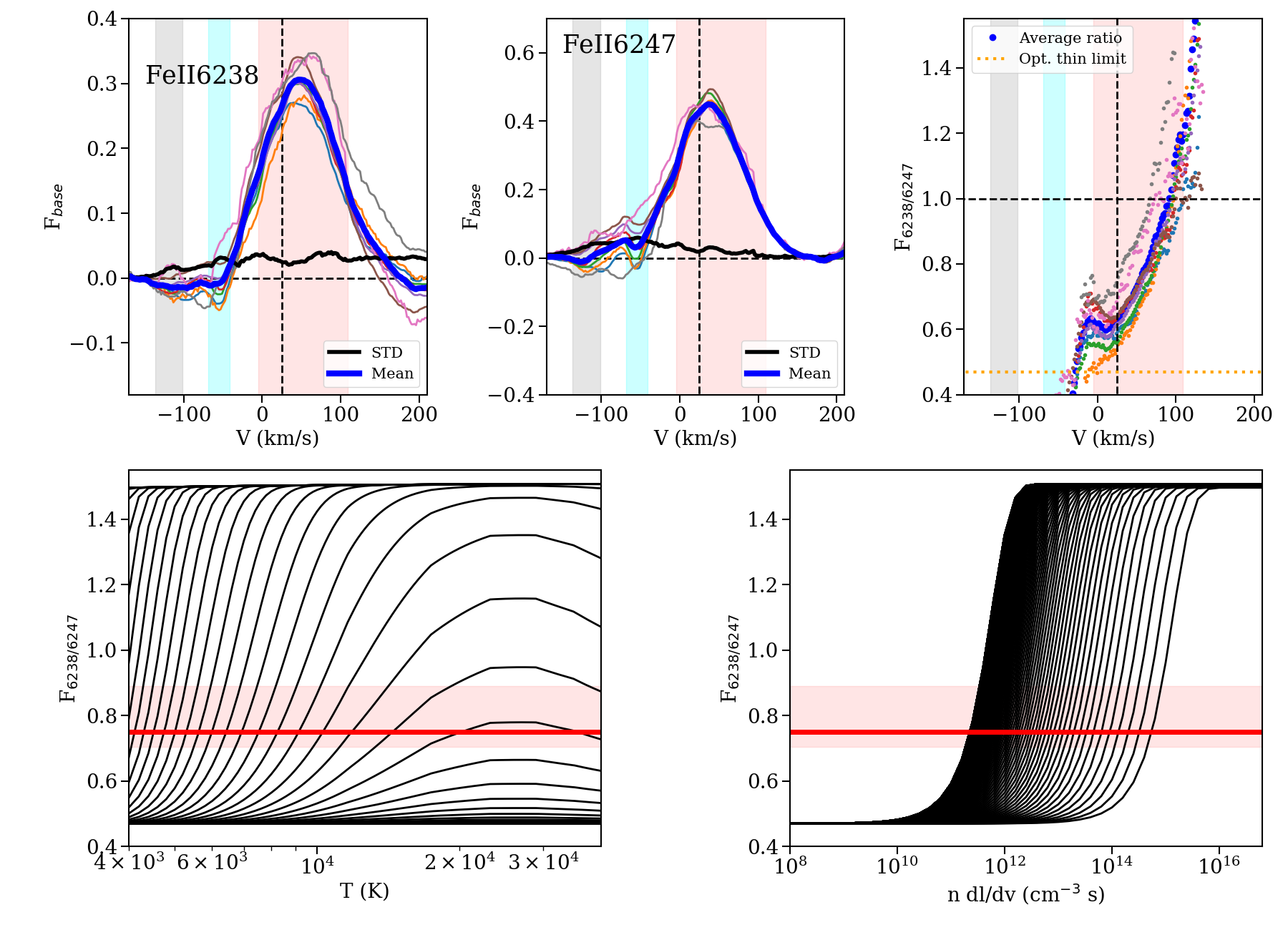} &
\includegraphics[width=5.5cm]{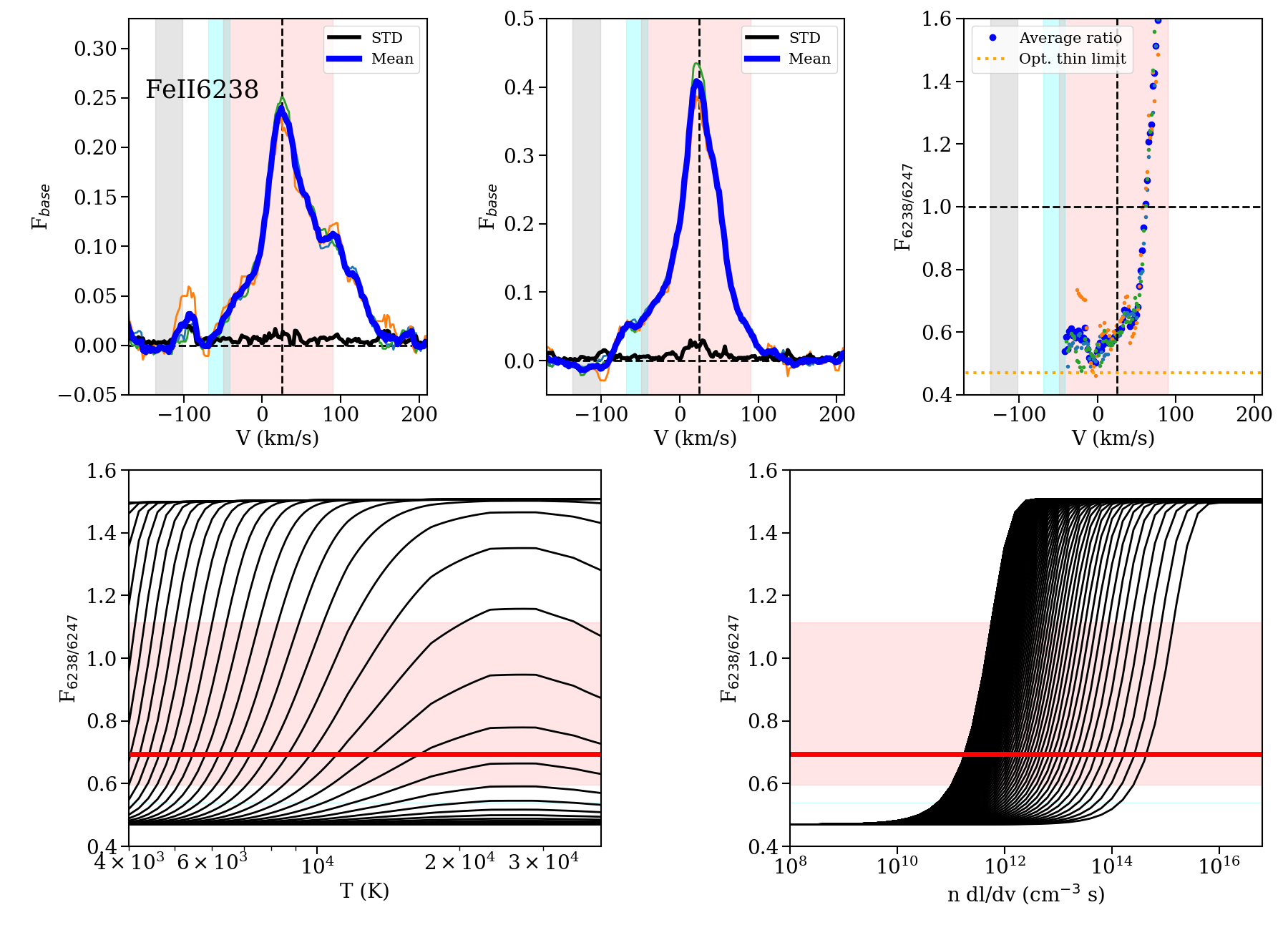} \\
\includegraphics[width=5.5cm]{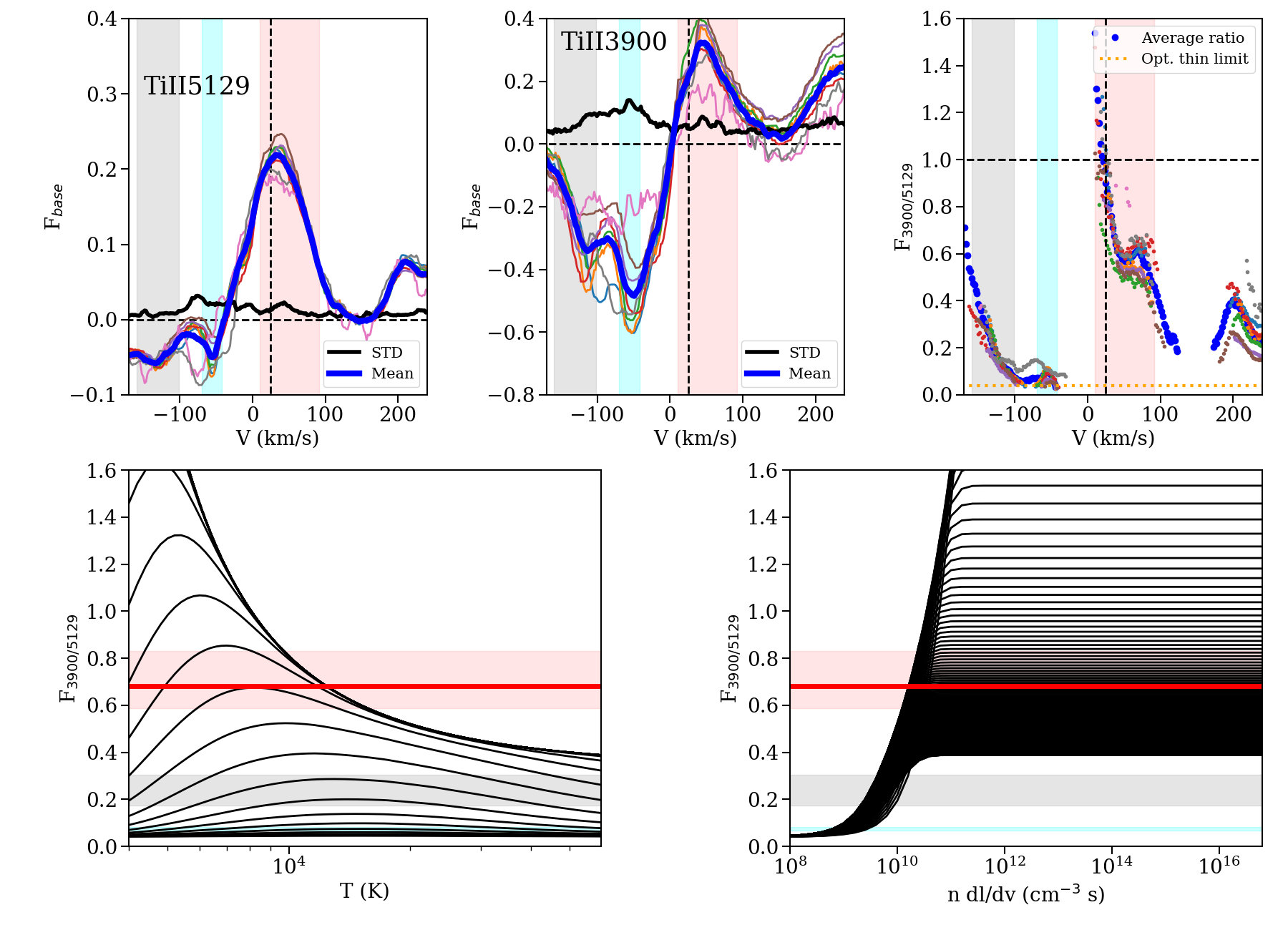} &
\includegraphics[width=5.5cm]{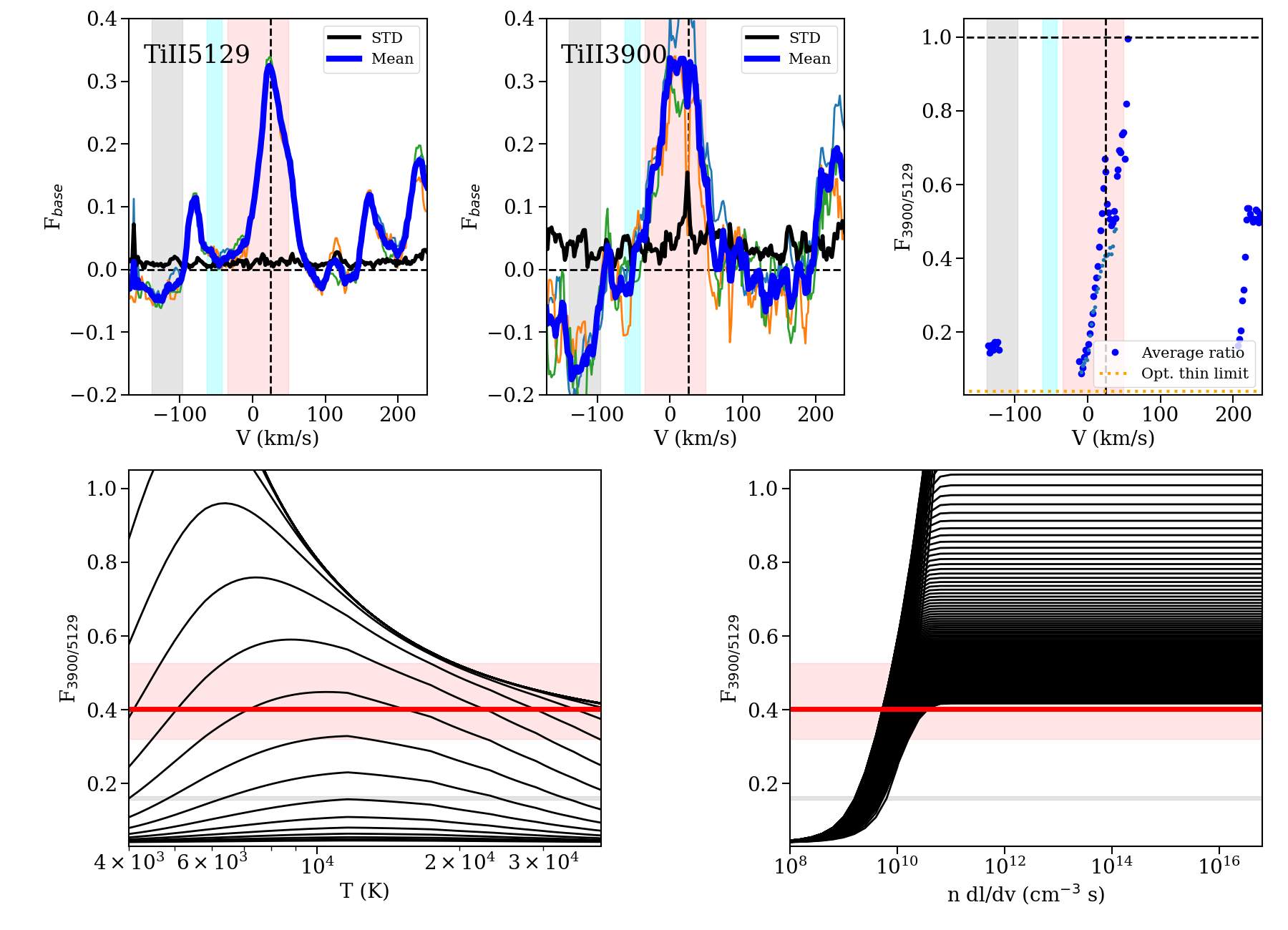} \\
\end{tabular}
\caption{Analysis of the line ratio for the FeII and TiII lines with PCygni profiles from 
transitions with upper common
levels (Figure \ref{Windratio-fig}, continued). The upper left panels show the
line emission in outburst; the upper right panel shows the quiescence data. 
The ratios observed are compared to the theoretical calculations for a range of temperatures and column density-velocity gradients in the lower panels.
In the velocity panels, the red area, cyan, and the gray area show the regions with significant redshifted emission, and slow and fast wind absorption, respectively. The shaded colored
areas in the temperature and density planes show the regions that are compatible with the line ratios observed 
in the emission and the two wind component parts with their 1$\sigma$ error.
}
\label{Windratio2-fig}%
\end{figure*}

Using nearby lines (with wavelengths within a $\leq$500$\AA$ range) is
preferred to avoid effects of underlying variable continuum. We note that the main uncertainty 
related to the atomic parameters  comes from  the B$_{ij}$ parameters, thus the main uncertainties lie in the derived
line opacities. This exercise was done for several velocity ranges along the line profile (see Section \ref{lineratio-sect}).
Some lines are shown in the main text, with the remaining lines shown in Figures \ref{FeIratio2-fig} and \ref{Windratio2-fig}.

\subsection{Disk analysis by brightness decomposition \label{bdecomp-app}}

To explore the location of the material producing the double-peaked emission
lines, we used a procedure similar to \citet{acke06} to convert line intensity in each
velocity bin into a relative brightness per unit area at each velocity. 
Assuming Keplerian rotation,
we can convert the velocities into radial distances. 
One main limitation comes from 
gaseous disks being partly supported by their own gas pressure, so the rotation
of hot atomic gas is
sub-Keplerian. The thermal deviation from Keplerian rotation depends on the
temperature, and this quantity ranges from $\sim$11 km s$^{-1}$ for a temperature of 10$^4$ K, to
$\sim$3.5 km s$^{-1}$ for a temperature of 10$^3$ K. This is a small fraction of the
total velocity span observed in the lines, justifying the approximation.
We fit the ESPaDOnS data only, since the OHP data are too noisy for the fit.

For the brightness decomposition, we assume a disk inclination of 30 degrees and
the outburst stellar values (R$_*$=1.93 R$_\odot$, M$_*$=16 M$_\odot$, L$_*$=8$\times$2400 L$_\odot$). A different inclination or the stellar mass can be
easily rescaled in the brightness decomposition plots, as long as the luminosity is kept 
constant\footnote{ A different
inclination $i$ would change the derived radii by sin$^2 i$/sin$^2(30)$, and a different stellar mass M$_*$ would change these derived radii
by a factor of M$_*$/16M$_\odot$, but the relative brightness would remain the same.}.
With these values we can estimate an effective temperature T$_{eff}$ , assuming that the stellar radius
does not change  during outburst with respect to the quiescent value, 
that reproduces the observed luminosity. We can also estimate the temperature at various distances
from the star, which depends on the heating mechanism and disk structure. 
We assume that the temperature of the innermost disk rim (the part producing
the highest observed velocities) is set by direct irradiation, and for the rest of the 
disk it follows a power law with radius.
Although the temperature estimate would be smaller if 
the emitting zone is larger (e.g., if the innermost disk is the emitting region),
at larger distances the only quantity that matters is the total luminosity (or R$_{emit}^2$T$_{emit}^4$)
as long as the emitting region is located at a smaller radius than
the detected disk emission\footnote{We note that we do not see much emission at velocities that would 
correspond to Keplerian rotation at 
radii below 0.5$\times$M$_*$/16M$_\odot$ au, although 
such emission may be masked by other effects including the innermost parts having a relatively 
small area compared to the more extended disk, or if there is temperature 
inversion (e.g., due to viscous dissipation).}.

\begin{figure*}
\centering
\begin{tabular}{cccccc}
 ~~~~~~~~~~~~~~~~~~~~~~~~~~~~~~ & {\bf Fe I 6400 \AA}& ~~~~~~~~~~~~~~~~~~~~~~~~~~~~~~ & ~~~~~~~~~~~~~~~~~~~~~~~~~~~~~~ & {\bf Fe I 6462 \AA} &  ~~~~~~~~~~~~~~~~~~~~~~~~~~~~~~\\
\end{tabular}
\begin{tabular}{cccc}
$\alpha$=-3/4 & $\alpha$=-3/7 & $\alpha$=-3/4 & $\alpha$=-3/7\\
\includegraphics[height=3.9cm]{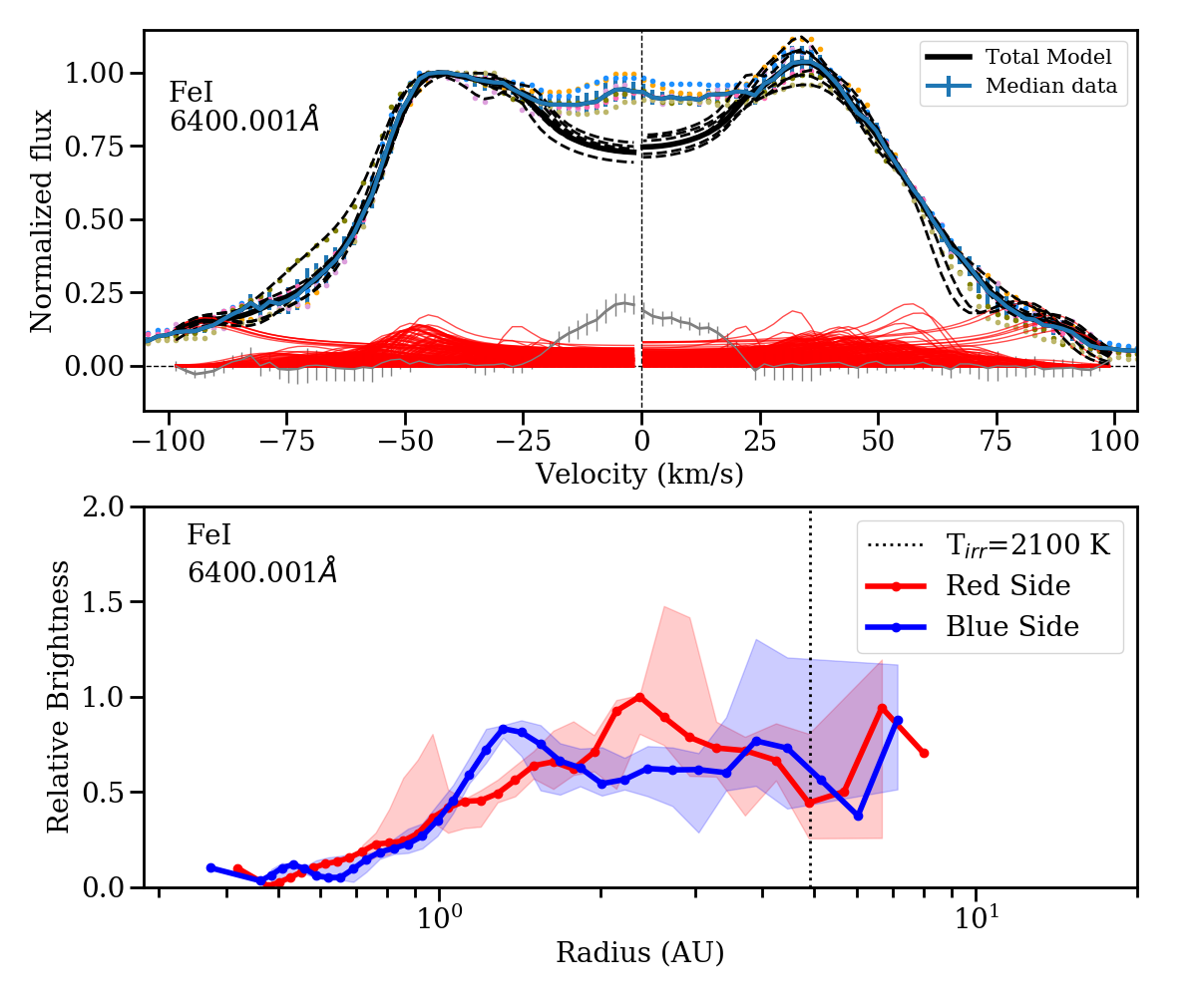} &
\includegraphics[height=3.9cm]{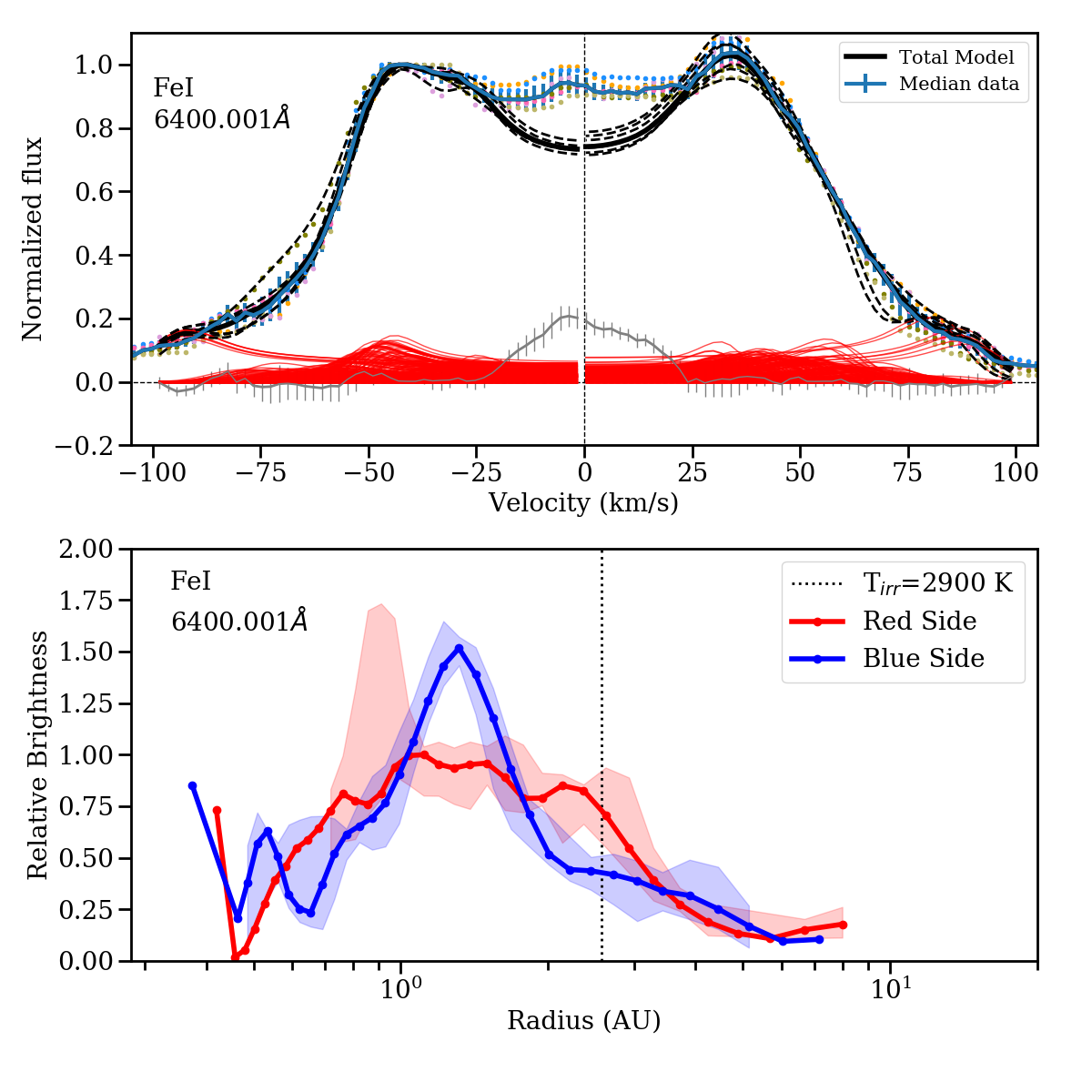} &
\includegraphics[height=3.9cm]{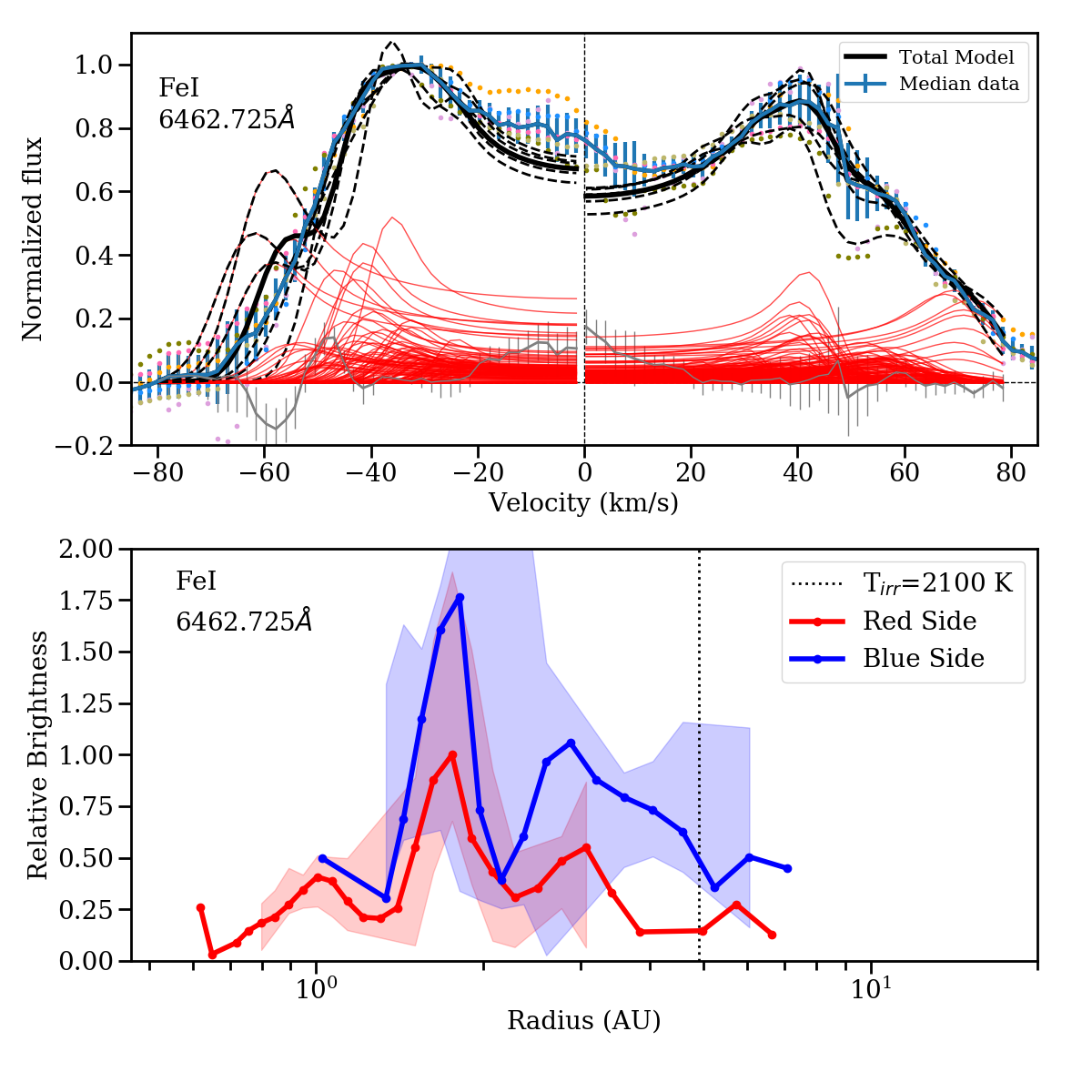} &
\includegraphics[height=3.9cm]{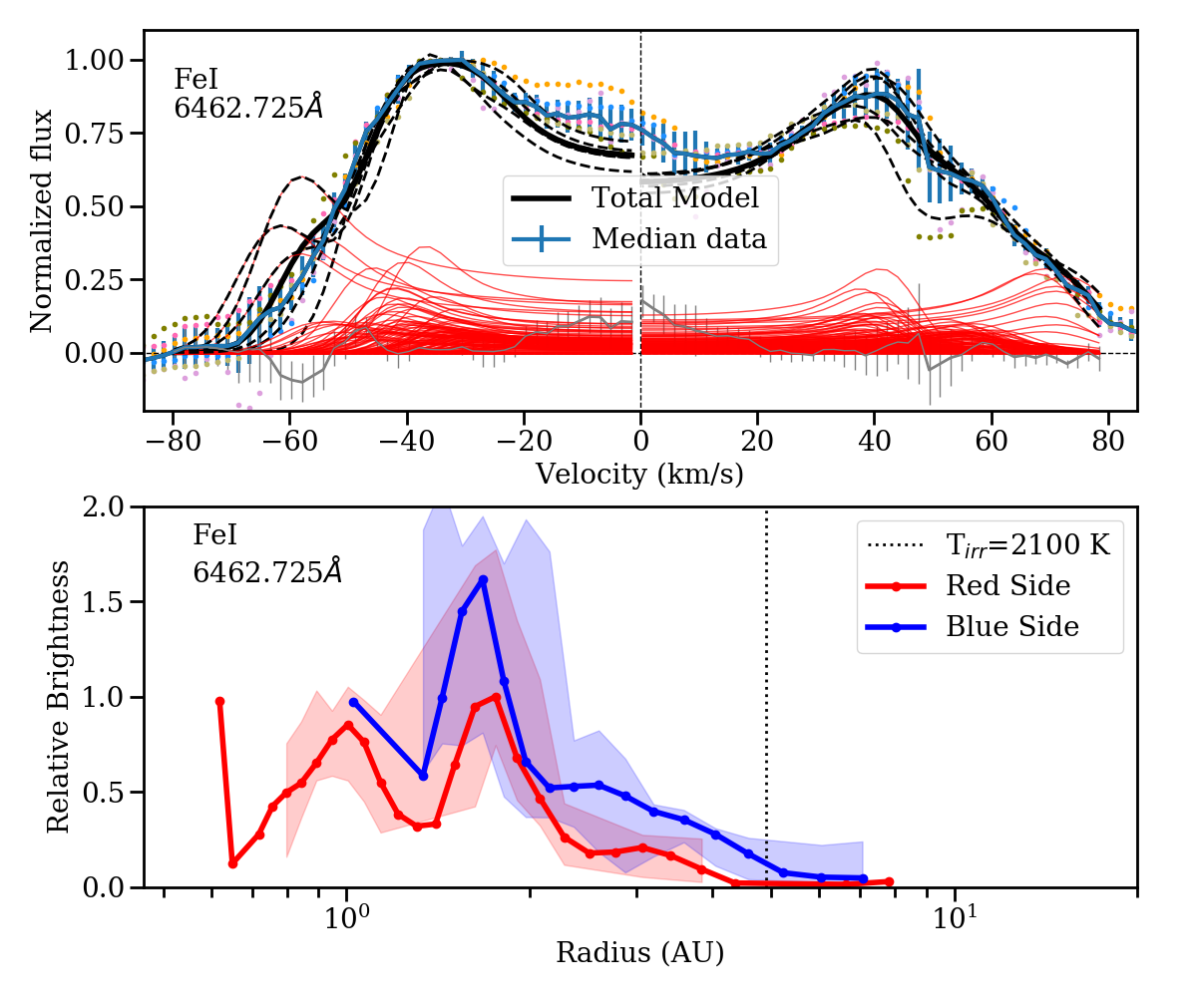} \\
\end{tabular}
\caption{Examples of how the temperature power law affects the radius derived from the brightness
profile. For each line, the left models show the results from a flat disk \citep[power-law exponent -3/4;][]{chiang97},
while the right models show the results assuming a flared disk \citep[power-law exponent -3/7;][]{dalessio99}. The difference is larger for lines with higher transition
probabilities (A$_{ki}$=9.3e6 s$^{-1}$ for FeI 6400\AA, A$_{ki}$=5.6e4 s$^{-1}$ for FeI 6462\AA) that are likely to be formed higher over the disk midplane.
Colors and symbols as in Figure \ref{brightness-fig}.}
\label{diskincli-fig}%
\end{figure*}

The temperature power law is poorly known, since we can have various sources of heating (e.g., 
irradiation, viscous dissipation) and gaseous disks are flared and have a vertical temperature structure. 
We tested several temperature versus radius power laws. These include the \citet{chiang97} 
structure for an irradiated flat disk, for which the temperature in
the disk 
is taken to be a power law of radius with exponent -3/4, which is also the
same power law than if we assume a self-luminous disk powered by viscous
dissipation\ \citep[e.g.,][]{hartmann98};
and the flared irradiated disk model, in which the temperature varies
as a power law  with an exponent -3/7 versus radius \citep[][]{dalessio99}. 
The result of using a steeper versus a 
shallower power law is that the strongest relative brightness moves to larger radii 
when the temperature decreases faster (i.e., if the
temperature is lower, a larger area is needed to reproduce the observed flux),
preserving the general shape of the profile and the blue/red (a)symmetry.
Figure \ref{diskincli-fig} shows
some examples of the effect in the derived brightness profiles depending on the
power law assumed. 
In general, the residuals of the flared disk models
are smaller than those of the flat disk fits, thus we conclude that an irradiated flared
disk model can fit better the velocity profiles observed, and this is what we
use through Section \ref{diskanalysis-sect}.

Finding Fe I emission in LTE is unlikely at temperatures below a thousand degrees. 
For a flat, irradiated disk and the stellar luminosity in outburst, 
T=1000 K is reached at about 20 au. Even though other sources of heating (e.g., 
thermal dissipation due to accretion) are likely to operate in this strongly
accreting system, we impose a cutoff at 35 au. Beyond this cutoff we deem that the 
emission becomes unreliable, especially for the strongest lines that often have
an emission core at low velocities that cannot be separated from the disk emission\footnote{Since
the spectral resolution is finite and the Keplerian velocity decreases with radius, this means
that the emitting radius (and associated brightness) becomes less and 
less well determined (and the area can become unrealistically
large) as we move to lower velocities, which
is an additional reason to impose a cutoff at large radii/low velocities}. We also estimated the radial velocity of the line using the 80\% level of the double peak of the line to estimate the line center,
which typically produces radial velocity values between 20-35 km s$^{-1}$, in agreement with
the narrow quiescence lines. A highly asymmetric line and/or
eccentric or asymmetric disk could lead to inaccurate values of the radial velocity.

Following \citet{acke06}, we decomposed the normalized line profile starting from the 
highest velocity at which the flux is $\geq$3$\sigma$. We assume that the emission 
at each velocity originates in a narrow ring
located at the position with the same Keplerian orbital velocity.
We generated simple disk model emission, assuming that each element of the velocity ring 
emits a line where the width is due to thermal broadening. 
The emission of the ring is proportional to the blackbody emission at the
relevant temperature and to the area of the emitting ring. We constructed the emission of the
ring by summing over all angles, breaking the ring into a 70 point grid in radius and
azimuthal angle. The ratio of the observed to theoretical flux at each 
velocity point
gives us the brightness associated with that ring. The resulting ring is subtracted from the total observed line, and
the procedure is repeated for the next highest velocity until the line has been completely
fitted or until we reach the 35 au cutoff. 

\begin{figure*}
\centering
\begin{tabular}{cccc}
\includegraphics[width=4.0cm,height=4.0cm]{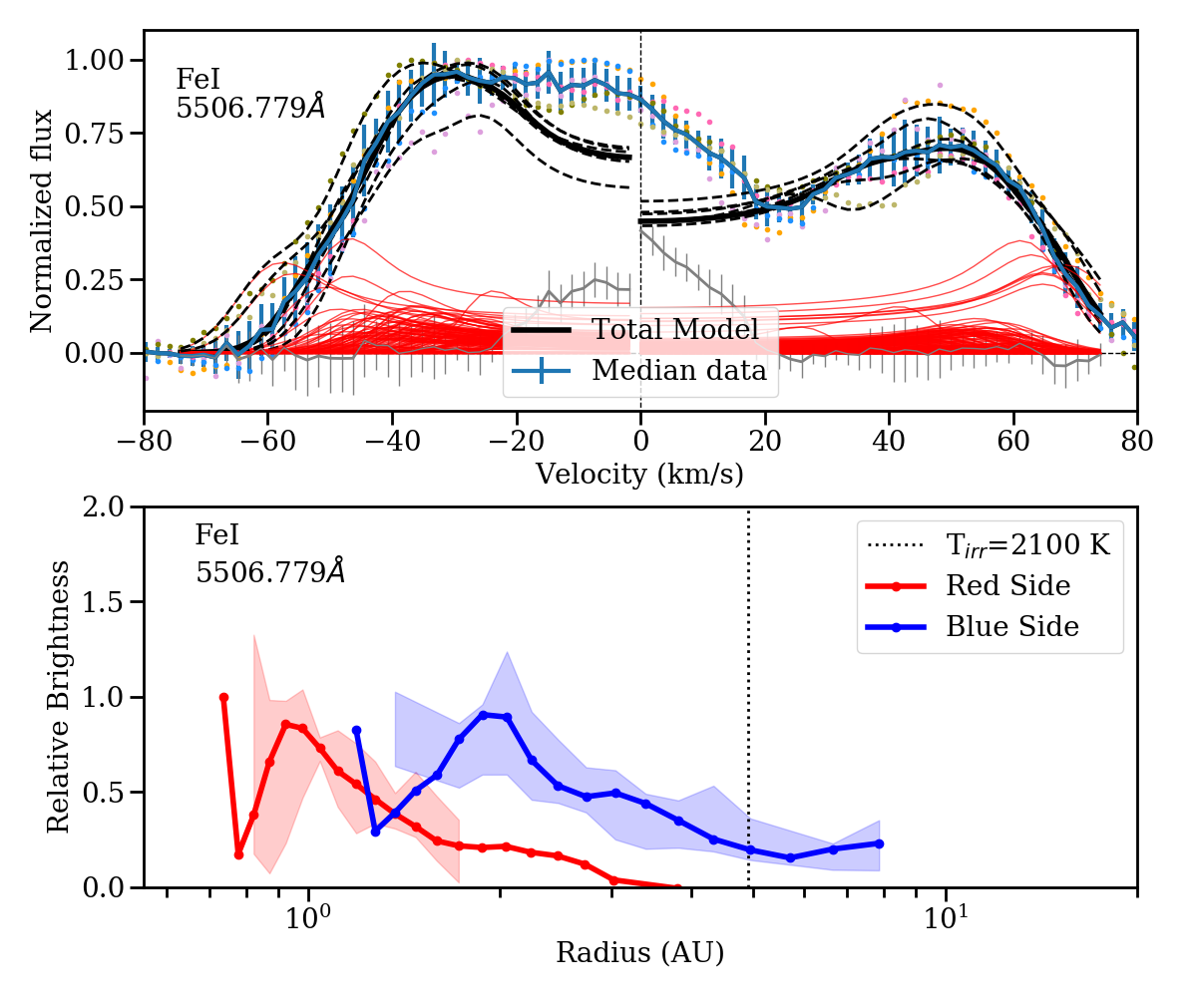} &
\includegraphics[width=4.0cm,height=4.0cm]{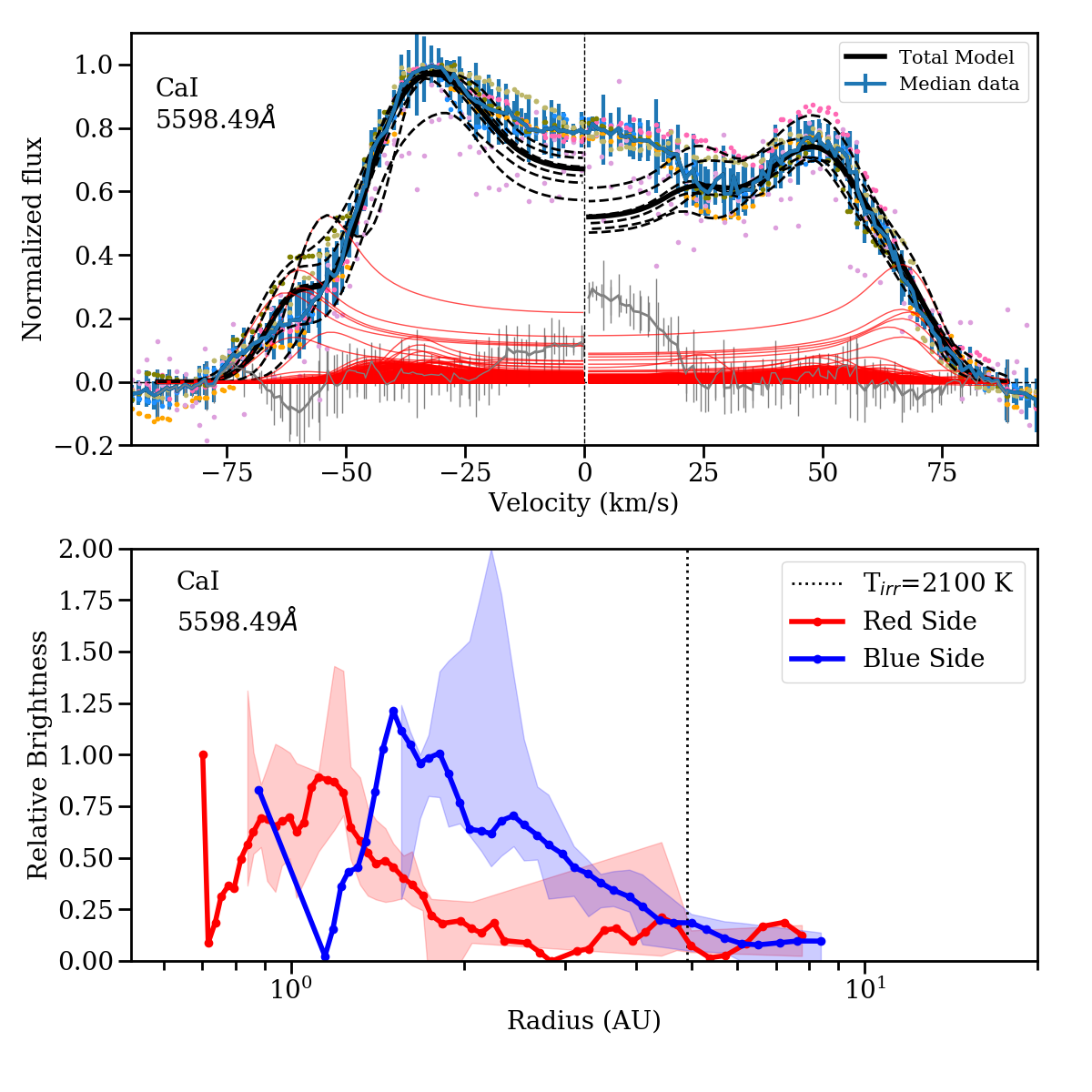} &
\includegraphics[width=4.0cm,height=4.0cm]{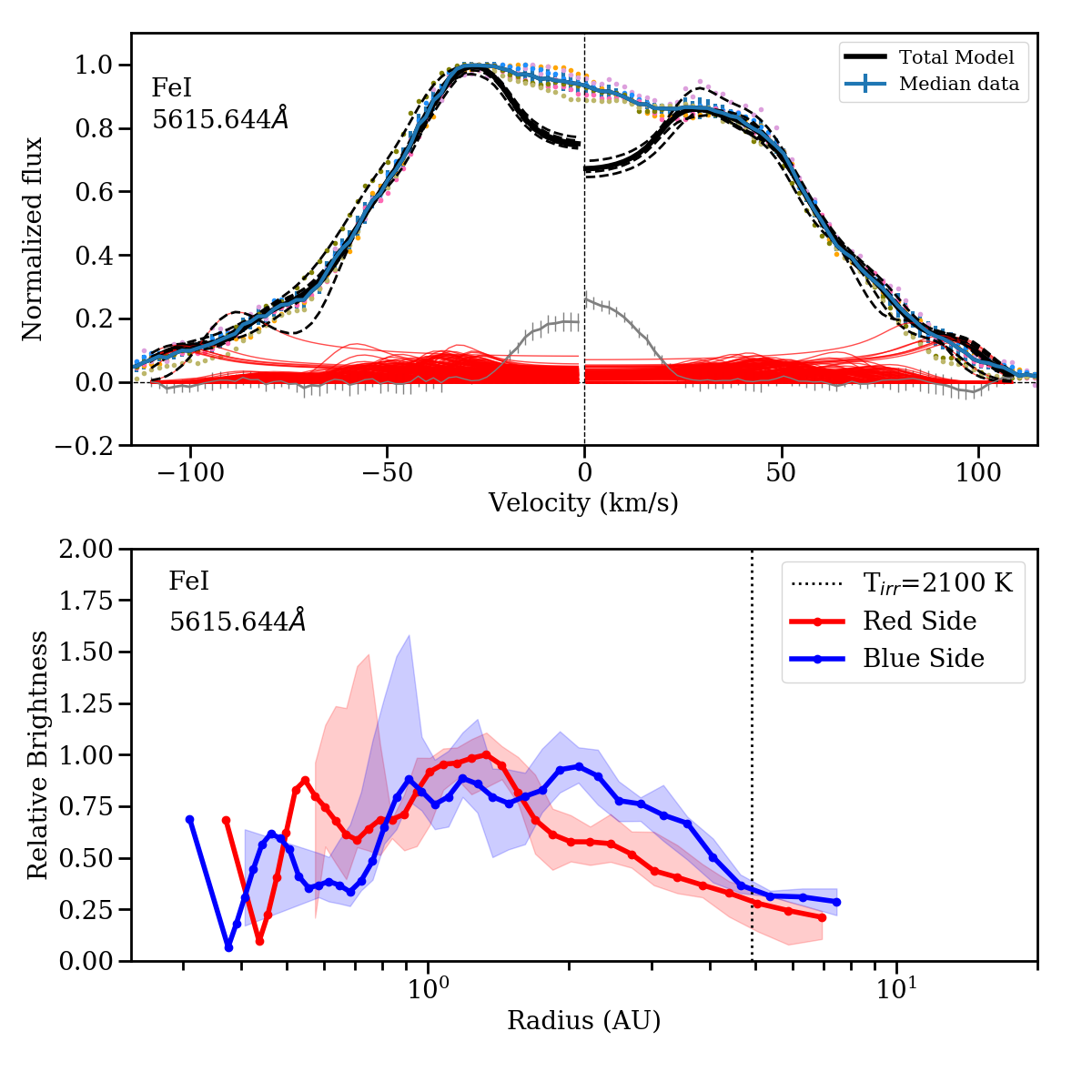} &
\includegraphics[width=4.0cm,height=4.0cm]{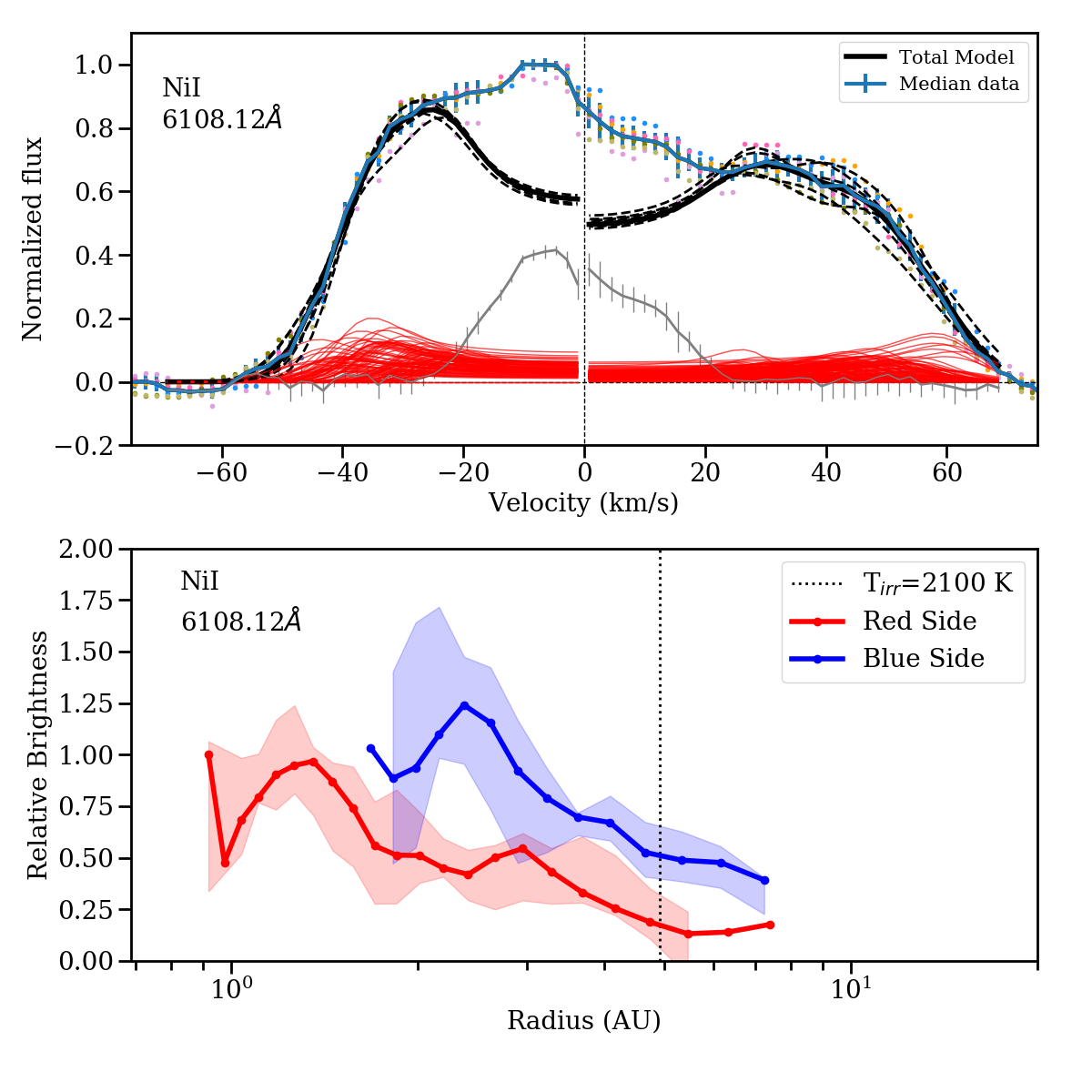} \\
\includegraphics[width=4.0cm,height=4.0cm]{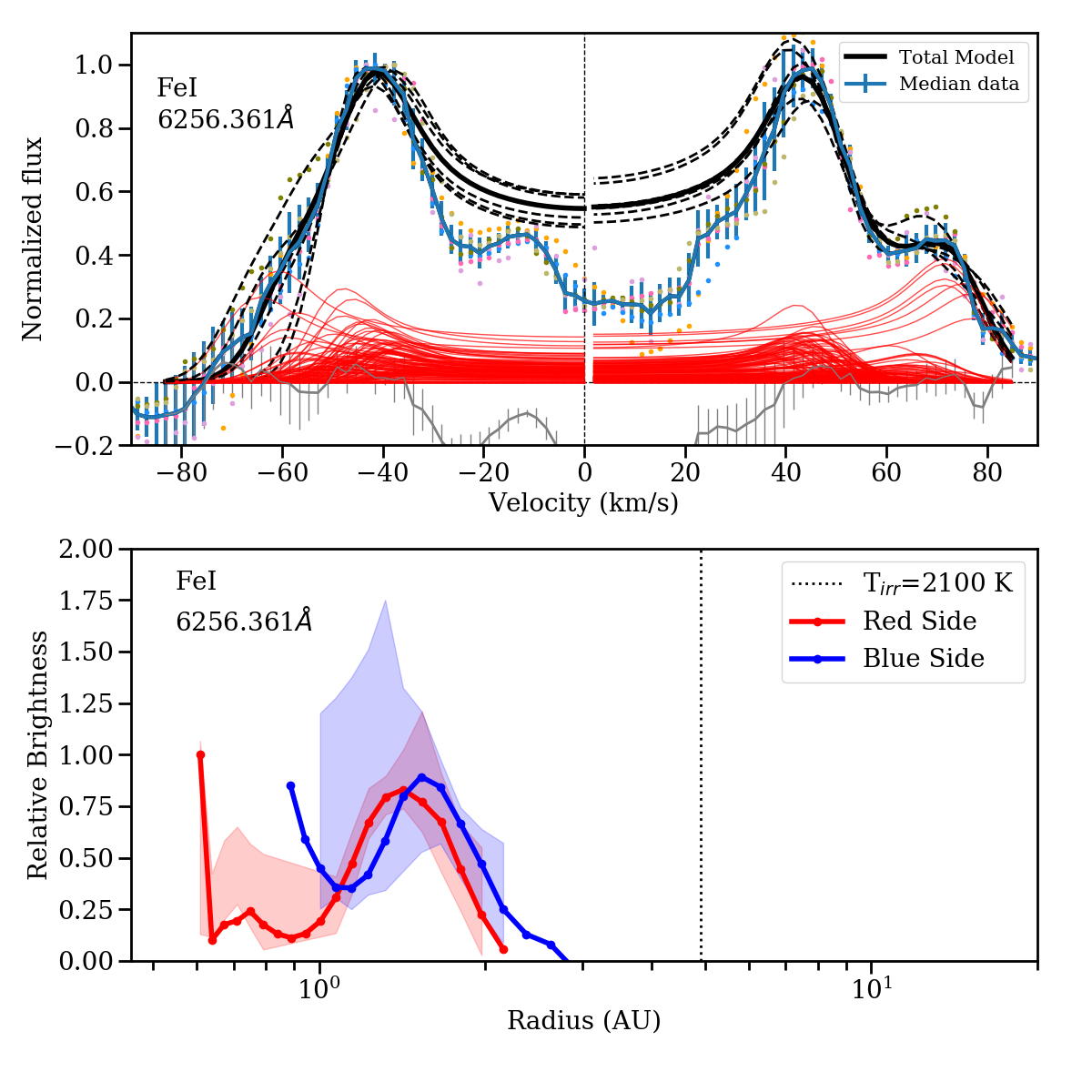} &
\includegraphics[width=4.0cm,height=4.0cm]{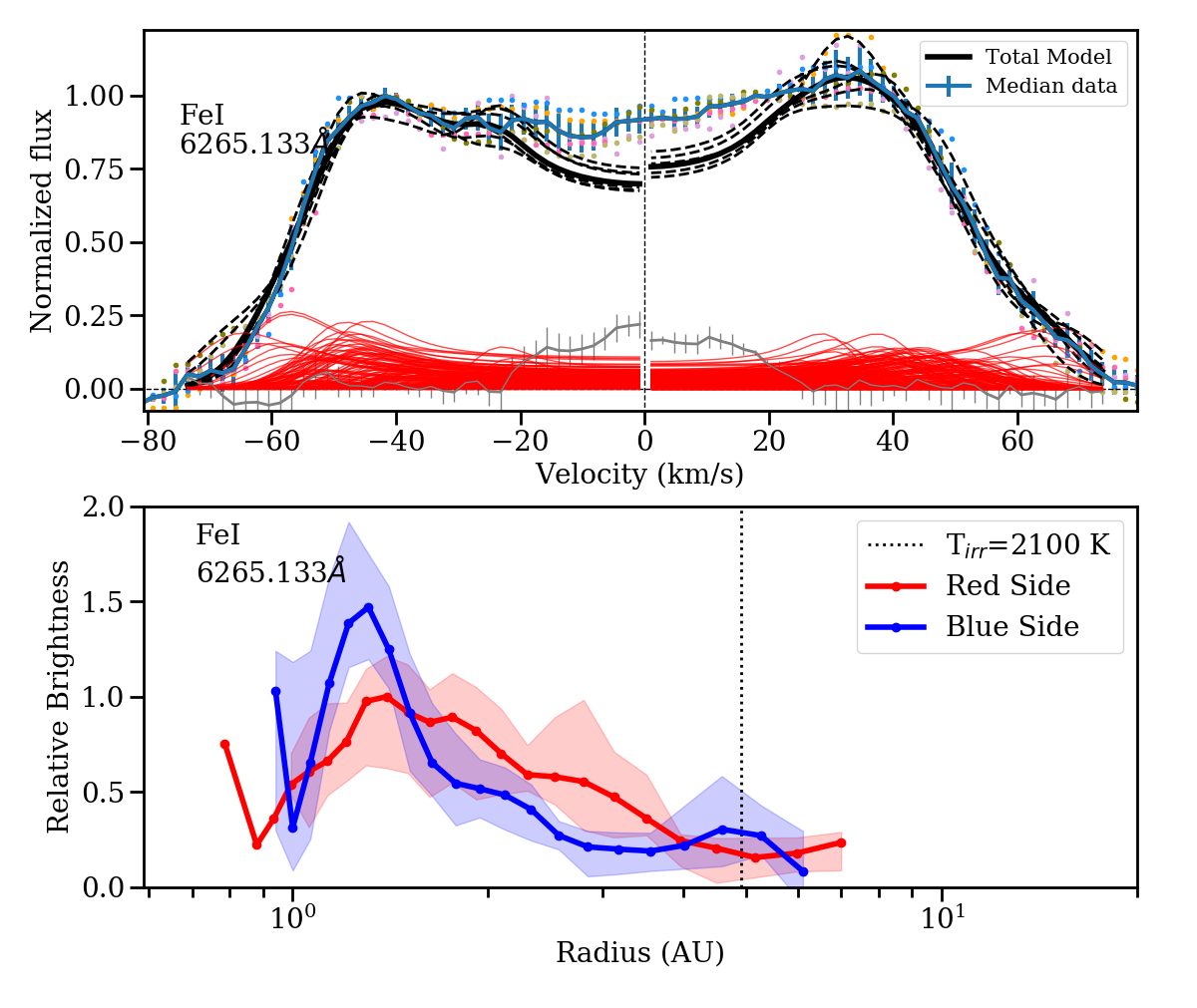} &
\includegraphics[width=4.0cm,height=4.0cm]{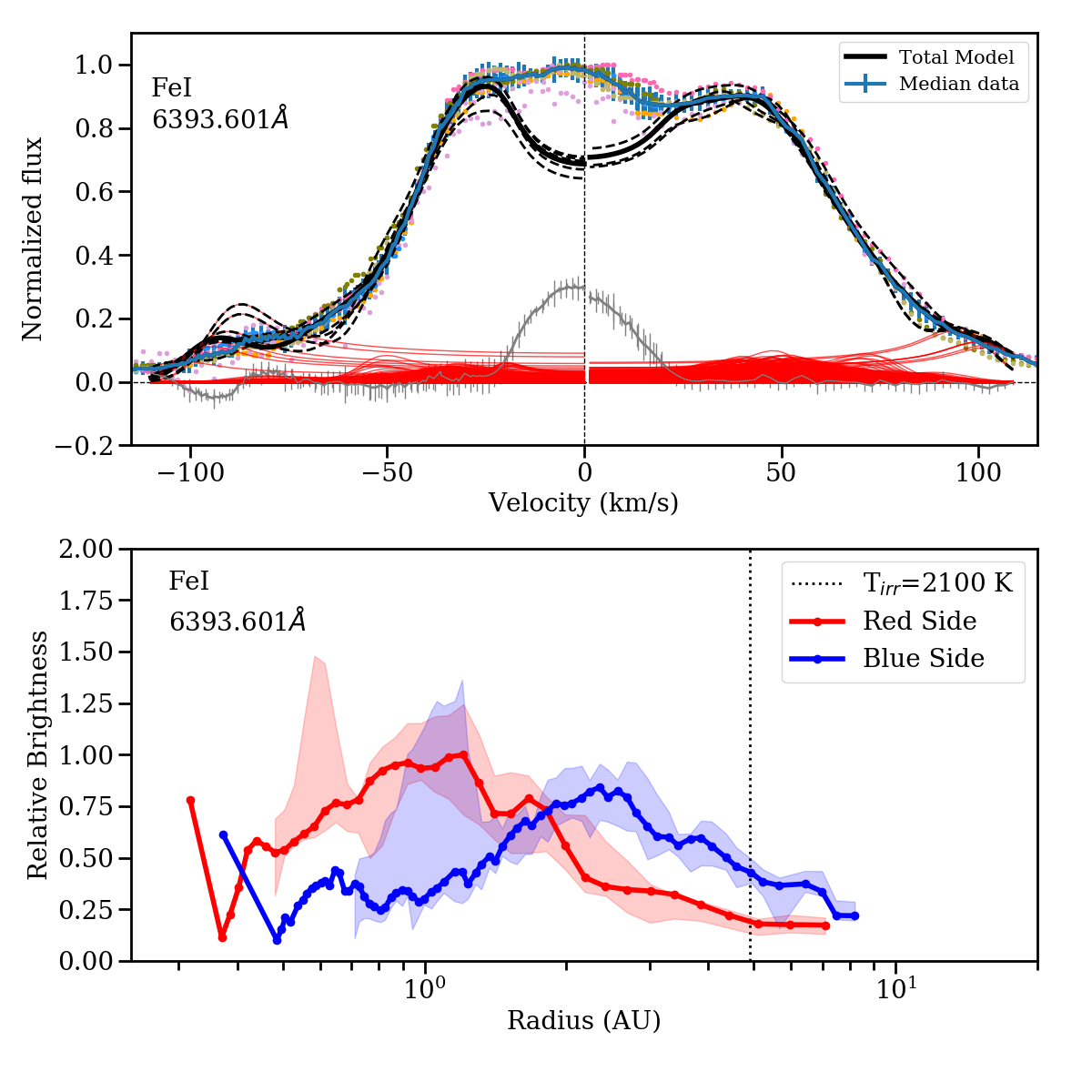} &
\includegraphics[width=4.0cm,height=4.0cm]{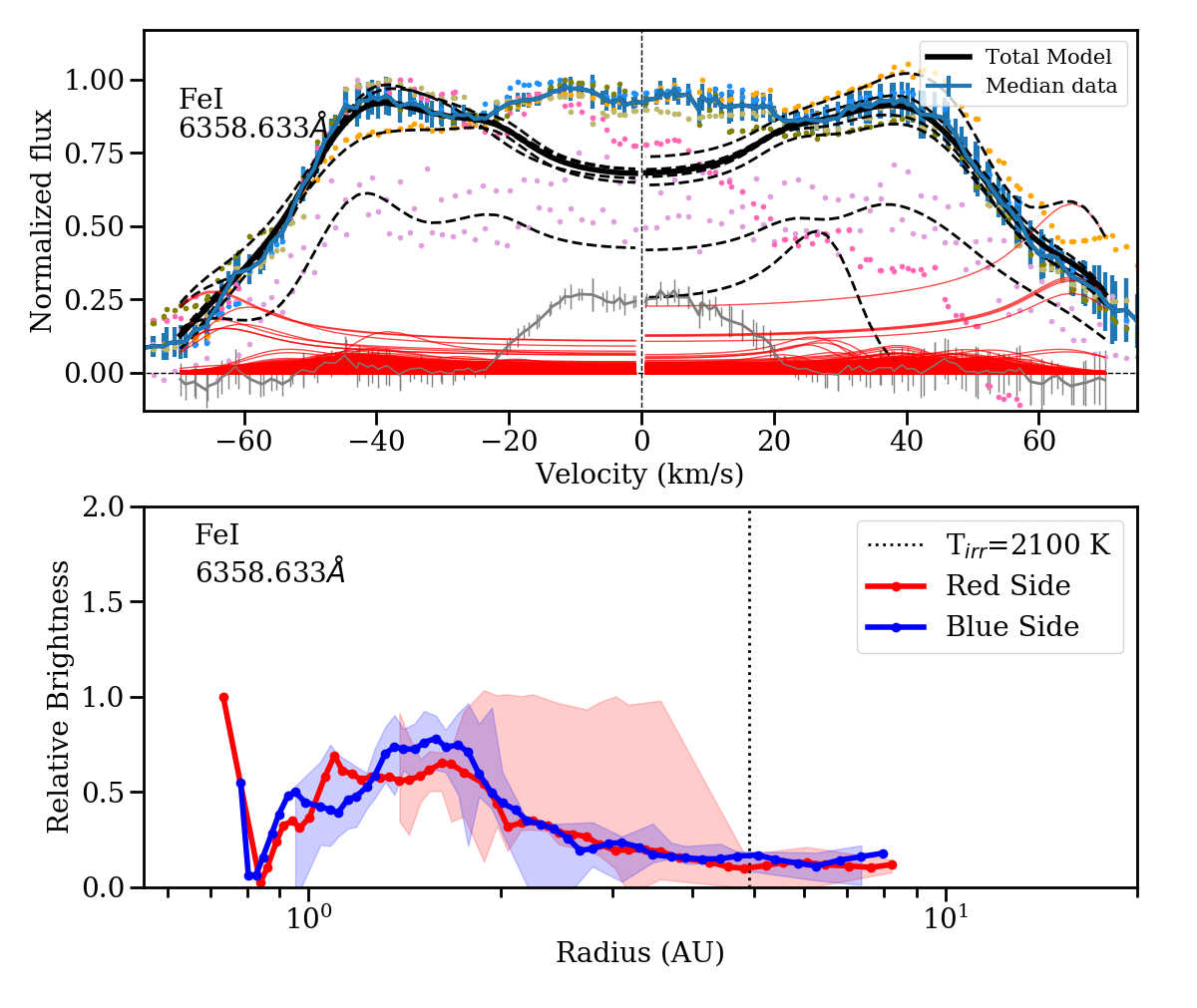} \\
\includegraphics[width=4.0cm,height=4.0cm]{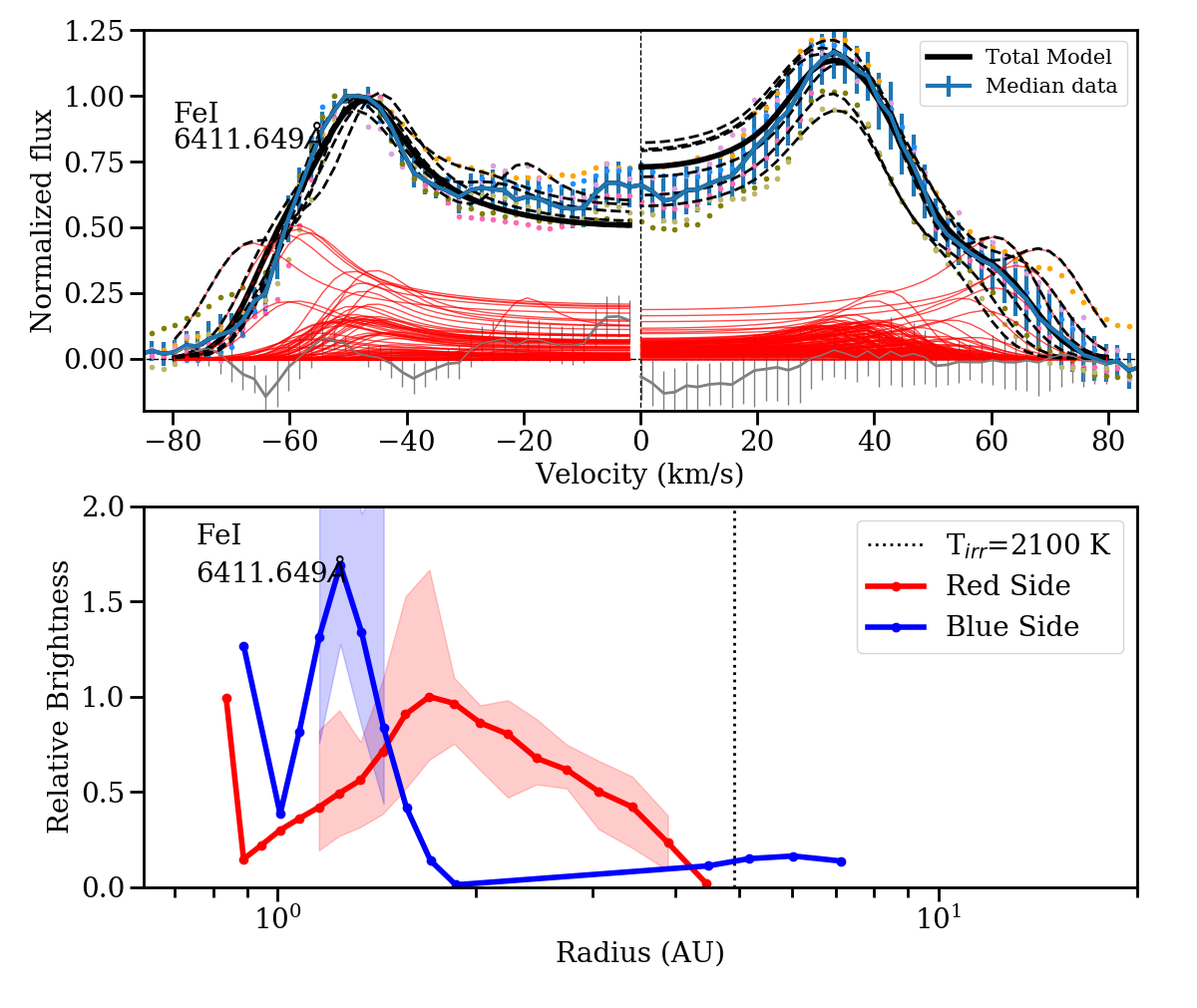} &
\includegraphics[width=4.0cm,height=4.0cm]{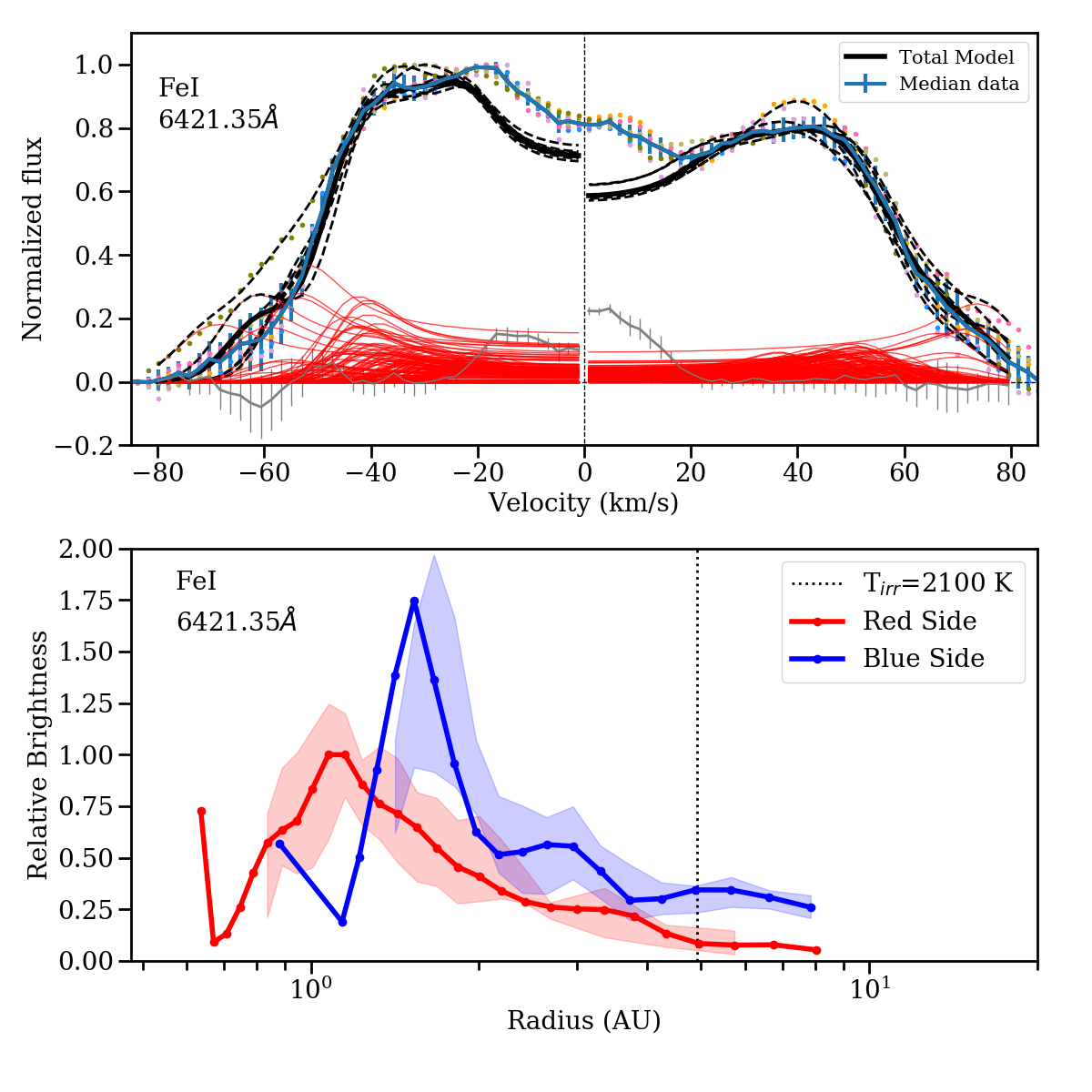} &
\includegraphics[width=4.0cm,height=4.0cm]{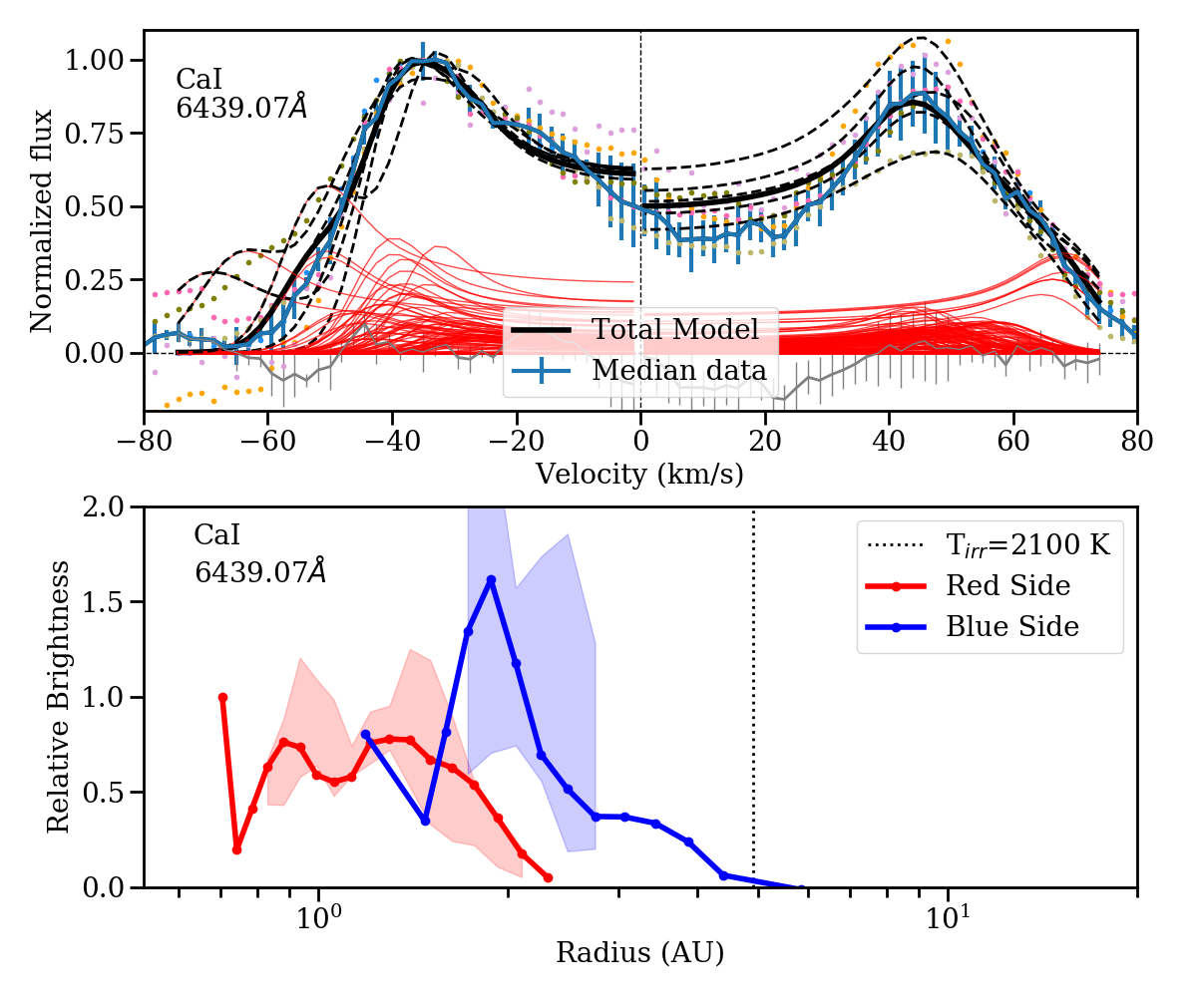}&
\includegraphics[width=4.0cm,height=4.0cm]{FeI_6462p725_v22irr_DA99flared_80_BD.png} \\
\includegraphics[width=4.0cm,height=4.0cm]{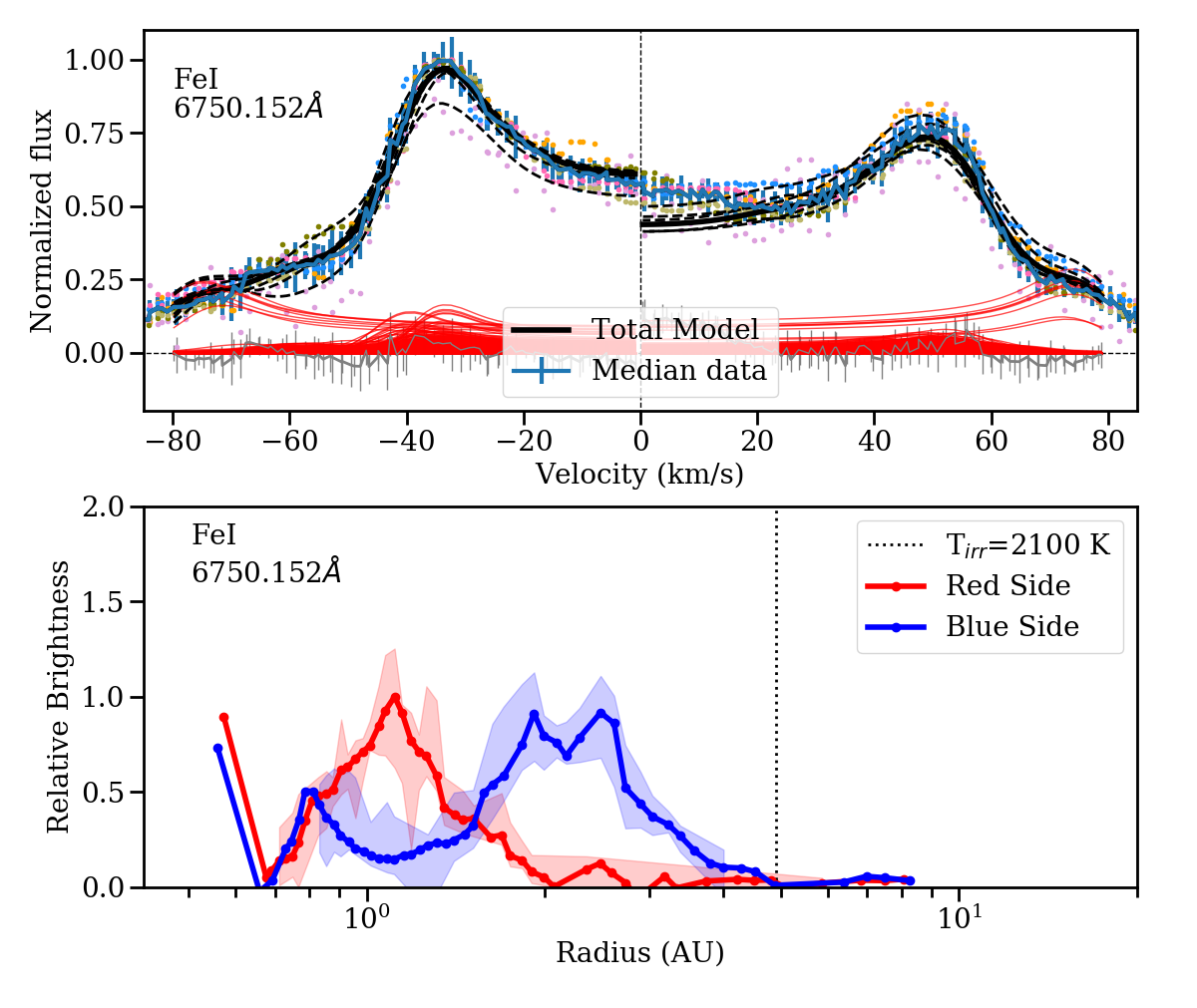} &
\includegraphics[width=4.0cm,height=4.0cm]{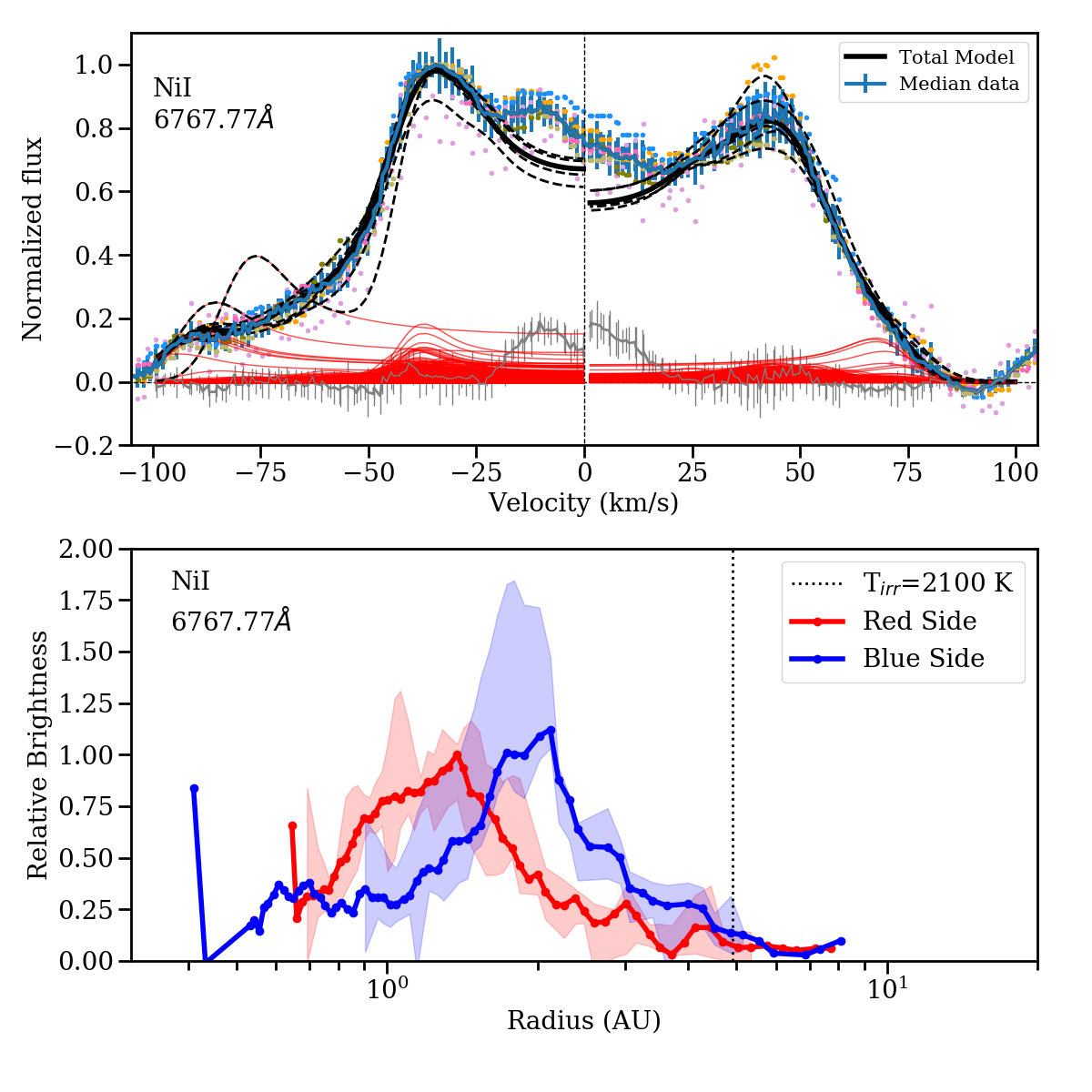} &
\includegraphics[width=4.0cm,height=4.0cm]{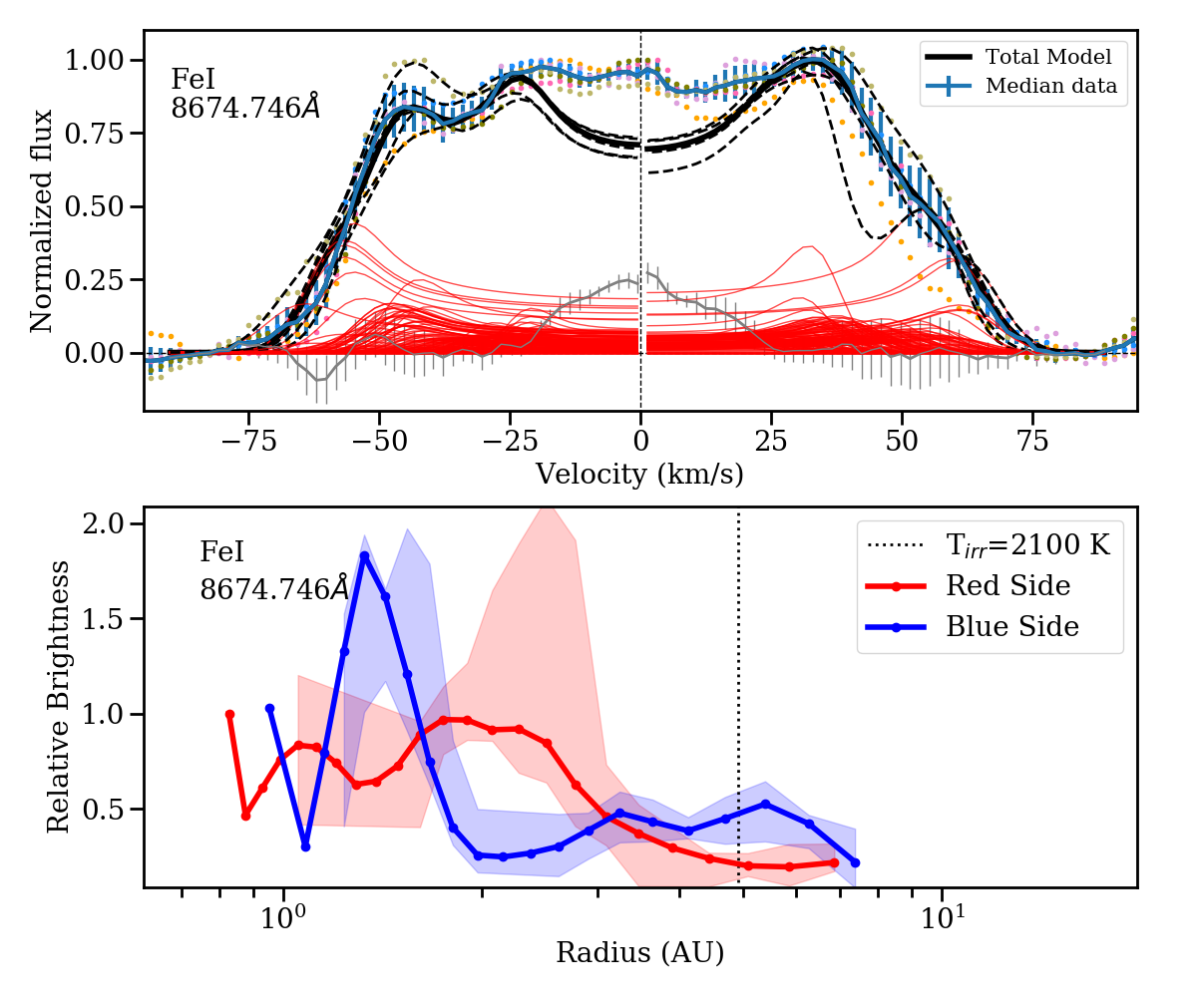} & 
\includegraphics[width=4.0cm,height=4.0cm]{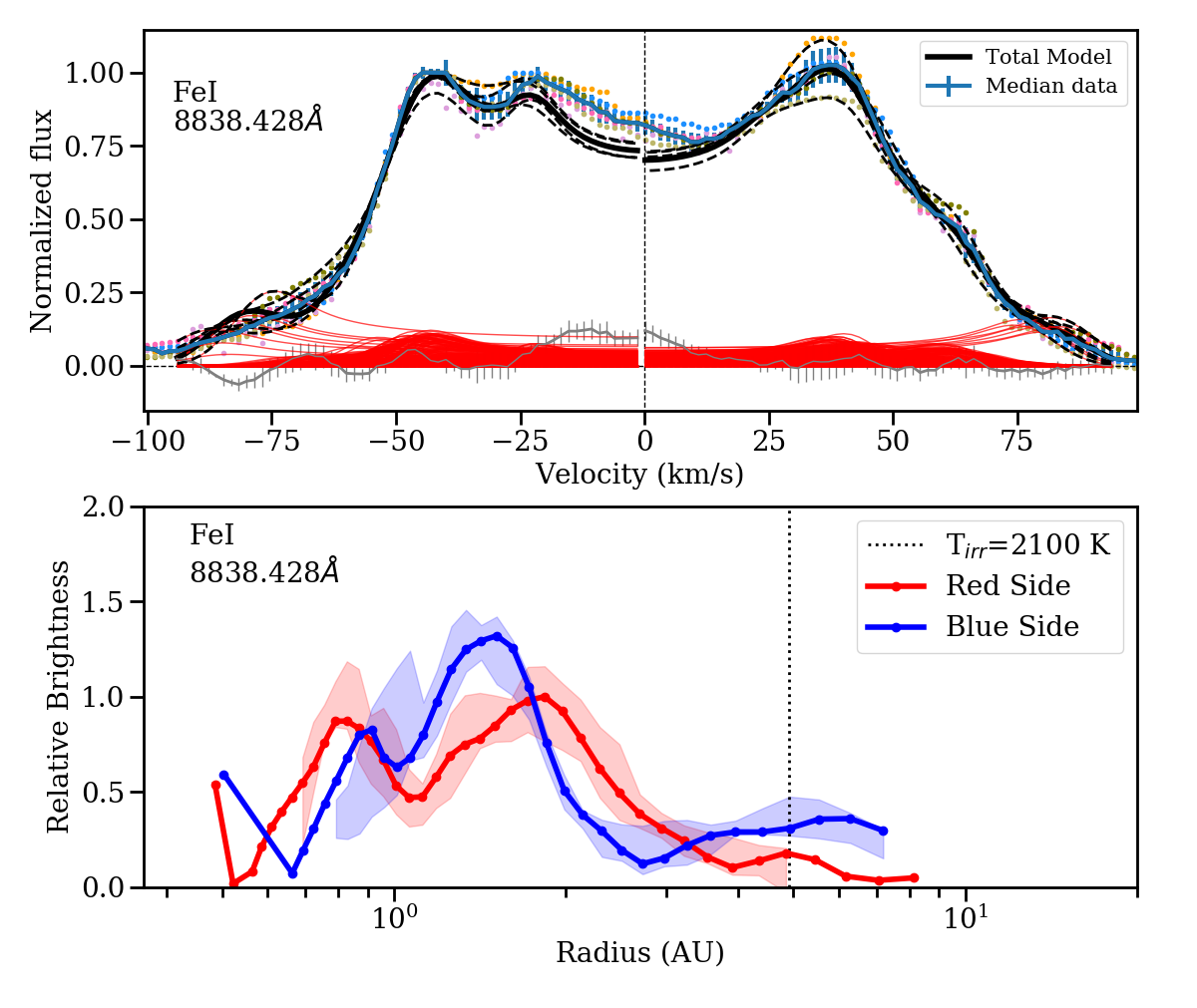} \\
\end{tabular}
\caption{Brightness profile decomposition for the disk-like lines not shown in the main text (Figure \ref{brightness-fig}, continued).
For each line, the upper panel shows the normalized, median line emission (blue), individual
data points (colored dots), and model fit. The black solid line represents the 
best fit to the
median data, while the dashed lines indicate the fits to the individual datasets that are used to
compute the errors. The zero velocity (measured with respect to the symmetry axis of the line) is shown as a dotted line. 
The individual ring fits are shown by thin red lines, and the residuals
are shown in gray. Since we impose a limit of 35 au to avoid fitting
potential narrow emission within the line, the center of the line is
not well fitted.  In the lower panel for each line, the red and blue thick lines show the best 
fit to the median spectrum for the red and blue sides, while the shaded areas show 
the range spanned by
the fits to the individual spectra. Because the noise in the single spectra is higher than in the median spectrum, there are no significant results at extreme velocities for some of the individual datasets. 
The vertical dotted line indicates the place where direct irradiation results in a temperature of 2100 K,
which is independent of the model assumptions and stellar mass.
All models are calculated for an inclination of 30 degrees and a stellar mass of 16 M$_\odot$. A different
inclination $i$ would change the derived radii by sin$^2 i$/sin$^2(30)$, and a different stellar mass M$_*$ would change the derived radii
by a factor of M$_*$/16M$_\odot$, but the relative brightnesses would remain the same.  }
\label{brightness2-fig}%
\end{figure*}

The comparison between the blue and the red sides
can provide information on the axisymmetry of the disk. In addition, this simple model allows to
check whether there is emission at locations more distant than those expected to have a temperature
of up to 2100 K (similar to the lower limit we obtained from Saha's equation) for a directly irradiated disk.
This would suggest that the disk is not (or not only) directly irradiated, but also heated by
other mechanisms such as accretion, but in general the emission is very low at larger radii.
The analysis is only valid   if the line is not contaminated by nearby features and
the continuum is relatively smooth. This results in only a subset of the lines with disk-like
profiles being appropriate for the brightness decomposition, which is discussed in Section \ref{diskanalysis-sect}.
A few examples of the brightness decomposition are shown in the main text; the rest of the  lines can
be found in Figure \ref{brightness2-fig}.

\clearpage
\onecolumn

\def\thetable{B.\arabic{table}}
\setcounter{table}{0}

\begin{scriptsize}
\begin{longtable}{lcccccccl}
\caption{Emission lines observed during outburst and quiescence. 
The lines are classified as ``double-peaked" or disk-like (Dsk), ``PCygni profiles" (PC). Those with complex,
multicomponent wind absorption signatures are denoted as 'PC+'. Potential
blends and contamination by atmospheric features are denoted by "Bl" and "Atm", respectively.
Those that could not be classified are labeled as "INDEF", and in some cases, a potential identification is given in the notes. The wavelengths provided 
are laboratory
measurements (air), except for non-identified lines, for which we list the observed wavelengths. If the line is not
seen, it is denoted as 'N', except if this part of the spectrum was not covered or if other features may have
difficulted the identification, in which case we denote it as '-'.
Lines denoted with vdA04 have been observed by \citet{vandenancker04} during the 1999 outburst.
Lines denoted with WS have been identified using Wahlgren \& Shore (S. Shore, private communication),
while the rest have been identified with the NIST database. Lines that are potentially pumped by other UV lines \citep{herczeg05}
are listed accordingly.
For some of the close blends without clear peaks, the species cannot be easily identified.
We note that the quiescence data had a lower spectral coverage, ending at $\sim$6940\AA. \label{alllines-table} }\\
\hline \hline
Species & $\lambda_{air}$  & A$_{ki}$  & E$_i$  & E$_k$   & Outb. & Quies. & Type & Notes \\
 &  (\AA) &  (s$^{-1}$) & (eV) & (eV)  &  &  &  &  \\
\hline
\endfirsthead
\caption{Continued.}\\
\hline \hline
Species & $\lambda_{air}$ & A$_{ki}$  & E$_i$  & E$_k$ & Outb. & Quies. & Type & Notes \\
 &  (\AA) &  (s$^{-1}$) & (eV) & (eV)  &  &  &  &  \\
\hline
\endhead
FeI & 3701.086 & 6.35E+07 & 2.998 & 6.347 & Y & - &    &     \\
FeI & 3704.461 & 1.42E+07 & 2.692 & 6.038 & Y & - &    &     \\
TiII: & 3706.230 & 3.10E+07 & 1.566 & 4.910 & Y & - &    &  WS. No TiII at 3819, 3833.5\AA.  \\
FeI & 3709.246 & 1.56E+07 & 0.915 & 4.256 & Y & - &    &     \\
MnI & 3708.856 & 5.20E+06 & 4.666 & 8.008 & Y & - &    & May be pumped by CIV   \\
HI & 3711.978 & 9.21e+03 & 10.199 & 13.538 & Y & - & Wk   &     \\
INDEF & 3715.7 & --- & --- & --- & Y & - &    &  VI 3714.9\AA?    \\
CaII: & 3736.900 & 1.70E+08 & 3.151 & 6.468 & Y & - & Wk &  \\
INDEF & 3742.5 & --- & --- & --- & Y & - &   &  VI 3741.5\AA? \\
HI & 3750.151 & 2.83e+04 & 10.199 &   13.504  & Y & - & PC,Bl   &     \\
FeI & 3761.408 & 1.04E+06 & 2.588 & 5.883 & Y & - &    &     \\
TiII & 3761.320 & 9.90E+07 & 0.574 & 3.869 & Y & - &    &     \\
HI & 3770.633 &  4.40e+04 & 10.199 &   13.486  & Y & - & PC   &     \\
FeI/TiII & 3774-6 & --- & --- & --- & Y & - & Bl & \\
NiI & 3783.530 & 3.30E+06 & 0.423 & 3.699 & Y & - &    &     \\
FeI/NiI & 3787 & --- & --- & ---  & Y & - & Bl & \\
HI & 3797.909  & 7.1225e+04 & 10.199 & 13.462 & Y & - & PC & \\
NiI & 3807.140 & 4.30E+06 & 0.423 & 3.678 & Y & - & Bl & \\
FeI & 3815.840 & 1.12E+08 & 1.485 & 4.733 & Y & - &    &     \\
FeI & 3821.178 & 5.54E+07 & 3.267 & 6.511 & Y & - &    &     \\
HI & 3835.395 & 1.22e+05 & 10.199 & 13.431 & Y & - & PC & \\
FeI & 3841.048 & 1.36E+08 & 1.608 & 4.835 & Y & - &    &     \\
FeI & 3849.966 & 6.05E+07 & 1.011 & 4.231 & Y & - &    &     \\
FeI & 3859.911 & 9.69E+06 & 0.000 & 3.211 & Y & - &    &     \\ 
CaI/FeI & 3873 & --- & --- & --- & Y & - & Bl & \\
FeI & 3879 & --- & --- & --- & Y & - & Bl & \\
HI & 3889.064 & 2.2148e+05 & 10.199 & 13.386 & Y & N & PC &  \\
TiII & 3900.540 & 1.60E+07 & 1.131 & 4.308 & Y & Y &  PC+  &  EXL-q   \\
INDEF & 3903 & --- & --- & --- & Y & Y & Wk   &     \\
FeII & 3905.626 & 4.00E-03 & 0.048 & 3.221 & Y & Y &    &     \\
FeI/CrI & 3913-6 & --- & --- & --- & Y & Y & Bl, Wk & PC, could be FeII \\
FeI & 3922.912 & 1.08E+06 & 0.052 & 3.211 & Y & N &    &  EXL-q   \\
CaII & 3933.660 & 1.47E+08 & 0.000 & 3.151 & Y & Y & PC+   &  vdA04   \\
FeII & 3938.290 & 6.10E+03 & 1.671 & 4.818 & Y & Y & PC   &     \\
FeI/CoI & 3944-5 & --- &--- & --- & Y & Y & Bl & \\
CoI & 3952.318 & 2.20E+05 & 0.432 & 3.568 & Y & Y &    &     \\
CaI/Fe & 3957.0 & --- & --- & --- & Y & Y & Bl & \\
CaII & 3968.470 & 1.40E+08 & 0.000 & 3.123 & Y & N &  PC  &  Blend with HI   \\
FeII & 3974.167 & 6.30E+03 & 2.704 & 5.823 & Y & N &    &     \\
FeI & 3977.741 & 6.41E+06 & 2.198 & 5.314 & Y & Y &    &  EXL-q   \\
MnI & 3982.157 & 3.50E+07 & 4.273 & 7.386 & Y & Y & Bl & \\
MnI/CrI/CoI & 3987-91 & --- & --- & --- & Y & N & Bl & \\
FeI/CoI & 3997.4-.9 & --- & --- & --- & Y & Y & Bl & EXL-q \\
FeI & 4001.662 & 7.47E+05 & 2.176 & 5.273 & N & Y & Wk & \\
MnI & 4003.260 & 1.10E+07 & 4.640 & 7.736 & Y & N &    &  Pumped by UV FeII?   \\
FeI: & 4005.242 & 2.04E+07 & 1.557 & 4.652 & Y & Y &    &  EXL-q   \\
INDEF   & 4013.0 &      &       &       & Y & N & PC & WS. TiII 4012.39\AA? VI 4012.495\AA?\\
FeI & 4013.782 & 4.19E+05 & 3.017 & 6.105 & Y & Y &    &     \\
INDEF: & 4023.7 & --- & --- & --- & N & Y & Wk &  WS. HeI 4023.98\AA?  \\
TiII & 4025.140 & 5.40E+05 & 0.607 & 3.687 & Y & Y &    &     \\
TiII & 4028.343 &5.14E+06 & 1.89 & 4.97 & Y & N & PC & WS \\
FeI/MnI & 4030.8-9 & ---& --- & --- & Y & Y & Bl & Several lines \\
FeI & 4033.186 & 8.40E+04 & 2.559 & 5.632 & Y & N &    &     \\
TiII & 4028.343 & 5.1e+06 & 1.892 & 4.969 & Y & Y & PC   &     \\
NiI & 4035.949 & 4.80E+05 & 3.655 & 6.726 & Y & N &    &     \\
FeI & 4045.812 & 8.62E+07 & 1.485 & 4.549 & Y & Y &    & 13db, EXL-q    \\
TiII & 4053.834 & 4.20E+06 & 1.893 & 4.950 & Y & Y &  PC  &     \\
FeI & 4062-3 & --- & --- & --- & Y & Y &  Bl  &  13db, EXL-q \\
FeI & 4067.271 & 2.19E+06 & 2.559 & 5.607 & Y & Y & Bl, Wk   &     \\
FeI/TiII & 4071-2 & --- & --- & --- & Y & Y & PC,Bl   &   \\
CrI & 4077.680 & --- & --- & --- & Y & Y &    &  13db, EXL-q   \\
FeI & 4083.549 & 1.84E+05 & 2.279 & 5.314 & Y & N &    &     \\
SiI & 4102.936 & 6.09E+04 & 1.909 & 4.930 & Y & Y &    &     \\
H$\delta$ & 4101.710 & 9.73e+05 & 10.199 & 13.220 & Y & Y & PC+   &     \\
CoI & 4110.540 & 5.50E+06 & 1.049 & 4.064 & Y & Y &    &     \\
MnI & 4113.867 & 1.50E+07 & 4.354 & 7.367 & Y & N & Bl & \\
MnI & 4114.384 & 1.50E+07 & 4.345 & 7.357 & Y & N & Bl & \\
FeI & 4118.886 & 1.90E+06 & 3.266 & 6.275 & Y & Y &    &     \\
FeI & 4123.728 & 6.04E+05 & 2.990 & 5.996 & Y & Y &    &     \\
FeII & 4128.748 & 2.60E+04 & 2.583 & 5.585 & Y & Y & PC   &     \\
FeI & 4132.058 & 1.18E+07 & 1.608 & 4.608 & Y & Y &    & 13db, EXL-q     \\
FeI & 4134.677 & 1.25E+07 & 2.832 & 5.829 & Y & Y &    &  EXL-q   \\
FeI & 4143.868 & 1.33E+07 & 1.557 & 4.549 & Y & Y &    &   EXL-q  \\
FeI: & 4150.249 & 7.70E+06 & 3.430 & 6.417 & Y & Y &  Bl   &  WS. Blend with FeII 4150\AA.  \\
FeI & 4154.805 & 1.40E+07 & 3.368 & 6.352 & Y & Y &    &     \\
FeI & 4156.798 & 1.20E+07 & 2.832 & 5.813 & Y & Y &    &     \\
FeI & 4161.077 & 9.50E+05 & 3.368 & 6.347 & Y & Y &    &     \\
TiII & 4163.348 &2.55E+07 & 2.590 & 5.567 & Y & N & PC & WS \\
FeI & 4172.122 & 9.80E+06 & 3.251 & 6.222 & N & Y & Bl & \\
FeII & 4173.461 & 4.43E+05 & 2.583 & 5.553 & Y & Y &    &  EXL-q   \\
FeI & 4175.636 & 1.14E+07 & 2.845 & 5.813 & N & Y & Wk & \\
FeI & 4177.594 & 3.71E+04 & 0.915 & 3.882 & N & Y & Bl & EXL-q \\
FeII & 4178.862 & 1.72E+05 & 2.583 & 5.549 & Y & Y &  PC  &  EXL-q   \\
FeI & 4181.754 & 2.32E+07 & 2.832 & 5.796 & Y & Y &    &     \\
FeI & 4184.891 & 1.03E+07 & 2.832 & 5.793 & Y & N &    &     \\
FeI & 4187.6-.8 & --- & --- & --- & Y & N & Bl & \\
FeI & 4191.4-.7 & --- & --- & --- & Y & Y & Bl,Wk   &  13db, EXL-o, EXL-q    \\
FeI & 4195.618 & 1.21E+06 & 3.017 & 5.972 & Y & Y &    &     \\
FeI & 4199.095 & 4.92E+07 & 3.047 & 5.999 & Y & Y &    &     \\
FeI & 4202.029 & 8.22E+06 & 1.485 & 4.435 & Y & Y &    &  EXL-o, EXL-q    \\
FeI & 4203-4 & --- & --- & --- & N & Y & Wk, Bl & \\
FeI & 4205.538 & 2.77E+06 & 3.417 & 6.364 & Y & Y &    &     \\
INDEF & 4209.646 &  -100 & --- & --- & Y & N &  Wk  & WS. Nearby {[FeII]} at 4208.82\AA   \\
FeI & 4216.184 & 1.84E+04 & 0.000 & 2.940 & Y & Y &    &  EXL-q   \\
FeII & 4233.167 & 7.22E+05 & 2.583 & 5.511 & Y & Y &  PC+  & 13db, EXL-o, EXL-q, vdA04  \\
CrI/FeI/FeII & 4238.8-.9 & --- & --- & --- & Y & N & Bl & EXL-q \\
CrII & 4246.409 & 3.73e+04 &  3.854 & 6.773 & Y & Y &  PC+   &  Redshifted   \\
FeI & 4250.787 & 1.02E+07 & 1.557 & 4.473 & Y & N &    &  EXL-q   \\
FeI & 4254.945 & 1.43E+05 & 3.018 & 5.931 & Y & Y &    &     \\
FeII & 4258.154 & 3.10E+04 & 2.704 & 5.615 & Y & Y &    & EXL-q. Low A$_{ki}$, weak PC.    \\
CrI & 4262.373 & 2.40E+06 & 3.079 & 5.987 & Y & Y & Wk   &     \\
FeII & 4273.326 & 9.10E+04 & 2.704 & 5.605 & Y & Y & PC,Bl   & 13db, EXL-q   \\
FeI & 4275.698 & 5.59E+04 & 2.559 & 5.458 & Y & N &    &     \\
FeI & 4278.231 & 1.00E+06 & 3.368 & 6.265 & Y & Y &    &  vdA04   \\
FeI & 4282.403 & 1.21E+07 & 2.176 & 5.070 & Y & Y &    &  EXL-q   \\
TiI: & 4287.400 & 1.46E+07 & 0.836 & 3.727 & N & Y & Wk, Bl & \\
TiII & 4290.230 & 4.6E+06 & 1.165& 4.054 & Y & Y &  PC  & vdA04   \\
TiII: & 4294.099 & 4.68E+06 & 1.084 & 3.971 & Y & Y &  PC,Bl  &  EXL-q,WS, vdA04. Blend with FeI/FeII   \\
FeII & 4296.572 & 7.00E+04 & 2.704 & 5.589 & Y & Y &  PC  &  EXL-q   \\
TiII & 4300.050 & 7.70E+06 & 1.180 & 4.063 & Y & Y & PC+   & EXL-q   \\
TiII & 4301.930 & 6.20E+06 & 1.161 & 4.042 & Y & Y &    &  EXL-q   \\
FeI & 4302.185 & 7.70E+05 & 3.047 & 5.928 & N & Y &    &     \\
FeII & 4303.176 & 2.20E+05 & 2.704 & 5.585 & Y & Y &  PC  &  EXL-q   \\
TiII & 4307.900 & 4.60E+06 & 1.165 & 4.042 & Y & Y &  PC  &   EXL-q, vdA04  \\
TiII & 4312.870 & 4.10E+06 & 1.180 & 4.054 & N: & Y & Bl & EXL-q\\
TiII & 4314.975 & 1.30E+07 & 1.161 & 4.033 & Y & Y &  PC,Bl  &     \\
TiII & 4320.960 & 2.40E+06 & 1.165 & 4.033 & Y & Y & PC   &  vdA04   \\
FeI/CrI & 4325 & --- & --- & --- & Y & Y & Bl &  13db, EXL-q \\
H$\gamma$ & 4340.472 & 2.53e+06 & 10.199 & 13.054 & Y & Y & PC & 13db, EXL-q, EXL-o, vdA04\\
TiII & 4344.290 & 7.20E+05 & 1.084 & 3.937 & Y & N & Wk   &     \\
FeII & 4351.769 & 4.66E+05 & 2.704 & 5.553 & Y & Y & PC   & WS, vdA04. Near FeI and {[FeII]}    \\
{[FeII]}/FeII & 4358-62 & --- & --- & --- & N & Y & Wk,Bl &   WS, vdA04  \\
TiII/FeI/FeII & 4367-9 & --- & --- & --- & Y & Y & Bl & EXL-q, vdA04 \\
FeI & 4375.930 & 2.95E+04 & 0.000 & 2.833 & N & Y &    & 13db, EXL-q. Blend with TiII    \\
TiII & 4374.815 & 3.00E+06 & 2.061 & 4.894 & Y & Y & Bl & vdA04. Blend with FeII  \\
FeI & 4383.545 & 5.00E+07 & 1.485 & 4.312 & N & Y &    &  13db, EXL-q   \\
FeII & 4385.387 & 4.50E+05 & 2.778 & 5.605 & Y & Y & Bl & EXL-q, vdA04 \\
TiII & 4386.844 & 2.40E+06 & 2.598 & 5.423 & N & Y & Bl &  \\
TiII & 4395.850 & 2.90E+05 & 1.243 & 4.063 & Y & Y & Bl,PC & Blend with FeII and TiII 4395\AA\\
ScII & 4400.389 & 1.52E+07 &  6.908 & 9.730 & Y & N &  PC+  &  WS. Pumped by UV H$_2$?   \\
FeI & 4404.750 & 2.75E+07 & 1.557 & 4.371 & Y & Y & Bl & EXL-q\\
FeI/MnI/TiII & 4407-9 & --- & --- & --- & Y & Y & Bl & Some lines unblended in quiescence\\
FeI/TiI/FeII & 4415-17 & --- & --- & --- & N & Y & Bl & 13db, EXL-q, vdA04 \\
TiII & 4418.330 & 3.00E+05 & 1.237 & 4.042 & Y & Y &  Bl  &     \\
FeI & 4422.568 & 8.72E+06 & 2.845 & 5.648 & Y & Y &    &     \\
FeI & 4427.310 & 3.41E+04 & 0.052 & 2.851 & Y & Y &    &     \\
FeI & 4430.614 & 7.45E+06 & 2.223 & 5.020 & Y & Y & Bl &  EXL-o, \\
NiI & 4431.033 & 4.10E+06 & 4.167 & 6.965 & Y & Y & Bl & \\
TiII & 4443.800 & 1.1E+07 & 1.080 & 3.869 & Y & Y & PC,Bl & Blend of TiII \\
TiII & 4444.558 & 3.90E+05 & 1.116 & 3.904 & Y & Y &  PC,Bl  &   Blend of TiII \\
FeI & 4447.130 & 8.91E+04 & 2.198 & 4.985 & Y & Y &  Bl  &  Unblended in quiescence  \\
FeI & 4447.717 & 5.11E+06 & 2.223 & 5.010 & Y & Y &  Bl  &   Unblended in quiescence    \\
TiII & 4450.490 & 2.00E+06 & 1.084 & 3.869 & Y & Y &   PC+ &   EXL-q. Shallower than TiII 4443\AA  \\
FeI & 4455.027 & 4.10E+06 & 3.882 & 6.664 & Y & Y & Bl & \\
FeI & 4459.117 & 2.52E+06 & 2.176 & 4.956 & Y & Y &    & EXL-q    \\
FeI & 4461.970 & 2.31E+06 & 3.603 & 6.380 & Y & Y &    &     \\
TiII & 4464.450 & 7.00E+05 & 1.161 & 3.937 & N & Y & Wk & \\
TiII & 4468.500 & 1.00E+07 & 1.131 & 3.904 & Y & Y & PC+   &  13db, EXL-q, vdA04    \\
INDEF &  4473.8  & --- & --- & --- & Y & N &    &  VI 4473\AA:   \\
FeI & 4476.018 & 1.01E+07 & 2.845 & 5.614 & N & Y &    &     \\
FeI/CrI & 4482 & --- &  --- & --- & Y & Y & Bl  &  13db:,  Several lines  \\
FeII & 4489.183 & 5.90E+04 & 2.828 & 5.589 & Y & Y & Bl & 13db, EXL-q, vdA04. Low A$_{ki}$, no PC\\
FeII & 4491.405 & 1.89E+05 & 2.856 & 5.615 & Y & Y & Bl & EXL-q, vdA04\\
FeI & 4494.563 & 3.45E+06 & 2.198 & 4.956 & Y & Y & Bl   &  EXL-q   \\
TiII & 4501.270 & 9.80E+06 & 1.116 & 3.869 & Y & Y & PC   &  EXL-q, vdA04   \\
FeII & 4508.288 & 7.30E+05 & 2.856 & 5.605 & Y & Y &  PC  &  EXL-q   \\
FeII    & 4515-6 & ---  &---     & ---  & Y & N & PC & FeII blend, WS, vdA04 \\
FeI & 4518.432 & 2.10E+04 & 3.237 & 5.980 & N & Y & Wk & \\
FeII & 4520.224 & 9.80E+04 & 2.807 & 5.549 & Y & Y &  PC  &   EXL-q, vdA04  \\
FeII & 4522.634 & 8.40E+05 & 2.844 & 5.585 & Y & Y &  PC  &   vdA04   \\
TiII & 4529.474 & 3.00E+05 & 1.572 & 4.308 & Y & Y &    &  13db. Weaker than other TiII   \\
TiII/FeII & 4534 & --- & --- & --- & Y & Y & Bl &  vdA04\\
FeII & 4549.474 & 1.00E+06 & 2.828 & 5.553 & Y & Y & Bl, PC+ & 13db, EXL-o, EXL-q, vdA04 \\
TiI & 4552.460 & 2.10E+07 & 0.836 & 3.559 & N & Y &    &     \\
CrI/FeI/FeII & 4556 &  &  &  & Y & Y & Bl & \\
CrII & 4558.660 & 8.80E+06 & 4.073 & 6.792 & Y & Y &  PC+   &  EXL-q, vdA04. Similar to TiII\\
TiII & 4563.770 & 8.80E+06 & 1.221 & 3.937 & Y & Y &  BC  &   EXL-o, EXL-q   \\
TiII & 4571.980 & 1.20E+07 & 1.572 & 4.283&  Y & Y &  PC+  &   vdA04 \\
FeII & 4576.340 & 6.40E+04 & 2.844 & 5.553 & Y & Y &    & EXL-q, vdA04    \\
CoI & 4580.131 & 2.70E+04 & 0.923 & 3.629 & N & Y & Wk & \\
FeII & 4583.829 & 7.22E+05 & 2.807 & 5.511 & Y & Y & Bl & vdA04 \\
CrII & 4588.199 & 1.20E+07 & 4.071 & 6.773 & N & Y & PC   &   vdA04 \\
TiII & 4589.950 & 1.30E+06 & 1.237 & 3.937 & Y & Y & Bl   &  EXL-q   \\
FeI/CrI/FeII & 4592 & --- & --- & --- & Y & N & Bl & \\
FeI & 4592.651 & 1.61E+05 & 1.557 & 4.256 & N & Y & Bl & EXL-q. Only in quiescence\\
FeI & 4603 &--- & --- & --- & Y & Y & Bl & Multiple blend\\
CrII & 4616.629 & 4.00E+06 & 4.072 & 6.757 & N & Y & Wk,Bl   &      \\
FeII/FeI/CrII & 4617-20 & --- & --- & --- & Y & Y & Bl   &     \\
FeII & 4629.339 & 1.72E+05 & 2.807 & 5.484 & Y & Y &  PC  &  13db, EXL-o, EXL-q, vdA04    \\
CrII & 4634.070 & 8.90E+06 & 4.072 &6.747& Y & Y & PC   & vdA04. Weak PC, near FeII \\
CrI & 4639.520 & 9.50E+06 & 3.111 & 5.782 & N & Y & Wk   &     \\
FeI & 4654.498 & 5.64E+04 & 1.557 & 4.220 & N & Y &    &     \\
CrI & 4656.809 & 1.10E+07 & 4.778 & 7.440 & Y & Y &  Bl  & May be pumped by [OIII]    \\
TiII & 4657.206 & 2.70E+05 & 1.243 & 3.904 & Y & Y &    &     \\
{[FeII]} & 4664.440 & 1.47E-01 & 0.121 & 2.778 & Y & N &    &  WS Blueshifted   \\
FeII & 4666.758 & 1.30E+04 & 2.828 & 5.484 & Y & Y &    & 13db, EXL-q. Low A$_{ki}$, no PC.    \\
FeII & 4670.182 & 3.20E+03 & 2.583 & 5.237 & Y & Y &    &  Low A$_{ki}$, no PC.  \\
FeI & 4691.411 & 1.01E+06 & 2.990 & 5.632 & Y & Y & Dsk & \\
CoI & 4699.177 & 7.40E+03 & 1.049 & 3.687 & Y & Y & Dsk & \\
MgI: & 4702.991 & 2.19E+07 & 4.346 & 6.981 & Y & N & Dsk &  EXL-o, EXL-q, offset \\
FeI & 4708.968 & 3.98E+05 & 3.640 & 6.272 & Y & Y & Dsk &  vdA04\\
FeI & 4710.283 & 1.05E+06 & 3.018 & 5.649 & N & Y &    &     \\
{[FeII]} & 4728.068 & 4.53E-01 & 0.083 & 2.704 & Y & Y &    &    \\
FeII & 4731.453 & 2.80E+04 & 2.891 & 5.511 & Y & Y & PC   &  13db, EXL-q, vdA04    \\
FeI & 4733.591 & 3.41E+04 & 1.485 & 4.103 & N & Y & Bl: & Only in quiescence \\
TiII & 4762.776 & 7.20E+04 & 1.084 & 3.687 & N & Y & Wk,PC & \\
CrI & 4764.626 & 7.30E+05 & 3.013 & 5.615 & Y & Y &    &     \\
TiII & 4779.985 & 6.20E+06 & 2.048 & 4.641 & Y & Y &  PC  &     \\
FeI & 4786.807 & 1.03E+06 & 3.017 & 5.607 & Y & Y & Dsk & \\
FeI & 4789.650 & 4.57E+06 & 3.547 & 6.134 & Y & N & Dsk & \\
FeI & 4791.246 & 3.56E+05 & 3.274 & 5.861 & Y & Y & Dsk & \\
FeI/TiII & 4798 & --- & --- & --- & Y & Y &  Bl  &     \\
TiII & 4805.100 & 1.10E+07 & 2.061 & 4.641 & Y & Y &  PC  &   EXL-o, vdA04    \\
FeI/TiI/CrII & 4811-4 & --- & --- & --- & Y & Y & Bl &  vdA04\\
CrII & 4824.127 & 1.70E+06 & 3.871 & 6.440 & Y & Y & PC+   &   EXL-o, EXL-q, vdA04. Strong PC. \\
FeII & 4833.197 & 4.60E+02 & 2.657 & 5.222 & Y & Y &    & Low A$_{ki}$, no PC FeII \\
FeI & 4835.868 & 1.10E+06 & 4.103 & 6.667 & Y & Y & Dsk & \\
FeI & 4839.544 & 3.90E+05 & 3.267 & 5.828 & Y & N & Dsk & \\
CrII & 4848.235 & 2.60E+06 & 3.864 & 6.421 & N & Y &    &  EXL-o, EXL-q    \\
H$\beta$ & 4861.350 & 8.42e+06 & 10.199 & 12.749 & Y & Y &    &  13db, EXL-o, EXL-q, vdA04    \\
CrII & 4876.473 & 1.60E+06 & 3.864 & 6.406 & - & Y &    &   EXL-q, vdA04. Blend with H$\beta$ in outburst.   \\
FeI & 4873.751 & 5.50E+04 & 3.301 & 5.844 & - & Y &    &  Blend with H$\beta$ in outburst   \\
CrII & 4884.607 & 5.80E+05 & 3.858 & 6.396 & Y & Y &    &     \\
FeI & 4891.492 & 3.08E+07 & 2.851 & 5.385 & Y & Y &    &     \\
FeII & 4893.820 & 2.50E+03 & 2.828 & 5.361 & Y & N &  Wk  &  Low A$_{ki}$, no PC, very weak   \\
NiI:s & 4900.971 & 5.00E+05 & 3.480 & 6.009 & Y & Y &    &     \\
FeI & 4903.310 & 6.58E+06 & 2.882 & 5.410 & Y & N & Dsk,Bl &  \\
TiII & 4911.193 & 3.20E+07 & 3.124 & 5.647 & Y & Y &    &   EXL-o   \\
FeII & 4923.921 & 4.28E+06 & 2.891 & 5.408 & Y & Y &  PC+  & 13db, EXL-q, vdA04     \\
INDEF & 4934.505 & --- & --- & --- & Y & Y &    &     \\
FeI: & 4939.686 & 1.39E+04 & 0.859 & 3.368 & Y & Y & Dsk,Bl & 13db \\
FeI & 4957.596 & 4.22E+07 & 2.808 & 5.308 & Y & Y & Dsk:   &  Dome-shaped\\
TiI & 4981.730 & 6.60E+07 & 0.848 & 3.337 & Y & Y & Bl & \\
FeI & 4985.253 & 1.48E+07 & 3.929 & 6.415 & Y & N & Bl & \\
FeII & 4989-90 & --- & --- & --- & N & Y & Bl, Wk   &     \\
FeII & 4993.358 & 6.90E+03 & 2.807 & 5.289 & Y & Y & Bl   &  13db, vdA04   \\
NiI & 5000.340 & 1.40E+07 & 3.635 & 6.114 & Y & Y & Bl & \\
FeI & 5001.863 & 3.70E+07 & 3.882 & 6.360 & Y & Y & Bl,Dsk & Maybe contaminated by TiI\\
FeII & 5018.434 & 2.00E+06 & 2.891 & 5.361 & Y & Y &  PC+  &  13db,  EXL-o, EXL-q, vdA04   \\
INDEF & 5031.9 & --- & --- & --- & Y & Y & Dsk & Likely FeI. Nearby ScII \\
TiI & 5036.470 & 3.94E+07 & 1.443 & 3.904 & Y & Y &    &     \\
TiI & 5038.400 & 3.87E+07 & 1.430 & 3.890 & Y: & Y &   &  Very weak in outburst    \\
FeI & 5041.071 & 4.29E+04 & 0.958 & 3.417 & Y & Y & Bl & 13db. Blended in outburst, near CaI\\
FeI & 5041.756 & 2.35E+05 & 1.485 & 3.943 & Y & Y & Bl & 13db. Blended in outburst  \\
FeI & 5049.819 & 1.65+06 & 2.279 & 4.733 & Y & Y & Bl & Blended in outburst\\
FeI & 5051.634 & 4.65E+04 & 0.915 & 3.368 & Y & Y & Dsk,Bl & 13db \\
FeI & 5065.192 & 8.86E+05 & 3.642 & 6.089 & Y & N & Dsk & \\
FeI & 5068.766 & 3.37E+06 & 2.940 & 5.385 & Y & Y & Dsk & \\
CrI & 5072.920 & 1.59E+05 & 0.941 & 3.385 & Y & Y & Dsk & \\
FeI & 5079.740 & 5.19E+04 & 0.990 & 3.430 & Y & Y & Dsk & Maybe contaminated by other Fe I lines \\
FeI & 5083.338 & 4.06E+04 & 0.958 & 3.397 & Y & Y & Dsk & 13db \\
FeI & 5098.698 & 4.83E+05 & 2.176 & 4.607 & Y & Y & Dsk, Bl & \\
INDEF &  5101.615 & --- & --- & --- & Y & Y &    &     \\
NiI & 5102.966 & 8.80E+04 & 1.676 & 4.105 & N & Y &    &     \\
FeI & 5107 & --- & --- & --- & Y & Y & Bl & Two lines \\
FeI: & 5107.447 & 4.18E+04 & 0.990 & 3.417 & N & Y &  Bl  &  Blend of several Fe I lines   \\
FeI & 5110.413 & 4.93E+03 & 0.000 & 2.425 & Y & Y & Dsk & 13db. Low-density line \\
FeII & 5120.352 & 2.59e+03 & 2.828 & 5.249 & Y & Y &    &  WS   \\
FeI & 5123.720 & 7.24E+04 & 1.011 & 3.430 & Y & Y &    & 13db     \\
TiII & 5129.150 & 1.00E+06 & 1.892 & 4.308 & Y & Y & PC   & 13db, vdA04     \\
FeI & 5131.468 & 2.58E+05 & 2.223 & 4.638 & N & Y &    &     \\
FeII & 5132.669 & 2.00E+03 & 2.807 & 5.222 & Y & Y &  & vdA04  \\
FeII: & 5136.802 & 2.80E+03 & 2.844 & 5.257 & Y & Y & Bl & More likely FeI\\
FeI & 5142.928 & 2.40E+04 & 0.958 & 3.368 & Y & Y & Dsk, Bl & \\
CoI & 5146.161 & 9.20E+03 & 1.740 & 4.149 & Y & Y &    &     \\
FeI & 5151.911 & 2.39E+04 & 1.011 & 3.417 & Y & Y & Bl   &  13db   \\
FeI & 5154.100 & 4.43E+05 & 3.882 & 6.286 & Y & Y &    &     \\
FeII & 5169.030 & 4.22E+06 & 2.891 & 5.289 & Y & N &  PC  &  EXL-q, vdA04   \\
MgI & 5172.684 & 3.37E+07 & 2.712 & 5.108 & Y & Y & Bl & 13db, EXL-q   \\
MgI & 5183.604 & 5.61E+07 & 2.717 & 5.108 & Y & Y &  PC  &  13db, EXL-q    \\
TiII & 5188.700 & 2.50E+06 & 1.582 & 3.971 & Y & Y & PC   &  vdA04     \\
FeI & 5194.941 & 2.87E+05 & 1.557 & 3.943 & N & Y &  Wk  &     \\
FeII & 5197.577 & 5.40E+05 & 3.230 & 5.615 & Y & Y & PC   &  EXL-q, vdA04   \\
FeI & 5202.336 & 5.11E+05 & 2.176 & 4.559 & Y & Y &  Wk  &     \\
CrI: & 5204.520 & 5.09E+07 & 0.941 & 3.323 & N & Y & Wk & 13db:, could be INDEF 5205\AA \\
CrI & 5206.040 & 5.14E+07 & 0.941 & 3.322 & Y & Y &    &     \\
FeI & 5208.594 & 6.23E+06 & 3.241 & 5.621 & Y & Y &    &     \\
TiI: & 5210 & --- & --- & --- & N & Y & Wk, Bl &  \\
TiI & 5211.222 & 3.10E+05 & 0.836 & 3.215 & Y & Y & Bl & Dome-shaped, blend\\
FeI & 5216.274 & 3.47E+05 & 1.608 & 3.984 & Y & Y & Dsk: & Dome-shaped \\
FeI & 5226.862 & 1.36E+07 & 3.038 & 5.410 & Y & Y &    &     \\
FeII: & 5227.487 & 1.22E+08 & 10.451 & 12.823 & Y & Y &  PC, Bl  &  Blueshifted, likely misclassified   \\
FeII & 5234.625 & 2.50E+05 & 3.221 & 5.589 & Y & Y &  PC  &  EXL-o, EXL-q, vdA04    \\
CrII & 5237.329 & 1.70E+06 & 4.073 & 6.440& Y & Y &    &   vdA04   \\
ScII & 5239.813 & 1.37E+07 &  1.455 & 3.821 & N & Y &  Bl  &  WS. Heavy blend \\
FeI & 5247.050 & 3.92E+02 & 0.087 & 2.450 & Y & N & Dsk & 13db,  \\
FeI & 5250.209 & 9.30E+02 & 0.121 & 2.482 & Y & Y &    &     \\
MnI & 5255.320 & 4.17E+06 & 3.133 & 5.491 & Y & Y &    &     \\
FeI/FeII & 5262 &  --- & --- & --- & Y & N &  Bl  &  vdA04   \\
NiI & 5265.712 & 3.70E+05 & 0.000 & 3.655 & Y & Y &    &     \\
FeI & 5269.537 & 1.27E+06 & 0.859 & 3.211 & Y & Y &    &  13db, EXL-q, vdA04   \\
CaI & 5270.270 & 5.00E+07 & 2.526 & 4.878 & Y & Y &    &  13db, EXL-q    \\
FeI & 5273.373 & 5.55E+05 & 2.484 & 4.835 & N & Y & Wk & \\
FeII & 5276.002 & 3.76E+05 & 3.199 & 5.549 & Y & Y &  PC  &   EXL-o, vdA04    \\
FeI & 5280.362 & 7.27E+05 & 3.642 & 5.989 & Y & Y & Dsk & Maybe affected by nearby FeII\\
FeII & 5284.109 & 1.90E+04 & 2.891 & 5.237 & Y & Y &    &   EXL-o, EXL-q, vdA04. No PC     \\
FeII: & 5306.180 & 3.28E+07 & 10.523 & 12.859 & N & Y &  Wk  &  Very weak if FeII, could be CrII   \\
FeII & 5316.615 & 3.89E+05 & 3.153 & 5.484 & Y & Y & PC+ & 13db, EXL-o, EXL-q, vdA04. Complex PC   \\
FeI & 5324.179 & 2.06E+07 & 3.211 & 5.539 & N & Y &    &     \\
FeI/FeII & 5325-6 &--- & --- & --- & Y & Y & Bl   &  vdA04    \\
FeI & 5328.038 & 1.15E+06 & 0.915 & 3.241 & Y & Y & Bl   &   13db, EXL-q, vdA04   \\
TiII & 5336.810 & 5.80E+05 & 1.582 & 3.904 & Y & Y &    &  Small or no PC, vdA04  \\
FeI & 5341.024 & 5.21E+05 & 1.608 & 3.929 & Y & Y &    &  13db    \\
MgI+CrII & 5346 & --- & --- & --- & Y & Y & Bl & Several lines \\
FeI & 5349.737 & 1.40E+06 & 4.386 & 6.703 & Y & N & Dsk &  \\
NiI: & 5353.391 & 1.20E+05 & 1.951 & 4.266 & Y & Y &    &   Offset, uncertain  \\
FeII & 5362.860 & --- & --- & --- & Y & Y &  PC  &   vdA04   \\
FeI & 5371.489 & 1.05E+06 & 0.958 & 3.266 & Y & Y &    &  EXL-o, EXL-q, vdA04    \\
TiII & 5381.015 & 3.20E+05 & 1.566 & 3.869 & Y & Y &  PC:  & Weak PC    \\
FeI & 5393.167 & 4.91E+06 & 3.241 & 5.539 & Y & N & Dsk, Bl & Blended with unknown line\\
FeI & 5397.128 & 2.58E+05 & 0.915 & 3.211 & Y & Y &    &  13db, EXL-q, vdA04    \\
FeI & 5405.350 & 8.45E+05 & 4.386 & 6.680 & Y & Y &    &   vdA04   \\
MnI & 5407.420 & 5.15E+05 & 2.143 & 4.435 & N & Y &    &     \\
FeI & 5415.199 & 7.67E+07 & 4.386 & 6.675 & Y & Y &    &     \\
TiII/CrII & 5419-20 & --- & --- & --- & Y & Y & Bl   &     \\
CrII & 5420.922 & 5.00E+05 & 3.758 & 6.044 & N & Y & Wk   &     \\
FeII & 5425.257 & 9.20E+03 & 3.199 & 5.484 & Y & Y & Bl & EXL-q\\
FeI & 5429.696 & 4.27E+05 & 0.958 & 3.241 & Y & Y & Bl & 13db, EXL-q, vdA04. Nearby FeII \\
FeI & 5432.948 & 4.30E+06 & 4.446 & 6.727 & N & Y &    &  vdA04    \\
FeI & 5434.524 & 1.70E+06 & 1.011 & 3.292 & Y & Y & Bl & vdA04 \\
NiI & 5435.870 & 1.90E+05 & 1.986 & 4.266 & N & Y &    &     \\
INDEF & 5448.014 & --- & --- & --- & Y & Y &    &     \\
FeI & 5455.609 & 6.05E+05 & 1.011 & 3.283 & Y & Y & Dsk:   &  13db, EXL-o, EXL-q, vdA04. Dome-shaped   \\
FeI & 5461.550  & 4.20E+05 & 4.446 & 6.715 & N & Y & Bl   & Nearby TiI, MgII   \\
FeI & 5464.279 & 1.77E+06 & 4.143 & 6.411 & Y & N & Dsk & \\
FeI & 5466.987 & 1.45E+05 & 3.573 & 5.840 & Y & N & Dsk, Wk & Low S/N\\
FeI & 5473.164 & 3.50E+05 & 4.191 & 6.456 & Y & N &    &     \\
CoI & 5477.078 & 6.80E+06 & 3.713 & 5.976 & Y & Y & Dsk & \\
CoI & 5483.340 & 9.00E+05 & 1.710 & 3.971 & N & Y &    &     \\
TiI & 5490.840 & 1.40E+04 & 0.048 & 2.305 & Y & Y & Bl & \\
FeI & 5497.516 & 6.25E+04 & 1.011 & 3.266 & Y & Y & Dsk & 13db, vdA04 \\
CrII & 5502.067 & 2.80E+05 & 4.168 & 6.421 & Y & Y &    &     \\
FeI & 5506.779 & 5.01E+04 & 0.990 & 3.241 & Y & Y & Dsk &  vdA04\\
CrII & 5508.606 & 2.80E+05 & 4.156 & 6.406 & Y & Y &    &  Nearby FeI   \\
MnI & 5510.191 & 8.00E+03 & 3.135 & 5.384 & Y & N & Bl & \\
FeI & 5512.256 & 9.90E+05 & 4.371 & 6.620 & N & Y &    &     \\
NiI & 5514.793 & 4.50E+05 & 3.847 & 6.095 & N & Y &    &     \\
{[FeII]} & 5527.54/.61 & 2.8/1.2E-01 & --- & --- & Y & Y &    &  WS   \\
FeII & 5534.847 & 3.00E+04 & 3.245 & 5.484 & Y & Y & PC   & 13db, EXL-o, EXL-q, vdA04     \\
FeI: & 5543.935 & 3.40E+06 & 4.218 & 6.453 & Y & N & Dsk, Bl & \\
FeI & 5546.506 & 1.00E+06 & 4.371 & 6.606 & Y & N &    &     \\
FeI & 5569.618 & 2.34E+07 & 3.417 & 5.642 & N & N &    &     \\
FeI & 5572.842 & 2.28E+07 & 3.397 & 5.621 & Y & N & Dsk & \\
{[OI]}: & 5577.340 & 1.26E+00 & 1.967 & 4.190 & Y & Y & Bl, Atm &  vdA04  \\
FeI/NiI/CrI & 5586-8 & --- & --- & --- & Y & Y & Bl & \\
NiI & 5592.280 & 1.10E+05 & 1.951 & 4.167 & N & Y &    &     \\
CaI & 5598.490 & 4.30E+07 & 2.521 & 4.735 & Y & N & Dsk & \\
FeI & 5615.644 & 2.64E+07 & 3.332 & 5.539 & Y & Y & Dsk &  EXL-o  \\
FeI & 5624.542 & 7.41E+06 & 3.417 & 5.621 & Y & Y & Dsk & \\
FeII: & 5627.497 & 2.93E+03 & 3.387 & 5.589 & Y & Y & Dsk:   &  WS. Potential blend FeII/[FeII]   \\
FeI & 5641.434 & 3.00E+06 & 4.256 & 6.453 & Y & Y & Dsk: & Dome-shaped \\
FeI/ScII & 5658 & --- & --- & --- & Y & Y & Bl & Several lines, vdA04 \\
FeI & 5662.938 & 2.45E+05 & 3.695 & 5.883 & Y & Y & Dsk & \\
FeI/ScII & 5667-8 & --- & --- & --- & Y & Y & Bl & WS \\
SiI & 5684.484 & 2.60E+06 & 4.954 & 7.134 & Y & Y & Dsk & \\
FeI & 5701.544 & 1.78E+05 & 2.559 & 4.733 & Y & N & Dsk, Bl & \\
FeI & 5702.347 & 6.30E+04 & 3.640 & 5.813 & Y & N & Dsk, Bl &  \\
FeI & 5709.378 & 2.13E+06 & 3.368 & 5.539 & Y & Y & Dsk, Bl & \\
FeI & 5708.966 & 1.50E-01 & 0.052 & 2.223 & Y & Y & Dsk, Bl & \\
FeI & 5711.848 & 1.50E+06 & 4.283 & 6.453 & Y & Y & Bl & Blend with unknown line\\
FeI & 5712.131 & 2.99E+05 & 3.417 & 5.587 & Y & Y & Bl & \\
VI & 5727.780 & --- & --- & --- & Y & N &  Bl,Wk  &  13db. Very weak    \\
FeI/FeII & 5747-8 & --- & --- & --- & Y & N & Bl &  Several lines\\
SiI & 5754.220 & 1.23E+06 & 4.954 & 7.108 & Y & N & Bl & {[N II]} nearby \\
FeI & 5762.413 & 1.30E+05 & 3.642 & 5.793 & Y & N & Bl,Dsk & \\
INDEF & 5764.681 & --- & --- & --- & Y & N &  Atm &     \\
INDEF & 5768.800 & --- & --- & --- & Y & N & Atm &     \\
INDEF & 5770.413 & --- & --- & --- & Y & N & Atm &     \\
NiI & 5846.994 & 2.40E+04 & 1.676 & 3.796 & N & Y &    &     \\
FeI & 5853.680 & --- & --- & --- & Y & Y & Dsk & 13db \\
HeI & 5875.62 & 7.07E+07 & 20.96 & 23.070 &  Y & Y & PC   &  EXL-o, EXL-q, vdA04   \\
NaI & 5889.951 & 6.16E+07 & 0.000 & 2.104 & Y & Y & PC+   & 13db, vdA04. Very complex profile.    \\
NaI & 5895.924 & 6.14E+07 & 0.000 & 2.102 & Y & Y & PC+   &  13db, vdA04. Very complex profile.   \\
FeI & 5929.677 & 1.10E+06 & 4.549 & 6.639 & Y & N &    &     \\
FeI & 5952.718 & 1.50E+06 & 3.984 & 6.066 & Y & N & Dsk,Wk & \\
FeII & 5991.376 & 4.2E+03 & 3.153 & 5.222 & Y & N & & vdA04. No PC, low A$_{ki}$. Box-like.\\
NiI & 6053.685 & 2.20E+06 & 4.236 & 6.283 & Y & N & Wk & \\
FeI & 6065.482 & 1.07E+06 & 2.609 & 4.652 & Y & Y & Dsk & 13db, EXL-o,  \\
FeII & 6084.111 & 3.00E+03 & 3.199 & 5.237 & Y & Y & Dsk: &  EXL-o, vdA04. No PC, low A$_{ki}$. Box-like. \\
FeI & 6103.293 & 1.52E+06 & 4.733 & 6.764 & Y & N & Dsk, Bl & \\
NiI & 6108.120 & 1.30E+05 & 1.676 & 3.706 & Y & Y & Dsk & \\
NII/FeII & 6113-6 & --- & --- & --- & Y & N &    & 13db, vdA04. Part of big blend. \\
CaI & 6122.220 & 2.87E+07 & 1.886 & 3.910 & Y & Y & Dsk: &  EXL-o. Box-like profile. \\
FeII/CoI/CrII/NiI & 6129-30 & --- & --- & --- & Y & Y & Bl &  vdA04\\
FeI & 6137.691 & 1.00E+06 & 2.588 & 4.608 & Y & Y & Bl & \\
FeI & 6141.732 & 1.23E+06 & 3.603 & 5.621 & Y & Y &    &     \\
FeI/FeII & 6147-8 & --- & --- & --- & N & Y & Bl & \\
FeII & 6149.238 & 1.30E+05 & 3.889 & 5.905 & Y & Y &    & vdA04    \\
CaI & 6162.170 & 4.77E+07 & 1.899 & 3.910 & Y & N & Bl & Part of big blend \\
CaI & 6169.1-.5 & --- & --- & --- & Y & N & Bl & \\
FeI & 6191.558 & 7.41E+05 & 2.433 & 4.435 & Y & Y & Dsk & 13db, EXL-o, vdA04  \\
FeI: & 6200.312 & 9.06E+04 & 2.609 & 4.608 & Y & N &  Wk  &  13db   \\
TiI: & 6220.490 & 1.80E+07 & 2.677 & 4.669 & Y & Y & Dsk & Uncertain, blueshifted. \\
FeI & 6230.723 & 9.99E+05 & 2.559 & 4.549 & Y & Y & Bl,Dsk: & \\
FeII & 6238.375 & 7.50E+04 & 3.889 & 5.876 & Y & Y &    &   EXL-o, EXL-q, vdA04   \\
FeII & 6247.562 & 1.60E+05 & 3.892 & 5.876 & Y & Y & PC:   &  13db, EXL-o, EXL-q, vdA04. Weak PC   \\
FeI & 6252.555 & 3.19E+05 & 2.404 & 4.386 & Y & Y & Dsk & \\
FeI & 6254.258 & 2.13E+05 & 2.279 & 4.260 & N & Y &    &  13db   \\
FeI & 6256.361 & 7.40E+04 & 2.453 & 4.435 & Y & Y & Dsk,Bl & \\
TiI & 6258.100 & 8.36E+06 & 1.443 & 3.424 & N & Y & Bl & \\
TiI & 6258.700 & 8.90E+06 & 1.460 & 3.441 & N & Y & Bl & \\
TiI & 6261.100 & 8.07E+06 & 1.430 & 3.409 & N & Y & Bl & \\
FeI & 6265.133 & 6.84E+04 & 2.176 & 4.154 & Y & Y & Dsk & \\
FeI & 6270.225 & 1.39E+05 & 2.858 & 4.835 & Y & Y & Bl,Dsk & \\
FeI & 6271.278 & 3.05E+04 & 3.332 & 5.308 & Y & N & Bl,Dsk & \\
{[OI]} & 6300.304 & 5.63E-03 & 0.000 & 1.967 & Y & Y &    &  vdA04. Components: 0 and -400 km s$^{-1}$   \\
FeI & 6318.017 & 3.75E+05 & 2.453 & 4.415 & Y & Y & Atm & 13db, vdA04 \\
FeI & 6335.330 & 1.58E+05 & 2.198 & 4.154 & Y & Y & Dsk,Atm & \\
SiII & 6347.100 & 5.84E+07 & 8.121 & 10.074 & Y & N & PC+   &   EXL-o, EXL-q, vdA04   \\
FeI & 6353.836 & 7.87E+00 & 0.915 & 2.865 & Y & Y & Atm & \\
FeI & 6358.633 & 5.19E+05 & 4.143 & 6.092 & Y & Y & Dsk & \\
{[OI]} & 6363.776 & 1.82E-03 & 0.020 & 1.967 & Y & Y &    &   vdA04. Components: 0 and -400 km s$^{-1}$   \\
FeII & 6369.462 & 1.40E+04 & 2.891 & 4.837 & Y & Y &  PC  &   vdA04. Near SiII line.   \\
FeI & 6393.601 & 4.81E+05 & 2.433 & 4.371 & Y & Y & Dsk & 13db, EXL-o, vdA04  \\
FeI & 6400.001 &  9.27E+06 & 3.603 & 5.539 & Y & Y & Dsk &  13db, EXL-o. Blend with [FeI] 6400 in quiescence?\\
FeI & 6408.018 & 3.12E+06 & 3.686 & 5.621 & Y & N & Dsk, Bl & 13db, near FeII \\
FeI & 6411.649 & 4.43E+06 & 3.654 & 5.587 & Y & N & Dsk & 13db, \\
FeII & 6416.905 & 3.60E+04 & 3.892 & 5.823 & Y & Y &    &   vdA04. Low A$_{ki}$, no PC  \\
FeI & 6421.350 & 3.04E+05 & 2.279 & 4.209 & N & Y & Dsk & 13db \\
FeII & 6432.680 & 8.50E+03 & 2.891 & 4.818 & Y & Y &    &    13db, vdA04. Low A$_{ki}$, no PC     \\
CaI & 6439.070 & 5.30E+07 & 2.526 & 4.451 & N & N & Dsk & \\
CaI & 6450.860 & --- & --- & --- & N & Y &    &  13db  \\
FeII & 6456.376 & 1.70E+05 & 3.903 & 5.823 & Y & Y &  PC   &  13db, EXL-o, EXL-q, vdA04    \\
FeI & 6462.725 & 5.60E+04 & 2.453 & 4.371 & Y & Y & Dsk & \\
FeII/{[FeII]}: & 6482-3 & --- & --- & ---  & Y & N & Bl,Atm &  vdA04. May include {[NII]}.\\
FeI & 6494 & --- & --- & --- & Y & Y & Atm, Bl & \\
FeII & 6516.053 & 8.30E+03 & 2.891 & 4.793 & Y & Y &  Atm  & 13db, EXL-o, EXL-q     \\
H$\alpha$ & 6562.570 & 5.39E+07 & 10.199 & 12.090 &  Y & Y & PC+   &  13db, EXL-o, EXL-q, vdA04   \\
FeI & 6593.870 & 5.28E+04 & 2.433 & 4.312 & Y & Y & Bl, Dsk: & Very close to H$\alpha$\\
VI &      6643.786 & 2.2e+05  &  1.950 &   3.815 & Y & Y & Dsk & EXL-o. May be Ni I 6643.64\AA.  \\
FeI & 6663.442 & 4.98E+05 & 2.424 & 4.284 & Y & N &    &     \\
HeI & 6678.150 & 6.37e+07 & 21.22 & 23.070 &  Y & Y &  PC  &  EXL-o, EXL-q, vdA04. May include FeII.    \\
LiI & 6707.76/.91 & 3.69e+07 & 0 & 1.847 & Y & - &    & EXL-o. Has absorption. Nearby VI, FeI, FeII.  \\
{[SII]} & 6716.44 & 4.3E-4/3.6E-5 & 0 & 1.845 & : & Y &  & vdA04. Uncertain in outburst due to blends.\\
CaI: & 6717.690 & 1.20E+07 & 2.709 & 4.554 & Y & - & Dsk &  EXL-o. Redshifted, could be misclassified. \\
{[SII]} & 6730.816 & 2.7E-4/1.56E-4 & 0 & 1.841 & Y & Y &  & vdA04\\
TiI & 6743.120 & 6.90E+05 & 0.900 & 2.738 & N & - &    &     \\
FeI & 6750.152 & 1.17E+05 & 2.424 & 4.260 & Y & - & Dsk & 13db \\
NiI & 6767.770 & 3.30E+05 & 1.826 & 3.658  & Y & - & Dsk: & \\
FeI & 6769.110 & --- & --- & --- & Y & - &    &    \\
INDEF & 6795.712 & --- & --- & --- & Y & - & Dsk &     \\
FeI & 6828.591 & 3.70E+06 & 4.638 & 6.453 & Y & - & Dsk: & Dome-shaped \\
FeI & 6999.884 & 4.70E+05 & 4.103 & 5.874 & Y & - & Atm & \\
INDEF & 7144.951 & --- & --- & --- & Y & - &  &     \\
FeII & 7155.157 & 1.46E-01 & 0.232 & 1.964 & Y & - & Atm & \\
{[CaII]} & 7291.47 & 1.30E+00 & 0 & 1.700 & Y & - & Atm & WS, vdA04 \\
{[CaII]} & 7323.890 & 1.300E+00& 0  & 1.692 & Y & - & Atm &   WS, vdA04  \\
FeII & 7449.335 & 1.68E+04 & 3.889 & 5.553 & Y & - &    &  vdA04   \\
FeII & 7462.380 & 2.70E+04 & 3.892 & 5.553 & Y & - &    &  13db, EXL-o, vdA04. Low A$_{ki}$, no PC  \\
FeII: & 7479.694 & 3.50E+03 & 3.892 & 5.549 & Y & - & Dsk & Unclassified FeI? \\
INDEF: & 7494.872 & --- & --- & --- & Y & - & Dsk: &     \\
INDEF: & 7496.917 & --- & --- & --- & Y & - & Dsk: &     \\
FeI & 7511.020 & 1.35E+07 & 4.178 & 5.828 & Y & - & Dsk, Atm & Complex background \\
FeII & 7515.832 & 8.10E+03 & 3.903 & 5.553 & Y & - &    &  vdA04   \\
INDEF & 7533.904 & --- & --- & --- & Y & - &  &  Strong   \\
KI & 7698.965 & 3.73E+07 & 0 & 1.610 & Y & - & Atm & 13db, vdA04. Line at 7665\AA\ masked by atm.\\  
FeII & 7711.710 &  4.94E+04 & 3.903 &5.511 & Y & - & &  vdA04 \\
INDEF: & 7749.000 & --- & --- & --- & Y & - & Atm   &  13db, EXLup   \\
OI & 7771.940 & 3.69e+07 & 9.15 & 10.740 & Y & - &  PC+  &  EXL-o, EXL-q, vdA04. Pumped by UV H$_2$?   \\
OI & 7774.170 & 3.69e+07 & 9.15 & 10.740 &  Y & - & PC+   &  EXL-o, EXL-q, vdA04. Pumped by UV H$_2$?    \\
OI & 7775.390 & 3.69e+07 & 9.15 & 10.740 &  Y & - &  PC+  &  EXL-o, EXL-q, vdA04. Pumped by UV H$_2$?   \\
NiI & 7788.940 & 8.40E+04 & 1.951 & 3.542 & Y & - &    &     \\
INDEF & 7790.6 & --- & --- & --- & Y & - &  Dsk  &  13db, Unclassified FeI?    \\
INDEF & 7832-4 & --- & --- & --- & Y & - &  Dsk:  &  Unclassified FeI?   \\
HI & 8249.990 & --- & 12.088 & 13.590 & Y & - &  Atm  &  EXL-q. Blended wit CaII 8248\AA.   \\
FeI & 8387.772 & 6.09E+05 & 2.176 & 3.654 & Y & - &    &  vdA04   \\
HI & 8413.318 & 1.96e+3 & 12.088 & 13.561 & Y & - &    &  13db, vdA04   \\
OI & 8446.25 &  3.22e+07 & 9.52 &  10.99 & Y & - & PC+ & EXLup, vdA04. Pumped by UV H$_2$? \\
FeI & 8468.407 & 2.63E+05 & 2.223 & 3.686 & Y & - & Atm & \\
CaII & 8498.020 & 1.11E+06 & 1.692 & 3.151 & Y & - & PC+   & 13db, EXL-o, EXL-q, vdA04    \\
FeI & 8514-5 & --- & --- & --- & Y & - & Dsk:, Bl & vdA04\\
CaII & 8542.090 & 9.90E+06 & 1.700 & 3.151 & Y & - & PC+   &   13db, EXL-o, EXL-q, vdA04   \\
FeI & 8598.830 & 6.69E+05 & 4.386 & 5.828 & Y & - &    &     \\
CaII & 8662.140 & 1.06E+07 & 1.692 & 3.123 & Y & - &  PC+  &   13db, EXL-o, EXL-q, vdA04   \\
FeI & 8674.746 & 6.17E+05 & 2.832 & 4.260 & Y & - &  Dsk:  & Variable emission   \\
FeI & 8688.625 & 7.74E+05 & 2.176 & 3.603 & Y & - &    &   vdA04  \\
INDEF & 8728-9 & --- & --- & --- & Y & - & Bl &  vdA04. Includes NI   \\
HI & 8750.46 & 2.02e+04 & 12.088 & 13.504 & Y & - & PC & 13db, EXL-o, vdA04   \\
FeI & 8757.187 & 3.52E+05 & 2.845 & 4.260 & Y & - & Dsk, Bl   &  13db   \\
INDEF & 8790-5 & --- & --- & --- & Y & - & Bl   &  Several lines   \\
INDEF & 8764 & --- & --- & --- & Y & - &  &     \\
MgI & 8806.757 & 1.27E+07 & 4.346 & 5.753 & Y & - &    &  13db,vdA04. Very strong, boxy/IPC profile. \\
FeI & 8824.220 & 3.53E+05 & 2.198 & 3.603 & Y & - & Dsk & 13db, EXL-o,vdA04\\
FeI & 8838.428 & 3.83E+05 & 2.858 & 4.260 & Y & - & Dsk & Complex profile\\
HI & 8862.782 & 3.16e+4 & 12.088 & 13.518 & Y & - &  PC+  &  vdA04   \\
CaII & 8912.07 & --- & --- & --- & Y & - &    &   vdA04  \\
Ca II & 8927.36 & --- & --- & ---  & Y & - & Atm  & vdA04    \\
HI & 9015.300 & 5.1558e+04 & 12.088 & 13.462 & Y & - & Atm,PC & vdA04\\
CaII & 9854.74 & 1.9E+07 & 7.505 & 8.763 & Y  & - &  & vdA04. Pumped by UV FeII? \\
FeII & 9891.55 & --- & 12.893 & 14.146 & Y & - & Bl   &     \\
FeII & 9932.383 & 8.6e+05  & 11.447 & 12.695 & Y & - & Bl   & Pumped by UV CIII?     \\
FeI & 9951.158 & --- & 5.393 & 6.639 & Y & - &    &     \\
TiI & 9957.550 & --- & 3.409 & 4.654 & Y & - &    &     \\
FeII & 9997.57 & 7.60E+04 &  5.484 &  6.724 & Y & - &    &   vdA04  \\
FeII & 10125.14 & --- & --- & --- & Y & - &  &  vdA04.  \\
TiI & 10179.924 & --- & 3.890 & 5.107 & Y & - &    &     \\
FeI:m & 10218.976 & --- & 5.478 & 6.691 & Y & - & Bl   & Multiple blend  \\
FeII & 10246.410 & 9.e+04 & 11.441 & 12.650 & Y & - &    &     \\
FeII/FeI/CI & 10327-8 & --- & --- & --- & Y & - &    &     \\
FeI & 10365.169 & --- & 5.587 & 6.783 & Y & - & Bl   &  More than one line possible   \\
FeI & 10371.687 & --- & 3.640 & 4.835 & Y & - & Dsk:,Wk   &  Likely disky, but noisy and weak  \\
INDEF & 10389.176 & --- & --- & --- & Y & - &  &     \\
TiI/FeII/FeI & 10397-402 & --- & --- & --- & Y & - & Bl &     \\
FeI & 10423.030 & --- & 2.692 & 3.882 & Y & - &    &     \\
FeI: & 10456.941 & --- & 5.539 & 6.725 & Y & - & Bl   & Multiple FeI/FeII blend   \\
FeI & 10469.654 & --- & 3.884 & 5.067 & Y & - &    &     \\
Fe II:  & 10473.269 & --- &  10.480  &    11.664 & Y & - & Bl &     \\
\hline
\end{longtable}
\end{scriptsize}

\end{appendix}

\end{document}